\documentclass[cernpreprint,texlive=2016,txfonts,UKenglish]{latex/atlasdoc}
\pdfoutput=1

\usepackage{latex/atlasphysics}
\usepackage{latex/atlasunit}
\usepackage[biblatex=false]{latex/atlaspackage}
\usepackage{latex/atlascontribute}
\usepackage{subfig}
\usepackage{url}
\usepackage{hyperref}
\usepackage{array}
\usepackage{multirow}
\usepackage{lineno}
\usepackage{subfig}
\usepackage{mathtools}
\usepackage{epstopdf}
\usepackage{epsfig}
\usepackage{float}
\usepackage{cite}

\usepackage{clusternote-defs}

\graphicspath{ {./logos/} {./figures/} }

\AtlasTitle{Topological cell clustering in the ATLAS calorimeters and its performance in LHC Run 1}\AtlasRefCode{PERF-2014-07}

\AtlasDate{\today}

\AtlasJournal{\EPJC}

\AtlasJournalRef{\EPJC 77 (2017) 490}
\AtlasDOI{10.1140/epjc/s10052-017-5004-5}

\PreprintIdNumber{CERN-PH-EP-2015-304}

\begin{document}

\AtlasAbstract{%
 The reconstruction of the signal from hadrons and jets emerging from the \pp{} collisions at the Large Hadron Collider (\LHC) and entering the \ATLAS{} calorimeters is based on a three-dimensional topological clustering of individual calorimeter cell signals. 
The cluster formation follows cell signal-significance patterns generated by electromagnetic and hadronic showers. 
In this, the clustering algorithm implicitly performs a topological noise suppression by removing cells with insignificant signals which are not in close proximity to cells with significant signals. 
The resulting \emph{topological cell clusters} have shape and location information, which is exploited to apply a local energy calibration and corrections depending on the nature of the cluster. 
Topological cell clustering is established as a well-performing calorimeter signal definition for jet and missing transverse momentum reconstruction in \ATLAS.
}

\maketitle

\tableofcontents

\section{Introduction}

The detectable final state emerging from the \pp{} collisions at the Large Hadron Collider (LHC) consists of particles and jets which are reconstructed with high precision for physics analyses. In the \ATLAS{} experiment \cite{DetectorPaper}, 
clusters of topologically connected calorimeter cell signals (\topos) are 
employed as a principal signal definition for use in the reconstruction of the (hadronic) part of the final state comprising isolated hadrons, jets and hadronically decaying $\tau$-leptons. In addition, \topos{} are also used to represent the energy flow from softer particles, which is needed for the reconstruction of full-event observables such as the missing transverse momentum. 

The algorithm building the \topos{} explores the spatial distribution of the cell signals in all three dimensions to establish connections between neighbours in an attempt to reconstruct the energy and directions of the incoming particles. 
The signals from cells determined to be connected are summed, and are used together with the cell locations to calculate direction, location, and shapes of the resulting clusters. 
Calorimeter cells with insignificant signals found to not be connected to neighbouring cells with significant signals are considered noise and discarded from further jet, particle and missing transverse momentum reconstruction.
   
The \topos, while well established in deep inelastic scattering experiments such as H1 \cite{Abt:1996hi} at HERA and in electron--positron collider experiments such as ALEPH \cite{Decamp:1990jra} at LEP and BaBar \cite{Aubert:2000bz} at PEP-II, are used here in an innovative implementation 
as fully calibrated three-dimensional objects representing the calorimeter signals in the complex final-state environment of hadron--hadron collisions. 
A similar application in this particular environment, previously developed by the \DZero{} Collaboration, implements the topological clustering in the two dimensions spanned by pseudorapidity and the azimuthal angle, thus applying the noise-suppression strategy inherent in this algorithm for jet reconstruction \cite{Abazov:2013hda}. 
Several features and aspects of the \ATLAS{} \topo{} algorithms and their validations have previously been presented in \citMultiRef{Cojocaru:2004jk,Lampl:2008zz,Barillari:2009zza,Pinfold:2012ac}. 
 
Some of the complexity of the final state in hadron--hadron collisions is introduced by particles from the underlying event generated by radiation and multiple parton interactions in the two colliding hadrons producing the hard-scatter final state. 
Other detector signal contributions from the collision environment, especially important for higher intensity operations at the \LHC, arise from \emph{\pu} generated by diffuse particle emissions produced by the additional \pp{} collisions occurring in the same bunch crossing as the hard-scatter interaction (\ipu). 
Further \pu{} influences on the signal are from signal remnants from the energy flow in other bunch crossings in the \ATLAS{} calorimeters (\opu).

This paper first describes the \ATLAS{} detector in \secRef{sec:atlas}, together with the \ds s used for the performance evaluations.
The motivations and basic implementation of the \topo{} algorithm are presented in \secRef{sec:topos}. 
The computation of additional variables associated with \topos{} including geometric and signal moments is described in \secRef{sec:moments}. 
The various signal corrections applied to \topos{} in the the context of the local hadronic calibration are presented in \secRef{sec:lcw}. 
\SecRef{sec:perf} summarises the performance of the \topo{} signal in the reconstruction of isolated hadrons and jets produced in the \pp{} collisions at \LHC.
Performance evaluations with and without \pu{} are discussed in this section, together with results from the corresponding Monte Carlo (\MC) simulations. The paper concludes with
a summary and outlook in \secRef{sec:concl}.

\newcommand{\baselabel}{sec:atlas}
\newcommand{\thislabel}{\baselabel}
\section{The \ATLAS{} experiment} \label{sec:atlas}

In this section the basic systems forming the \ATLAS{} detector are described in \secRef{\thislabel:det}, followed in \secRef{\thislabel:data} by a description of the datasets considered in this paper and the corresponding run conditions in data. The \MC{} simulation setup for final-state generation and 
the simulation of the calorimeter response to the incident particles is described in \secRef{\thislabel:mc}.  

\renewcommand{\thislabel}{\baselabel:det}
\subsection{The \ATLAS{} detector}\label{sec:atlas:det}

The \ATLAS{} experiment features a multi-purpose detector system with a forward--backward symmetric cylindrical geometry. It 
provides nearly complete and hermetic coverage of the solid angle around the \pp{} collisions at the LHC. A detailed description of the \ATLAS{} experiment
can be found in Ref.~\cite{DetectorPaper}.

\subsubsection{The \ATLAS{} detector systems}\label{\thislabel:det}
The detector closest to the \pp{} collision vertex 
is the inner tracking detector (ID). 
It has complete azimuthal coverage and spans the pseudorapidity\footnote{ATLAS uses a right-handed coordinate system with its origin at the nominal interaction point (IP) in the centre of the detector and the $z$-axis along the beam pipe. The $x$-axis points from the IP to the centre of the LHC ring, and the $y$-axis points upward. Cylindrical coordinates $(r,\phi)$ are used in the transverse plane, $\phi$ being the azimuthal angle around the beam pipe. The pseudorapidity is defined in terms of the polar angle $\theta$ as $\eta=-\ln\tan(\theta/2)$.} 
region $|\eta|<2.5$. It consists of a silicon pixel detector,  a silicon micro-strip detector, and a straw-tube transition
radiation tracking detector covering  $|\eta|<2$. The ID is immersed into a uniform axial magnetic field of \unit{2}{\text{T}}{} provided by a thin superconducting solenoid magnet. 

\begin{figure}[!th] \centering
	\includegraphics[width=\figfullwidth]{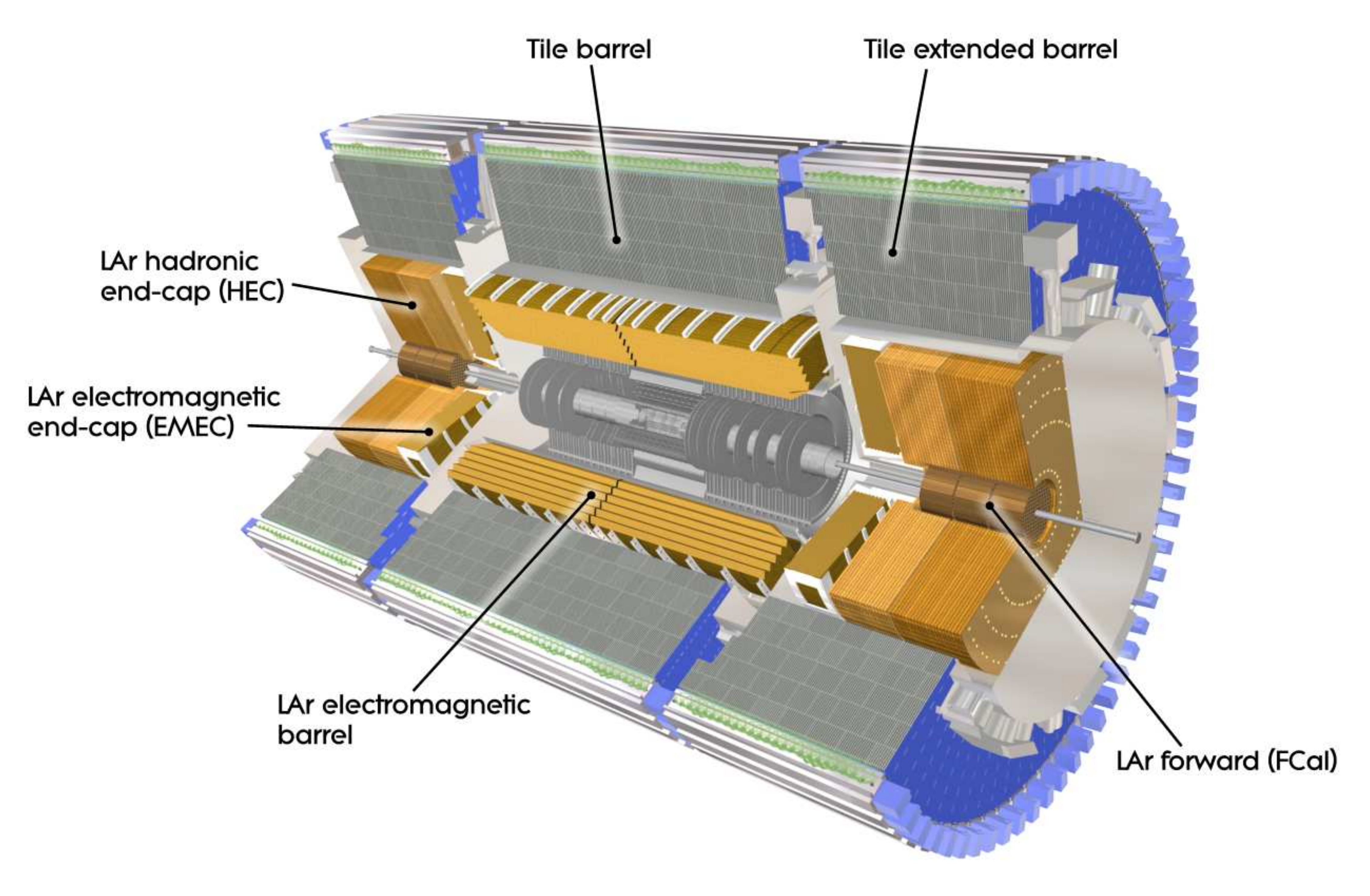}
	\caption{Cutaway view on the \ATLAS{} calorimeter system.}\label{fig:ATLAS_calos}
\end{figure}

The ATLAS calorimeter system is illustrated in \figRef{fig:ATLAS_calos}. It comprises several calorimeters with various read-out granularities and with different technologies.
The electromagnetic calorimeter (\LArEM) surrounding the ID is a high-granularity liquid-argon sampling calorimeter (\LAr), using lead as an absorber. It is divided into one barrel (\LArEMB; $|\eta|<1.475$) and two \EndCap{} (\LArEME; $1.375<|\eta|<3.2$) regions. 

The barrel and \EndCap{} regions also feature pre-samplers mounted between the cryostat cold wall and the calorimeter modules. The barrel pre-sampler (\LArPreSamplerB) covers $|\eta| < 1.52$, while the \EndCap{} pre-sampler (\LArPreSamplerE) covers $1.5 < |\eta| < 1.8$.  

The hadronic calorimeters are divided into three distinct sections.  The most central section contains the central barrel region ($|\eta|<0.8$) and two extended barrel regions ($0.8<|\eta|<1.7$). These regions are instrumented with scintillator-tile/steel hadronic calorimeters (\Tile). 
Each barrel region consists of $64$ modules with individual azimuthal  ($\phi$) coverages of $\pi/32$ 
rad. 
The two hadronic \EndCap{} calorimeters (\LArHEC; $1.5<|\eta|<3.2$) feature liquid-argon/copper calorimeter modules. The two forward calorimeters (\LArFCAL; $3.1<|\eta|<4.9$) are instrumented with liquid-argon/copper and liquid-argon/tung\-sten modules for elec\-tro\-mag\-ne\-tic and 
hadronic energy measurements, respectively.

The \ATLAS{} calorimeters have a highly granular lateral and longitudinal segmentation. Including the pre-samplers, there are seven sampling 
layers in the combined central calorimeters (\LArPreSamplerB, three in \LArEMB{} and three in \Tile) and eight sampling layers in the \EndCap{} region (\LArPreSamplerE, three in \LArEME{} and four in \HEC). The three \FCal{} modules provide three sampling layers in the forward region.  
Altogether, the calorimeter system has about $188\,000$ \readout{} channels. 
The \LArEM{} calorimeter are between $24$ radiation lengths (\radl) and \unit{27}{\radl}{} deep.
The combined depth of the calorimeters for hadronic energy measurements is more than 10 hadronic interaction lengths ($\lambda$) nearly everywhere across the full detector acceptance ($|\eta| \leq 4.9$). 
The amount of inactive material in front of the calorimeters depends on $\eta$. 
It varies from about \unit{2}{\radl}{} at $\eta = 0$ to about \unit{4}{\radl}{} at $|\eta| \approx 1.8$, when measured from the nominal interaction point in \ATLAS{} to the first active sampling layer (including \LArPreSamplerB{} and \LArPreSamplerE).
It can increase to more than \unit{6}{\radl}{} in the transition region between central and \EndCap{} calorimeters ($|\eta| \approx 1.45$ and $|\eta| \approx 1.7$).
The amount of inactive material for hadrons is approximately \unit{1}{\lambda}{} across the full covered $\eta$-range, with spikes going up to more than \unit{2}{\lambda}{} in transition regions and in regions with complex cryostat structures and beam line services ($|\eta| \approx 4$). 

The absorption power of the \ATLAS{} calorimeters and their segmentation allow for very precise energy-flow reconstruction based on the \topos{} described in this paper, with considerable exploitation of the \topo{} shapes for signal calibration purposes. For more details of the calorimeter \readout{} structures, absorption characteristics, inactive material distributions, and cell signal formation, see \citRef{DetectorPaper}. 
The segmentation of the \readout{} structure in the various calorimeter sampling layers, each named by a dedicated identifier (\sampid), is shown in \tabRef{tab:readout}.

\begin{table}[!tp] \centering
\caption{The \readout{} granularity of the \ATLAS{} calorimeter system \cite{DetectorPaper}, given in terms of $\Delta\eta\times\Delta\phi$ with the exception of the forward calorimeters, where it is given in linear measures $\Delta x \times \Delta y$, due to the non-pointing \readout{} geometry of the \LArFCAL. For comparison, the \LArFCAL{} granularity is approximately $\Delta\eta\times\Delta\phi = 0.15 \times 0.15 (0.3 \times 0.3)$ at $\eta = 3.5(4.5)$.
The total number of \readout{} cells, including both ends of the calorimeter system, with (without) pre-samplers is $187\,652$ ($178\,308$).}\label{tab:readout}
\renewcommand{\arraystretch}{1.1}
\begin{small}
\begin{tabular}{|p{0.2\textwidth}|p{0.25\textwidth}rp{0.2\textwidth}p{0.15\textwidth}|}
	\hline \hline
	\multicolumn{1}{|l|}{Calorimeter} & \multicolumn{1}{l}{Module \hspace{\fill}Sampling (\sampid)} 
		& $N_{\text{cells}}$ & $\eta$-coverage & \multicolumn{1}{c|}{$\Delta\eta\times\Delta\phi$} \\
	\hline
	\multirow{2}{0.2\textwidth}{\textit{Electromagnetic calorimeters}}  & \LArEMB  &  $109\,568$ & $|\eta|<1.52$ & \\
	 & \multicolumn{1}{r}{\LArPreSamplerB} & $7\,808$ & $|\eta|<1.52$ & $0.025\times\pi/32$ \\
	 & \multicolumn{1}{r}{\LArEMBN{1}}       & & $|\eta|<1.4$ & $0.025/8\times\pi/32$ \\
	 &                                                                     & & $1.4<|\eta|<1.475$ & $0.025\times\pi/128$ \\
	 & \multicolumn{1}{r}{\LArEMBN{2}}       & & $|\eta|<1.4$ & $0.025\times\pi/128$ \\
	 &                                                                     & & $1.4<|\eta|<1.475$ & $0.075\times\pi/128$ \\
	 & \multicolumn{1}{r}{\LArEMBN{3}}       & & $|\eta|<1.35$ & $0.050\times\pi/128$ \\
	 & \LArEME & $63\,744$ & $1.375 < |\eta| < 3.2$ & \\
	 & \multicolumn{1}{r}{\LArPreSamplerE} & $1\,536$ & $1.5<|\eta|<1.8$ & $0.025\times\pi/32$\\
	& \multicolumn{1}{r}{\LArEMEN{1}} & & $1.375<|\eta|<1.425$ & $0.050\times\pi/32$\\
	 &                                                              & & $1.425<|\eta|<1.5$    & $0.025\times\pi/32$ \\
	 &                                                              & & $1.5<|\eta|<1.8$    & $0.025/8\times\pi/32$ \\
	 &                                                              & & $1.8<|\eta|<2.0$    & $0.025/6\times\pi/32$ \\
	 &                                                              & & $2.0<|\eta|<2.4$    & $0.025/4\times\pi/32$ \\
	 &                                                              & & $2.4<|\eta|<2.5$    & $0.025\times\pi/32$ \\
	 &                                                              & &  $2.5<|\eta|<3.2$    & $0.1\times\pi/32$ \\
	& \multicolumn{1}{r}{\LArEMEN{2}} & & $1.375<|\eta|<1.425$ & $0.050\times\pi/128$ \\
	  &                                                             & & $1.425<|\eta|<2.5$ & $0.025\times\pi/128$ \\
	  &                                                             & & $2.5<|\eta|<3.2$ & $0.1\times\pi/128$ \\
	& \multicolumn{1}{r}{\LArEMEN{3}} & & $1.5<|\eta|<2.5$ & $0.050\times\pi/128$ \\ \hline
	\multirow{1}{0.2\textwidth}{\textit{Hadronic calorimeters}}  & \Tile{} (barrel)  & $2\,880$ 
	                                                                                                                  &  $|\eta|<1$ & \\
	 & \multicolumn{1}{r}{\TileBarN{0/1}}          &  &  & $0.1\times\pi/32$ \\
	 & \multicolumn{1}{r}{\TileBarN{2}}             &   & & $0.2 \times\pi/32$\\ 
	 & \Tile{} (extended barrel) & $2\,304$
	                                         & $0.8 < |\eta| < 1.7$ & \\
	 & \multicolumn{1}{r}{\TileExtN{0/1}}             &  & & $0.1\times\pi/32$ \\
	 & \multicolumn{1}{r}{\TileExtN{2}}             &  & & $0.2\times\pi/32$ \\ 
	 & \LArHEC                                             & $5\,632$ & $1.5<|\eta|<3.2$ & \\
	 & \multicolumn{1}{r}{\LArHECN{0/1/2/3}}  &   & $1.5<|\eta|<2.5$ & $0.1\times\pi/32$ \\
	  &                                                                & & $2.5<|\eta|<3.2$ & $0.2\times\pi/16$ \\ \hline
	 \multirow{1}{0.2\textwidth}{\textit{Forward calorimeters}} & \LArFCAL & $3\,524$ & $3.1<|\eta|<4.9$ & \multicolumn{1}{c|}{$\Delta x \times \Delta y$} \\ \cline{5-5}
	 & \multicolumn{1}{r}{\LArFCALN{0}} & & $3.1<|\eta|<3.15$ & $\unit{1.5}{\cm}\times\unit{1.3}{\cm}$ \\
	 &                                                                & & $3.15<|\eta|<4.3$ & $\unit{3.0}{\cm}\times\unit{2.6}{\cm}$ \\
	 &                                                                & & $4.3<|\eta|<4.83$ & $\unit{1.5}{\cm}\times\unit{1.3}{\cm}$ \\
	 & \multicolumn{1}{r}{\LArFCALN{1}} & & $3.2<|\eta|<3.24$ & $\unit{1.7}{\cm}\times\unit{2.1}{\cm}$ \\
	 &                                                                & & $3.24<|\eta|<4.5$ & $\unit{3.3}{\cm}\times\unit{4.2}{\cm}$ \\
	 &                                                                & & $4.5<|\eta|<4.81$ & $\unit{1.7}{\cm}\times\unit{2.1}{\cm}$ \\
	 & \multicolumn{1}{r}{\LArFCALN{2}} & & $3.29<|\eta|<3.32$ & $\unit{2.7}{\cm}\times\unit{2.4}{\cm}$ \\
	 &                                                                & & $3.32<|\eta|<4.6$ & $\unit{5.4}{\cm}\times\unit{4.7}{\cm}$ \\
	 &                                                                & & $4.6<|\eta|<4.75$ & $\unit{2.7}{\cm}\times\unit{2.4}{\cm}$ \\ \hline
	  \hline
\end{tabular}
\end{small}
\end{table}

The muon spectrometer surrounds the \ATLAS{} calorimeters. A system of three large 
air-core toroids, a barrel and two \EndCap s with eight coils each, generates a magnetic field in the
pseudorapidity range of $|\eta| < 2.7$. The muon spectrometer measures the full momentum of muons based on their tracks reconstructed
with three layers of precision tracking chambers  in the toroidal field. It is also instrumented with separate trigger chambers.

\subsubsection{The \ATLAS{} trigger} \label{\thislabel:trigger}

The trigger system for the \ATLAS{} detector in \runone{} consisted of a hardware-based Level~1~(\Lone) trigger 
and a software-based High Level Trigger~(\HLT) \cite{Aad:2012xs}.  
For the evaluation of the \topo{} reconstruction performance, samples of minimum-bias (MB) triggered events, samples of events selected by jet triggers, and samples of events with hard objects such as muons, which are not
triggered by the calorimeter, are useful.
 
The \ATLAS{} \MB{} trigger \cite{Martin:2010zz} used signals from a dedicated system of scintillators (\MBTS{} \cite{Aad:2010af}; $2.1 < |\eta| < 3.8$) at \Lone{} in 2010 and 2011 data-taking. 
Depending on the run period, it required one hit in either of the $\eta$ hemispheres, or one hit in each $\eta$ hemisphere. 
In 2012, the MB samples were triggered by a zero-bias trigger. 
This trigger unconditionally accepted events from bunch crossings occurring a fixed number of  \LHC{} cycles after a high-energy electron or photon was accepted by the \Lone{} trigger.
The \Lone{} trigger rate for these hard objects scales linearly with luminosity, thus the collision environment generated by the luminosity-dependent additional \pp{} interactions discussed in \secRef{sec:atlas:data:pu} is well reflected in the \MB{} samples.

For triggering on collision events with jets at \Lone, jets are first built from coarse-granu\-la\-rity calorimeter towers using a sliding-window algorithm (\Lone-jets). The events are accepted if they have \Lone-jets passing triggers based on (1) the transverse momentum (\pT)  of individual \Lone-jets (single-jet triggers) or on (2) the detection of several such jets at increasing transverse momenta (multi-jet triggers).
Those events accepted by \Lone{} are then subjected to refined jet-trigger decisions based on jet \pT{} and multi-jet topology in the \HLT, now using jets that are reconstructed from calorimeter cell signals with algorithms similar to the ones applied in the offline precision reconstruction \cite{Tamsett:2013rya}.  

A \Zboson{} boson sample is collected from muon triggers at \Lone. 
Since the trigger rate and the reconstruction of the decay properties of the accepted \Zmumu{} events are basically unaffected by \pu,
this sample is not only unbiased in this respect but also with respect to other possible biases introduced by the \ATLAS{} calorimeter signals.

\renewcommand{\thislabel}{\baselabel:data}
\subsection{Dataset} \label{sec:atlas:data}

\begin{figure}[t!] \centering
\subfloat[Peak luminosities in data runs]{\includegraphics[width=0.95\textwidth]{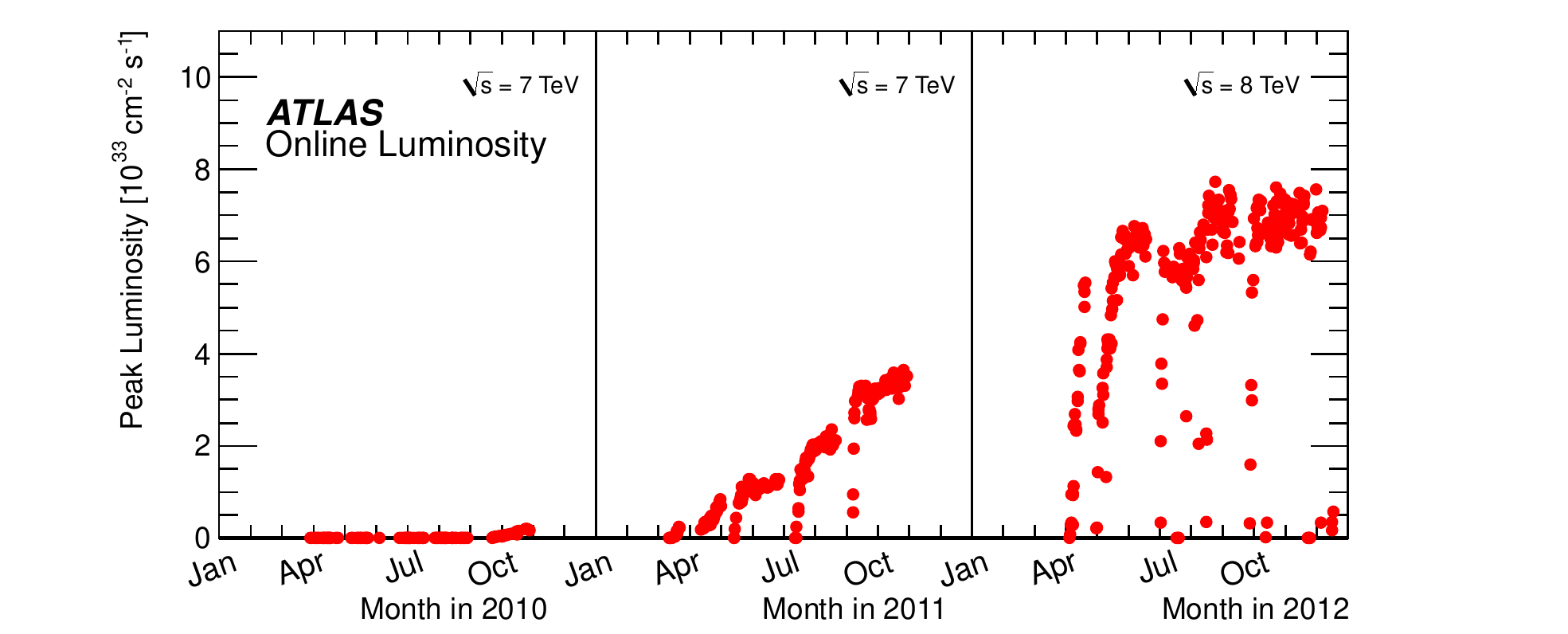}\label{fig:lhc:lumivstime}} \\
\subfloat[Mean number of additional \pp{} interactions per bunch crossing]{\includegraphics[width=0.95\textwidth]{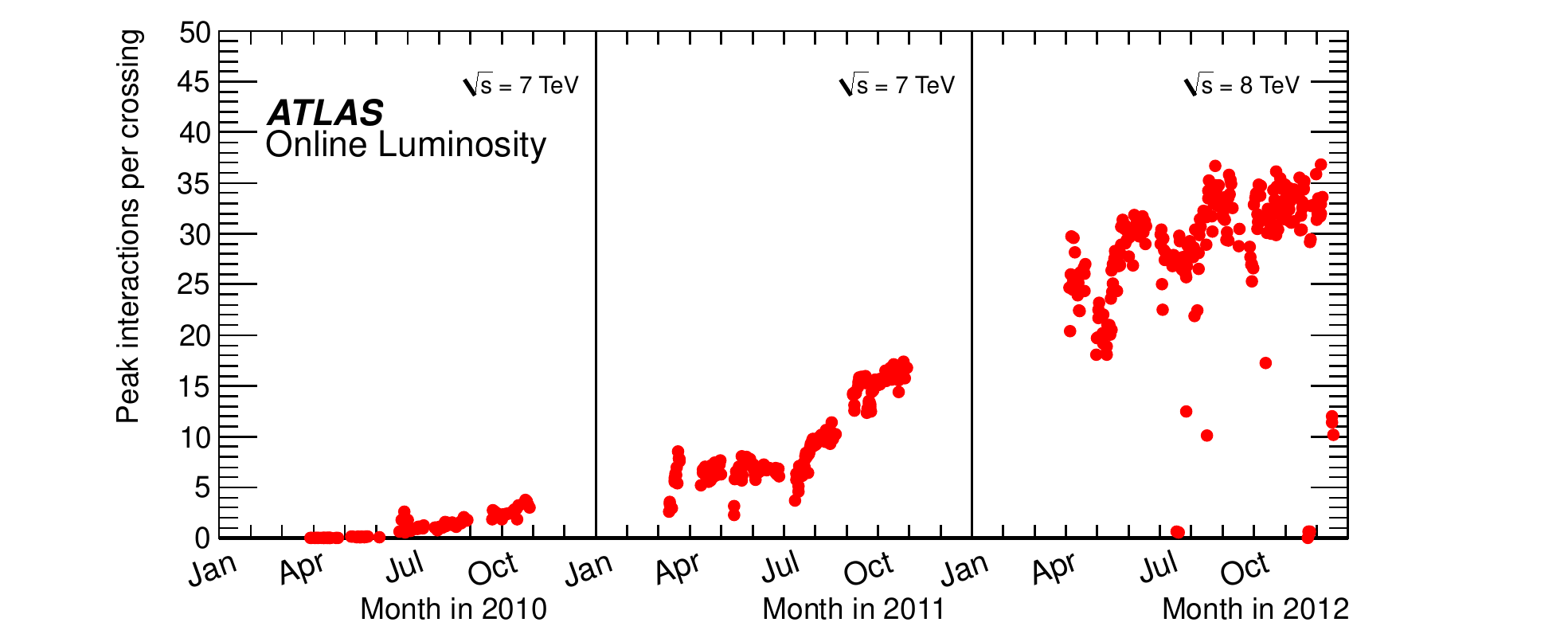}\label{fig:lhc:muvstime}} 
\caption[]{The peak luminosities measured by the \ATLAS{} online luminosity monitor system throughout the run years are shown in \subref{fig:lhc:lumivstime}. The mean number of additional \pp{} interactions at the beginning of each \LHC{} fill is shown in \subref{fig:lhc:muvstime} for the same period in time.
\label{fig:lhc}}
\end{figure}

The data used for the evaluation of the \topo{} reconstruction performance are selected from \pp{} collision events at a \cms{} of \sqrts{7}, recorded with the \ATLAS{} detector in 2010, and at \sqrts{8} in 2012. 
The overall amount of high-quality data recorded at those times corresponds to \unit{\sim\!45}{\invpb}{} in 2010, and \unit{\sim\!20.3}{\invfb}{} in 2012.  
Peak instantaneous luminosities reached in the first three years of \LHC{} running (\LHC{} \runone) are shown in \figRef{fig:lhc}\subref{fig:lhc:lumivstime}. 
Some early data recorded during the very first \pp{} collisions in the \LHC{} in 2009 are considered for the studies of the \topo{} reconstruction performance as well. 
The corresponding events are extracted from approximately  $540\,000$ \pp{} collisions at $\sqrt{s} = \unit{900}{\GeV}$, recorded during stable beam conditions and corresponding to about \unit{12}{\invmb}.
Occasional references to 2011 run conditions, where protons collided in the \LHC{} with \sqrts{7}{} and \ATLAS{} collected data corresponding to \unit{\sim\!5.1}{\invfb}, are provided to illustrate the evolution of the operational conditions during \LHC{} \runone{} relevant to \topo{} reconstruction. 
The specific choice of 2010 and 2012 data for the performance evaluations
encompasses the most important scenarios with the lowest and highest luminosity operation, respectively.  

\subsubsection{\PU{} in data} \label{\thislabel:pu}
One important aspect of the contribution from additional \pp{} interactions (\pu) to the calorimeter signal in data is the sensitivity of the \ATLAS{} liquid-argon calorimeters to this \pu{} as a function of the instantaneous luminosity, and as a function of the signal history from previous bunch crossings. 

\begin{figure}[t!] \centering
	\includegraphics[width=\fighalfwidth]{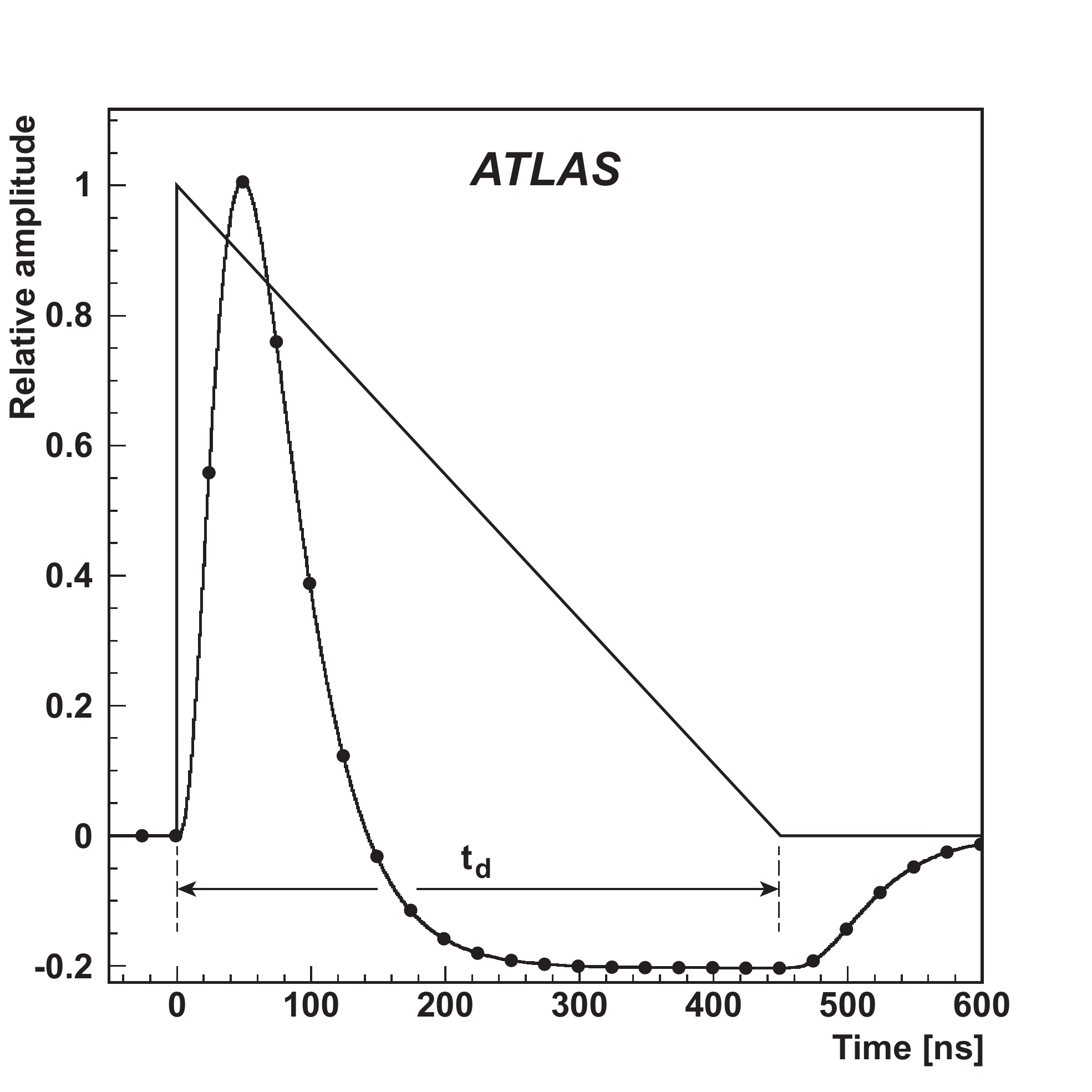}
	\caption[Liquid argon calorimeter pulse shape]{The pulse shape in the \ATLAS{} \LAr{} calorimeters. The unipolar triangular pulse is the current pulse in the liquid argon generated by fast ionising particles. Its characteristic time is the drift time (charge collection time) $t_{\text{d}}$, with $t_{\text{d}} \approx \unit{450}{\ns}$ in the example for the central \LArEMB{} calorimeter shown here.
 The shaped pulse is superimposed, with a characteristic duration of $\tau_{\text{signal}} \approx \unit{600}{\ns}$. The full circles on the shaped pulse indicate the nominal bunch crossings at \unit{25}{\ns} intervals. The figure has been adapted from Ref. \cite{Aad:2014una}.}\label{fig:larpulse}     
\end{figure}

In the initial phase of data-taking in 2010 the proton beam intensities at \LHC{} were relatively low. 
The recorded events contain on average three additional \pp{} interactions, as shown in \figRef{fig:lhc}\subref{fig:lhc:muvstime}. 
In addition, the initial bunch crossing interval of $t_{\text{BX}} = \unit{750}{\ns}$ was larger than the window of sensitivity of the \LAr{} calorimeter,
which is defined by the duration $\tau_{\text{signal}}$ of the shaped signal, with $\tau_{\text{signal}} \approx \unit{600}{\ns}$, as depicted  in \figRef{fig:larpulse} for the typical charge collection time of $t_{\text{d}} = \unit{450}{\ns}$ in this detector.
In later data-taking periods in 2010 the bunch crossing interval was reduced to $t_{\text{BX}} = \unit{175}{\ns}$, which is within the sensitivity of the \LAr{} calorimeter signal formation ($t_{\text{BX}} < \tau_{\text{signal}}$). Nevertheless, the still-low instantaneous luminosity reduced the amount of energy scattered into the calorimeter in the other bunch crossings to a negligible contribution with little effect on the signal history. 

Throughout operations in 2011 and 2012, the proton beam intensities in the \LHC{} were significantly increased, leading to the corresponding increases in the number of \pu{} interactions per bunch crossing shown in \figRef{fig:lhc}\subref{fig:lhc:muvstime}.
At the same time, $t_{\text{BX}}$ 
was reduced to \unit{50}{\ns}. These two changes in the run conditions introduced 
a sensitivity of the \LAr{} calorimeter signal to the signal residuals from \pp{} interactions occurring in $N_{\text{BX}}^{\text{PU}} \approx 12$ preceding bunch crossings at the \LHC{}
(\opu), in addition to \pu{} interactions in the current bunch crossing (\ipu). 
The \opu{} effect on the cell signal depends on $N_{\text{BX}}^{\text{PU}} \approx \tau_{\text{signal}}/t_{\text{BX}}$ and the energy deposited in each of the $N_{\text{BX}}^{\text{PU}}$ bunch crossings.

The bipolar shape of the \LAr{} calorimeter signal shown in \figRef{fig:larpulse} reduces the overall effect of \pu, because it
features a net-zero integral over time. 
This leads to 
can\-cel\-lation on average of \ipu{} signal contributions by \opu{} signal residuals in any given calorimeter cell. By design of the shaping amplifier, and the choice of digitally sampling the shaped pulse amplitude in time with a frequency of \unit{40}{\text{MHz}}{}
in the \readout, the most efficient suppression is achieved for \unit{25}{\ns}{} bunch spacing in the \LHC{} beams.
It is fully effective in the limit where for each bunch crossing contributing to \opu{} about the same amount of energy is deposited
in a given calorimeter cell. 
A small loss of efficiency is observed for \unit{50}{\ns}{} bunch spacing, due to the less frequent injection of energy by the fewer previous bunch crossings.

Approximately the first ten bunch crossings in each \LHC{} bunch train at \unit{50}{\ns}{} bunch spacing are characterised by different \opu{} contributions from the collision history. 
This history gets filled with signal remnants from an increasing number of past bunch crossings with \pp{} interactions the 
larger the time difference between the bunch crossing and the beginning of the train becomes.   
The remaining bunch crossings in a train, about 26 of a total of 36 in 2011 and 62 of a total of 72 in 2012, have an \opu{} signal contribution which is stable within the 
bunch-to-bunch fluctuations in the beam intensity. 
In 2012 data  a dedicated cell-by-cell correction is applied in the offline cell signal reconstruction to compensate for the corresponding variations  in the \opu.
Further details of the \ATLAS{} liquid-argon calorimeter \readout{} and signal processing can be found in Ref.~\cite{Aad:2010ai}.

Even with a constant proton bunch intensity and apart from the bunch train effects,  
the efficiency of \pu{} suppression by signal shaping is reduced by the large fluctuations in the number of additional interactions from bunch crossing to bunch crossing, and by the different energy-flow patterns of the
individual collisions in the time window of sensitivity $\tau_{\text{signal}}$ in the \LAr{} calorimeters. 
Consequently, the signal shows a principal sensitivity 
to \pu,  even after shaping and digital filtering in the \readout. 
This is evident from the residual event-by-event deviation of the cell-signal baseline, which depends on the specific \pu{} condition at the time of the triggered event, from the (average zero) baseline expected from the signal shaping.  
These baseline fluctuations can lead to relevant signal offsets once the noise suppression is applied, which is an important part of the calorimeter signal extraction strategy using \topos{} presented in \secRef{sec:topos}. 

The \Tile{} calorimeter shows very little sensitivity to \pu{} since most of the associated (soft particle) energy flow is absorbed in the \LAr{} calorimeters in front of it. Moreover, \opu{} is suppressed by a shorter signal collection time and a short pulse shaping time, reducing the sensitivity of the signal to only about three bunch crossings at \unit{50}{\ns}{} intervals \cite{Aad:2010af}.

\subsubsection{Effect on calorimeter noise} \label{\thislabel:noise}

\begin{figure*}[t!]
\centering
\subfloat[$\sigma_{\mathrm{noise}}(\left|\eta\right|)$ in $2010$ ($\mu = 0$)]{\includegraphics[width=\fighalfwidth]{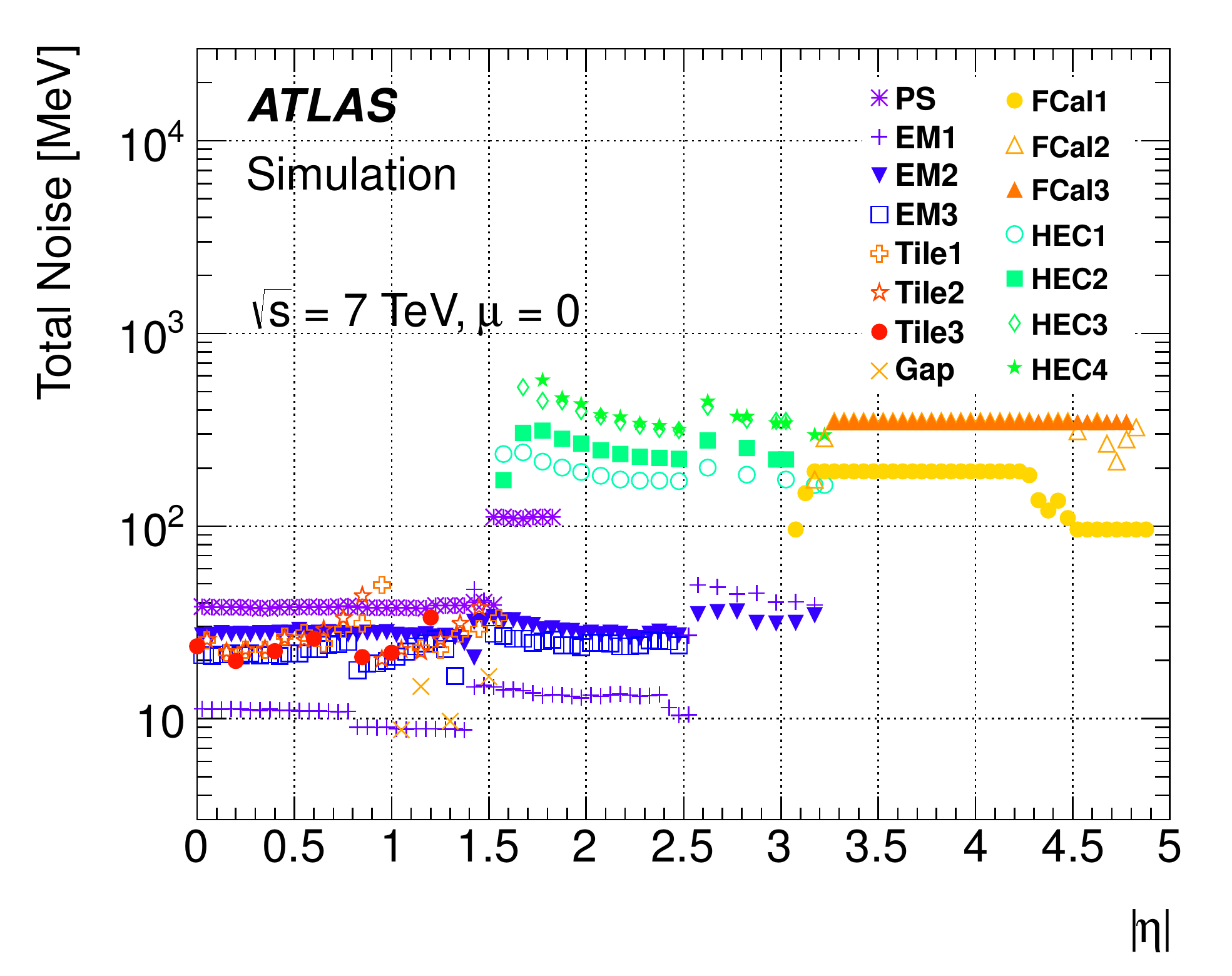}\label{fig:noise:2010}}
\subfloat[$\sigma_{\mathrm{noise}}(\left|\eta\right|)$ in $2011$ ($\mu = 8$)]{\includegraphics[width=\fighalfwidth]{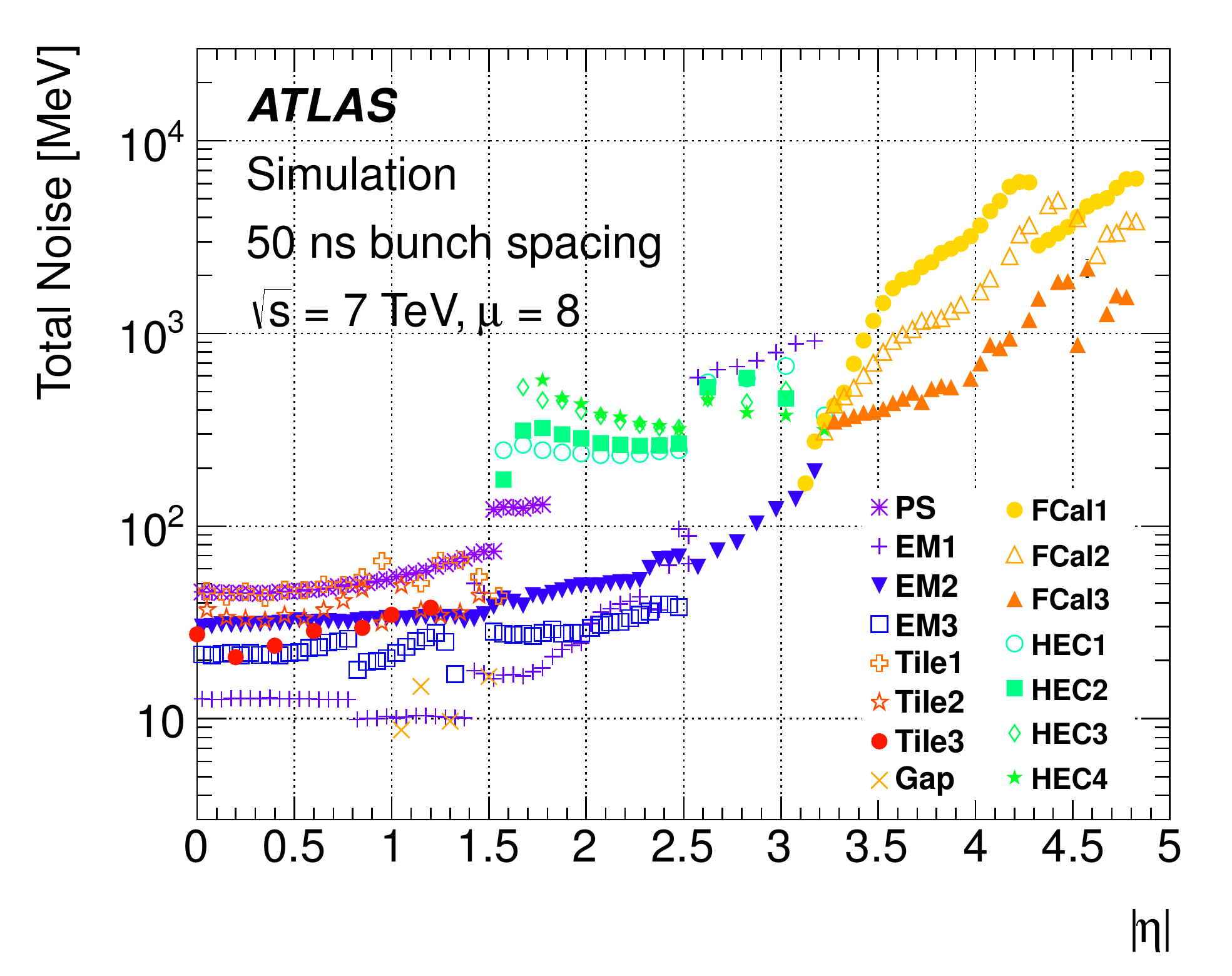}\label{fig:noise:2011}}\qquad
\subfloat[$\sigma_{\mathrm{noise}}(\left|\eta\right|)$ in $2012$ ($\mu = 30$)]{\includegraphics[width=\fighalfwidth]{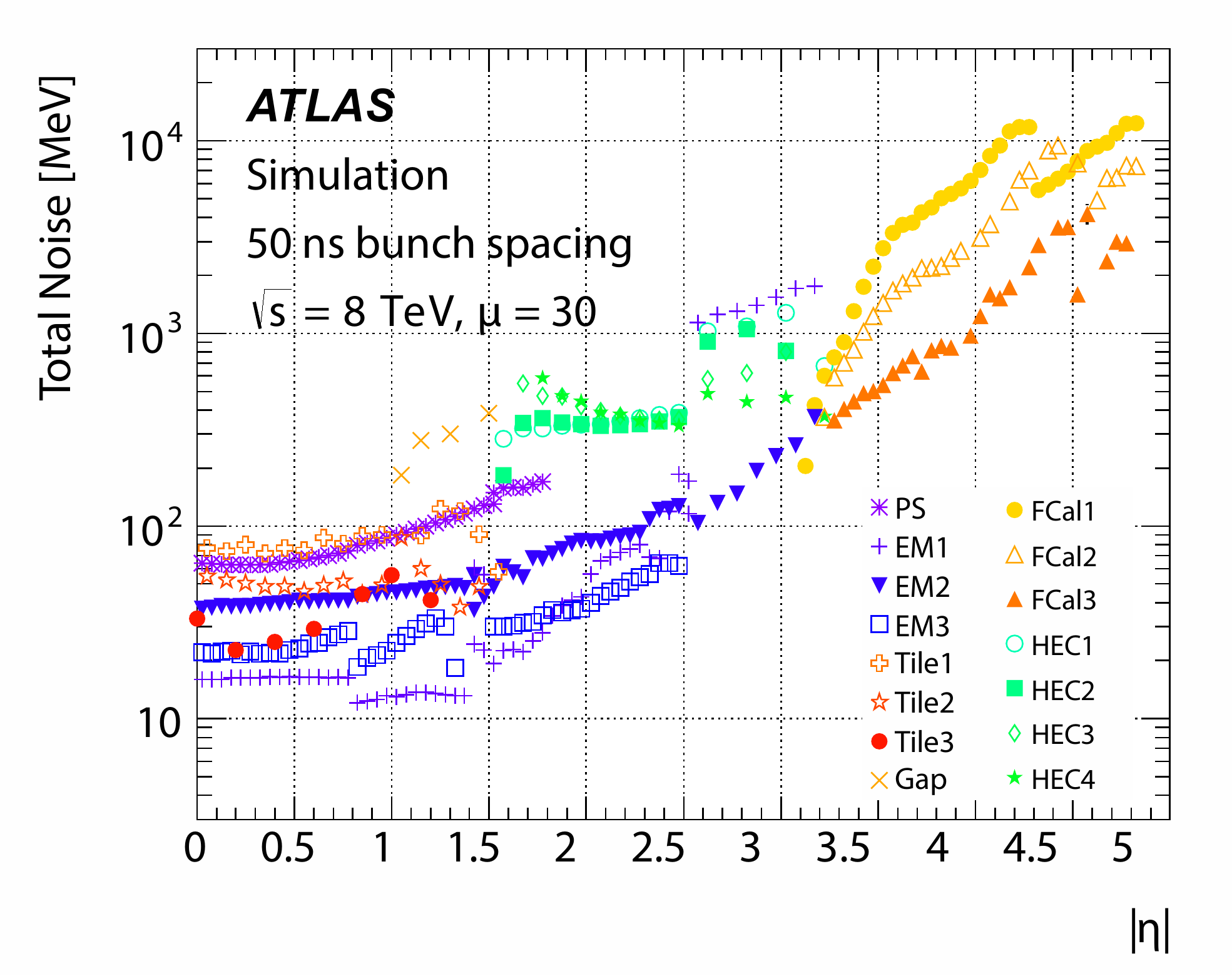}\label{fig:noise:2012}}
\caption[]{The energy-equivalent cell noise in the \ATLAS{} calorimeters on the electromagnetic (\EM) scale 
as a function of the direction $\left|\eta\right|$ in the detector, for \subref{fig:noise:2010} the $2010$ configuration 
with  $\mu = 0$, \subref{fig:noise:2011} the $2011$ configuration with $\mu = 8$ (both plots from \citRef{Aad:2014bia}), and \subref{fig:noise:2012} the $2012$ configuration with  $\mu = 30$. The various colours
indicate the noise in the pre-sampler (\texttt{PS}) and the three layers 
of the \LAr{} \LArEM{} calorimeter, 
the three layers of the \Tile{} calorimeter, the four
layers of the hadronic end-cap (\LArHEC) calorimeter, and the three 
modules of the forward (\LArFCAL) calorimeter. 
\label{fig:noise}}
\end{figure*}   

In \ATLAS{} operations prior to 2011 the cell noise was dominated by electronic noise. 
The short bunch crossing interval  and higher instantaneous luminosity in 2011 and 2012 \LHC{} running added additional and dominant noise contributions 
from the cell-signal baseline fluctuations introduced by \pu, as
discussed in \secRef{\thislabel:pu}.
These fluctuations, even though not perfectly following a Gaussian distribution,\footnote{Selected examples of the actual distributions taken from data are shown in the context of the \topo{} formation discussed in \secRef{sec:topos:formation:collect}.} can nevertheless be expressed as noise measured by the standard deviation of their distribution, taken from simulated \MB{} events and scaled to the expected number of \pu{} interactions.   
The cell noise thresholds steering the \topo{} formation described in \secRef{sec:topos} thus needed to be increased from those used in 2010 to accommodate this \pu-induced noise.
This is done by adjusting the nominal energy-equivalent noise \sigTot{} according to 
\begin{equation}
        \sigTot = \left\{\begin{array}{lr}
	\sigEle & \quad\text{(2010 operations)}, \\[6pt]
	\sqrt{\left(\sigEle\right)^{2} + \left(\sigPU\right)^{2}}   &
                                 \quad\text{(2011 and 2012 operations)}.
                                 \end{array}\right. 
\label{eq:noise}
\end{equation} 
Here, \sigEle{} is the electronic noise, and \sigPU{} the noise from \pu, 
corresponding to an average of eight additional \pp{} interactions per bunch crossing ($\mu = 8$) in 2011, and $\mu = 30$ in 2012. 
These configurations are choices based on the expected average $\langle\mu\rangle$ for the run year.
They needed to be made before the respective data-taking started, to allow for a fast turn-around reconstruction of the collected data. 
As $\mu$ changes with the decrease of
the instantaneous luminosity $L_{\text{inst}}$  through-out the \LHC{} proton fill, \sigPU{} is only optimal for the small subset of data recorded when $L_{\text{inst}}$ generated the   
nominal (\textit{a priori} chosen) $\mu$ \pu{} interactions on average. 
\LHC{} operations at lower $\mu$ lead to slightly reduced calorimeter sensitivity to relevant small signals, as \sigPU{} is too large. 
For data-taking periods with higher than nominal $\mu$ the noise suppression is not optimal, leading to more noise contributions to the \topo{} signals. 
 
The change of the total nominal noise \sigTot{} and its dependence on the
calorimeter region in \ATLAS{} can be seen by comparing \figMultiRefLabel~\ref{fig:noise}\subref{fig:noise:2010}, \ref{fig:noise}\subref{fig:noise:2011}, and \ref{fig:noise}\subref{fig:noise:2012}. In most calorimeter regions,
the total noise rises significantly above the electronic noise with increasing \pu{} activity, as expected. This increase is largest in the forward calorimeters, where  $\sigPU \gg \sigEle$ by more than one order of magnitude, already under 2011 run conditions.

\renewcommand{\thislabel}{\baselabel:mc}
\subsection{Monte Carlo simulations} \label{sec:atlas:mc}
The energy and direction of particles produced in \pp{} collisions
are simulated using various \MC{} event generators. An overview of these
generators for \LHC{} physics can be found in Ref.~\cite{Buckley:2011ms}.
The samples for comparisons to 2010 data are produced at \sqrts{7}, while the \MC{} samples for 2012 analysis are generated at \sqrts{8}. Some configuration details for the inclusive jet and inclusive \Zboson{} boson \MC{} samples and the simulated \MB{} samples are given below. 

\subsubsection{Monte Carlo simulations of signal samples} \label{\thislabel:signal}

Simulated signal samples include inclusive jet-production, which is generated using \pythia{} \cite{Sjostrand:2006za} version 6.425  for 2010 analyses, and \pythia 8 \cite{Sjostrand:2007gs} version 8.160 for 2012 analysis. 
Both generators model the hard sub-process in the final states of the generated \pp{} collisions using a \twototwo{} matrix element at leading order in the strong coupling \alphas. 
Additional radiation is modelled in the leading-logarithmic (LL) approximation by \ptordered{} parton showers \cite{Corke:2010zj}.
Multiple parton interactions (MPI) \cite{Sjostrand:2004ef}, as well as fragmentation and hadronisation 
based on the Lund string model \cite{Andersson:1983ia}, are also generated.

For comparisons with 2012 data, samples of \Zboson{} bosons with \Zmumu{} are generated. The next-to-leading-order (NLO) \powheg{} \cite{Nason:2004rx,Frixione:2007vw} model is used, 
with the final-state partons showered by \pythia 8 using the \CTTEN{} NLO parton distribution function (PDF) \cite{Lai:2010vv} and the \ATLAS{} \AUTWO{}  \cite{ATLAS:2012uec} set of tuned parton shower and other soft underlying event generation parameters. 
\pythia 8 also provides the MPI, fragmentation and hadronisation for these events. 

\subsubsection{Minimum-bias samples and \pu{} modelling} \label{\thislabel:mb}

The \MB{} samples for 2012 running conditions are generated using \pythia 8 with the \ATLAS{} \AMTWO{} \cite{ATLAS:2012uec} set of tuned soft interaction parameters  and the \MSTW{} PDF set \cite{Martin:2009iq}. 
A single, fully simulated event for that run year is built  by overlaying a number $N_{\text{PU}}$ of generated \MB{} events onto one generated hard-scatter event. The actual $N_{\text{PU}}$ is drawn from a Poisson distribution around the average number $\langle\mu\rangle$ of additional \pp{} collisions per bunch crossing. 
The value of $\langle\mu\rangle$ is measured by the experiment as an average over one 
\emph{luminosity block}, which can last as long as two minutes, with its actual duration depending on the central data acquisition configuration at the time of the data-taking. 
The measurement of $\langle\mu\rangle$ is mainly based on single $\eta$-hemisphere hit counting as well as counting coincidental hits in both $\eta$-hemispheres with the fast \ATLAS{} luminosity detectors consisting of two small Cerenkov counter (\texttt{LUCID}; $5.6 < |\eta| < 6.0$) and two sets of small diamond sensors forming two beam conditions monitors (\texttt{BCM}; $|\eta| = 4.2$).  
Details of these detectors and the measurement are given in \citRef{Aad:2013ucp}.
The distribution of the measured $\langle\mu\rangle$ over the whole run period is taken into account in the \pu{} simulation.

The \LHC{} bunch train structure with 72 proton bunches per train and \unit{50}{\ns} spacing between the bunches in 2012,  
is also modelled by organising the simulated collisions into four such trains. 
This allows the inclusion of \opu{} effects driven by the distance of the hard-scatter events from the beginning of the bunch train, as discussed in \secRef{sec:atlas:data:pu}. A correction depending on the bunch position in the train is applied to data and \MC{} simulations to mitigate these effects. 
Bunch-to-bunch intensity fluctuations in the \LHC{} are not included in the \MC{} modelling. These are corrected in the data by the correction depending on the position of the bunch in the train.

\subsubsection{Minimum-bias overlay samples for 2012}\label{\thislabel:ovly}

In addition to the fully generated and simulated \MC{} samples described earlier, samples with events mixing data and \MC{} simulations are used to study the \topo{} reconstruction performance. 
These samples are produced by overlaying one event from the \MB{} samples collected by the zero-bias trigger described in \secRef{sec:atlas:det:trigger}
and a hard-scatter interaction from the \MC{} generator \cite{Rimoldi:2011zz,Haas:2012,Marshall:2014mza}. 
The generated hard-scatter event is simulated using  the detector simulation described in \secRef{\thislabel:det}, but without any noise effects included. The recorded and simulated raw electronic signals are then overlaid prior to the digitisation step in the simulation. This results in modelling both the detector noise and the effect of \pu{} from data with the correct experimental conditions on top of the simulated event. Theses samples are useful for
detailed comparisons of \topo{} signal features in 2012, as they do not depend on limitations in the soft-event modelling introduced by any of the generators. 

\subsubsection{Detector simulation} \label{\thislabel:det}

The \geant{} software toolkit \cite{Agostinelli:2002hh} within the \ATLAS{} simulation framework \cite{Aad:2010ah} propagates the stable 
particles\footnote{Stable particles are those with laboratory frame lifetimes  $\tau$ defined by $c \tau > 10$ \mm.}
produced by the event generators through the \ATLAS{} detector and simulates their interactions with the detector material and the signal formation. Hadronic showers are simulated with the quark-gluon-string-plasma  model employing a quark--gluon string model \cite{Folger:2003sb} at high energies and the Bertini intra-nuclear cascade model \cite{Guthrie:1968ue,Bertini:1970zs,Bertini:1971xb} at low energies (QGSP\_BERT). 
There are differences between the detector simulation used in 2010 and in 2012. A newer version of \geant{} (version 9.4) is employed in 2012, together
with a more detailed description of the \LAr{} calorimeter absorber structure. These geometry changes introduce an increase of about $2\%$ in the calorimeter response to pions with energies of less than \unit{10}{\GeV}.

\renewcommand{\thislabel}{\baselabel:jetreco}
\subsection{Hadronic final-state reconstruction in \ATLAS{}} \label{sec:atlas:jetreco}
The fully reconstructed final state of the \pp{} collisions in \ATLAS{} includes identified individual particles comprising electrons, photons, muons, and $\tau$-leptons, in addition to jets and missing transverse momentum (\met). 
Calorimeter signals contribute to all objects, except for muons. The \topos{} introduced in detail in \secRef{sec:topos} are primarily used for the reconstruction of isolated hadrons, jets and \met.  
 
Jets are reconstructed using \topos, with their energies either reconstructed on the basic (electromagnetic) scale presented in \secRef{sec:topos:kinematics}, or on the fully calibrated and corrected (hadronic) scale described in \secRef{sec:lcw}.    

Additional refinement of the jet energy scale (\JES) may 
include reconstructed charged-particle tracks from the ID. More details of jet reconstruction and calibration can be found in \citMultiRef{Aad:2014bia,Aad:2011he}.

Jets used in the studies presented here are reconstructed in data and \MC{} simulations using the  \antikt{}  jet algorithm \cite{Cacciari:2008gp} as implemented in the \FJ{} package \cite{Cacciari:2011ma}. 
The jet size is defined by the radius parameter $R$ in the jet algorithm, where both $R = 0.4$ and $R = 0.6$ are used. 
Full four-mo\-men\-tum recombination is used, restricting the input \topo{} signals to be positive for a meaningful jet formation. 
The jets are fully calibrated and corrected after formation, including a correction for \pu{} signal contributions. 
For 2012, the \pu{} correction employs the reconstructed median transverse momentum density in the event and the area of the jet to subtract the \pT{ } contribution from \pu{}, following the suggestions in \citRef{Cacciari:2007fd}. In addition, an \MC{} simulation-based residual correction is applied \cite{Aad:2015ina}.

\renewcommand{\baselabel}{sec:topos}
\renewcommand{\thislabel}{\baselabel}
\section{Topological cluster formation and features} \label{sec:topos}
The collection of  the calorimeter signals of a given collision event into clusters of topologically connected cell signals is an attempt to extract the significant signal from a background of electronic noise and 
other sources of fluctuations such as \pu.
This strategy is most effective in a highly granular calorimeter system such as the one employed by \ATLAS. 
Finely segmented lateral \readout{} together with longitudinal sampling layers allows the resolution of energy-flow structures generating these spatial signal patterns, thus 
retaining only signals important for particle and jet reconstruction while efficiently removing insignificant signals induced by noise.    
The signal extraction is guided by reconstructing three-dimensional ``energy blobs" from particle showers in the active calorimeter volume. 
Individual \topos{} are not solely expected to contain the entire response to a single particle all of the time. Rather, depending on the incoming particle types, energies, spatial separations and cell signal formation, individual \topos{} represent the full or fractional response to a single particle (full shower or shower fragment), the merged response of several particles, or a combination of merged full and partial showers.

\subsection{\Topo{} formation} \label{\thislabel:formation}

The collection of calorimeter cell signals into \topos{} follows spatial signal-significance patterns generated by particle showers. 
The basic observable controlling this cluster formation is the cell signal significance \cellsig, which is defined as the ratio of the cell signal to the average (expected) noise \sigTotEMCell{} in this cell, as estimated for each run year according to \eqRef{eq:noise} (with $ \sigTotEMCell = \sigTot$),
\begin{equation}
\cellsig = \frac{\ecellem}{\sigTotEMCell} .
\label{eq:significance}
\end{equation}
Both the cell signal \ecellem{} and \sigTotEMCell{} are measured on the electromagnetic (\EM) energy scale.
This scale reconstructs the energy deposited by electrons and photons correctly but does not include any corrections for the loss of signal for hadrons due to the non-compensating character of the \ATLAS{} calorimeters.

\Topos{} are formed by a growing-volume algorithm starting from a calorimeter cell with a highly significant seed signal.
The seeding, growth, and boundary features of \topos{} are in this algorithm controlled by the three respective parameters $\{S,N,P\}$, which define signal thresholds in terms of \sigTotEMCell{} and thus apply selections based on \cellsig{} from \eqRef{eq:significance},
\begin{align}
\left|\ecellem\right| > S \sigTotEMCell &\Rightarrow \left|\cellsig\right| > S && (\text{primary seed threshold, default\ } S = 4);   \label{eq:toposeed} \\[6pt]
\left|\ecellem\right| > N \sigTotEMCell &\Rightarrow\left|\cellsig\right| > N  && (\text{threshold for growth control, default\ } N = 2); \label{eq:topogrowth} \\[6pt]
\left|\ecellem\right| > P \sigTotEMCell &\Rightarrow\left|\cellsig\right| > P  && (\text{principal cell filter, default\ } P = 0).  \label{eq:topofilter}
\end{align}
Useful configurations employ a $S > N \ge P$ rule, as reflected in the default configuration for \ATLAS{} indicated above. The default values are derived from optimisations of the response and the relative energy resolution for charged pions in test-beam experiments using \ATLAS{} calorimeter prototypes \cite{Speckmayer:2008zz}. 

\subsubsection{Collecting cells into \topos} \label{\thislabel:formation:collect}

\Topo{} formation is a sequence of \emph{seed and collect} steps, which are repeated until all topologically connected cells passing the criteria given in \eqMultiRef{eq:toposeed}{and}{eq:topogrowth} and their direct neighbours satisfying the condition in \eqRef{eq:topofilter} are found. 
The algorithm starts by selecting all cells with signal significances \cellsig{} passing the threshold defined by $S$ in \eqRef{eq:toposeed}
from calorimeter regions which are allowed to seed clusters.\footnote{Calorimeter cells marked as having \readout{} or general signal extraction problems in the actual run conditions are not considered as seeds.} These seed cells are then ordered in decreasing \cellsig.

Each seed cell forms a \emph{\prclus}. The cells neighbouring a seed and satisfying \eqRef{eq:topogrowth} or \eqRef{eq:topofilter} are collected into the corresponding \prclus.    
Here \emph{neighbouring} is generally defined as two calorimeter cells being directly adjacent in a given sampling layer, or, 
if in adjacent layers, 
having at least partial overlap in the $(\eta,\phi)$ plane.
 This means that the cell collection for \topos{} can span modules within the same calorimeter as well as calorimeter sub-detector transition regions. 
Should a neigbouring cell have a signal significance passing the threshold defined by the parameter $N$ in \eqRef{eq:topogrowth},
its neighbours are collected into the \prclus{} as well. If a particular neighbour is a seed cell passing the threshold $S$ defined in \eqRef{eq:toposeed}, the two \prcluss{} are merged. If a neighbouring cell is attached to two different \prcluss{} and its signal significance is above the threshold defined by $N$, the two \prcluss{} are merged.  
This procedure is iteratively applied to further neighbours until the last set of neighbouring cells with significances passing the threshold defined by $P$ in \eqRef{eq:topofilter},  but not the one in \eqRef{eq:topogrowth}, is collected.
At this point the formation stops.

\begin{figure}[t!] \centering
\subfloat[First sampling \LArEMBN{1} ($0.2 <|\etacell| < 0.4$)]{\includegraphics[width=0.46\textwidth]{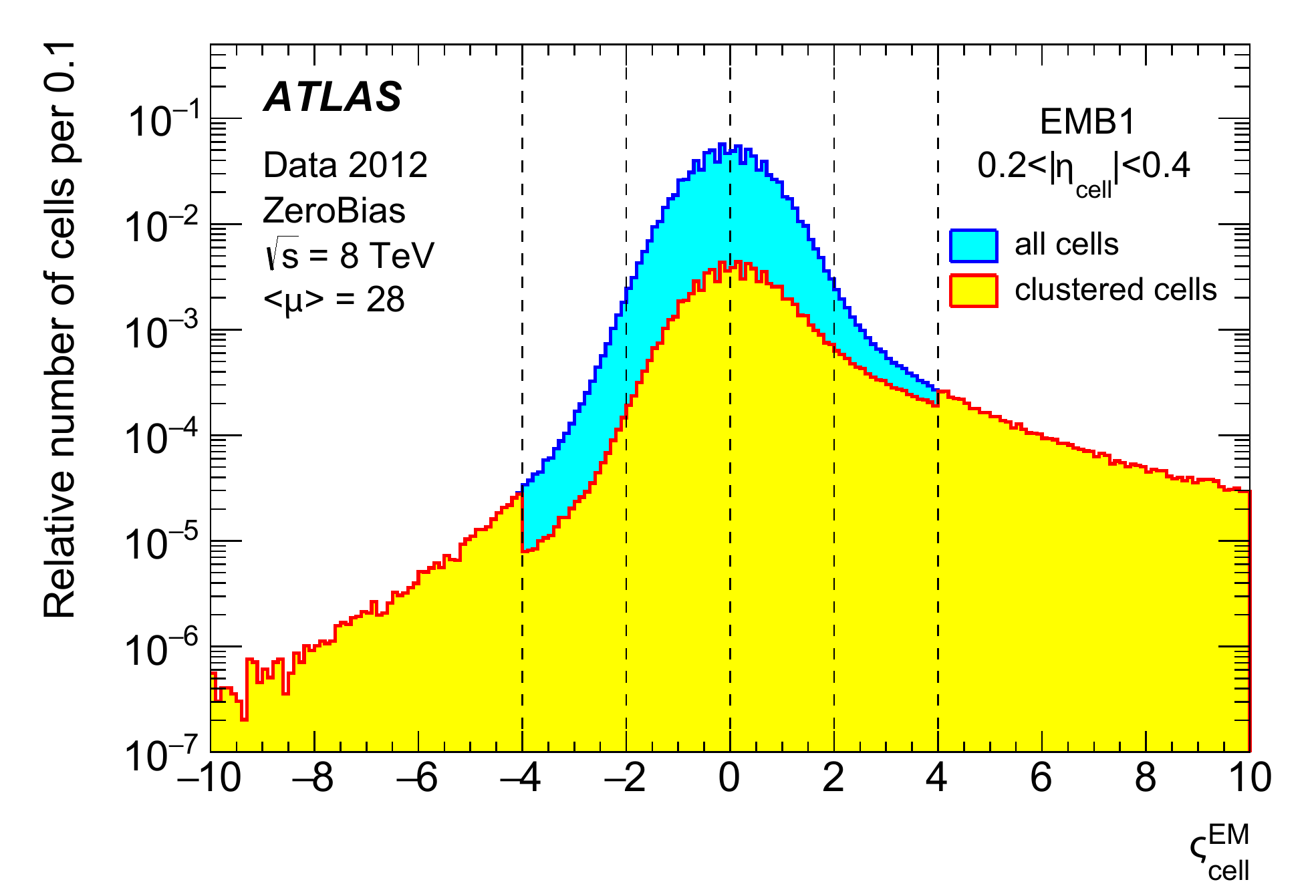}\label{fig:cellsignif:EMB}} 
\subfloat[First sampling \LArEMEN{1} ($1.6 <|\etacell| <1.8$)]{\includegraphics[width=0.46\textwidth]{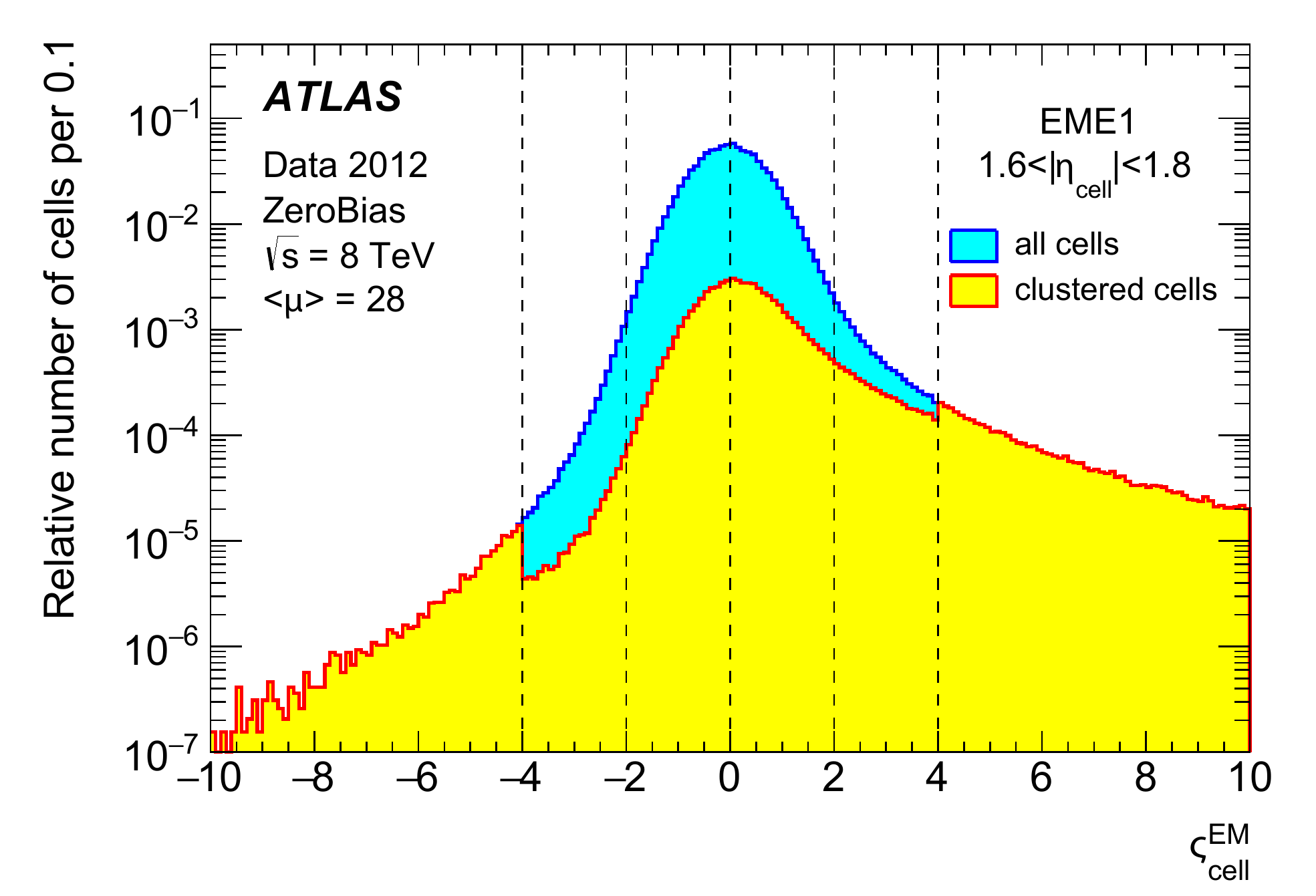}\label{fig:cellsignif:EME}}\qquad
\subfloat[First sampling \LArHEC{0} ($1.6 <|\etacell| < 1.8$)]{\includegraphics[width=0.46\textwidth]{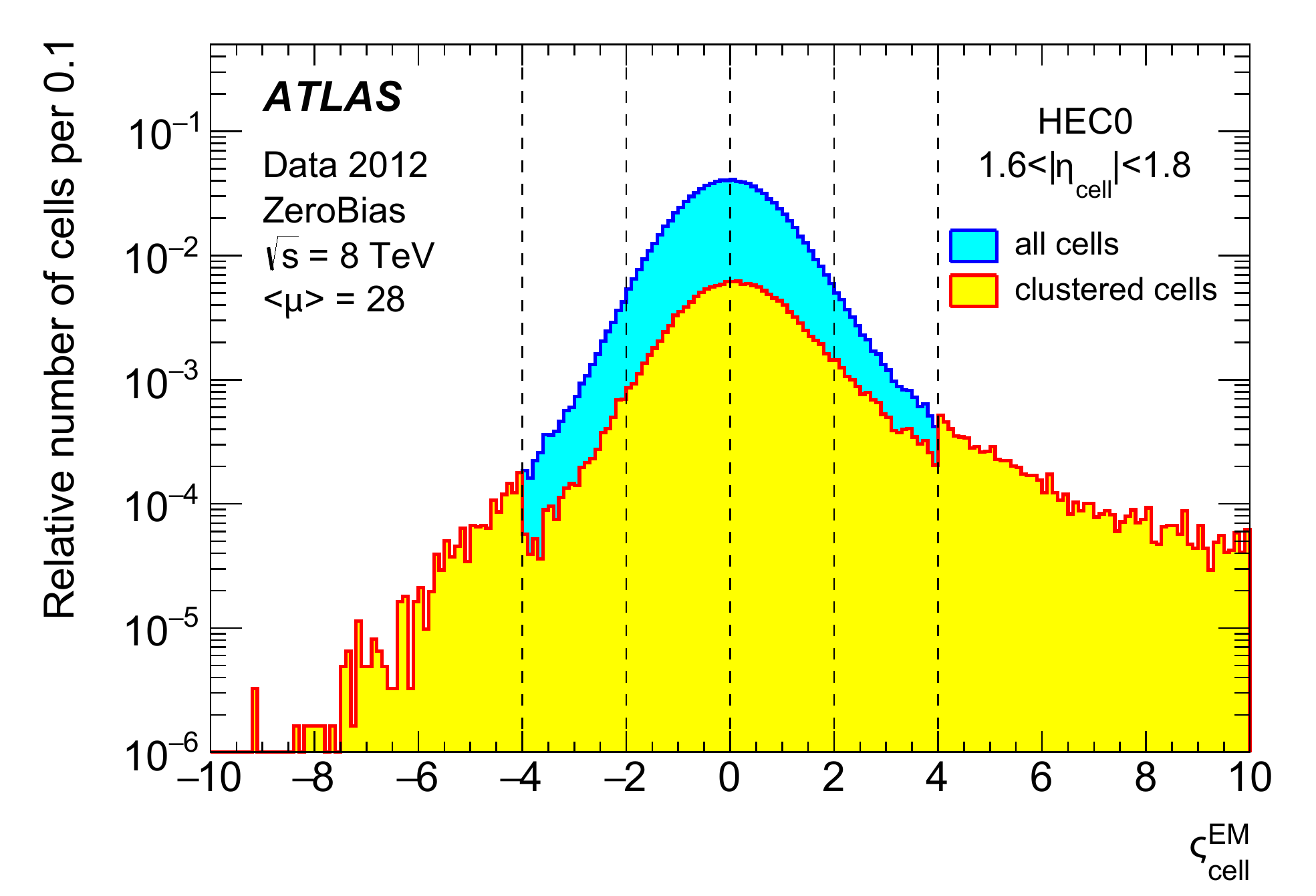}\label{fig:cellsignif:HEC}}
\subfloat[First module \LArFCAL{0} ($3.6 < |\etacell| < 3.8$)]{\includegraphics[width=0.46\textwidth]{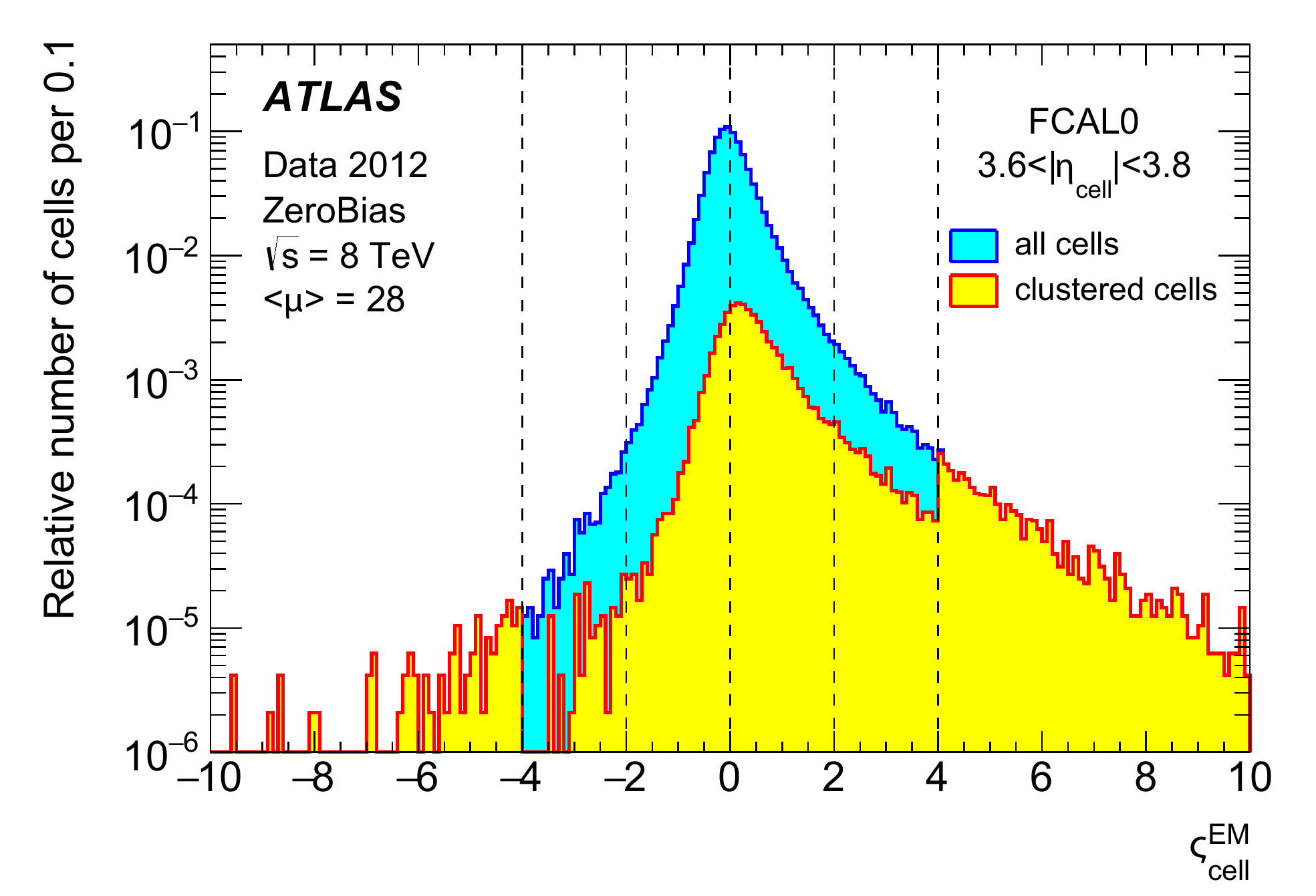}\label{fig:cellsignif:FCAL}}
\caption[]{Signal significance (\cellsig) distributions for all cells (blue/cyan) and for cells after the noise suppression in the topological cell clustering is applied (red/yellow), in selected sampling layers of the \LAr{} calorimeters: \subref{fig:cellsignif:EMB} the first sampling of the central  electromagnetic \LAr{} calorimeter (\LArEMB), \subref{fig:cellsignif:EME} the first sampling of the electromagnetic \LAr{} end-cap calorimeter (\LArEME), \subref{fig:cellsignif:HEC} the first sampling of the hadronic \LAr{} end-cap calorimeter (\LArHEC), and \subref{fig:cellsignif:FCAL} the first module of the \LAr{} forward calorimeter (\LArFCAL). The spectra are extracted from 2012 zero-bias data at \sqrts{8}{} with an average number of \pu{} interactions $\langle\axing\rangle = 28$. The dashed lines indicate $S = \pm 4$, $N = \pm 2$, and $P = 0$.}
\label{fig:cellsignif}
\end{figure}
 
The resulting \prclus{} is characterised by a core of cells with highly significant signals.
This core is surrounded by an envelope of cells with less significant signals.
The configuration optimised for \ATLAS{} hadronic final-state reconstruction 
is $S = 4$, $N = 2$, and $P = 0$, as indicated in \eqMultiRef{eq:toposeed}{to}{eq:topofilter}.
This particular configuration with $P = 0$ means that any cell neighbouring a cell with signal significance passing the threshold given by $N$ in \eqRef{eq:topogrowth} 
is collected into a \prclus{}, independent of its signal. Using the correlations between energies in adjacent cells in this way allows the retention of cells with signals that are close to the noise levels while preserving the noise suppression feature of the clustering algorithm.

The implicit noise suppression implemented by the \topo{} algorithm discussed above leads to significant improvements in various aspects of the calorimeter performance, such as the energy and spatial resolutions in the presence of pile-up. 
Contributions from large negative and positive signal fluctuations introduced by
\pu{} can survive in a given event, though, and thus contribute to the sensitivity to \pu{} observed in e.g.
the jet response \cite{Aad:2015ina}, in addition to the cell-level effects mentioned in \secRef{sec:atlas:data:pu}. 
Examples of the effect of this noise suppression on the cells contributing to zero-bias events recorded with \ATLAS{} in 2012 are shown in the cell signal-significance spectra in \figMultiRefLabel~\ref{fig:cellsignif}\subref{fig:cellsignif:EMB}, \ref{fig:cellsignif}\subref{fig:cellsignif:EME}, \ref{fig:cellsignif}\subref{fig:cellsignif:HEC}, and \ref{fig:cellsignif}\subref{fig:cellsignif:FCAL} for four different \LAr{} calorimeters in \ATLAS.

\subsubsection{Treatment of negative cell signals} \label{\thislabel:formation:negative}

Negative cell signals in the \ATLAS{} calorimeters are the result of 
fluctuations introduced predominantly by \pu{} and, to a lesser extent, by electronic noise, as discussed in \secMultiRef{sec:atlas:data:pu}{and}{sec:atlas:data:noise}.
The thresholds in \eqMultiRef{eq:toposeed}{to}{eq:topofilter} are applied in terms of the absolute value of \cellsig. This means that not only large positive cell signals can seed a cluster, but also those with large negative signals. 
In addition, cells with negative signals can also contribute to the cluster growth control and are added to the envelope around the \topo{} core. 

The use of cells with $\ecellem < 0$ as \topo{} seeds provides a diagnostic tool for the amount of noise  in the overall calorimeter signal for a given event.   At the fixed noise value given in \eqRef{eq:noise} and used in \eqRef{eq:toposeed}, the luminosity-dependent actual noise in the event is reflected in the number of \topos{} reconstructed with negative seeds. This number serves as an estimator mainly for \opu. 

\Topos{} with negative seeds often have a total energy $\eclusem < 0$ as well, especially when $|\cellsig| \gg P$.
This is due to the dominance of the negative seed and the correlation between this seed signal and signals in the neighbouring cells, which likely also have $\ecellem < 0$. 
If a negative seed signal is generated by \opu, it is induced by a particle injected into the calorimeter more than \unit{100}{\ns}{} before the event. Its residual signal trace is scaled by the negative undershoot of the shaping function shown in \figRef{fig:larpulse}.
This particle also injected significant energy in the neighbouring cells at the same time, due to its electromagnetic or hadronic shower, which leads to $\ecellem < 0$ in these cells at the time of the event. For the same reasons, \topos{} from \opu{} seeded by $\ecellem >0$ often yield $\eclusem > 0$, because they are typically generated by particles injected in past bunch crossings closer in time (within \unit{100}{\ns}).
The \topos{} with $\eclusem < 0$ can be used to provide an average global cancellation of contributions of clusters seeded by positive fluctuations in \opu{} in full event observables including \met~\cite{Aad:2012re}. 

Clustering cells with $\ecellem < 0$ in any \topo, including those containing and seeded by large positive signals,  improves noise suppression due to the local cancellation of random positive (upward) noise fluctuations by negative (downward) fluctuations within this cluster. 
Allowing only positive signals to contribute introduces a bias in the cluster signal, while the random cancellation partially suppresses this bias.  

To reconstruct physics objects such as jets from \topos, only those clusters with a net energy $\eclusem > 0$ are considered. The expectation is that clusters with net negative energy have no contribution to the signal of the reconstructed object, as there is no correlation of the corresponding downward fluctuation mainly induced by the energy flow in previous bunch crossings with the final state that is triggered and reconstructed.

\subsubsection{Cluster splitting} \label{\thislabel:formation:split}

The \prcluss{} built as described in \secRef{\thislabel:formation:collect} can be too large to provide a good measurement of the energy flow from the particles generated in the recorded event. 
This is true because spatial signal structures inside those clusters are not explicitly taken into account in the formation. In particular, local signal maxima indicate the presence of two or more particles injecting energy into the calorimeter in close proximity.

To avoid biases in jet-finding and to support detailed jet substructure analysis as well as a high-quality \met{}{}
reconstruction, \prcluss{} with two or more local maxima are split between the corresponding signal peaks in all three spatial dimensions. 
A local signal maximum is defined by $\ecellem > \unit{500}{\MeV}$, in addition to the topological requirements for this cell to have at least four neighbours and that none of the neighbours has a larger signal. 
Also, the location of cells providing local maxima is restricted to cells in the \EM{} sampling layers \LArEMBN{2}, \LArEMBN{3}, \LArEMEN{2}{} and \LArEMEN{3}, and to \LArFCALN{0}.  
This means that for a \prclus{} located completely inside the electromagnetic calorimeters, or extending from the electromagnetic to the hadronic calorimeters, splitting is guided by the spatial cell signal distributions in the highly granular electromagnetic calorimeters. 
The cluster splitting is refined in an additional step, where
signal maxima can be provided by cells from the thin \EM{} sampling layers \LArEMBN{1}{} and \LArEMEN{1} with a highly granular $\eta$-strip \readout{} geometry, all sampling layers in the hadronic calorimeters (\LArHECN{0}{} to \LArHECN{3}, \TileN{0}{} to \TileN{2}), and the hadronic forward calorimeter modules \LArFCALN{1}{} and \LArFCALN{2}.\footnote{Signals in the pre-samplers and gap scintillators are not considered at all in guiding the \topo{} splitting (see \citRef{DetectorPaper} for a detailed description of the \ATLAS{} calorimeters).}
The use of \LArEMBN{1}{} and \LArEMEN{1}{} in the \topo{} splitting improves the photon separation in $\pi^{0}\to\gamma\gamma$. 

\newlength{\figspecwidth}\setlength{\figspecwidth}{0.9\figthreequarterwidth}
\begin{figure}[tb!]\centering
\subfloat[Cells passing selection in \eqRef{eq:toposeed}]{\includegraphics[width=0.9\fighalfwidth]{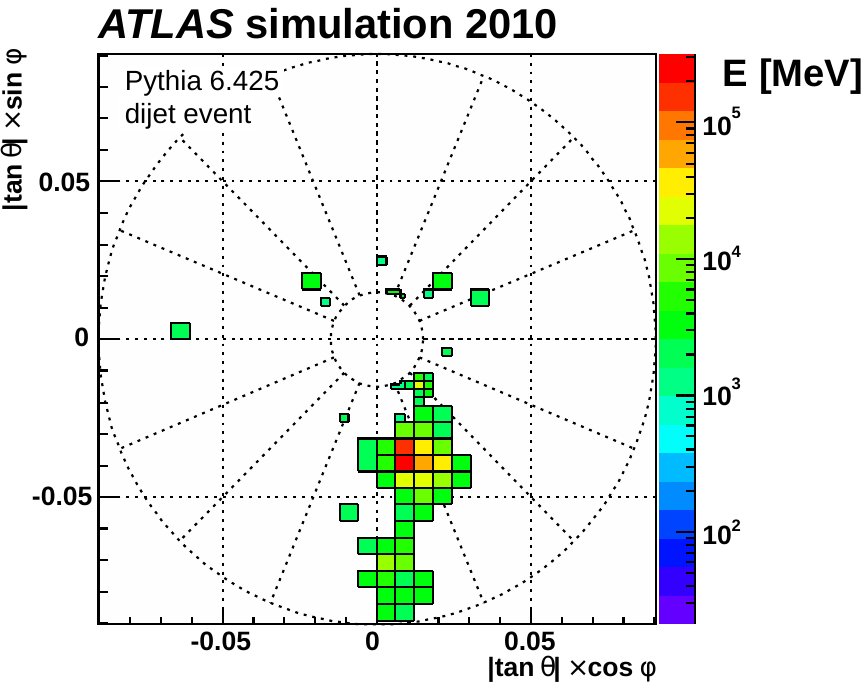}\label{fig:fcaltopos:sig4}}\quad
\subfloat[Cells passing selection in \eqRef{eq:topogrowth}]{\includegraphics[width=0.9\fighalfwidth]{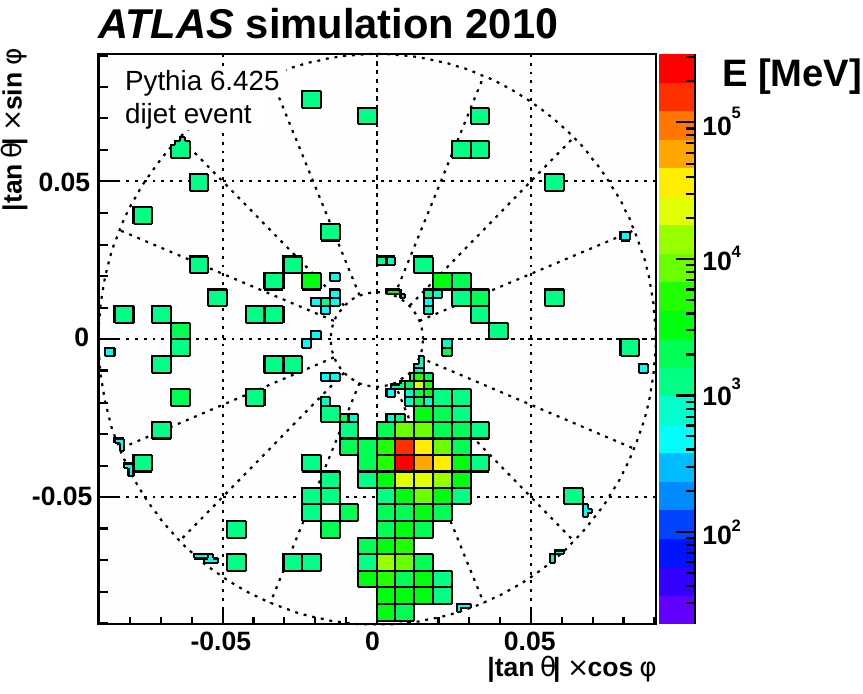}\label{fig:fcaltopos:sig2}}
\qquad
\subfloat[All clustered cells]{\includegraphics[width=\figspecwidth]{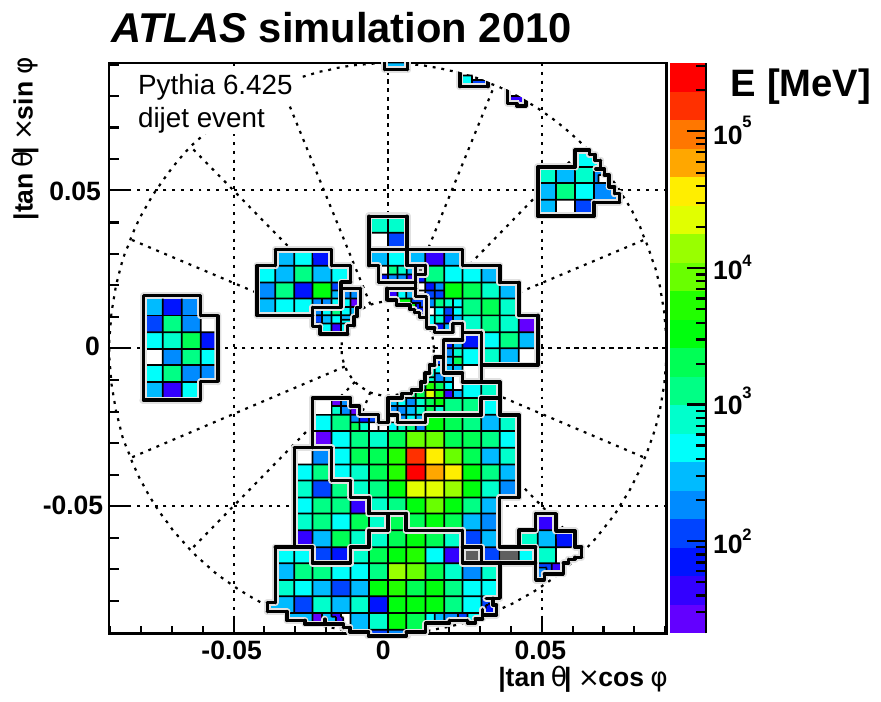}\label{fig:fcaltopos:all}}
\caption[]{Stages of \topo{} formation in the first module (\LArFCALN{0}) of the \LArFCAL{} calorimeter for a simulated dijet event with at least one jet entering this calorimeter. Shown in \subref{fig:fcaltopos:sig4} are cells with signal significance $|\cellsig| > 4$ that can seed \topos, in \subref{fig:fcaltopos:sig2} cells with $|\cellsig| > 2$ controlling the \topo{} growth, and in \subref{fig:fcaltopos:all} all clustered cells and the outline of \topos{} and \topo{} fragments in this module. All clusters shown in \subref{fig:fcaltopos:all} which do not contain a seed cell from this module are seeded in other modules of the \LArFCAL, or in other calorimeters surrounding it. \PU{} is not included in this simulation, but electronic noise is modelled. Cells not colour coded but inside a \topo{} have a negative signal, while cells shaded grey are completely surrounded by clustered cells but not part of a \topo{} themselves.  The cell and cluster boundaries are displayed on a dimensionless grid using  
the polar angle $\theta$ and the azimuthal angle $\phi$. This view maintains the cell shapes and proportions.  
For the definition of the cell signal significance \cellsig{} see \eqRef{eq:significance}.}
\label{fig:fcaltopos}
\end{figure}

The cluster splitting algorithm can find cells which are neighbours to two or more signal maxima. In this case, the cell is assigned to the two highest-energy clusters after splitting of the original \topo{} it is associated with. This means that each cell is only shared once at most, and, even then, is never shared between more than two clusters.   

The sharing of its signal between the two clusters with respective energies \eclusemi{1}{} and \eclusemi{2}{} is expressed in terms of two geometrical weights \wcellgi{1}{} and \wcellgi{2}. 
These weights are calculated from the distances of the cell to the \cog{} of the two clusters  ($d_{1}$, $d_{2}$),  measured in units of a typical electromagnetic shower size scale in the \ATLAS{} calorimeters,\footnote{This scale is motivated by the Moli\`{e}re radius of the electromagnetic shower, which in good approximation is set to \unit{5}{\cm}{} for all calorimeters.} and the cluster energies, 
\begin{eqnarray}
\wcellgi{1} & = & \dfrac{\eclusemi{1}}{\eclusemi{1} + r \eclusemi{2}}\,,    \label{eq:geoweight:w1} \\[6pt]
\wcellgi{2} & = & 1 - \wcellgi{1}\,,                                                            \label{eq:geoweight:w2} \\[6pt]
r & = & \exp(d_{1}-d_{2})\,.                                                                  \label{eq:geoweight:r}
\end{eqnarray}
The geometrical weights reflect the splitting rule that each cell can only appear in two \prcluss{} at most, as $\wcellgi{1} + \wcellgi{2} = 1$. After splitting, the final \prcluss{} are the \topos{} used for further reconstruction of the recorded or simulated final state. 

\FigRef{fig:fcaltopos} shows an example of \topos{} generated by an \MC{} simulated jet in the first module of the \ATLAS{} forward calorimeter under 2010 run conditions (no \pu). Possible seed cells, as defined in \eqRef{eq:toposeed}, are shown in \figRef{fig:fcaltopos}\subref{fig:fcaltopos:sig4}. 
Cells with signal significances above the threshold $N$
specified in \eqRef{eq:topogrowth} are displayed in \figRef{fig:fcaltopos}\subref{fig:fcaltopos:sig2}. The cells from this module included in any \topo{} are shown in \figRef{fig:fcaltopos}\subref{fig:fcaltopos:all}. This display shows the effectiveness of cluster splitting in tracing signal structures. 
Comparing \figMultiRefLabel{} \ref{fig:fcaltopos}\subref{fig:fcaltopos:sig4} and \ref{fig:fcaltopos}\subref{fig:fcaltopos:all} clearly shows the survival of cells with $|\cellsig| < 2$ in the vicinity of more significant signals, even if those are not in the same module (or sampling layer). 

\begin{figure}[t!] \centering
\subfloat[]
{\includegraphics[width=\fighalfwidth]{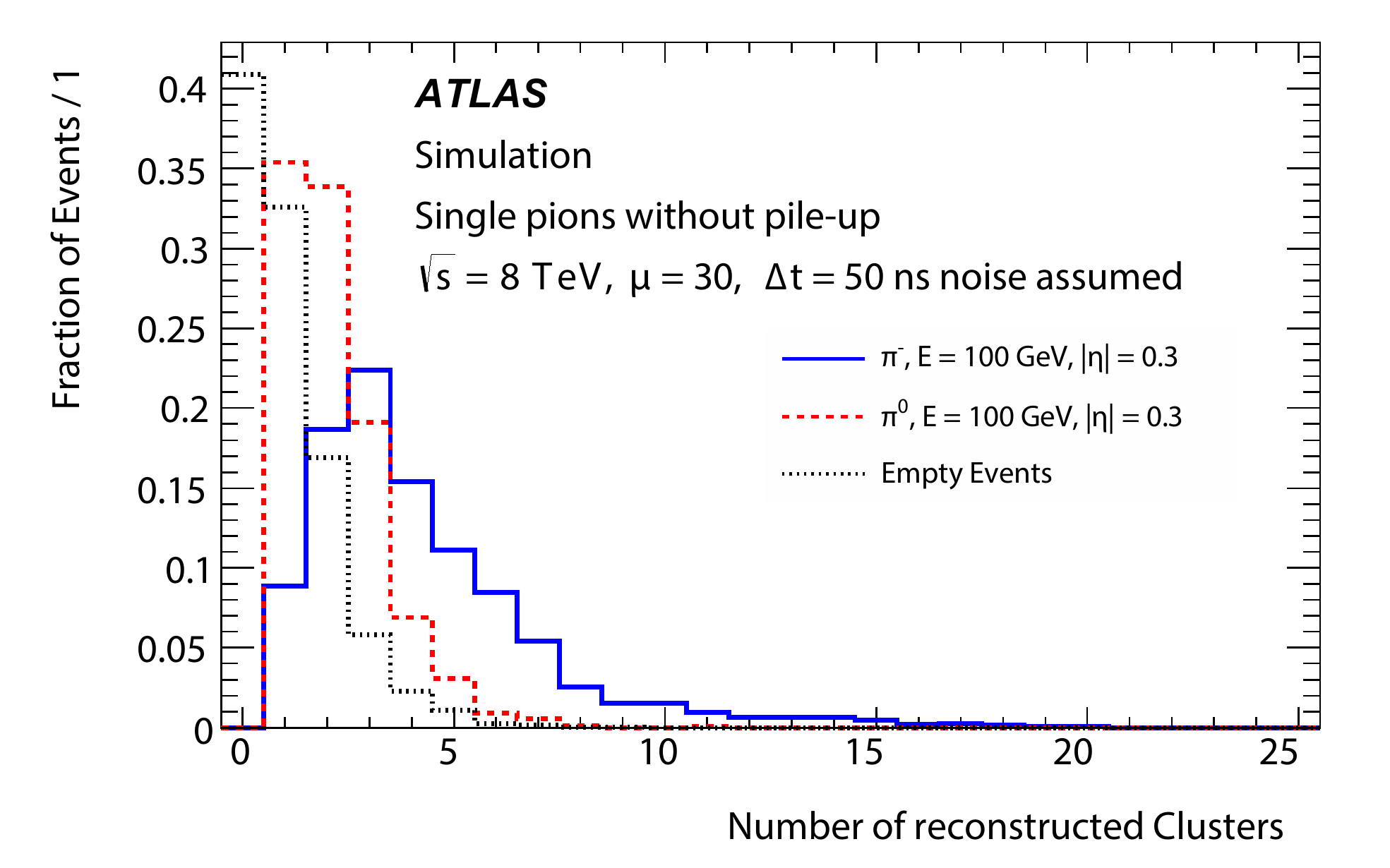}\label{fig:nclusters_barrel}}
\subfloat[]
{\includegraphics[width=\fighalfwidth]{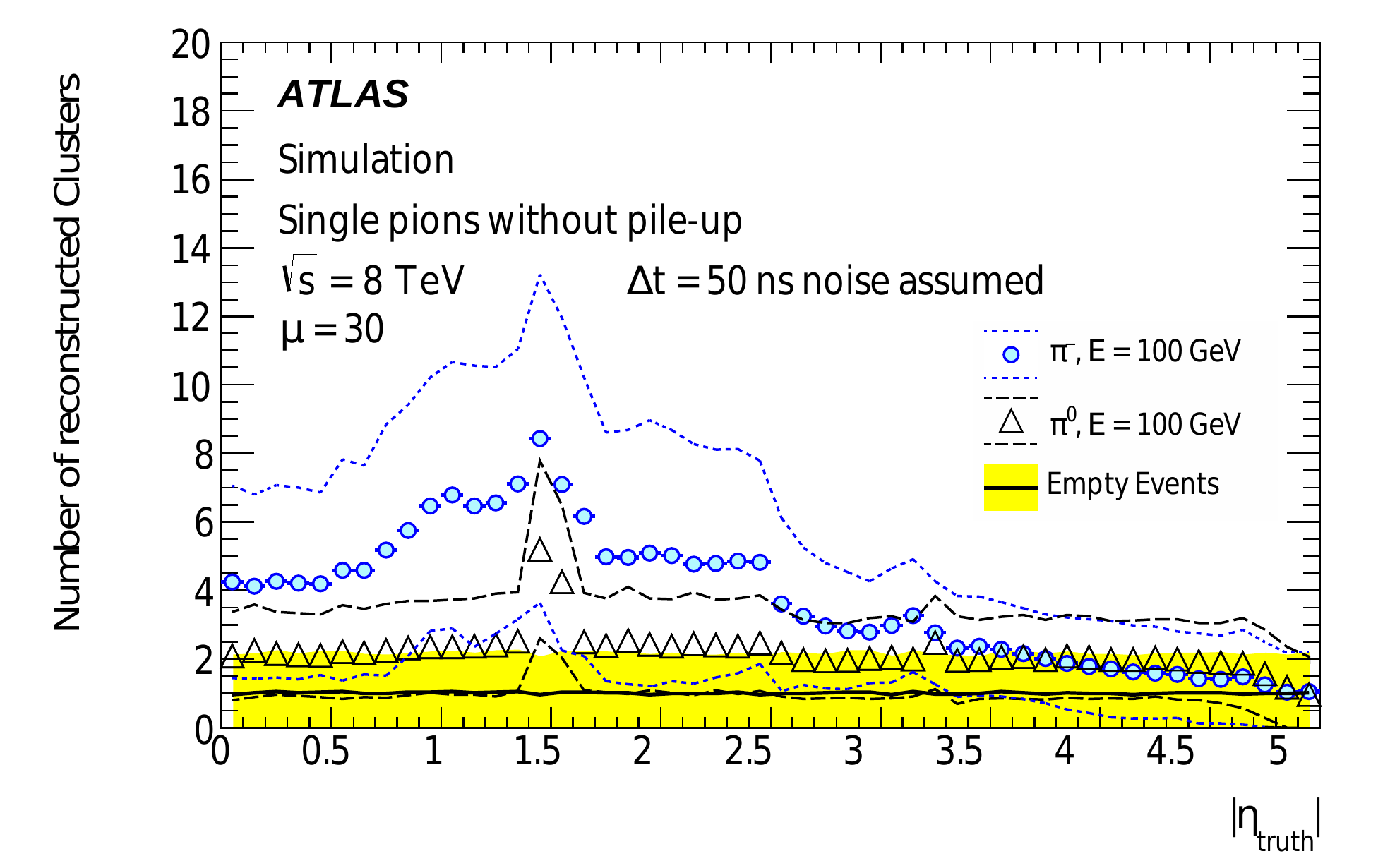}\label{fig:nclusters_vs_eta}}
\caption[]{The number of reconstructed clusters for simulated charged and neutral single pions without actual \pu{} added but with nominal \pu{} noise used in the reconstruction. In \subref{fig:nclusters_barrel} the distribution of the number of clusters \Nclus{} is shown for neutral and charged pions injected into the \ATLAS{} calorimeters at $|\eta| = 0.3$ with an energy of $E = \unit{100}{\GeV}$, together with the \Nclus{} distribution for empty events (\topos{} generated by electronic noise only). The distributions are individually normalised to unity. 
The dependence of the average $\langle\Nclus\rangle$ on the generated $\eta_{\text{gen}}$ is shown in \subref{fig:nclusters_vs_eta}, again for $\pi^{0}$, $\pi^{-}$ and empty events. The shaded area and the dashed lines indicate the spread (in terms of RMS) around the central value.}
\label{fig:nclusters} 
\end{figure}

\subsubsection{Cluster multiplicities in electromagnetic and hadronic showers}\label{\thislabel:formation:multi}

One of the original motivations behind any cell clustering is to reconstruct single-particle showers with the highest possible precision in terms of energy and shape. 
The immediate expectation is that the clustering algorithm should be very efficient in reconstructing one cluster for each particle entering the calorimeter. 
While this view is appropriate for dense and highly compact electromagnetic showers with relatively small shower-to-shower fluctuations in their longitudinal (along the direction of flight of the incoming particle) and lateral (perpendicular to the direction of flight) extensions, hadronic showers are subject to much larger intrinsic fluctuations leading to large shower-to-shower variations in their shapes and compactness. 
Hadrons generated in inelastic interactions in the course of the hadronic shower can even travel significant distances and generate sub-showers outside the direct neighbourhood of the calorimeter cell containing the initial hadronic interaction. 
This means that \topos{} can contain only a fraction of the hadronic shower.
 
The distributions of the \topo{} multiplicity \Nclus{} for single particles which primarily generate electromagnetic showers ($\pi^{0}$) and hadronic showers ($\pi^{-}$) in the central (barrel) calorimeter region are shown in \figRef{fig:nclusters}\subref{fig:nclusters_barrel}. The dependence of the average \Nclus{} on the pseudorapidity $\eta$ is displayed in \figRef{fig:nclusters}\subref{fig:nclusters_vs_eta}. 

Neutral pions with $E_{\pi^{0}} = \unit{100}{\GeV}$ injected into the detector at a fixed direction often generate only one \topo{} from largely overlapping electromagnetic showers, as the angular distance between the two photons from \twoprong{\pi^{0}}{\gamma\gamma}{} is small.
This is demonstrated by the \Nclus{} distribution for \topos{} generated by $\pi^{0}$ at $|\eta| = 0.3$  in \ATLAS{} in \figRef{fig:nclusters}\subref{fig:nclusters_barrel} peaking at $\Nclus = 1$, with a probability only slightly larger than the one for $\Nclus =2$. 
 In the latter case the two \topos{} from the $\pi^{0}$ are generated by (1) resolving the two photon-induced showers, (2) a possible residual imperfect signal collection and \prclus{} splitting in the \topo{} algorithm, or by (3) accidental inclusion of additional  \topo(s) generated by electronic noise.  
While the particular reason for the second cluster depends on effects introduced by local features including the calorimeter \readout{} granularity and cell noise levels at a given direction $\eta$, hypothesis (1) is found to be least likely as it is observed that the energy sharing between the two \topos{} is typically very asymmetric. The leading \topo{} generated by $\pi^{0}$ at \unit{100}{\GeV}{} contains very close to \unit{100}{\%}{} of the total energy in this calorimeter region, indicating that the second and any further \topos{} arise from hypotheses (2) and (3).     

\FigRef{fig:nclusters}\subref{fig:nclusters_vs_eta} shows the average \Nclus{} as a function of the generated particle direction $\eta = \eta_{\text{gen}}$. 
Especially around transition regions at $|\eta| \approx 1.4$ (central to \EndCap{} calorimeters) and $|\eta| \approx 3.2$ (\EndCap{} to forward calorimeters), which both have reduced calorimetric coverage, \Nclus{} can significantly increase due to reduction or loss of the core signal of the showers. 

The number of clusters generated by $\pi^{-}$ with $E = \unit{100}{\GeV}$ injected at $\eta = 0.3$ peaks at $\Nclus = 3$ and has a more significant tail to higher multiplicities, as shown in \figRef{fig:nclusters}\subref{fig:nclusters_barrel}. 
This is expected for hadronic showers, where the distance between two inelastic interactions with significant energy release is of the order of the nuclear interaction length \lamnucl, typically $\mathcal{O}(\unit{10}{\cm})$. This can lead to several well-separated \topos. 
For example, at \unit{100}{\GeV}{} incident energy the leading \topo{} generated by $\pi^{-}$ contains on average \unit{85}{\GeV}, while the next-to-leading \topo{} contains about \unit{10}{\GeV}{} on average. The remaining energy is distributed among one or more low-energy \topos.    

The wider hadronic shower spread introduces a higher sensitivity of \Nclus{} to the calorimeter \readout{} granularities and transition regions, as can be seen in \figRef{fig:nclusters}\subref{fig:nclusters_vs_eta}. 
The transition regions at $|\eta| \approx 0.8\text{--}1.0$, $|\eta| \approx 1.4$ and $|\eta| \approx 3.2$  affect the \topo{} formation more than in the case of electromagnetic showers, not only in terms of the peak \Nclus{} but also in terms of the range in $\eta$. In particular the region around $|\eta| \approx 0.8\text{--}1.0$ has a larger effect on \Nclus{} for hadrons than for electromagnetic interacting particles, as this is the transition from the central to the extended \Tile{} calorimeter introducing reduced calorimetric coverage for hadrons. The central electromagnetic calorimeter provides hermetic coverage here, without any effect on \Nclus. The sharp drop of \Nclus{} for $\pi^{-}$ at $|\eta| = 2.5$ corresponds to the reduction in calorimeter cell granularity by a factor of approximately four. 

\subsection{Cluster kinematics} \label{\thislabel:kinematics}

The cluster kinematics are the result of the recombination of cell energies and directions. The presence of cells with $\ecellem < 0$ requires a special recombination scheme to avoid directional biases.

The cluster directions are calculated as signal-weighted barycentres $(\etaclus,\phiclus)$. Using $\ecellem < 0 $ in this scheme leads to distortion of these directions, even projecting them into the wrong hemispheres. Ignoring the contribution of cells with negative signals, on the other hand, biases the cluster directions with contributions from upward noise fluctuations. To avoid both effects, the cluster directions are calculated with absolute signal weights $|\ecellem|$,
\begin{eqnarray}
\etaclus & = & \dfrac{\sum_{i=1}^{\ncell} \wcellgi{i} \cdot |\ecellemi{i}|\cdot\etacelli{i}}
	                       {\sum_{i=1}^{\ncell} \wcellgi{i} \cdot |\ecellemi{i}|} 
\label{eq:etabaryem} \\[6pt]
\phiclus & = & \dfrac{\sum_{i=1}^{\ncell}  \wcellgi{i} \cdot |\ecellemi{i}|\cdot\phicelli{i}}
	                       {\sum_{i=1}^{\ncell} \wcellgi{i} \cdot |\ecellemi{i}|} . 
\label{eq:phibaryem}
\end{eqnarray} 
Here \ncell{} is the number of cells in the cluster, and \wcellgi{i}{} are the geometrical signal weights introduced by cluster splitting, as given in \eqMultiRef{eq:geoweight:w1}{to}{eq:geoweight:r} in \secRef{\thislabel:formation:split}. The direction of each cell is given by $(\etacell,\phicell)$, calculated from its location with respect to the centre of \ATLAS{} at $(x=0,y=0,z=0)$ in the detector reference frame. The cluster directions are therefore reconstructed with respect to this nominal detector centre.

The total cluster signal amplitude \eclusem{} reflects the correct signal contributions from all cells, 
\begin{equation}
\eclusem = \sum_{i=1}^{\ncell} \wcellgi{i} \ecellemi{i}\,,
\label{eq:eclusem}
\end{equation} 
and is calculated using the signed cell signals \ecellemi{i} and taking into account the geometrical signal weights.
In general, all clusters with $\eclusem > 0$ are used for the reconstruction of physics objects in the \ATLAS{} calorimeters, including the very few ones seeded by cell signals $\ecellem <0$.  

Each \topo{} is interpreted as a massless \emph{\ppart} in physics object reconstruction. The energy and momentum components on the \EM{} scale are calculated from the basic reconstructed kinematic variables $(\eclusem,\etaclus,\phiclus)$ as 
\begin{equation}
\Pclusem = \eclusem \cdot \left(1,\sin\thetaclus\cos\phiclus,\sin\thetaclus\sin\phiclus,\cos\thetaclus\right) 
                             = \left(\eclusem,\pvecclusem\right)
	\label{eq:clusfourmom}
\end{equation}
with terms involving \thetaclus, the polar angle calculated from \etaclus, and \phiclus.
 
The massless \ppart{} interpretation 
is appropriate as there is no physically meaningful 
cluster mass without a specific and valid particle hypothesis for the origin of the signal. Such a hypothesis seems to be impossible to obtain from the calorimeter signals alone, especially for hadrons or hadronically decaying particles, where particle identification often requires a measurement of the charge.
A \topo{} mass could in principle be reconstructed from the cell signals and their spatial distribution, but this observable is dominated by lateral shower spreading, which does not represent a physically meaningful mass. It is also highly affected by the settings for the noise thresholds, which control the lateral and longitudinal spread of the cluster in a given \pu{} environment (see \secRef{\thislabel:formation:collect}).  

In addition, hadronic showers tend to be split more often into two or more \topos, as discussed in \secRef{\thislabel:formation:multi} for single particles.
Also, it is very likely in the \pp{} collision environment at the \LHC{} that a given \topo{} contains signals from several particles, especially when located inside a jet, as a mix of electromagnetic and hadronic showers or shower fragments. These issues make a physical particle hypothesis very unlikely, and any cluster mass measurement would be very hard to interpret or validate in relation to a \lq\lq real\rq\rq{} particle.

\renewcommand{\baselabel}{sec:moments}
\renewcommand{\thislabel}{\baselabel}
\section{\Topo{} moments} \label{sec:moments}

\begin{figure}[ttb!] \centering
\includegraphics[width=\figfullwidth]{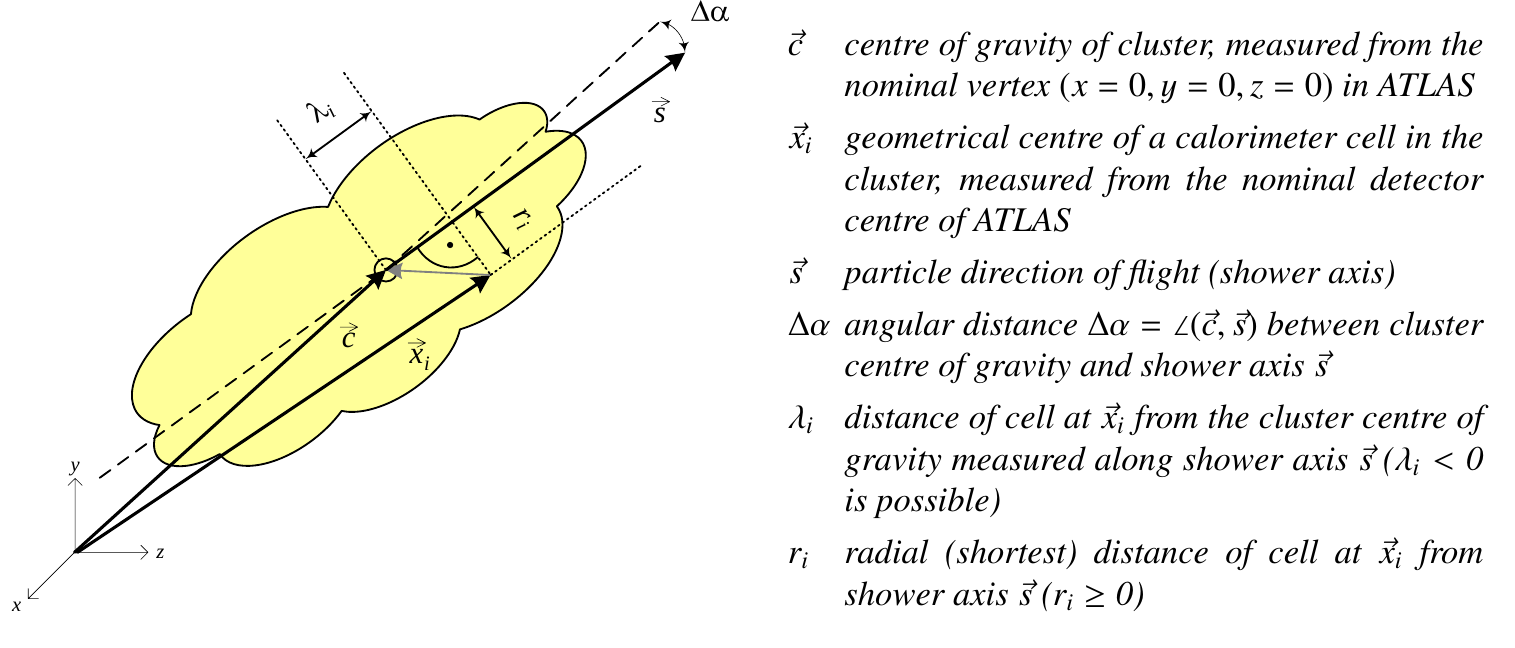}
\caption[]{Schematic view of geometrical moments for \topos.}
\label{fig:moments}
\end{figure}

The shape of a \topo{} and its internal signal distribution contain valuable information for signal characterisation with respect to its origin, and therefore cluster-based calibrations. 
The list of reconstructed observables (\lq\lq cluster moments\rq\rq) is long. 
In this section the focus is on moments used to evaluate the signal quality in data, to determine the cluster location and size, and  
to calibrate each cluster. The geometry relevant to some of the moments is depicted in \figRef{fig:moments}. 
Moments which are useful for purely technical reasons, such as those related to the information about the true energy deposited in the calorimeter in \MC{} simulations, are not discussed in this paper.

Most moments are defined at a given order $n$ for a given calorimeter cell variable \varcell{} as
\begin{equation}
\langle \varcellj{n} \rangle = \dfrac{\sum_{\{i\,|\ecellemi{i} >0\}} \wcellgi{i} \ecellemi{i} \varcellij{i}{n}}
	{\sum_{\{i\,|\ecellemi{i} > 0\}} \wcellgi{i} \ecellemi{i}} \,. 
\label{eq:momcalc}
\end{equation} 
All moments use the \EM{} scale cell signals \ecellem, thus they do not depend on any refined calibration. 
The moment calculation is further restricted to in-time signals, meaning only cells with $\ecellem > 0$ are considered. 
Even though higher-order moments can be reconstructed, only centroids ($n = 1$) and spreads ($n = 2$) are used.


\subsection{Geometrical moments} \label{\thislabel:geometry}

Each \topo{} with at least three cells with $\ecellem > 0$ has a full set of geometrical moments. Simple directional moments (barycentres in $(\eta,\phi)$ space) and locations (\cogs) are available for all clusters.
Not all geometrical moments can be evaluated in a meaningful way for all \topos, mostly due to lack of relevant information in clusters with few cells. In this case, a default value specific to each of these moments is provided.

\subsubsection{Location} \label{\thislabel:geometry:location}

The location of a \topo{} is defined by its \cog{} $\vec{c}$ in three-dimensional space, as shown in \figRef{fig:moments}. This centre is calculated from the first moments of the three Cartesian coordinates specifying the calorimeter cell centres, following the definition given in \eqRef{eq:momcalc}. These locations are provided in the nominal detector frame of reference defined by the interaction point (IP) being located at $(x=0,y=0,z=0)$. 

In addition to the absolute location measured by the \cog, the distance \lamctr{} of the \cog{} from the calorimeter front face, determined along the shower axis (see below and \figRef{fig:moments}), is calculated for each \topo. 

\subsubsection{Directions} \label{\thislabel:geometry:directions}

The direction of a \topo{} is given by $(\etaclus,\phiclus)$, reconstructed as given in \eqMultiRef{eq:etabaryem}{and}{eq:phibaryem}. 
In addition, the first- and second-order directional moments using \etacell{} and \phicell{} are calculated using \eqRef{eq:momcalc} with $n =1$ and $n =2 $, respectively.\footnote{The first directional moment in $\eta$($\phi$) is only identical to \etaclus(\phiclus) for \topos{} without negative signal cells, because negative signal cells are omitted from its calculation while they contribute to the \etaclus(\phiclus) reconstruction.}
%
The reference for these direction measures is the IP discussed above.

The shower axis is a measure of the direction of flight of the incoming particle. It 
is defined by a principal value analysis of the energy-weighted spatial correlations between cells with $\ecellem > 0$ with respect to the cluster centre in Cartesian coordinates, 
\begin{equation}
	\matij{C}{u}{v} = \dfrac{1}{\mathcal{W}} \sum_{\{i\,|\ecellemi{i} > 0\}} \left(\wcellgi{i}\ecellemi{i}\right)^{2} (u_{i} - \langle u \rangle)(v_{i} - \langle v \rangle) \,,
\label{eq:principal_axis} 
\end{equation} 
with all permutations of $u,v \in \{x,y,z\}$. The normalisation $\mathcal{W}$ is given by
\begin{equation}
	\mathcal{W} = \sum_{\{i\,|\ecellemi{i} > 0\}} \left(\wcellgi{i}\ecellemi{i}\right)^{2} \,.
\label{eq:pa_norm}
\end{equation}
The \matij{C}{u}{v}{} fill a symmetric $3 \times 3$ matrix $\mat{C} = \left[\matij{C}{u}{v}\right]$. The eigenvector of \mat{C}{} closest to the direction $\vec{c}$ from the IP to the \cog{} of the \topo{} is taken to be the shower axis $\vec{s}$. If the angular distance $\Delta\alpha$ between $\vec{c}$ and $\vec{s}$ is $\Delta\alpha > \degree{20}$, $\vec{c}$ is used as the shower axis. \FigRef{fig:moments} depicts the geometry of the two axis definitions for \topos.

\subsubsection{Extensions and sizes} \label{\thislabel:geometry:size}

The size of the \topo{} is calculated with respect to the shower axis $\vec{s}$ and the \cog{} $\vec{c}$. For this, cells are first located with reference to $\vec{s}$ and $\vec{c}$. The distances of a cell at $\vec{x}_{i}$ to the shower axis and the \cog{} are then given by
\begin{align}
r_{i}            &= | (\vec{x}_{i} - \vec{c}\,) \times \vec{s}\,| && \text{(radial distance to shower axis)}; \label{eq:radial_mom}\\[6pt]
\lambda_{i} &= (\vec{x}_{i} - \vec{c}\,) \cdot \vec{s}         && \text{(longitudinal distance from shower \cog)}. \label{eq:long_mom}
\end{align}
The first moment $\langle\lambda\rangle$ calculated according to \eqRef{eq:momcalc} with $\varcelli{i} = \lambda_{i}$ and 
$n =1$ is $\langle\lambda\rangle = 0$ by definition. The same equation is used for the first moment $\langle r\rangle$ of $r_{i}$ ($\varcelli{i} = r_{i}$, $n = 1$). The longitudinal and lateral extensions of a \topo{} can then respectively be measured in terms of the second moments $\langle\lambda_{i}^{2}\rangle$ and $\langle r_{i}^{2}\rangle$, again using \eqRef{eq:momcalc}, but with $n = 2$. Specifying cluster dimensions in this fashion describes a spheroid with two semi-axes of respective lengths $\sqrt{\langle\lambda^{2}\rangle}$ and $\sqrt{\langle r^{2}\rangle}$.

As calorimeter technologies and granularities change as function of \etaDet{} in \ATLAS, measures representing the lateral and longitudinal extension of \topos{}
in a more universal and normalised fashion are constructed. These measures are defined in terms of second moments with value ranges from 0 to 1,  
\begin{align}
	\LAT &= \dfrac{\langle r^{2}\rangle_{\text{out}}}{\langle r^{2}\rangle_{\text{out}} + \langle r^{2}\rangle_{\text{core}}}  && \text{normalised lateral energy dispersion (width measure)}; \label{eq:radial2}\\[6pt]
	\LONG &= \dfrac{\langle\lambda^{2}\rangle_{\text{out}}}{\langle\lambda^{2}\rangle_{\text{out}} + \langle\lambda^{2}\rangle_{\text{core}}} && \text{normalised longitudinal energy dispersion (length measure)}\,. 
\label{eq:long2}
\end{align}
The $\langle r^{2}\rangle_{\text{out}}$ term is calculated using \eqRef{eq:momcalc} with $n =2$ and $\varcelli{i} = r_{i}$, but with $r_{i} = 0$ for the two most energetic cells in the cluster. 
The term $\langle r^{2}\rangle_{\text{core}}$ is calculated with the same equation, but now with a fixed $r_{i} = r_{\text{core}}$ for the two most energetic cells, and $r_{i} = 0$ for the rest. 
The calculation of the corresponding terms $\langle\lambda^{2}\rangle_{\text{out}}$ and $\langle\lambda^{2}\rangle_{\text{core}}$ for \LONG{} follows the same respective rules, now with $\varcelli{i} = \lambda_{i}$ in \eqRef{eq:momcalc} and $\lambda_{\text{core}}$ for the most energetic cells in $\langle\lambda^{2}\rangle_{\text{core}}$.\footnote{The constant parameters $\lambda_{\text{core}}$ and $r_{\text{core}}$  are introduced to ensure a finite contribution of the highest-energy cells to \LONG{} and \LAT, respectively, as  those can be very close to the principal shower axes. The specific choices $\lambda_{\text{core}} = \unit{10}{\cm}$ and $r_{\text{core}} = \unit{4}{\cm}$ are motivated by the typical length of electromagnetic showers and the typical lateral cell size in the \ATLAS{} electromagnetic calorimeters.} 
    
The normalised moments \LONG{} and \LAT{} do not directly provide a measure of spatial \topo{} dimensions, rather they measure the energy dispersion in the cells belonging to the \topo{} along the two principal cluster axes. 
Characteristic values are $\LONG\rightarrow 0$ ($\LAT \rightarrow 0$) indicating few highly energetic cells distributed in close proximity along the longitudinal (lateral) cluster extension, and $\LONG \rightarrow 1$ ($\LAT \rightarrow 1$) indicating a longitudinal (lateral) distribution of cells with more similar energies. 
Small values of \LONG{} (\LAT) therefore mean short (narrow) \topos, while larger values are indicative of long (wide) clusters. 

The effective size of the \topo{} in $(\eta,\phi)$ space can in good approximation be  estimated as\footnote{The $\sigma_{\eta}$ and $\sigma_{\phi}$ in this equation represent the energy-weighted root mean square (RMS) of the respective cell directions \etacell{} and \phicell. Correspondingly, the full width at half maximum estimates for the \topo{} are closer to $2.35 \sigma_{\eta}$ and $2.35 \sigma_{\phi}$.}
\begin{align}
	\sigma_{\eta} \simeq \sigma_{\phi} \simeq \atan\left(\dfrac{\sqrt{\langle r^{2}\rangle}}{|\vec{c}\,|}\right) \times \cosh(\eta)\,.
	\label{eq:cluswidth}
\end{align} 
The fact that this approximation holds for both the cluster size in $\eta$ ($\sigma_{\eta}$) and $\phi$ ($\sigma_{\phi}$) is due to the particular granularity of the \ATLAS{} calorimeters. 

\subsection{Signal moments} \label{\thislabel:signal}

\Topo{} moments related to the distribution of the cell signals inside the cluster are useful in determining the density and compactness of the underlying shower, the significance of the cluster signal itself, and the quality of the cluster reconstruction. 
These moments thus not only provide an important input to the calibrations and corrections discussed in \secRef{sec:lcw}, but also support data quality driven selections in the reconstruction of physics objects. 
Additional \topo{} signal quality moments related to instantaneous, short term, and long term detector defects introducing signal efficiency losses are available but very technical in nature, and very specific to the \ATLAS{} calorimeters. 
Their discussion is outside of the scope of this paper. 

\subsubsection{Signal significance} \label{\thislabel:signal:signif}

The significance of the \topo{} signal is an important measure of the relevance of a given cluster contribution to the reconstruction of physics objects. 
Similar to the cell signal significance \cellsig{} given in \eqRef{eq:significance} in  \secRef{sec:topos:formation},  it is measured with respect to the total noise \sigTotEMClus{} in the \topo. 
The definition of \sigTotEMClus{} assumes incoherent noise in the cells contributing to the \topo,\footnote{\OPU{} introduces a coherent component into the calorimeter cell noise due to the correlation of signals in adjacent cells in showers generated by past energy flow. This contribution is reflected on average in the value for \sigTotEMCell, but cannot explicitly be evaluated for any given cell due to its highly stochastic and beam-conditions dependent nature.}     
\begin{align}
\sigTotEMClus = \sqrt{\sum_{i=1}^{N_{\cell}} \left(\sigTotEMCelli{i}\right)^{2}}\,.
\label{eq:cluster_noise}
\end{align}
Here $N_{\cell}$ is the number of cells forming the cluster, including the ones with $\ecellem < 0$. 
As discussed in \secRef{sec:atlas:data:noise}, the individual overall cell noise \sigTotEMCelli{i} is set according to the nominal \pu{} condition for a given data taking period.
The \topo{} signal significance \clussig{} is then measured using \sigTotEMClus{} and \eclusem,  
\begin{align}
	\clussig = \dfrac{\eclusem}{\sigTotEMClus} \,.
\label{eq:clus_signif}
\end{align}
In addition to \clussig, \cellsig{} of the cell with the highest significant signal (the original cluster seed) is available to further evaluate the \topo.
A highly significant seed is a strong indication of an important cluster signal, even if \clussig{} may be reduced by inclusion of a larger number of less significant cell signals. 

\subsubsection{Signal density} \label{\thislabel:signal:density}

The signal density of the \topo{} is indicative of the nature of the underlying particle shower. 
It can be evaluated in two different approaches. First,  
\eclusem{} can be divided by the volume the cluster occupies in the calorimeter. 
This volume is the sum of volumes of all cells contributing to the cluster. 
The signal density reconstructed this way is subject to considerable instabilities introduced by signal fluctuations from noise, as large volume cells can be added with a very small signal due to those fluctuations.  

The default for \topo{} calibration is the second and more stable estimate of the \topo{} signal density measured by the cell-energy-weighted first moment $\rhoclus = \langle\rhocell\rangle$ of the signal densities $\rhocelli{i} = \ecellemi{i}/V_{\cell,i}$ of cells $i = 1 \ldots \ncell$ forming the cluster.
Here $V_{\cell,i}$ is the volume of cell $i$.
The \rhoclus{} variable is calculated using \eqRef{eq:momcalc} with $\varcelli{i} = \rhocelli{i}$ and $n =1$. 
It is much less sensitive to the accidental inclusion of large volume cells with small signals into the cluster, and is used in the context of \topo{} calibration.  
The corresponding second moment is calculated using \eqRef{eq:momcalc} with $n = 2$. It indicates the spread of cell energy densities in the \topo, thus its compactness.

\subsubsection{Signal timing} \label{\thislabel:signal:timing}

The \topo{} signal timing is a sensitive estimator of its signal quality. It is particularly affected by large signal remnants from previous bunch crossings contributing to the cluster, or even exclusively forming it, and can thus be employed as a tag for \topos{} indicating \pu{} activity.   

The reconstructed signal \ecellem{} in all calorimeter cells in \ATLAS{} is derived from the reconstruction of the peak amplitude of the time-sampled analogue signal from the calorimeter shaping amplifiers. 
In the course of this reconstruction the signal peaking time \tcell{} with respect to the \unit{40}{\text{MHz}}{} LHC bunch crossing clock is determined as well. The timing \tclus{} of a \topo{} is then calculated from 
\tcelli{i}{} of the clustered cells $i = 1,\ldots \ncell$ according to
\begin{align}
	\tclus = \dfrac{\sum_{\{i|\cellsigi{i}>2\}} \left(\wcellgi{i}\ecellemi{i}\right)^{2} \tcelli{i}}{\sum_{\{i|\cellsigi{i}>2\}} \left(\wcellgi{i}\ecellemi{i}\right)^{2}}\,,
	\label{eq:clustime} 
\end{align} 
where only cells with a signal significance \cellsigi{i} 
sufficient to reconstruct \ecellemi{i}{} and \tcelli{i}{} are used ($\cellsigi{i} > 2$). The particular weight of the contribution of \tcelli{i}{} to \tclus{} in \eqRef{eq:clustime} is found to optimise the cluster timing resolution \cite{Cojocaru:2004jk}.

\subsubsection{Signal composition} \label{\thislabel:signal:shapes}

The signal distribution inside a \topo{} is measured in terms of the energy sharing between the calorimeters contributing cells to the cluster, and other variables measuring the cell signal sharing. The energy sharing between the electromagnetic and hadronic calorimeters is expressed in terms of the signal ratio $f_{\text{emc}}$, and can be used as one of the characteristic observables indicating an underlying electromagnetic shower. The signal 
fraction $f_{\text{max}}$ carried by the most energetic cell in the cluster is a measure of its compactness.
The signal fraction $f_{\text{core}}$ of the summed signals from the highest energetic cell in each longitudinal calorimeter sampling layer contributing to the \topo{} can be considered as a measure of its \emph{core signal strength}. It is
sensitive not only to the shower nature but also to specific features of individual hadronic showers.  
These fractions are calculated for each \topo{} with $\eclusem > 0$ as follows (\EMC{} denotes the electromagnetic calorimeters\footnote{For the purpose of this calculation, the \EMC{} consists of sampling layers \LArEMBN{1} to \LArEMBN{3}, \LArEMEN{1} to \LArEMEN{3}, and \LArFCALN{0}.} in \ATLAS),
\begin{align}
	f_{\text{emc}} &= \dfrac{1}{\eclusemi{\text{pos}}}\sum_{\{i\,\in\,\text{\EMC};\ecellemi{i} > 0\}} \wcellgi{i}\ecellemi{i} && (\text{\EMC{} signal fraction in cluster}); \label{eq:f_emc} \\[6pt]
	f_{\text{max}} &= \dfrac{1}{\eclusemi{\text{pos}}}\max\left\{\wcellgi{i}\ecellemi{i}\right\}                    && (\text{most energetic cell signal fraction in cluster}); \label{eq:f_max} \\[6pt]
	f_{\text{core}} &= \dfrac{1}{\eclusemi{\text{pos}}}\sum_{s\,\in\,\{\text{samplings}\}} \max_{i\,\in\,s}\left\{\wcellgi{i}\ecellemi{i}\right\} && (\text{core signal fraction in cluster}). \label{eq:f_core}
\end{align}
The index $s$ steps through the set of calorimeter sampling layers with cells contributing to the \topo. 
Only cells with $\ecellem > 0$ are used in the calculation of these fractions.  
Correspondingly, they are normalised to \eclusemi{\text{pos}}{} given by
\begin{align}
	\eclusemi{\text{pos}} = \sum_{\{i|\ecellemi{i}>0\}} \wcellgi{i}\ecellemi{i}\,.
\label{eq:eclus_pos_norm}
\end{align}
All these moments have a value range of $[0,1]$. 

One of the variables that can be considered for further evaluation of the relevance of the cluster signal in the presence of \pu{} is
the ratio of \eclusemi{\text{pos}}{} to \eclusem. It is sensitive to the negative energy content of a given \topo{} which is largely injected by \opu{} dominated by the negative tail of the bipolar signal shaping function discussed in \secRef{sec:topos:formation:negative}.

\subsubsection{Topological isolation} \label{\thislabel:signal:isolation}

The implicit noise suppression in the topological clustering algorithms leads to signal losses affecting the calorimeter response to particles, as further discussed in \secRef{sec:lcw:ooc}. As these signal losses appear at the boundary of the \topo, corresponding corrections need to be sensitive to whether the lost signals may be included in another close-by cluster or if they are lost for good. This is particularly important for jets, where the \topo{} density can be very high.   

The degree of isolation is measured by the isolation moment $f_{\text{iso}}$, with $0 \leq f_{\text{iso}} \leq 1$. 
A \topo{} with $f_{\text{iso}} = 1$ is completely isolated, while a cluster with $f_{\text{iso}} = 0$ is completely surrounded by others.
The isolation measures the sampling layer energy (\scriptij{E}{s}{\text{\EM}})-weighted 
fraction of non-clustered neighbour cells on the outer perimeter of the \topo. 
Here \scriptij{E}{s}{\text{\EM}} is defined as the sum of the energies \ecellem{} of all cells in a \topo{} located in a given sampling layer $s$ of the calorimeter. 

The isolation moment is reconstructed by first counting the number of calorimeter cells $N_{\text{cell},s}^{\text{noclus}}$ in sampling layer $s$ neighbouring a \topo{} but not collected into one themselves.
Second, the ratio $N_{\text{cell},s}^{\text{noclus}}/N_{\text{cell},s}^{\text{neighbour}}$ of this number 
to the number of all neighbouring cells ($N_{\text{cell},s}^{\text{neighbour}}$) for each $s$ contributing to the cluster is calculated. 
The per-cluster \scriptij{E}{s}{\text{\EM}}-weighted average of these ratios from all included $s$ is the isolation moment $f_{\text{iso}}$,    
\begin{align}
	f_{\text{iso}} = 
		\dfrac{\sum_{s \in \{\text{samplings with\ }\scriptij{E}{s}{\text{\EM}}>0\}} \scriptij{E}{s}{\text{\EM}} N_{\text{cell},s}^{\text{noclus}}/N_{\text{cell},s}^{\text{neighbour}}}{\sum_{s \in \{\text{samplings with\ }\scriptij{E}{s}{\text{\EM}}>0\}}  \scriptij{E}{s}{\text{\EM}}}\,.
\label{eq:iso_mom}
\end{align}

\renewcommand{\baselabel}{sec:lcw}
\renewcommand{\thislabel}{\baselabel}
\section{Local hadronic calibration and signal corrections} \label{sec:lcw}

The motivation for the calibration scheme described in this section arises from the intention to provide a calorimeter signal for physics object reconstruction in \ATLAS{} which is calibrated outside any particular assumption about the kind of object. 
This is of particular importance for final-state objects with a significant hadronic signal content, such as jets and, to a lesser degree, $\tau$-leptons. 
In addition to these discrete objects, the precise reconstruction of the missing transverse momentum requires well-calibrated hadronic signals even outside hard final-state objects, to e.g. avoid deterioration of the \met{} resolution due to highly fluctuating (fake) \pT-imbalances introduced by the non-linear hadronic response on the \EM{} scale.

The \topo{} moments provide information sensitive to the nature of the shower generating the cluster signal. 
This information can be explored to apply moment-dependent calibrations cluster-by-cluster, and thus correct for the effects of the non-compensating calorimeter response to hadrons, accidental signal losses due to the clustering strategy, and energy lost in inactive material in the vicinity of the \topo.
The calibration strategy discussed in some detail in the following is \emph{local} because it attempts to calibrate highly localised and relatively small (in transverse momentum flow space) \topos.\footnote{As cells and clusters are localised in the calorimeters, the preferred variables for this space are the azimuth $\phi$ and the pseudorapidity $\eta$, rather than the rapidity $y$. As \topos{} are reconstructed as massless \pparts{} (see \secRef{sec:topos:kinematics}), $y = \etaclus$ for the complete object.}
As the local hadronic calibration includes cell signal weighting, the calibration based on \topos{} is referred to as
\lq\lq local hadronic cell weighting\rq\rq{} (\LCW) calibration.

All calibrations and corrections are derived using \MC{} simulations of 
single pions (charged and neutral) at various energies in all \ATLAS{} calorimeter regions. This fully simulation-based approach requires good agreement between data and these \MC{} simulations for the \topo{} signals and moments used for any of the applied corrections in terms of distribution shapes and averages. 
Reconstructed observables which are not well-modelled by simulation are not considered. The \datatomc{} comparisons for most used observables are shown in the context of the discussion of the methods using them.

\subsection{General \topo{} calibration strategy} \label{\thislabel:overview}

The \LCW{} calibration aims at the cluster-by-cluster reconstruction of the calorimeter signal on the appropriate (electromagnetic or hadronic) energy scale. In this, 
the cluster energy resolution is expected to improve by using other information in addition to the cluster signal in the calibration. 
The basic calorimeter signal inefficiencies that this calibration must address are given below.
\begin{description}
\item[\textbf{Non-compensating calorimeter response:}]  All calorimeters employed in \ATLAS{} are non-compensating, meaning their signal for hadrons
is smaller than the one for electrons and photons depositing the same energy ($e/\pi >1$).
Applying corrections to the signal locally so that $e/\pi$ approaches unity on average  improves the linearity of the response as well as the resolution for
jets built from a mix of electromagnetic and hadronic signals. It also improves the reconstruction of full event observables 
such as \met,
which combines signals from the whole calorimeter system and requires balanced electromagnetic and hadronic responses in and outside signals from (hard) particles and jets.  
\item[\textbf{Signal losses due to clustering:}] The \topo{} formation applies an intrinsic noise suppression, as discussed in detail in \secRef{sec:topos:formation}. Depending on the \pu{} conditions and the corresponding noise thresholds, a significant amount of true signal can be lost this way, in particular at the margins of the \topo. This requires corrections to allow for a more uniform and linear calorimeter response.
\item[\textbf{Signal losses due to energy lost in inactive material:}] This correction is needed to address the limitations in the signal acceptance in active calorimeter regions due to energy losses in nearby inactive material in front, between, and inside the calorimeter modules.
\end{description}

\begin{figure}[tb!]\centering
\includegraphics[width=0.98\textwidth]{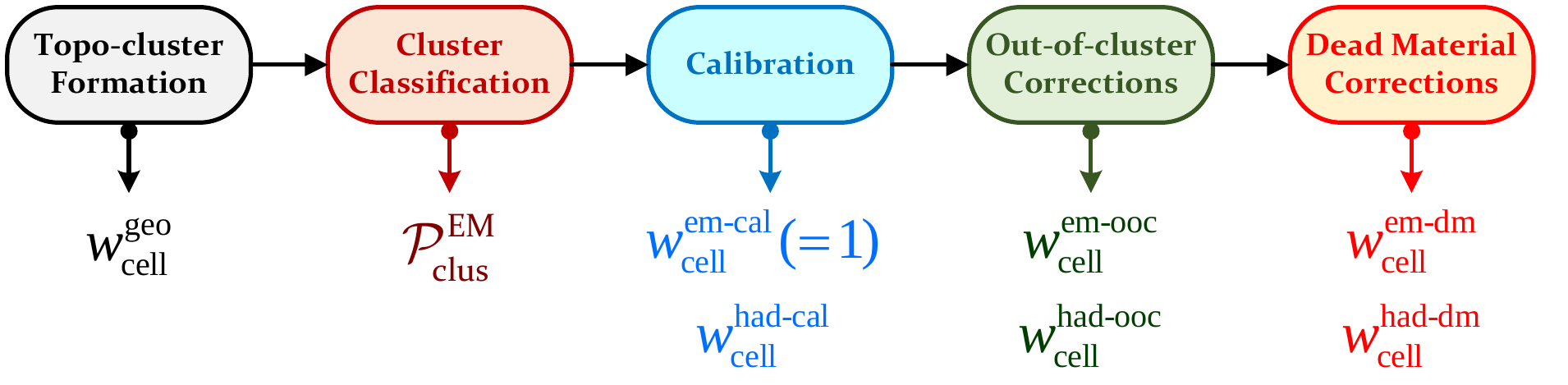}
\caption[]{Overview of the local hadronic cell-weighting (\LCW) calibration scheme for \topos. Following the \topo{} formation, the likelihood for a cluster to be generated by electromagnetic energy deposit (\EMLike) is calculated. After this, the sequence of calibration and corrections indicated in the schematics is executed, each yielding cell signal weights for the two possible interpretations of the cluster signals. These weights are indicated in the figure. They are then used together with \EMLike{} to calculate the \topo{} energy and barycentre from the contributing calorimeter cells, as described in the text.}
\label{fig:calib_scheme}
\end{figure} 

The corrections collected in the \LCW{} calibration address these three main sources of signal inefficiency. 
The specifics of the calibrations and corrections applied to correct for these signal inefficiencies depend on the nature of the energy deposit -- hadronic (\HAD) or electromagnetic (\EM). 
Therefore, the first step of the \topo{} calibration procedure is to determine the probability $0 \leq \EMLike\leq 1$ that a given \topo{} is generated by an electromagnetic shower. This approach provides straightforward dynamic scales (cluster-by-cluster) for the application of specific electromagnetic (\EMLike) and hadronic ($1-\EMLike$) calibrations and corrections. 
For \topos{} with $\EMLike = 1$, it suppresses the application of a hadronic calibration mostly addressing the non-compensating response to hadrons, and applies the electromagnetic-signal-specific corrections for the losses introduced by clustering and inactive material mentioned above. Reversely, very hadronic \topos{} with $\EMLike = 0$ receive the appropriate hadronic calibration and hadronic-signal-specific signal loss corrections.   

The main differences in the hadronic and electromagnetic calibration of \topos{} are  the magnitudes of the applied corrections, which in the \EM{} case
are significantly smaller than for \HAD. Applying an exclusive categorisation based on the probability distributions described in \secRef{\thislabel:classification} can lead to inconsistent calibrations especially for low-energy or small (few cells only) clusters, as misclassification for these kinds of \topos{} is more likely than for clusters with higher energies or larger sizes. 
To allow for smooth transitions and reduce the dependency on the classification, the signal weights \wcellc{} applied to cell signals in the \topo{} at any  of the calibration and correction steps are calculated as
\begin{equation}
	\wcellc = \EMLike\cdot\wcellcem + ( 1 - \EMLike )\cdot\wcellchad \,. 
	\label{eq:effweight}
\end{equation}
The weights \wcellcem{} and \wcellchad{} represent the factors applied by the \EM{} or \HAD{} calibration to the cell signal. The effective representation of all calibration steps in terms of these cell-level signal weights implements a consistent approach independent of the nature of the actual correction applied at any given step. As detailed in \secMultiRef{\thislabel:hadcal}{to}{\thislabel:dmc}, the weights can depend on the cell signal itself, thus yielding a different weight for each cell. 
They can also represent cluster-level corrections generating the same weight for all cells, or a subset of cells, of the \topo. 
This cell weighting scheme therefore provides not only the corrected overall cluster energy after each calibration step by weighted cell signal re-summation, but also the corresponding (possibly modified) cluster barycentre. Thus the cumulative effect on the \topo{} energy and direction can be validated after each step.
The steps of the general \LCW{} calibration are schematically summarised in \figRef{fig:calib_scheme}, and the individual steps are described in more detail below.

The \EM{} calibrations and corrections and their respective parameters are determined with single-particle \MC{} simulations of neutral pions for a large set of energies distributed uniformly in terms of $\log(E)$ between \unit{200}{\MeV} and \unit{2}{\TeV}, at various directions \etaDet. 
The same energy and \etaDet{} phase space is used for the corresponding simulations of charged pions to determine the \HAD{} calibrations and corrections.   
The signals in these simulations are reconstructed with thresholds corresponding to the nominal \sigTotEM{} for a given run period, which reflects the \pu{} conditions according to \eqRef{eq:noise} in \secRef{sec:atlas:data:noise}. 
Only electronic noise is added into the signal formation in the \MC{} simulation, so that the derived calibrations and corrections effectively correct for signal losses introduced by the clustering itself. 
In particular, additional signal from \pu{} and modifications of the true signal by \opu{}  are not considered, as these are expected to cancel on average.

\subsection{Cluster classification} \label{\thislabel:classification}

\begin{figure}[t!] \centering
\includegraphics[width=\figthreequarterwidth]{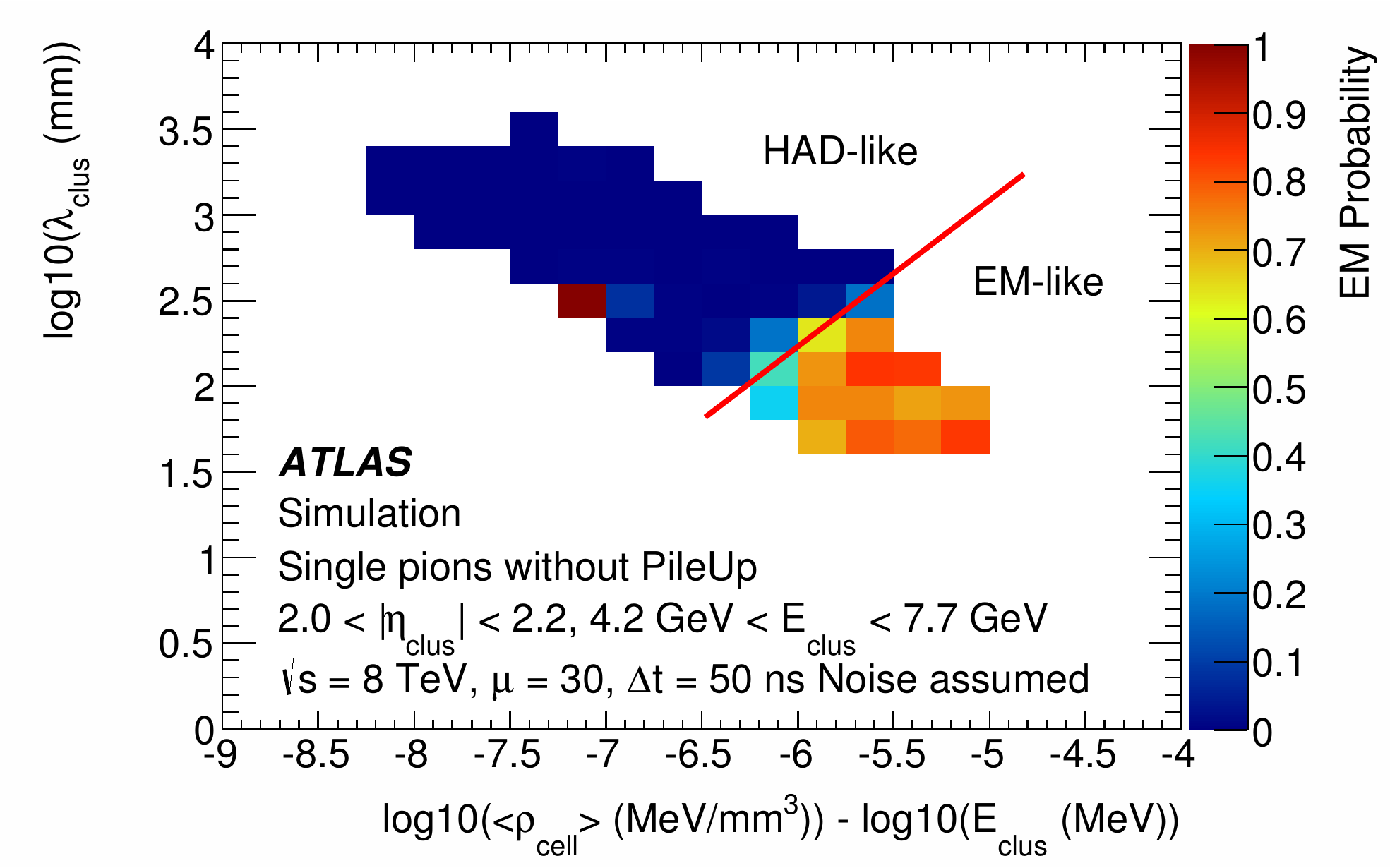}
\caption[]{Distribution of the likelihood $\EMLike(\rhoclus/\eclusem,\lamctr)$ for reconstructed \topos{} to originate from an electromagnetic shower as a function of the shower depth \lamctr{} and the normalised cluster signal density $\rhoclus/\eclusem$, with $\rhoclus = \AVE{\rhocell}$ being the energy-weighted average of \rhocell. The shown distribution is determined as described in the text, in a selected bin of the cluster energy \eclusem{} and the cluster direction \etaclus. The red line indicates the boundary of the $\EMLike > \unit{50}{\%}$ selection, below which the \topo{} is classified as mostly electromagnetic (\lq\lq\EM-like\rq\rq) and above  which it is classified as mostly hadronic (\lq\lq HAD-like\rq\rq). 
The small \EM-like area at the edge of the \HAD-like region stems from neutral pions showering late. 
These areas are typical in regions of the detector where the second layer of the \EM{} calorimeter is thinner and substantial parts of the shower are deposited in its last layer. 
The larger volume of the cells in this last layer leads to the reduced energy density while the position at the back of the EM calorimeter means a larger \lamctr.}\label{fig:emlike}
\end{figure}

As discussed in \secRef{sec:moments}, most \topos{} provide geometrical and signal moments 
sensitive to the nature of the shower producing the cluster signal.
In particular, electromagnetic showers with their compact shower development, early starting point and relatively small intrinsic fluctuations can generate cluster characteristics very different from those generated by hadronic showers. 
The latter are in general subjected to larger shower-by-shower fluctuations in their development and can be located deeper into the calorimeter. 
In addition, the hadronic showers show larger variations of their starting point in the calorimeter. 
A classification of each \topo{} according to its likely origin 
determines the most appropriate mix of \EM{} and \HAD{} calibration and correction functions to be applied.

\begin{figure}[t!] \centering
	\subfloat[\lamctr{} distribution for \EM{} clusters]{\includegraphics[width=\fighalfwidth]{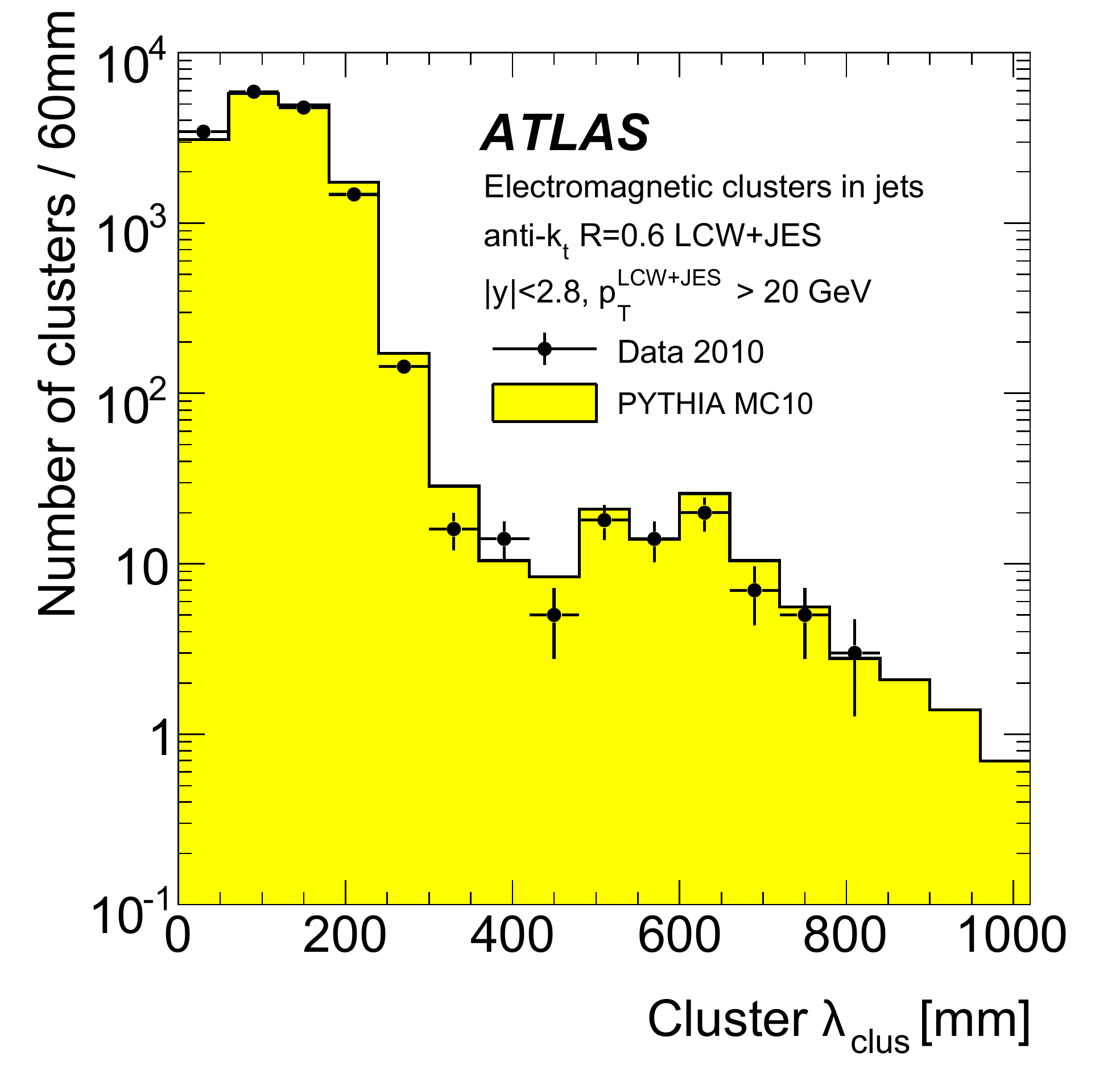}\label{fig:lambda:distem}}
	\subfloat[\lamctr{} distribution for \HAD{} clusters]{\includegraphics[width=\fighalfwidth]{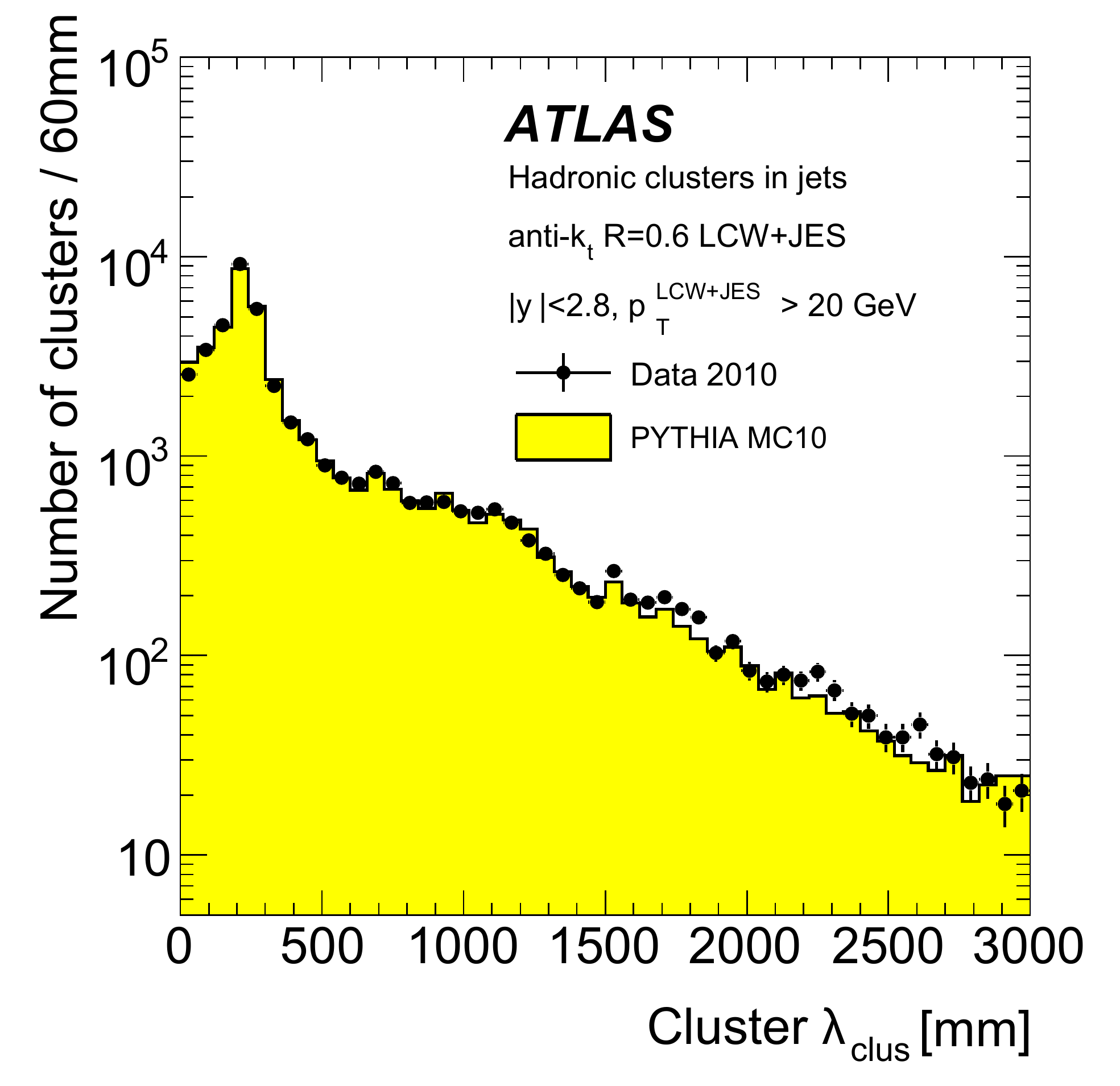}\label{fig:lambda:disthad}} \\
	\subfloat[$\langle\lamctr\rangle(\eclusem)$ for \EM{} clusters]{\includegraphics[width=\fighalfwidth]{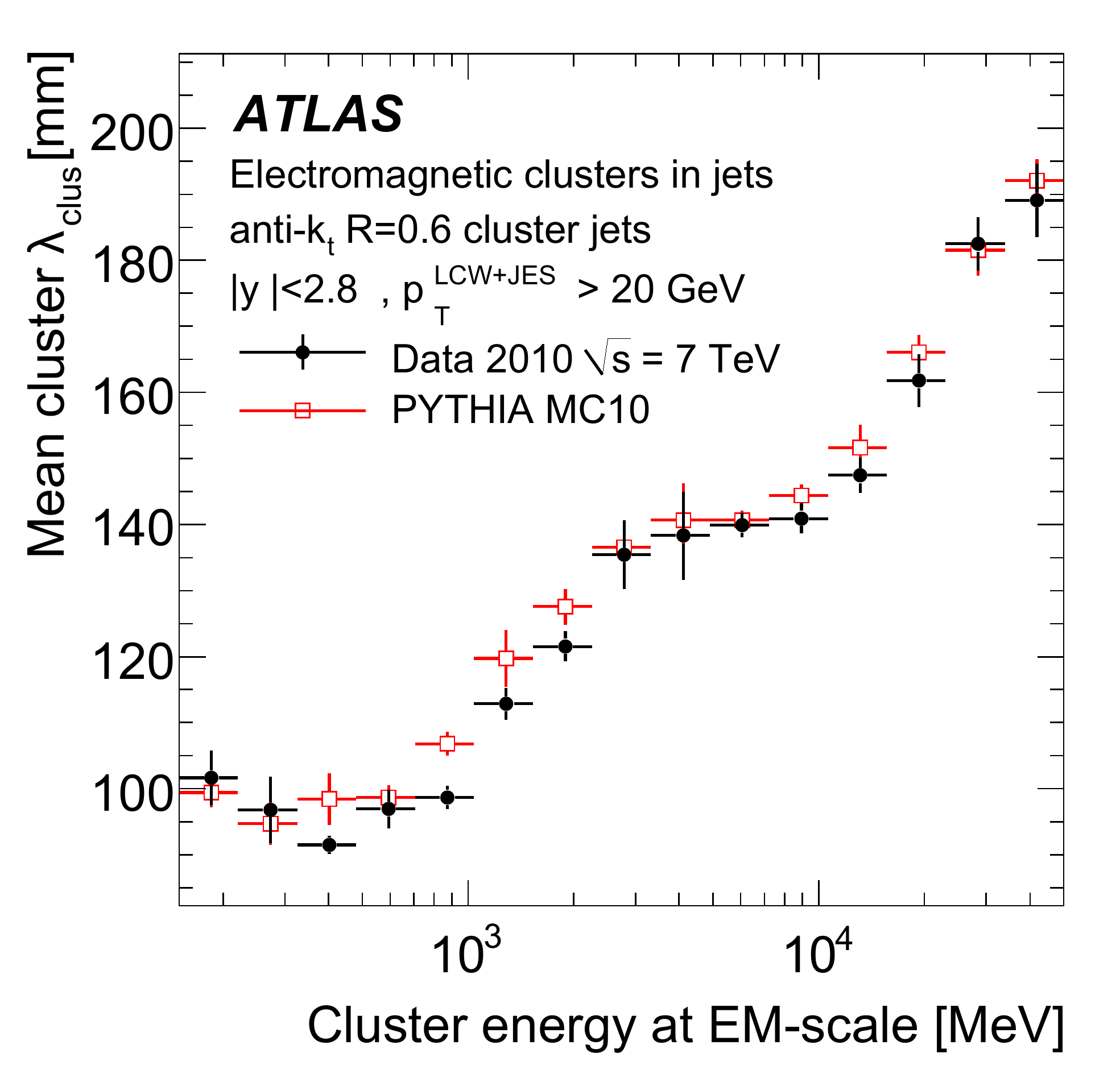}\label{fig:lambda:aveem}}
	\subfloat[$\langle\lamctr\rangle(\eclusem)$ for \HAD{} clusters]{\includegraphics[width=\fighalfwidth]{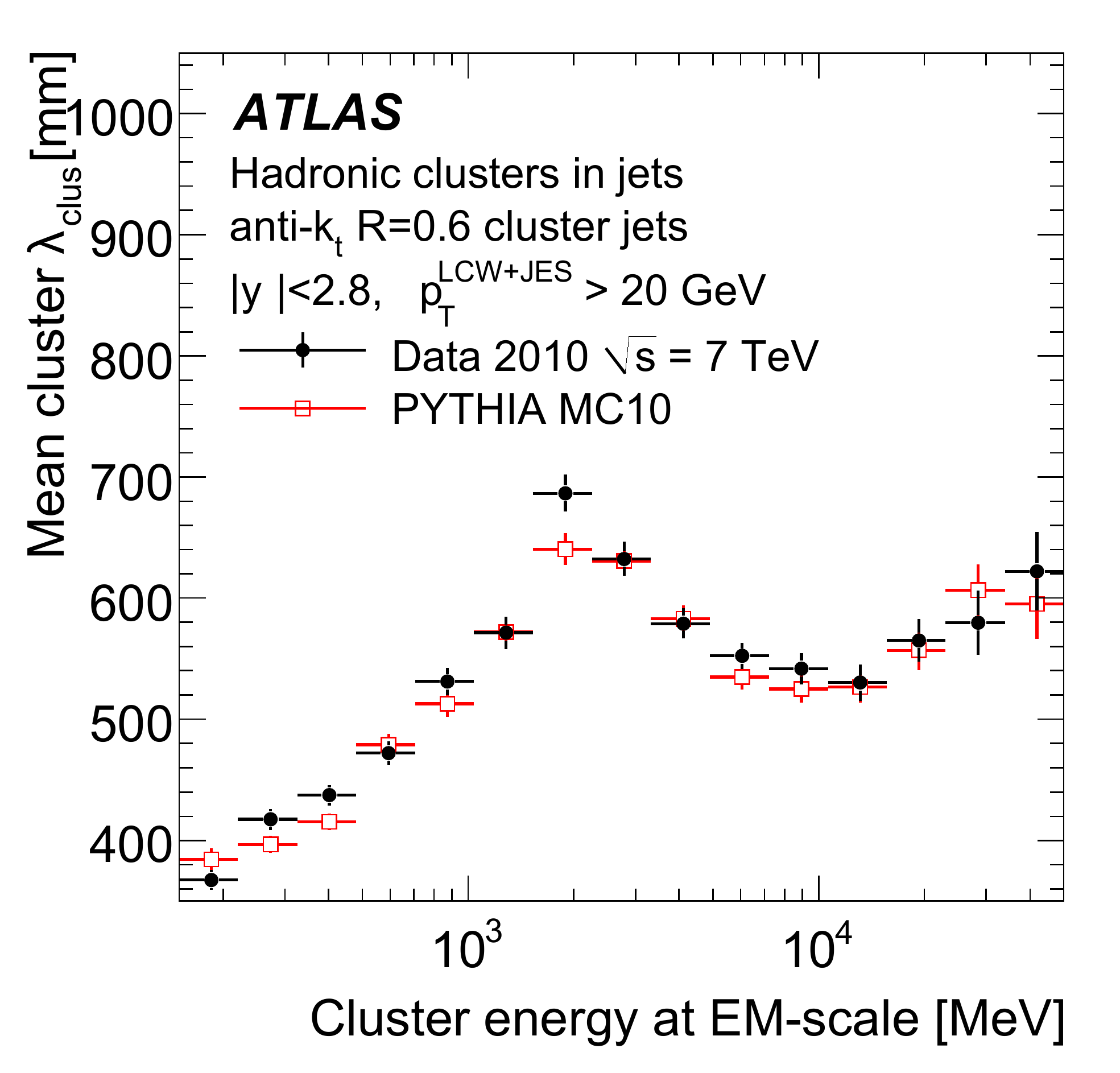}\label{fig:lambda:avehad}}
	\caption[]{The distribution of the longitudinal depth \lamctr{} of \topo{} inside \antikt{} jets with $R = 0.6$, $|y|<2.8$, and $\pT > 20$ \GeV, for clusters classified as \subref{fig:lambda:distem}  electromagnetic (\EM) and \subref{fig:lambda:disthad} hadronic (\HAD), in 2010 data and \MC{} simulations (no \pu). Also shown is the average \topo{} depth $\langle\lamctr\rangle$ as function of the cluster energy \eclusem{} for the same \topos{} classified as \subref{fig:lambda:aveem} \EM{} and \subref{fig:lambda:avehad} \HAD, respectively. The figures are adapted from \citRef{Aad:2011he}.}
\label{fig:lambda}
\end{figure}

The depth  \lamctr{} of the \topo{} (\secRef{sec:moments:geometry:location}) and its average cell signal density \rhoclus{} (\secRef{sec:moments:signal:density}), both determined in bins of the cluster energy \eclusem{} and the cluster direction \etaclus, are found to 
be most efficient in classifying the \topos. 
Using the \MC{} simulations of single charged and neutral pions entering the calorimeters at various pseudorapidities and at various momenta, the probability for a cluster to be of electromagnetic origin (\EMLike) is then determined by measuring the efficiency for detecting an \EM-like cluster in bins of 
four \topo{} observables,
\begin{equation}
	\obsset{\clus}{\text{class}} = \left\{\;\eclusem,\;\etaclus,\;\log_{10}(\rhoclus/\rho_{0})-\log_{10}(\eclusem/E_{0}),\log_{10}(\lamctr/\lambda_{0})\;\right\}\,,
	\label{eq:clusobs} 
\end{equation}
in this sequence mapped to bin indices $\mathit{ijkl}$ in the full accessible phase space.
The density scale is $\rho_{0} = \unit{1}{\MeV\,\mm^{-3}}$, the signal normalisation is $E_{0} = \unit{1}{\MeV}$, and longitudinal depth is measured in terms of $\lambda_{0} = \unit{1}{\mm}$.  
Here the density \rhoclus{} is divided
by the cluster signal \eclusem. This provides a necessary reference scale for its evaluation. As an absolute measure, \rhoclus{} is less powerful in separating electromagnetic from hadronic energy deposits, as the same densities can be generated by electromagnetically and hadronically interacting particles of different incident energies.

The likelihood \EMLike{} is defined in each bin $\mathit{ijkl}$ as
\begin{equation}
	\EMLike(\eclusem,\etaclus,\rhoclus/\eclusem,\lamctr) \mapsto \EMLikei{\mathit{ijkl}} = \dfrac{\varepsilon^{\pi^{0}}_{\mathit{ijkl}}}{\varepsilon ^{\pi^{0}}_{\mathit{ijkl}}+2\varepsilon^{\pi^{\pm}}_{\mathit{ijkl}}}\,,
        \label{eq:clus_class_like}                                                   
\end{equation}
with $0 \leq \EMLikei{\mathit{ijkl}} \leq 1$. The efficiencies $\varepsilon_{\mathit{ijkl}}^{\pi^{0}(\pi^{\pm})}$ are calculated as 
\begin{align}
	\varepsilon_{\mathit{ijkl}}^{\pi^{0}(\pi^{\pm})} = \dfrac{N_{\mathit{ijkl}}^{\pi^{0}(\pi^{\pm})}}{N_{\mathit{ij}}^{\pi^{0}(\pi^{\pm})}}\,.
        \label{eq:clus_class_eff} 
\end{align}
\noindent 
Here $N_{\mathit{ijkl}}^{\pi^{0}(\pi^{\pm})}$ is the number of \topos{} from $\pi^{0}$ ($\pi^{\pm}$) in a given bin $\mathit{ijkl}$, while $N_{\mathit{ij}}^{\pi^{0}(\pi^{\pm})}$ is the number of $\pi^{0}$ ($\pi^{\pm}$) found in bin $\mathit{ij}$ of the $(\eclusem,\etaclus)$ phase space. 
On average there is no detectable difference in the development of $\pi^{+}$ and $\pi^{-}$ initiated hadronic showers affecting the \topo{} formation. 
The distributions of the observables in \obsset{\clus}{\text{class}} as well as the correlations between them are the same.
Therefore \topos{} from $\pi^{+}$ and $\pi^{-}$ showers occupy the same bins in the \obsset{\clus}{\text{class}}{} phase space, yielding $N_{\mathit{ijkl}}^{\pi^{\pm}} = N_{\mathit{ijkl}}^{\pi^{+}} = N_{\mathit{ijkl}}^{\pi^{-}}$, $N_{\mathit{ij}}^{\pi^{\pm}} = N_{\mathit{ij}}^{\pi^{+}} = N_{\mathit{ij}}^{\pi^{-}}$, and $\varepsilon^{\pi^{-}}_{\mathit{ijkl}} + \varepsilon^{\pi^{+}}_{\mathit{ijkl}} = 2\varepsilon^{\pi^{\pm}}_{\mathit{ijkl}}$ in the definition of \EMLike{} in \eqRef{eq:clus_class_like}.
This normalisation reflects the use of all three pion charges at equal probability in \MC{} simulations, thus maintaining the correct isospin-preserving ratio.

For performance evaluation purposes, any \topo{} with the set of observables \obsset{\clus}{\text{class}}{} from \eqRef{eq:clusobs} located in a bin $\mathit{ijkl}$ with $\EMLikei{\mathit{ijkl}} \geq 0.5$ is classified as \EM{} and with $\EMLikei{\mathit{ijkl}} < 0.5$  is classified as \HAD. 
In the rare case where a \topo{} has too few cells or too little signal to meaningfully reconstruct the observables in \obsset{\clus}{\text{class}}, the cluster is likely generated by noise or insignificant energy deposits and is thus neither classified nor further corrected or calibrated. 
%
An example of a \EMLike{} distribution in a given phase space bin $\mathit{ij}$ is shown in \figRef{fig:emlike}. All distributions and their bin contents are accessed as lookup tables to find \EMLike{} for a given cluster. 

The distributions of \lamctr{} for \topos{} in jets reconstructed with the \antikt{} algorithm  with $R = 0.6$ are shown for clusters respectively classified as electromagnetic or hadronic, in 2010 data and \MC{} simulations (no \pu) in \figMultiRefLabel~\ref{fig:lambda}\subref{fig:lambda:distem} and \ref{fig:lambda}\subref{fig:lambda:disthad}. The specific structure of each distribution reflects the longitudinal segmentation of the electromagnetic and hadronic calorimeters in \ATLAS. 
The average cluster depth \AVE{\lamctr}{} as a function of the cluster energy is shown in \figMultiRefLabel~\ref{fig:lambda}\subref{fig:lambda:aveem} and \ref{fig:lambda}\subref{fig:lambda:avehad} for the same \EM{} and \HAD{} \topos, respectively. 
The \EM{} \topos{} show the expected linear dependence of \AVE{\lamctr}{} on $\log{\eclusem}$ in \figRefLabel~\ref{fig:lambda}\subref{fig:lambda:aveem}, with some modulations introduced by the \readout{} granularity of the \EMC{}. 
The \AVE{\lamctr}{} dependence on \eclusem{} shown for \HAD{} \topos{} in \figRefLabel~\ref{fig:lambda}\subref{fig:lambda:avehad} features a similar shape up to $\eclusem \approx \unit{2}{\GeV}$. 
This energy range is dominated by \topos{} from low-energy hadrons, in addition to clusters from less-energetic hadronic shower fragments created by the splitting algorithm described in \secRef{sec:topos:formation:split}. For $\eclusem > \unit{2}{\GeV}$ the average \lamctr{} is increasingly dominated by higher-energy clusters produced by splitting and located in the electromagnetic calorimeter, thus pulling it to lower values. 
The rise of \AVE{\lamctr}{} for \topos{} with $\eclusem \gtrsim \unit{10}{\GeV}$ reflects increasing contributions from energetic hadrons with dense showers generating high-energy clusters deeper in the hadronic calorimeter.
The good agreement between data and \MC{} simulations for both classes of \topos{} supports the use of \lamctr{} for the cluster classification derived from \MC{} simulations for data \cite{Aad:2011he}.

\subsection{Hadronic calibration} \label{\thislabel:hadcal}

\begin{figure}[t!] \centering
\subfloat[\ecellem{} in \LArPreSamplerB]{\includegraphics[width=\fighalfwidth]{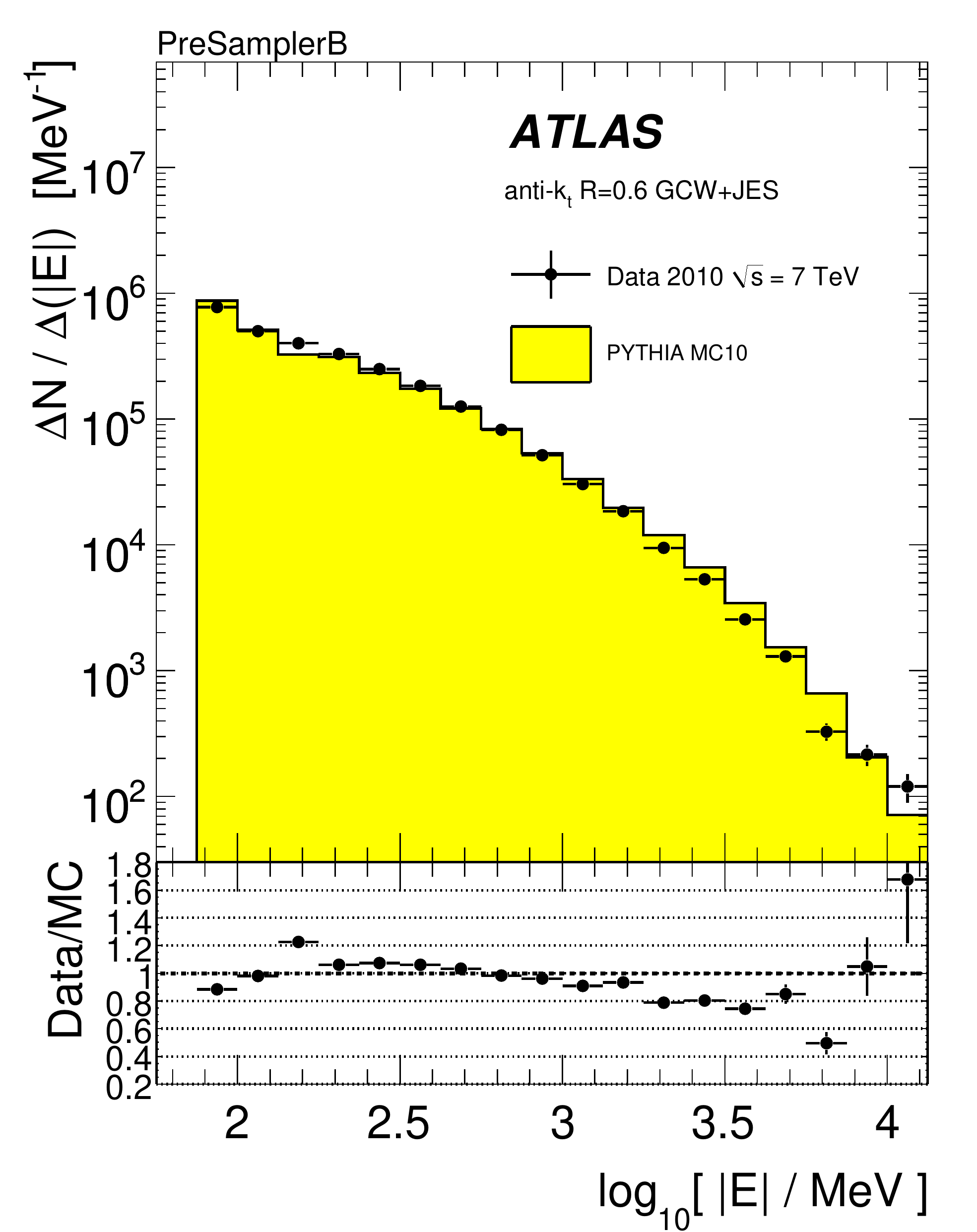}\label{fig:celldensEMB:PreSampler}}
\subfloat[\rhocell{} in \LArEMBN{2}]{\includegraphics[width=\fighalfwidth]{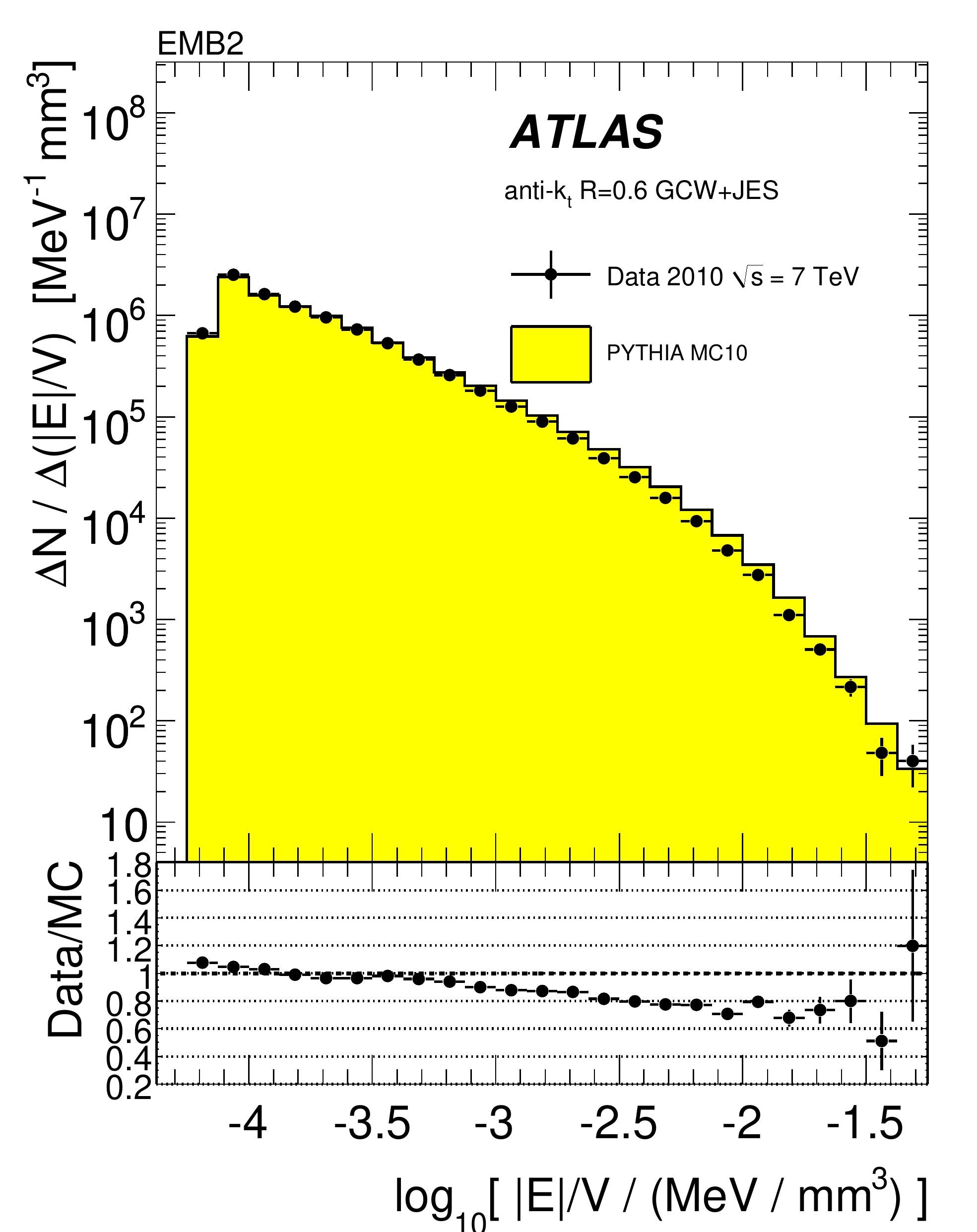}\label{fig:celldensEMB:EMB2}}
\caption[]{Distributions of the cell energy \ecellem{} in the \subref{fig:celldensEMB:PreSampler} central pre-sampler (\LArPreSamplerB) and the cell energy density \rhocell{} in the second sampling of \subref{fig:celldensEMB:EMB2} the central (\LArEMBN{2}) electromagnetic calorimeter in \ATLAS, as observed inside \antikt{} jets with $R = 0.6$, calibrated with the global sequential (GCW+JES) calibration scheme described in \citRef{Aad:2011he}, in 2010 data (no \pu) and the corresponding \MC{} simulations. The \datatomc{} ratio of the spectra is shown below the corresponding distributions. The figure uses plots from \citRef{Aad:2011he}.}
\label{fig:celldensEMB}
\end{figure}

The hadronic calibration for \topos{} attempts to correct for non-compensating calorimeter response, meaning to establish an average $e/\pi = 1$ for the cluster signal. The calibration reference 
is the locally deposited energy in the cells of a given \topo, which is defined as the sum of all energies released by various shower processes in these cells. In each of the cells, the signal \ecellem{} from this deposited energy \ecelldep{} is reconstructed on the electromagnetic energy scale. This yields cell signal weights defined as
\begin{equation}
	\wcell{} = \dfrac{\ecelldep}{\ecellem}\,.
	\label{eq:calweight}
\end{equation}
In the case of electromagnetic signals, $\wcell{} = \wcellcem{} \equiv 1$ by construction of the electromagnetic scale. In hadronic showers, \ecelldep{} has contributions from energy loss mechanisms 
which do not contribute to the signal, including nuclear binding energies and escaping energy carried by neutrinos. In this case, $\wcell{} = \wcellchad \neq 1$ with $\wcellchad > 1$ for hadronic
inelastic interactions within the cell volume, and $\wcellchad < 1$ for deposits by ionisations.\footnote{This is because the electromagnetic energy scale reconstructs 
a signal larger than expected for the deposited energy in case of pure ionisation, due  to the lack of showering.}
The appropriate value of \wcellchad{} reflecting on average the energy loss mechanism generating \ecellem{} in a given cell is determined by the hadronic calibration as a function of a set of observables \obsset{\cell}{\calibhad}{} associated with the cell and the \topo{} it belongs to. 
It is then applied to \ecellem{} according to \eqRef{eq:effweight} in the signal reconstruction.

\begin{figure}[t!] \centering
\subfloat[\rhocell{} in \LArEMEN{2}]{\includegraphics[width=\fighalfwidth]{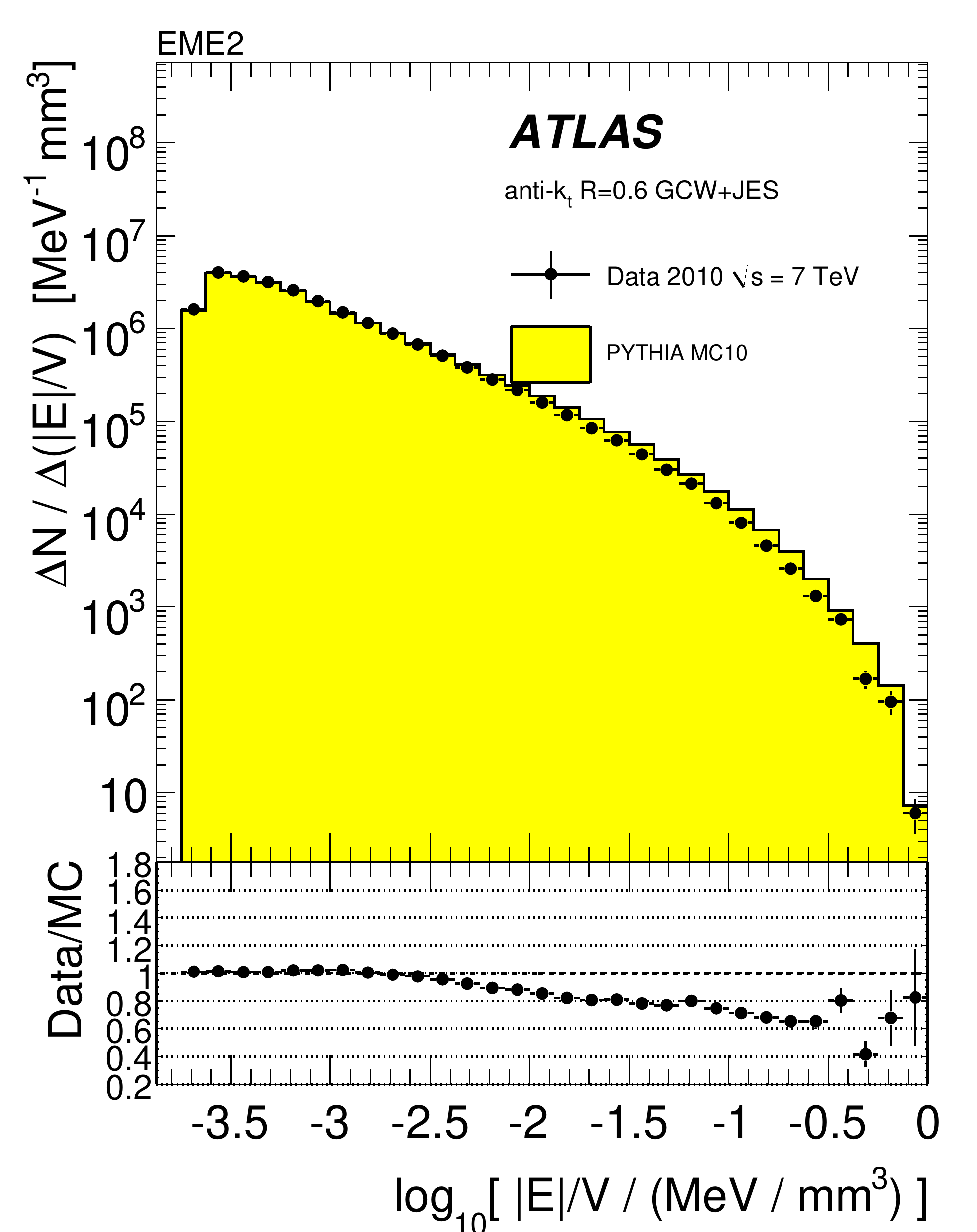}\label{fig:celldensEMC:EME2}}
\subfloat[\rhocell{} in \LArFCALN{0}]{\includegraphics[width=\fighalfwidth]{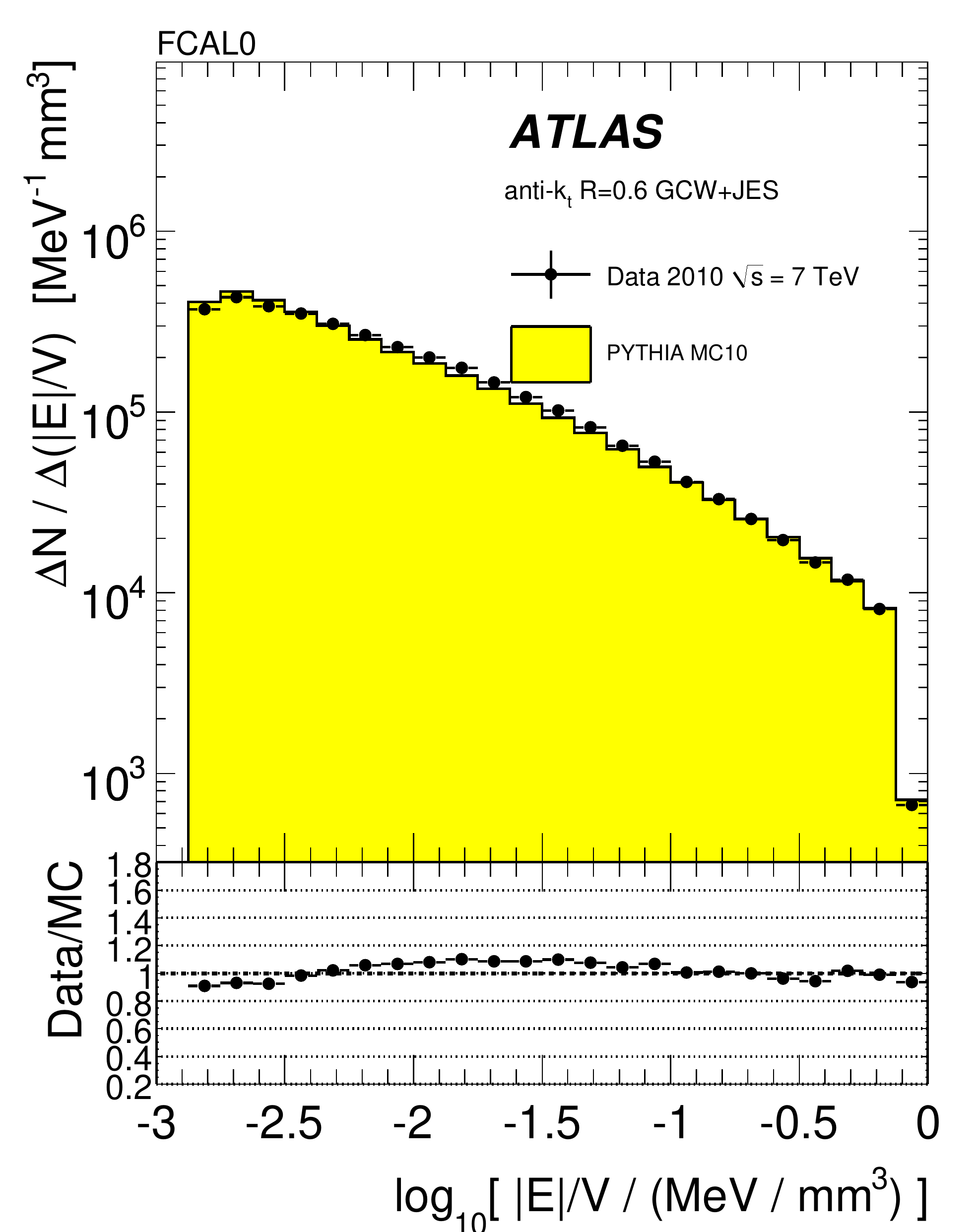}\label{fig:celldensEMC:FCAL0}}
\caption[]{Distributions of the cell energy density \rhocell{} in the \subref{fig:celldensEMC:EME2} second sampling of the \EndCap{}  (\LArEMEN{2}) electromagnetic calorimeter, and the \subref{fig:celldensEMC:FCAL0} first module of the forward calorimeter (\LArFCALN{0}) in \ATLAS, as observed inside \antikt{} jets with $R = 0.6$, calibrated with the GCW+JES scheme described in \citRef{Aad:2011he}, in 2010 data and \MC{} simulations (no \pu). The \datatomc{} ratio of the spectra is shown below the corresponding distributions. The figure uses plots from \citRef{Aad:2011he}.}
\label{fig:celldensEMC}
\end{figure}

Simultaneously using all simulations of charged single pions for all energies and directions, lookup tables are constructed from binned distributions relating \obsset{\cell}{\calibhad}, defined as
\begin{equation}
	\obsset{\cell}{\calibhad} = \left\{\;\sampid,\;\etacell,\;\log_{10}(\rhocell/\rho_{0}),\;\log_{10}(\eclusem/E_{0})\;\right\}\,,
	\label{eq:cellobs}
\end{equation} 
to the hadronic signal calibration weight \wcellchad. The cell location is defined by one of the sampling layer identifiers \sampid{} listed in \tabRef{tab:readout} in \secRef{sec:atlas:det} and the direction of the cell centre \etacell{} extrapolated from the nominal detector centre of \ATLAS. 
The cell signal density \rhocell{} is measured as discussed in \secRef{sec:moments:signal:density}, and \eclusem{} is the signal of the \topo{} to which the cell contributes to.
The lookup tables are binned in terms
of \obsset{\cell}{\calibhad}{} such that \wcellchad{} in each bin in the filled table is the average over all cells with observables fitting into this bin, with each contributing weight calculated as given in \eqRef{eq:calweight}. These
average weights are then retrieved for any cell in a \topo{}
as a function of \obsset{\cell}{\calibhad}.
The cluster signal and directions are re-summed as discussed in \secRef{\thislabel:full}.
The scales $\rho_{0}$ and $E_{0}$ in \eqRef{eq:cellobs} are the same as the ones used in \eqRef{eq:clusobs}.

\begin{figure}[t!] \centering
\subfloat[\rhocell{} in \TileN{1}]{\includegraphics[width=\fighalfwidth]{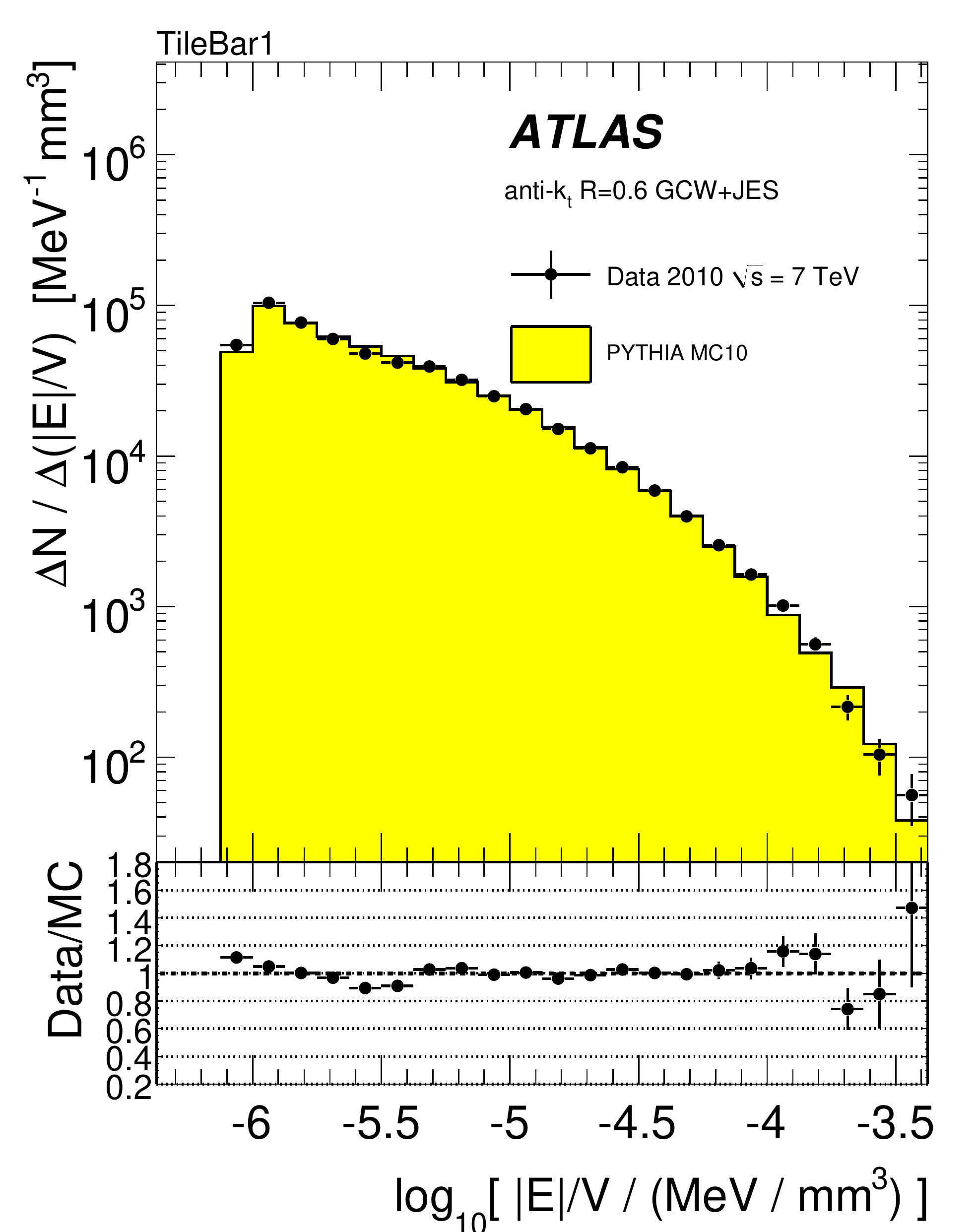}\label{fig:celldensHAC:TileBar1}}
\subfloat[\rhocell{} in \LArHECN{0}]{\includegraphics[width=\fighalfwidth]{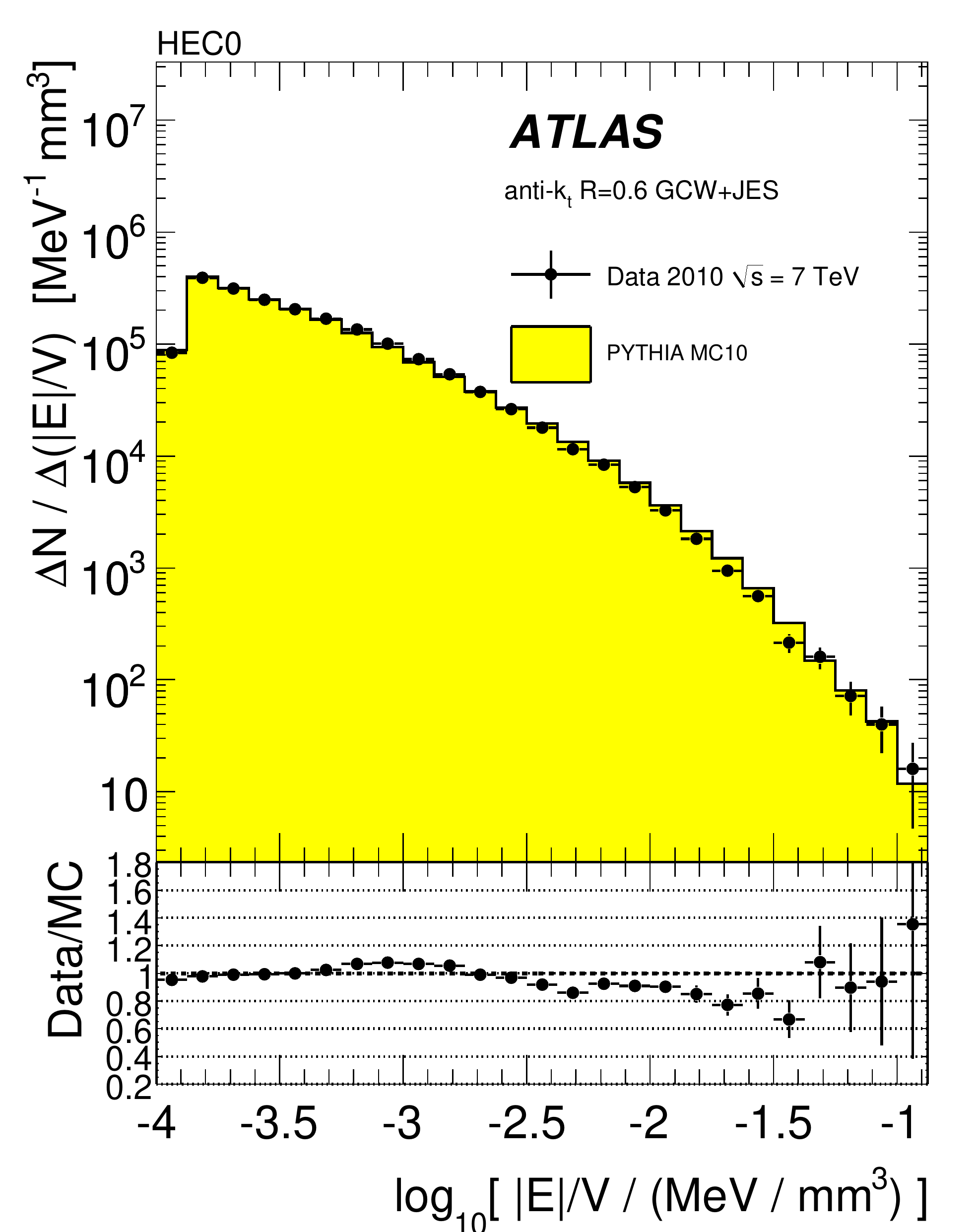}\label{fig:celldensHAC:HEC0}}
\caption[]{Distributions of the cell energy density \rhocell{} in the central \subref{fig:celldensHAC:TileBar1} and \EndCap{} \subref{fig:celldensHAC:HEC0} hadronic calorimeters in \ATLAS, as observed inside \antikt{} jets with $R = 0.6$ calibrated with the GCW+JES scheme described in \citRef{Aad:2011he}, in 2010 data and \MC{} simulations (no \pu). The \datatomc{} ratio of the spectra is shown below the corresponding distributions. The figure uses plots from \citRef{Aad:2011he}.}
\label{fig:celldensHAC}
\end{figure}

The \ecellem{} distribution in the \LArPreSampler{} and the \rhocell{} distribution in the the \LArEMBN{2} sampling of the central electromagmetic calorimeter are shown in \figRef{fig:celldensEMB} for cells in \topos{} inside jets reconstructed with the \antikt{} algorithm using a distance parameter $R = 0.6$. 
Discrepancies between data and \MC{} simulations mostly in the high-end tails of the distributions indicate more compact electromagnetic showers in the simulation.
This also seen in \figRefLabel{}~\ref{fig:celldensEMC}\subref{fig:celldensEMC:EME2} for the \rhocell{} distribution for the same kind of jets in the \LArEMEN{2} sampling of the electromagnetic \EndCap{} calorimeter. 
Better agreement between data and \MC{} simulations over the whole spectrum is observed for the \rhocell{} distributions in the first module (\LArFCALN{0}) of the 
forward calorimeter shown in \figRefLabel{}~\ref{fig:celldensEMC}\subref{fig:celldensEMC:FCAL0}, and in the second sampling of the central hadronic (\TileN{1}) 
and the first sampling of the hadronic \EndCap{} (\LArHECN{0}) calorimeters shown in \figRef{fig:celldensHAC}.
%
%
%
%
Overall, the quality of the modelling of the cell signal densities is sufficient for \topo{} calibration purposes. 
The figures are taken from \citRef{Aad:2011he}.

\subsection{Correction for \ooc{} signal losses} \label{\thislabel:ooc}

In the process of applying the noise suppression 
described in  \secRef{sec:topos:formation},
cells with small true deposited energy generated by 
\EM{} or \HAD{} showers may not be collected into a \topo, either due to lack of significance of 
their small
signal, or due to the absence of a neighbouring cell with a significant signal. The energy losses introduced by this effect are estimated using single-particle \MC{} simulations. A corresponding \emph{\ooc} correction is determined and applied to nearby \topos. The cells with true energy not included into clusters are referred to as \emph{lost cells}.

\begin{figure}[t!] \centering
	\includegraphics[width=\figfullwidth]{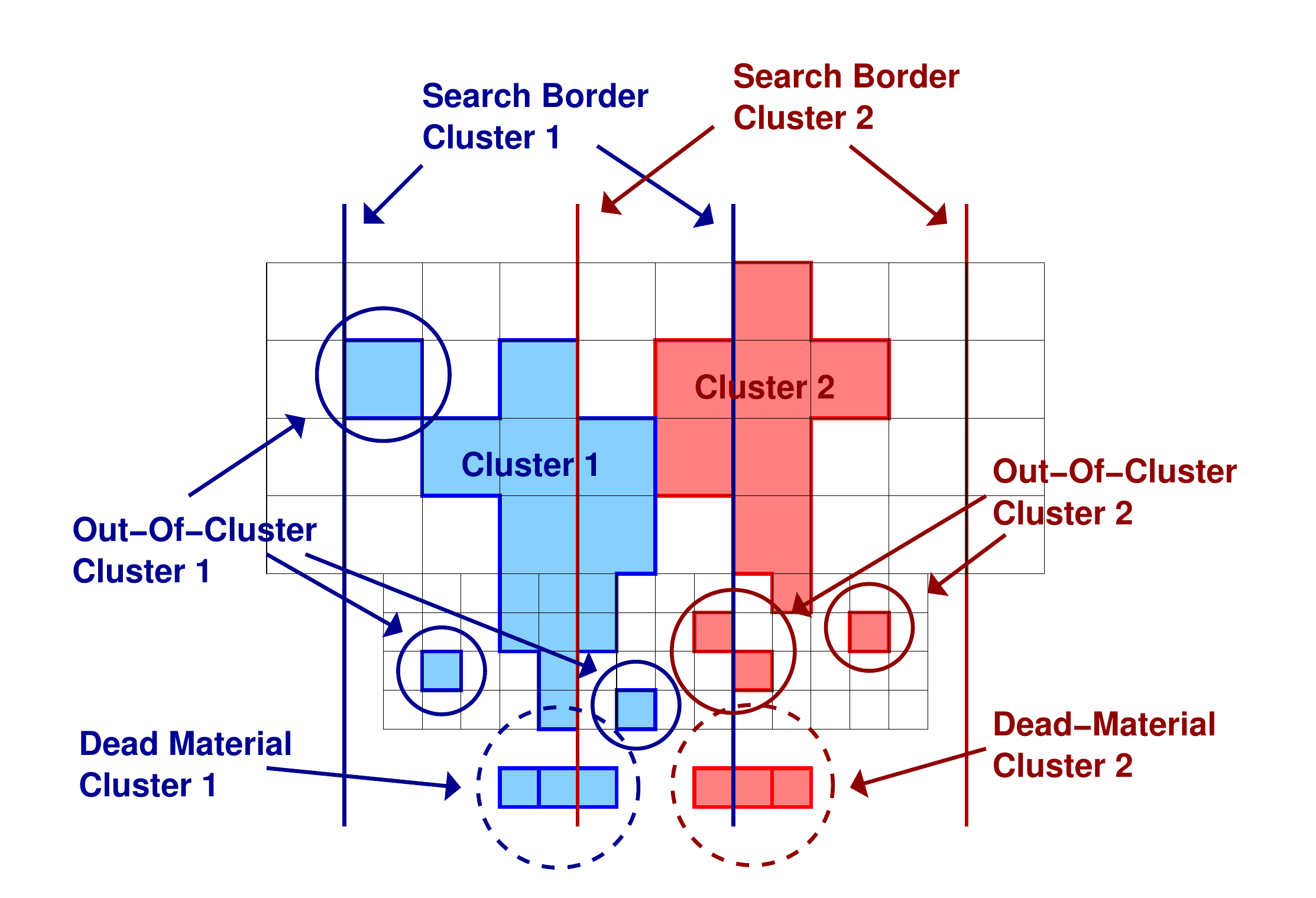}
	\caption[]{Illustration of the assignment scheme for cells inside the calorimeter with true signal not captured in a \topo{} in the context of the \ooc{} correction  (see \secRef{\thislabel:ooc}) and for 
	dead material cells outside the calorimeter for the \dm{} correction discussed in \secRef{\thislabel:dmc}. The deposited energy in cells inside the \topo{} is used to determine
	the hadronic calibration described in \secRef{\thislabel:hadcal}. A schematic depiction of a typical section of the \ATLAS{} \EndCap{} calorimeter with four highly granular electromagnetic samplings and four coarser hadronic samplings is shown in a view with $\eta$ as the horizontal and the depth $z$ as the vertical coordinate. The boxes at small $z$ in front of the \EM{} calorimeter symbolise upstream energy losses collected into dead material cells.}
	\label{fig:ooc_dm_scheme}
\end{figure}

The challenge in determining this correction is the assignment of the energy deposited in a lost cell to a certain cluster. As discussed in \secRef{sec:topos:formation:split} and seen in \figRef{fig:nclusters}, hadronic showers in particular 
can generate more than one \topo. 
An algorithm defining an \emph{\ooc{} neighbourhood} to search for the lost cells has been developed for this assignment. 
This is depicted schematically in \figRef{fig:ooc_dm_scheme}. 
The actual size of the neighbourhood for a given \topo{} is determined by the maximum angular distance between the cluster and the lost cells. 
This distance depends on \etaclus, and thus reflects granularity changes and shower size variations. It varies from approximately \unit{\pi/3}{\rad}{} ($60^{\circ}$) at $\etaclus = 0$ to \unit{7\pi/90}{\rad}{} ($14^{\circ}$) for $\etaclus > 3.2$. 
The energy \eclusooc{} deposited in all lost cells associated with a given \topo{} is then used to derive the \ooc{} correction factor \wcluso,
\begin{align}
	\wcluso = \dfrac{\eclusooc + \eclusdep}{\eclusdep} \;\;\;\;\;\text{and}\;\;\;\;\; \eclusooc = \sum_{i \in \text{\{lost cells\}}} \ecelldepi{\text{lost},i}\,.
	\label{eq:oocweight}
\end{align} 
Here \eclusdep{} is the summed deposited energy of all cells inside the cluster. The \ooc{} correction is a cluster-level correction featuring $\wcluso \ge 1$. 

\FigRef{fig:ooc_dm_scheme} shows that a lost cell can be located in the two overlapping \ooc{} neighbourhoods of two close-by \topos.
In this case \ecelldepi{\text{lost}} of this lost cell is assigned to both clusters, with a weight proportional to their respective deposited energies \eclusdepi{1(2)}. The \ooc{} correction takes into account shared and non-shared lost cells and is derived for each of the two clusters separately using \eqRef{eq:oocweight} with
\begin{align}
	\eclusooci{1(2)} = \underbracket{\vphantom{\dfrac{\eclusdepi{1(2)}}{\eclusdepi{1}+\eclusdepi{2}}\sum_{j \in \{\text{lost cells}\}} \ecelldepi{\text{lost},j}}\sum_{i \in \{\text{lost cells}\}} \ecelldepi{\text{lost},i}}_{\text{non-shared lost cells}} + 
		                    \underbracket{\dfrac{\eclusdepi{1(2)}}{\eclusdepi{1}+\eclusdepi{2}}\sum_{j \in \{\text{lost cells}\}} \ecelldepi{\text{lost},j}}_{\text{shared lost cells}} \,.
\label{eq:eclus_ooc_lost}
\end{align}
There are no spatial distance criteria applied to the sharing.

The scheme for the \ooc{} correction 
ignores lost energy deposited in inactive areas of the detector, outside calorimeter cells. This effect is corrected for later in the calibration sequence (see \secRef{\thislabel:dmc}) such that this component is not double-counted. 

The \ooc{} correction is different for electromagnetic and hadronic showers and is therefore separately determined with neutral and charged pion single-particle simulations.  
The three-dimensional set of observables \obsset{\clus}{\oocab}
\begin{equation}
	\obsset{\clus}{\oocab}  = \left\{\;\etaclus,\;\log_{10}(\eclusem/E_{0}),\;\log_{10}(\lamctr/\lambda_{0})\;\right\}
	\label{eq:ooc_obs}
\end{equation}
is used to bin \wcluso. 
The weight is applied to the signal of nearly all cells of the \topo{} receiving the \ooc{} correction such that $\wcello = \wcluso$. 
The exceptions are cells located in the \LAr{} pre-samplers \LArPreSamplerB{} and \LArPreSamplerE, and the \Tile{} scintillators located between the barrel and \EndCap{} cryostats, where $\wcello = 1$ always.
The normalisations $E_{0}$ and $\lambda_{0}$ in \eqRef{eq:ooc_obs} are the same as used in \eqRef{eq:clusobs}.

\begin{figure}[t!] \centering
\subfloat[$f_{\text{iso}}$ distribution for \EM-tagged clusters]{\includegraphics[width=\fighalfwidth]{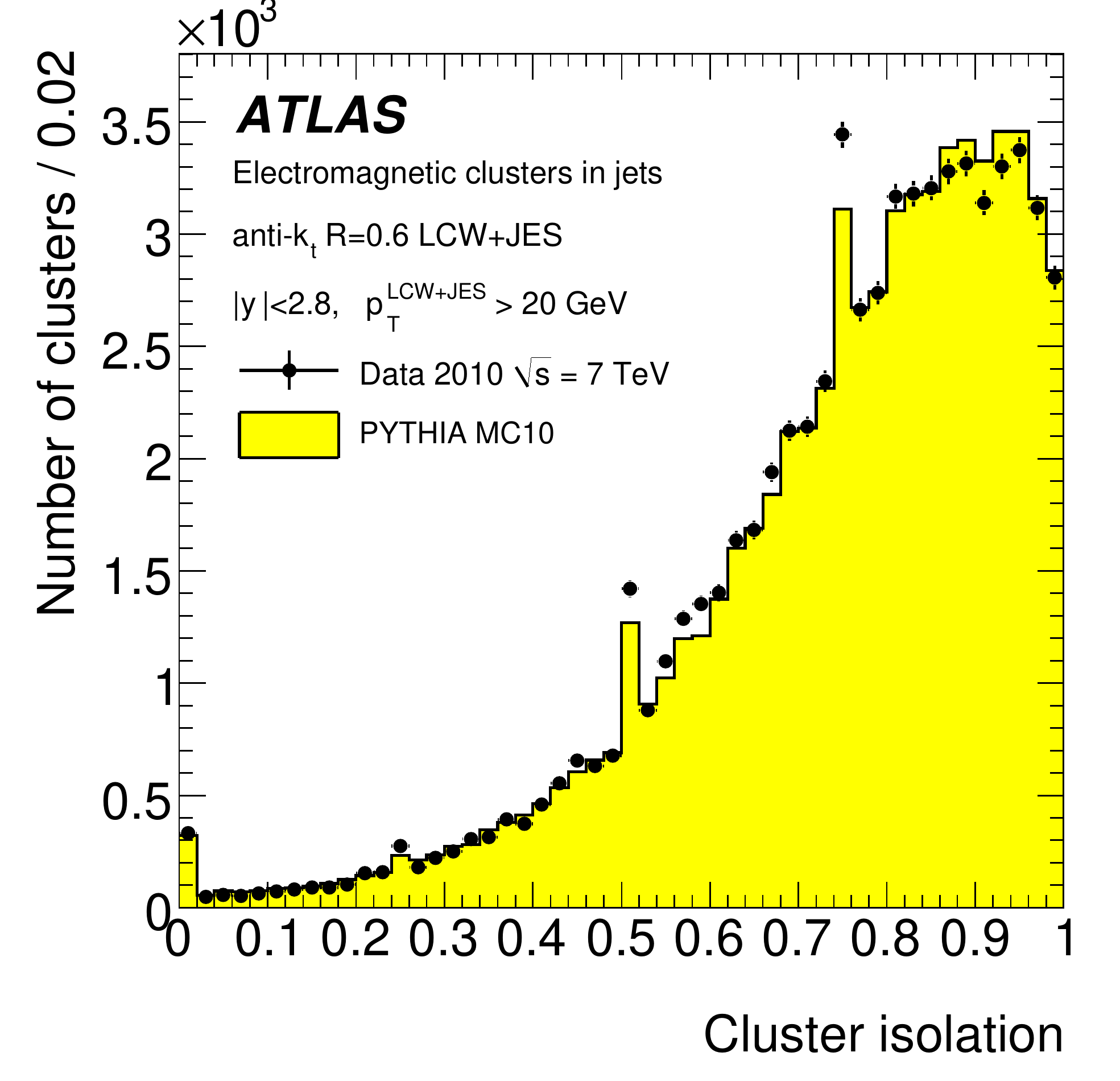}\label{fig:iso:emdist}}
\subfloat[$f_{\text{iso}}$ distribution for \HAD-tagged clusters]{\includegraphics[width=\fighalfwidth]{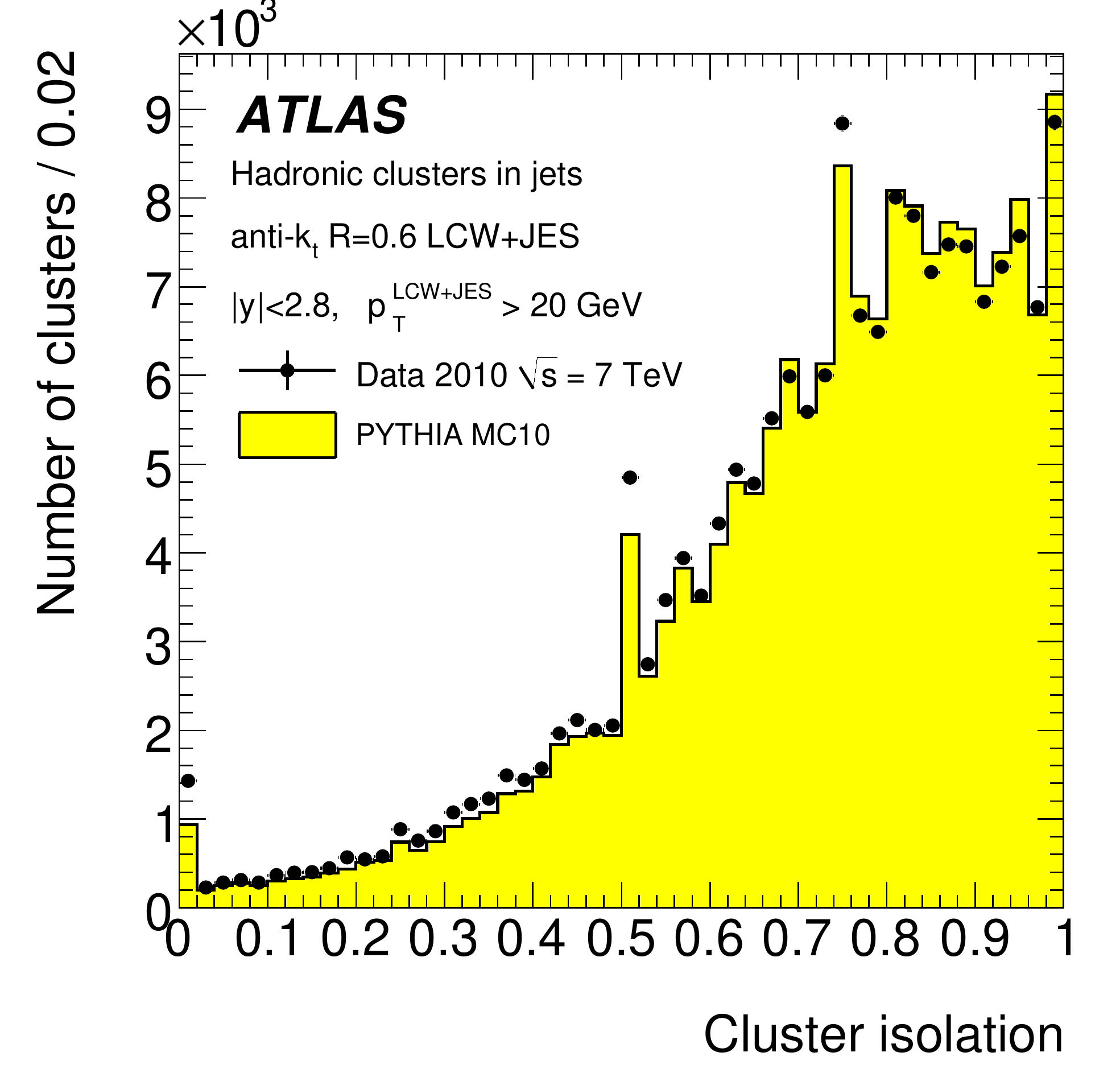}\label{fig:iso:haddist}}
\qquad
\subfloat[$\langle f_{\text{iso}}\rangle(\eclusem)$ for \EM-tagged clusters]{\includegraphics[width=\fighalfwidth]{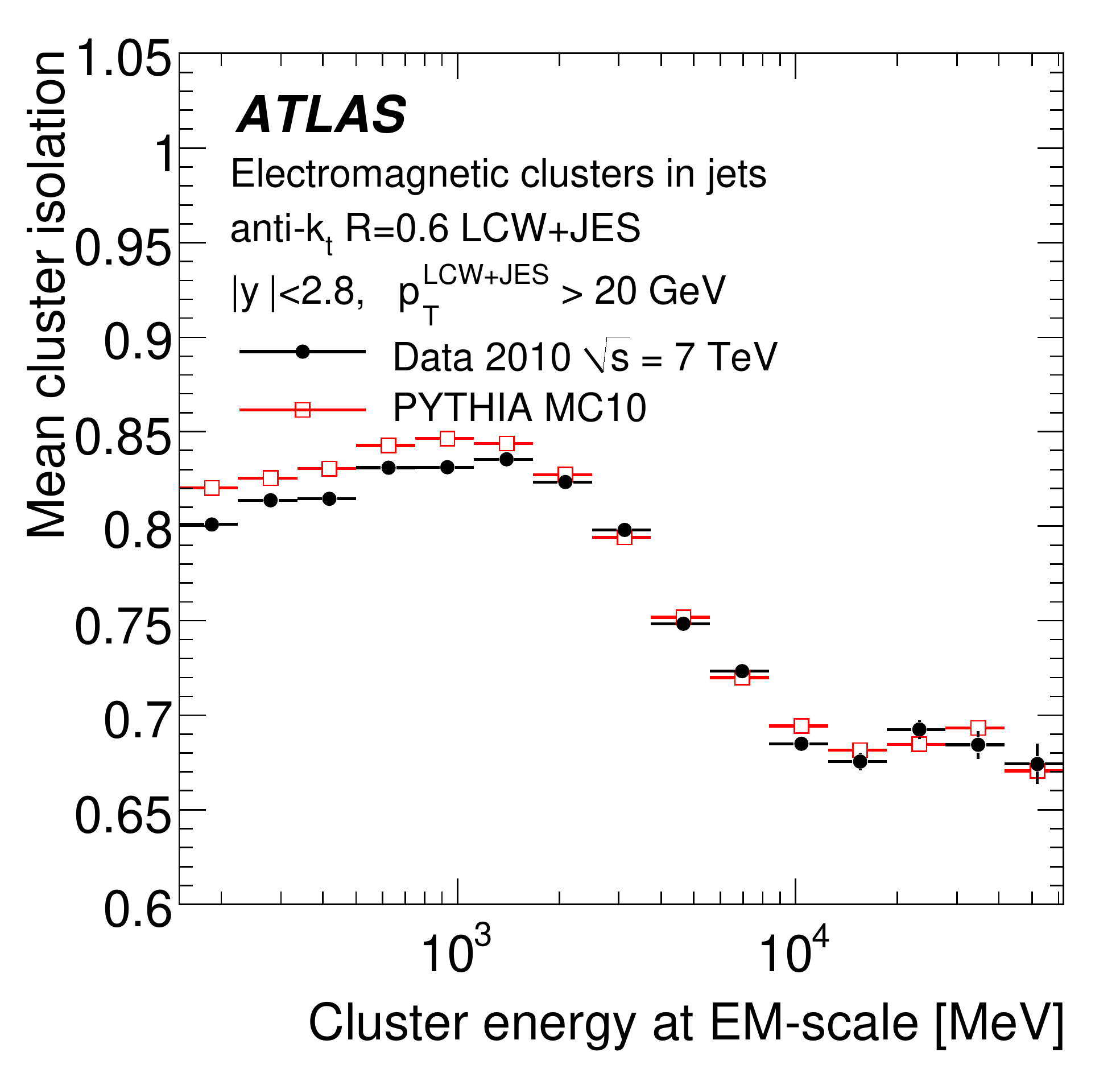}\label{fig:iso:emave}}
\subfloat[$\langle f_{\text{iso}}\rangle(\eclusem)$ for \HAD-tagged clusters]{\includegraphics[width=\fighalfwidth]{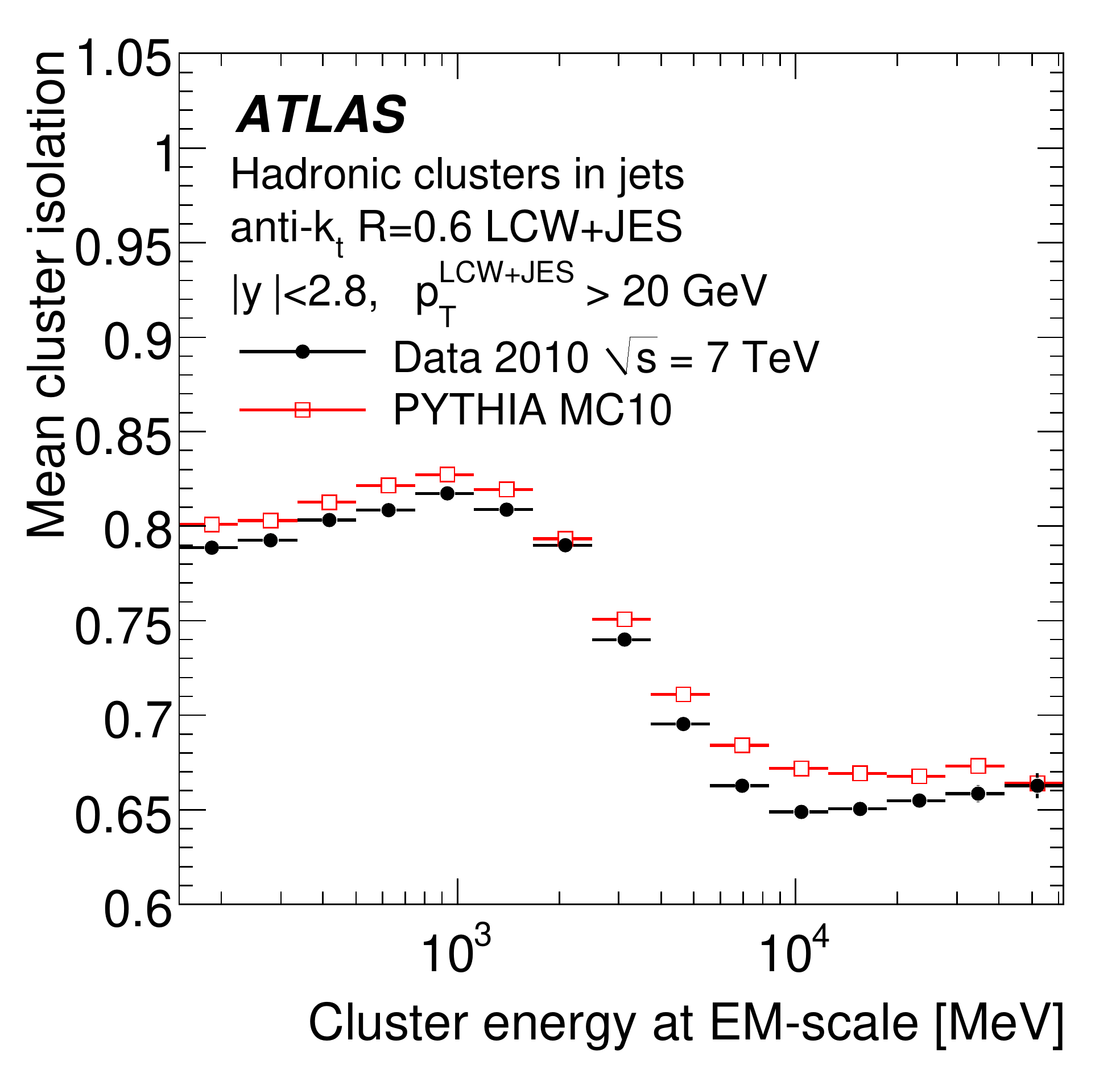}\label{fig:iso:hadave}} 
\caption[]{The distribution of the isolation moment $f_{\text{iso}}$ in \subref{fig:iso:emdist} clusters classified as electromagnetic, and \subref{fig:iso:haddist} clusters classified as hadronic. 
The average isolation $\langle f_{\text{iso}}\rangle$ as a function of the cluster signal \eclusem{} is shown in \subref{fig:iso:emave} for electromagnetic and in \subref{fig:iso:hadave} for hadronic \topos. The figures are taken from \citRef{Aad:2011he}.}
\label{fig:iso}
\end{figure}

While the determination of the \ooc{} correction depends on this assignment algorithm, the application of the correction is context dependent. 
A \topo{} in a jet is likely to have directly neighbouring 
clusters which can capture its \ooc{} signal loss. It is therefore expected that \topos{} in jets need less \ooc{} corrections than isolated \topos{} away from other clusters. The degree of
isolation is measured by the isolation moment  $f_{\text{iso}}$ introduced in \secRef{sec:moments:signal:isolation}. The \ooc{} correction is effectively $f_{\text{iso}} \wcluso(\obsset{\clus}{\oocab})$. 
This correction can change the barycentre and \cog{} of \topos{} containing cells from the \LAr{} pre-samplers or the \Tile{} scintillators. 

\FigRef{fig:iso} shows $f_{\text{iso}}$ for \topos{} classified as either electromagnetic or hadronic
in jets reconstructed with the \antikt{} algorithm and $R = 0.6$   \cite{Aad:2011he}. 
A good agreement between data and \MC{} simulations is observed, both for the details of the respective $f_{\text{iso}}$  in \figMultiRefLabel~\ref{fig:iso}\subref{fig:iso:emdist} and \ref{fig:iso}\subref{fig:iso:haddist} and the average as a function of \eclusem{} in \figMultiRefLabel~\ref{fig:iso}\subref{fig:iso:emave} and \ref{fig:iso}\subref{fig:iso:hadave}.  
The \eclusem{} dependence of $f_{\text{iso}}$ is very similar for both kinds of \topos. 

The peak structure in the $f_{\text{iso}}$ distributions shown in \figMultiRefLabel~\ref{fig:iso}\subref{fig:iso:emdist} and \ref{fig:iso}\subref{fig:iso:haddist} is indicative of \topos{} which have a large fraction of their energy in one sampling layer in the (regular) \ATLAS{} calorimeter \readout{} segmentation with at least 16 cells around the perimeter of clustered cells in a sampling layer. 
The isolation of this layer then dominates the overall $f_{\text{iso}}$, as given by \eqRef{eq:iso_mom} in \secRef{sec:moments:signal:isolation}. 
This dominance of just one sampling layer with the minimal number of cells is typical for \topos{} seeded by a cell barely above the seed threshold defined in \eqRef{eq:toposeed} and too little energy in the neighbouring samplings to further expand the cluster. 
Neighbouring cells then limit $f_{\text{iso}}$ to the multiples of $1/16$ visible in \figMultiRefLabel{}~\ref{fig:iso}\subref{fig:iso:emdist} and \subref{fig:iso:haddist}. 
Even multiples of $1/16$ occur more often than odd multiples since they can be produced more easily by \topos{} with a different number of neighbours. 
The fact that clusters close to the noise threshold are mainly responsible for the peaks explains the mismatch between data and \MC{} simulations observed in the peak heights, and points to non-perfect modelling of noise and very small signals.
The overall structure of the $f_{\text{iso}}$ spectrum in data is well reproduced in terms of the peak locations by \MC{} simulations. 

\subsection{\DM{} corrections} \label{\thislabel:dmc}

\begin{figure}[t!] \centering
	\includegraphics[width=\figfullwidth]{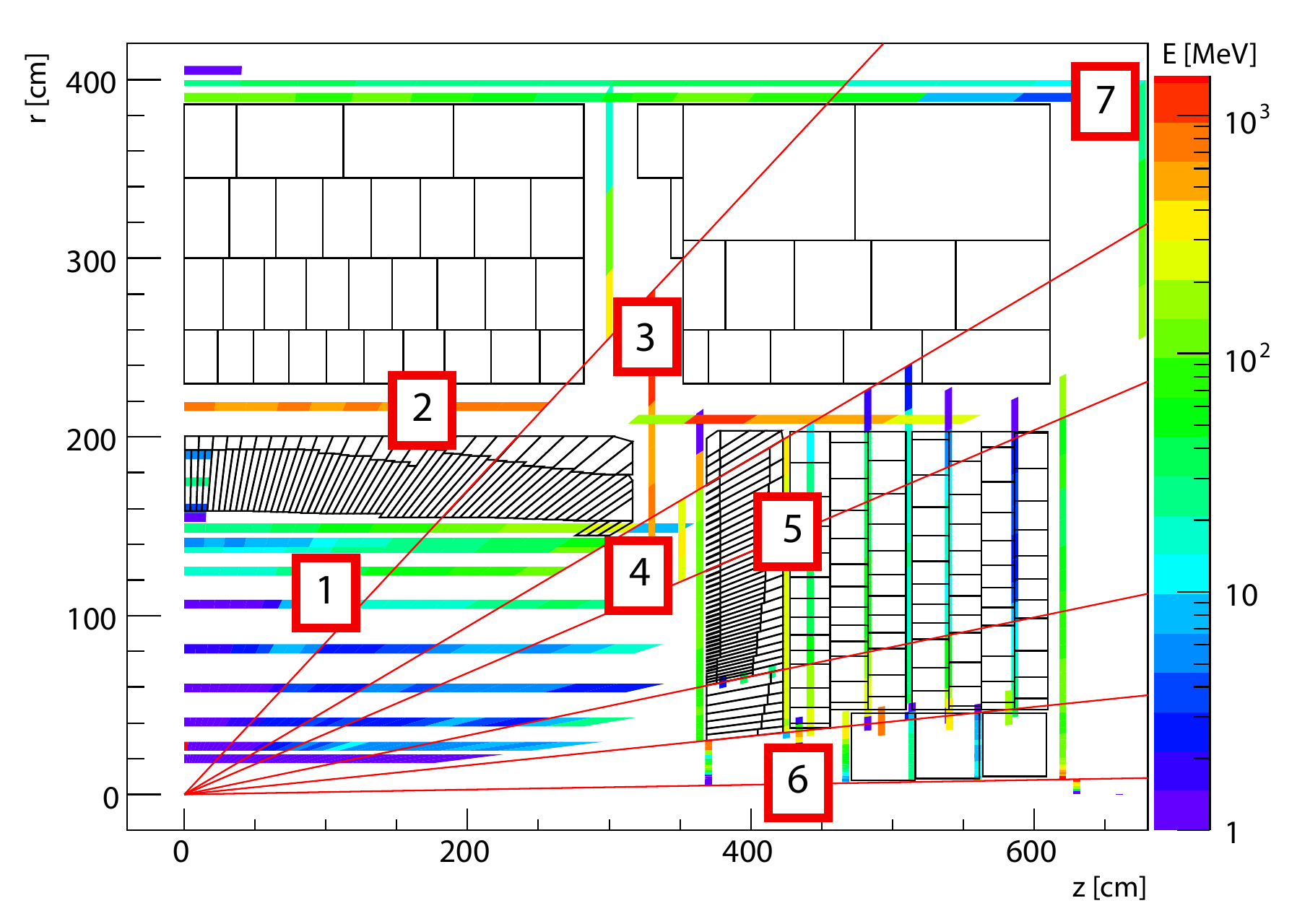}
	\caption[]{The average energy loss in the virtual \dm{} cells for charged 100 \GeV{} pions. The numbers 1 to 7 indicate the different regions, with region 8 (not displayed) being
	everywhere outside regions 1--7. The \dm{} cells are superimposed on a schematic $(r,z)$ view showing a quarter of the \ATLAS{} calorimeter system with its \readout{} segmentation.}
	\label{fig:dm_cells}
\end{figure}

Particles traversing the inactive (dead) material in front of or between calorimeter modules can deposit energy in it, thus reducing the measurable energy. 
This energy loss is addressed on average by the \dm{} correction. It is derived with single-particle \MC{} simulations, where the deposited energy in the dead material outside of the active calorimeter can be calculated. 
This material is divided into virtual cells with a pointing geometry in $(\eta,\phi)$. These cells are similar to the \ATLAS{} calorimeter cells, but typically larger in size. Depending on the particle's direction of flight, eight 
distinct regions are mapped out, as summarised in \tabRef{tab:dm_cells}. The energy deposited in the dead material cells is determined for charged and neutral pions at various energies and
directions, and almost everywhere correlated with measurable signals.

\FigRef{fig:dm_cells} shows a projection of the \dm{} cells  where energy loss is recorded to determine the \dm{} correction. The assignment to a \topo{} is 
based on the same search-border strategy used for the determination of the \ooc{} correction and illustrated in \figRef{fig:ooc_dm_scheme}, with a refinement of the  
assignment procedure specific for the determination of \dm{} corrections. 
Instead of using the full deposited energy \eclusdep{} in the \topo{} as input for sharing in \eqRef{eq:oocweight}, the energy $\eclusdep(s)$ deposited in a selected sampling layer $s$  is used to assign the \dm{} energy to \topos. 
For a given cluster $k$ out of $N_{\clus}$ \topos{} which have cells from $s$ included, the assignment weight $w$ is calculated using
\begin{align}
	w = \dfrac{\sqrt{\eclusdepi{k}(s)}\times\exp(-\Delta R_{k}/R_{0})}%
	               {\displaystyle\sum_{i=1}^{N_{\clus}}\sqrt{\eclusdepi{i}(s)}\times\exp(-\Delta R_{i}/R_{0})}\,,\text{\ \ with\ }\Delta R_{k(i)} = \sqrt{(\Delta\eta_{k(i)})^{2}+(\Delta\phi_{k(i)})^{2}} \text{\ \ and\ }R_{0} = 0.2\,.
	               \label{eq:assign_weight}
\end{align}
The choice of $s$ depends on the \dm{} regions indicated in \figRef{fig:dm_cells}. 
The distances $\Delta\eta$ and $\Delta\phi$ are measured between the \topo{} direction and the \dm{} cell direction. 
The normalisation of $w$ is calculated using all $N_{\clus}$ clusters such that $0 \leq w \leq 1$.
It is rare that two clusters are close to the same \dm{} cell,  most often $w = 1$ is found for the closest \topo, and $w = 0$ for the next closest ones. 
 
This weighted energy loss is collected as a function of observables of the associated \topo{} given in \tabRef{tab:dm_cells}. 
Lost energy deposited in front of the calorimeter is compensated for by applying a correction proportional to the pre-sampler signals in \topos{} which contain these signals. 
In the forward region the signal in the first module \LArFCALN{0}{} of the \LArFCAL{} is used for this purpose.

\begin{table}[t!] \centering
\renewcommand{\arraystretch}{1.25}
\caption[]{Overview of the signals used to correct for \dm{} losses in the various regions around the \ATLAS{} calorimeters. The numbered regions are shown in \figRef{fig:dm_cells}. The parameter values used for the dead material correction are extracted from lookup tables. Region 8 comprises all \dm{} volumes with energy loss outside regions 1 to 7. These are mostly small volumes located between and behind the active calorimeters.}
\label{tab:dm_cells}
\begin{tabular}{|p{0.1\textwidth}|p{0.4\textwidth}|p{0.4\textwidth}|}
	\hline \hline
	Regions & Description & Cluster signals for \dm{} correction \\ \hline
	\multicolumn{1}{|c|}{1}           & In front of \LArEMB                           & Energy in \LArPreSamplerB \\
	\multicolumn{1}{|c|}{2}           & Between \LArEMB{} and \Tile            & Energies in last layer of \LArEMB{} and first layer of \Tile \\
	\multicolumn{1}{|c|}{3}           & In front of \Tile{} gap scintillators            & Energy in \Tile{} gap scintillators \\
	\multicolumn{1}{|c|}{4}           & In front of \LArEME                           & Energy in \LArPreSamplerE \\
	\multicolumn{1}{|c|}{5}	     & Between \LArEME{} and \LArHEC & Energies in last layer of \LArEME{} and first layer of  \LArHEC \\ 
	\multicolumn{1}{|c|}{6}           & In front of \LArFCAL                          & Energy in first \LArFCAL{} module \\ \hline
	\multicolumn{1}{|c|}{7}           & Behind calorimeters                                & \multirow{2}{*}{\parbox[c]{0.4\textwidth}{Energy in last layer  of hadronic calorimeters and \obsset{\clus}{\dmab}{} given in \eqRef{eq:dm_obs}}} \\
	\multicolumn{1}{|c|}{8}           & Everywhere else                                     &  \\ 
	\hline \hline
\end{tabular}
\end{table}

Energy lost between an electromagnetic and a hadronic calorimeter module (regions 2 and 5 in \tabRef{tab:dm_cells} and \figRef{fig:dm_cells}) is found to be proportional to
	$\sqrt{E_{\text{l}}^{\text{\EM}} \cdot E_{\text{f}}^{\text{\EM}}}$,
where $E_{\text{l}}^{\text{\EM}}$ is the energy in the last sampling layer of the electromagnetic calorimeter, and $E_{\text{f}}^{\text{\EM}}$ is the energy in the first sampling layer of the hadronic calorimeter. Both $E_{\text{l}}^{\text{\EM}}$ and $E_{\text{f}}^{\text{\EM}}$ are reconstructed on the electromagnetic energy scale. This correction is only applied to \topos{} which span the material between the two calorimeters. 

\DM{} corrections for longitudinal leakage (region 7 in \tabRef{tab:dm_cells} and \figRef{fig:dm_cells}) are applied to \topos{} that contain cells from the very last (hadronic) calorimeter sampling layer. These corrections are calculated in three-dimensional bins of a set of observables \obsset{\clus}{\dmab}, with
\begin{equation}
	\obsset{\clus}{\dmab} = \left\{\;\etaclus,\;\log_{10}(\eclusem/E_0),\;\lamctr\;\right\}\,,
	\label{eq:dm_obs}
\end{equation} 
and $E_{0}$ from \eqRef{eq:clusobs} in \secRef{\thislabel:classification}. The same set of observables is used as input to correct \dm{} energy losses in \topos{} that are located in the direct neighbourhood of inactive material categorised as region 8 and that have no other \dm{} correction applied.

Like the \ooc{} correction, the \dm{} correction is a cluster-based correction. It is expressed in terms of a weight \wclusd, which is determined from the various correction functions or lookup tables. The corresponding cell signal weight  is the same for all cells of the given cluster ($\wcelld = \wclusd$). This correction therefore does not affect the \topo{} barycentre or \cog.   

\subsection{Fully calibrated cluster kinematics} \label{\thislabel:full}

The reconstructed and fully calibrated \topo{} energy \ecluslcw{} depends on the \EM{} likelihood of the cluster, as discussed in \secRef{\thislabel:classification}, and is characterised by $\ecluslcw \geq \eclusem$. The cluster direction changes due to the calibration, because it is calculated from energy-weighted cell directions using \eqMultiRef{eq:etabaryem}{and}{eq:phibaryem} with $\wcellg \to \wcellc$. 

The effective cell calibration weight \wcellc{} from \eqRef{eq:effweight} after any of the calibrations or corrections are applied yields the cluster energy \ecluscal{} after the calibration 
\begin{equation}
	\ecluscal = \sum_{i\,\in\text{\ cluster}} \wcellci{i} \ecellemi{i}\,.
	\label{eq:cluscalib}
\end{equation} 
While the signal weights determined for each calibration and correction are independently derived, the overall effect of the calibration sequence leads to a factorised accumulation of \wcellc{} in the reconstruction of the cell energies. This is summarised in \tabRef{tab:calsummary}. The overall weight \wcellclcw{} given in item \myref{ref:dmc} of the table is used cell-by-cell in \eqRef{eq:cluscalib} to calculate the final cluster energy \ecluslcw{} by setting $\wcellci{i} = \wcellclcwi{i}$ and thus yielding $\ecluslcw = \ecluscal$. As discussed earlier, \wcellclcw{} is also used to recalculate the cluster directions \etaclus{} and \phiclus.  The final fully calibrated four-momentum reconstructed for any \topo{} is given by replacing \eclusem{} in \eqRef{eq:clusfourmom} in \secRef{sec:topos:kinematics} with \ecluslcw.

\begin{table}[t!]\centering
\caption[]{Summary of the calibration and correction sequence applied to \topos{} from the \EM{} to the final \LCW{} scale.
}
\myarraystretch
\setcounter{myrefctr}{0}
\begin{tabular}{|l|c|c|}
\hline\hline
	\multicolumn{1}{|c|}{Procedure}     & Parameters          & Effective cell signal weight after each step \\ \hline
	\myrefstep{ref:form}{} Cluster formation  & \wcellg                & \wcellg                                   \\  \hline
	\myrefstep{ref:class}{} Classification        & \EMLike               & \wcellg                                    \\ \hline
        \myrefstep{ref:calib}{} Calibration            & $\wcellcem (= 1)$  & \multirow{2}{*}{$\wcellg\left[\EMLike\,\wcellcem + ( 1-\EMLike )\, \wcellchad\right]$}  \\
                                                                     & \wcellchad           &                                               \\ \hline
       \myrefstep{ref:ooc}{} \OOC                & \wcelloem & \multirow{2}{*}{$\displaystyle{\wcellg \prod_{\kappa\,\in\{\calib,\oocab\}} \left[\EMLike\cdot\wcell{\text{em}-\kappa}+ ( 1-\EMLike )\cdot\wcell{\text{had-}\kappa}\right]}$} \\
                                                                & \wcellohad &                                                   \\ 
\hline	
      \myrefstep{ref:dmc}{} \DM                  & \wcelldem  &  \multirow{2}{*}{$\wcellclcw = \displaystyle{\wcellg \prod_{\kappa\,\in\{\calib,\oocab,\dmab\}} \left[\EMLike\,\wcell{\text{em}-\kappa}+ ( 1-\EMLike )\,\wcell{\text{had-}\kappa}\right]}$} \\ 
                                                                & \wcelldhad &                                                   \\
\hline\hline               
\end{tabular}
\label{tab:calsummary}
\end{table}

\begin{figure}[t!]\centering
        \sfcompress       
	\subfloat[$\ecluscal/\eclusem$ versus \eclusem]{\includegraphics[width=\figsixpanelwidth]{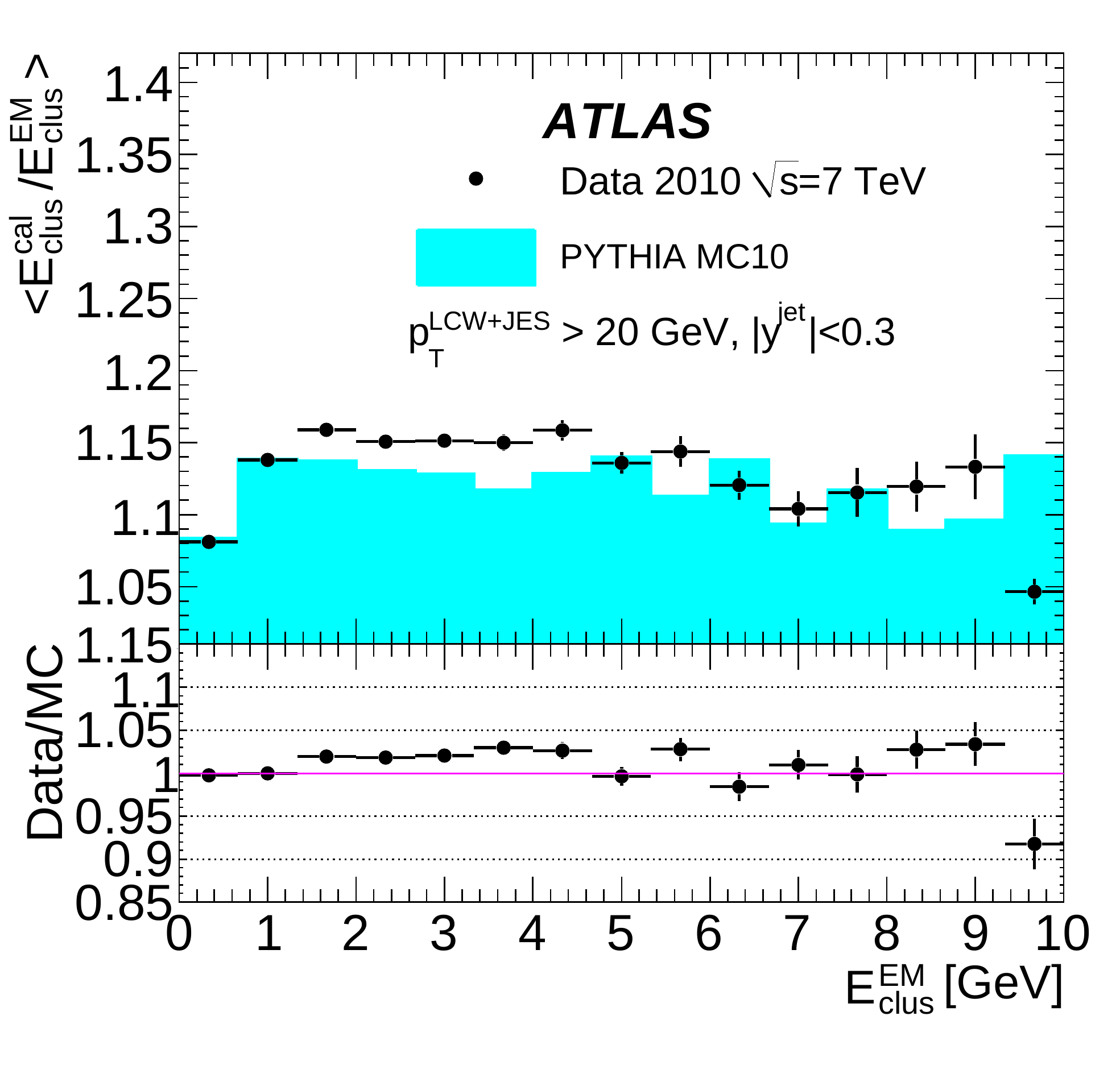}\label{fig:cluscal:e:cal}} \qquad
	\subfloat[$\ecluscal/\eclusem$ versus \etaclus]{\includegraphics[width=\figsixpanelwidth]{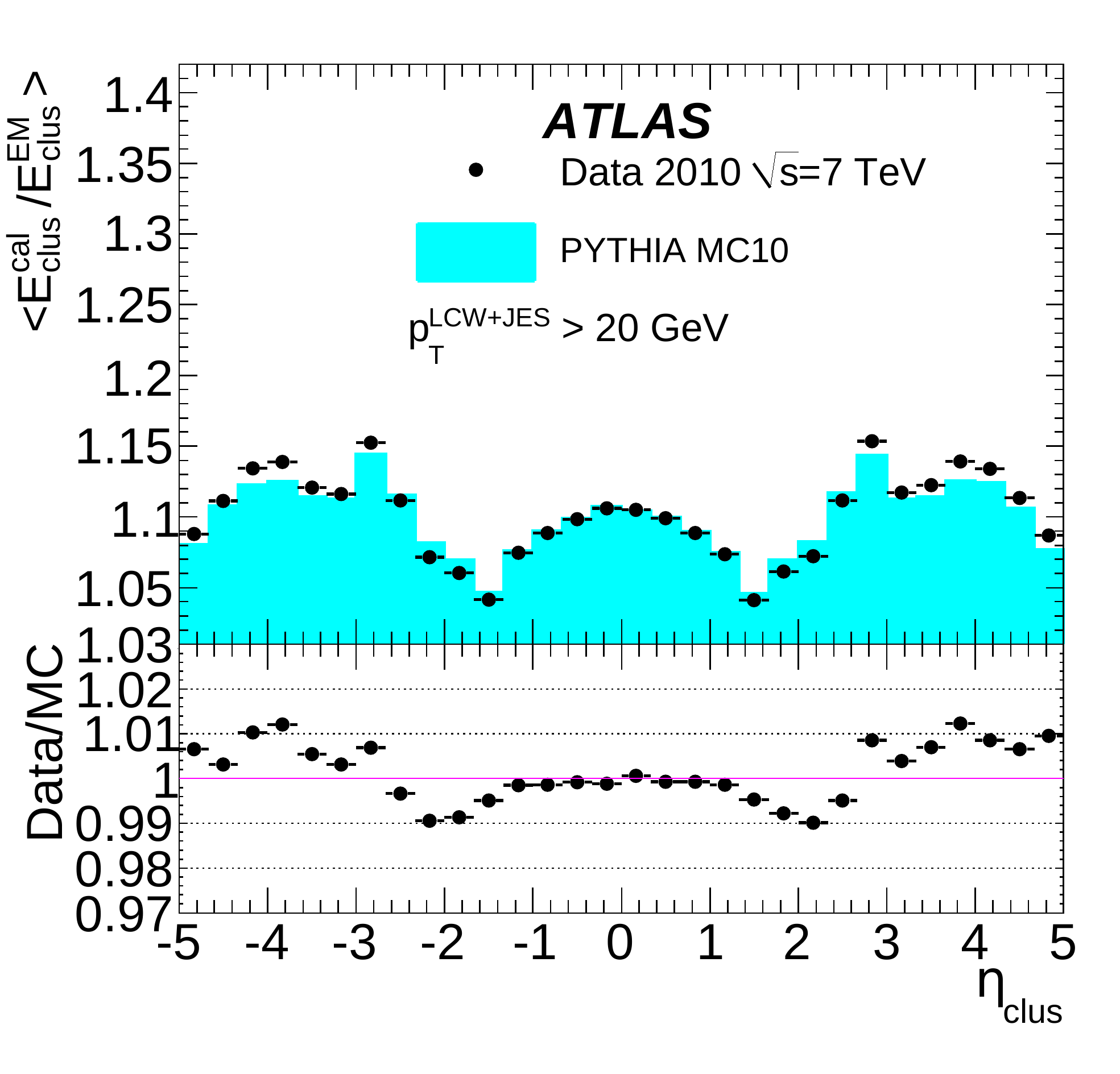}\label{fig:cluscal:eta:cal}}
        \\      
	\subfloat[$E^{\calib+\oocab}_{\clus}/\eclusem$ versus \eclusem]{\includegraphics[width=\figsixpanelwidth]{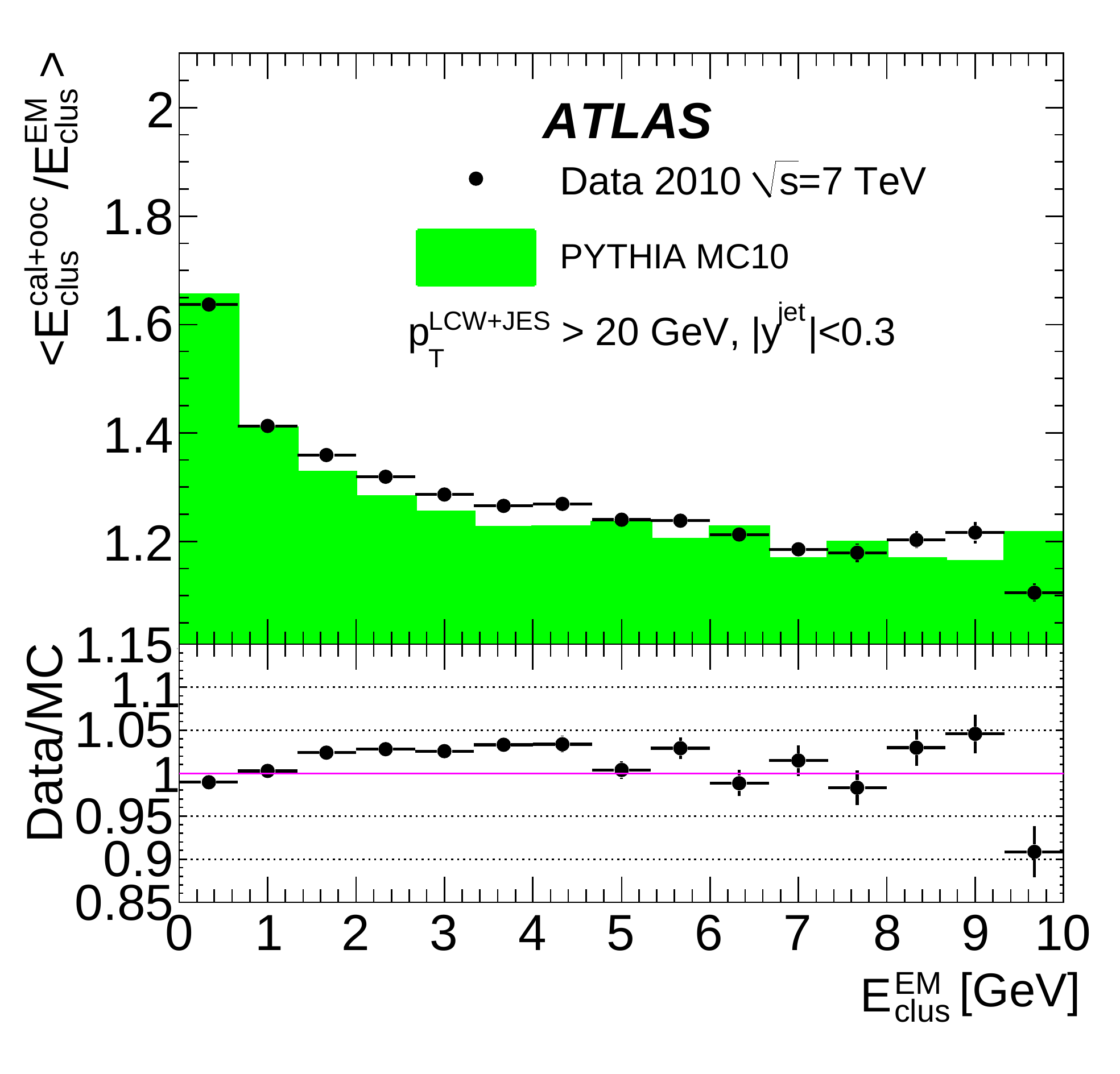}\label{fig:cluscal:e:cal_ooc}} \qquad
	\subfloat[$E^{\calib+\oocab}_{\clus}/\eclusem$ versus \etaclus]{\includegraphics[width=\figsixpanelwidth]{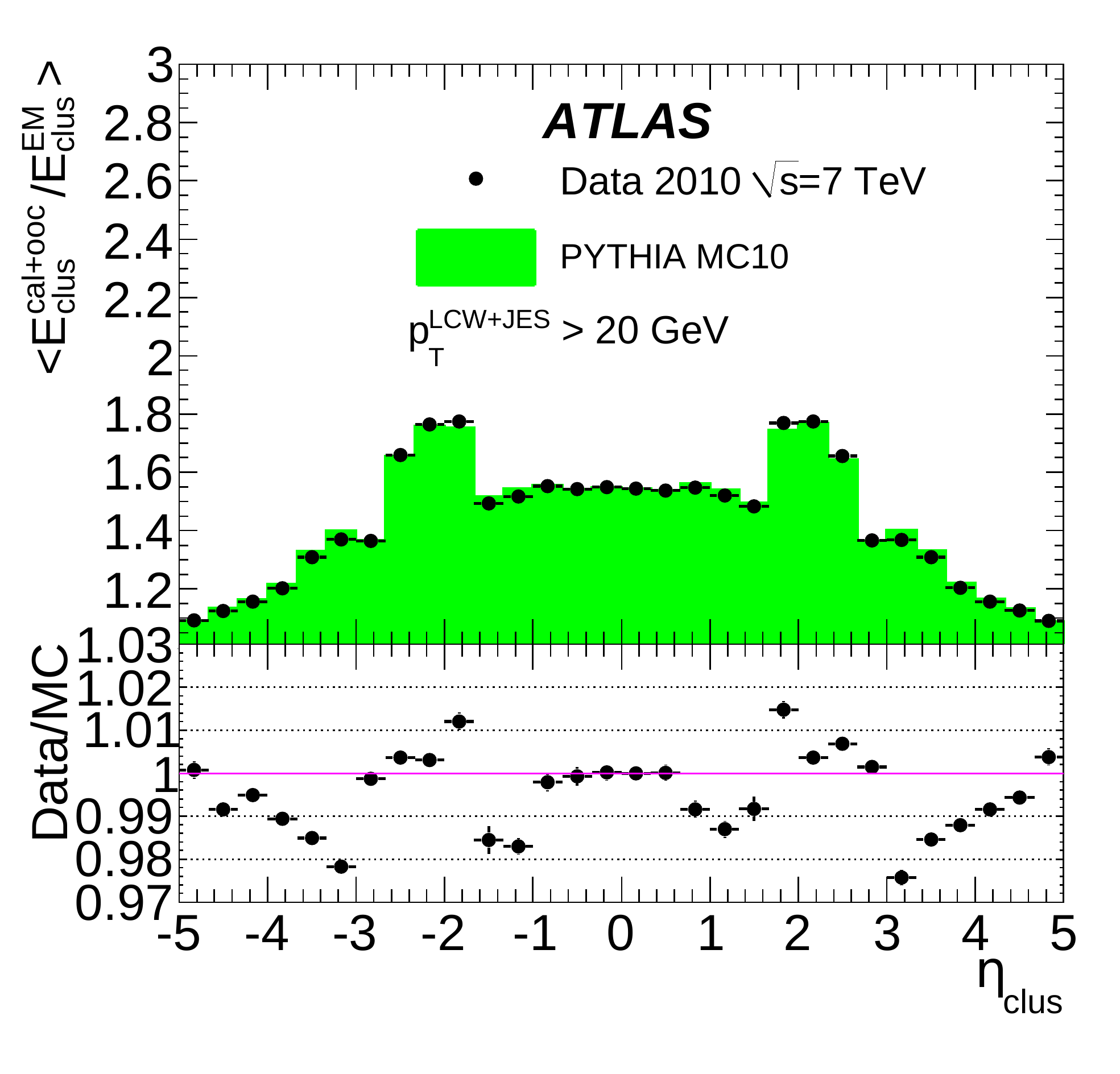}\label{fig:cluscal:eta:cal_ooc}}
        \\      
	\subfloat[$\ecluslcw/\eclusem$ versus \eclusem]{\includegraphics[width=\figsixpanelwidth]{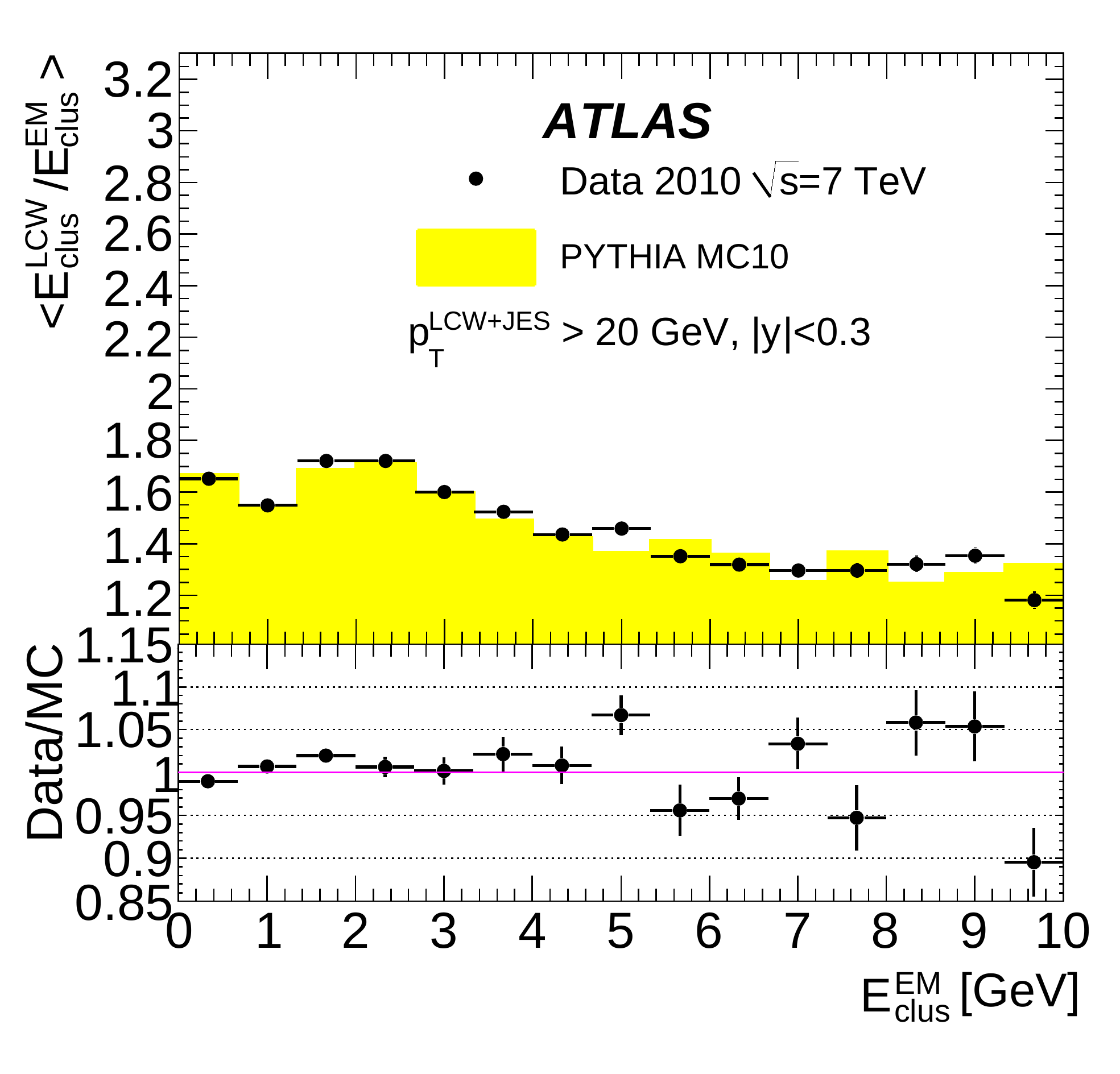}\label{fig:cluscal:e:lcw}}\qquad
	\subfloat[$\ecluslcw/\eclusem$ versus \etaclus]{\includegraphics[width=\figsixpanelwidth]{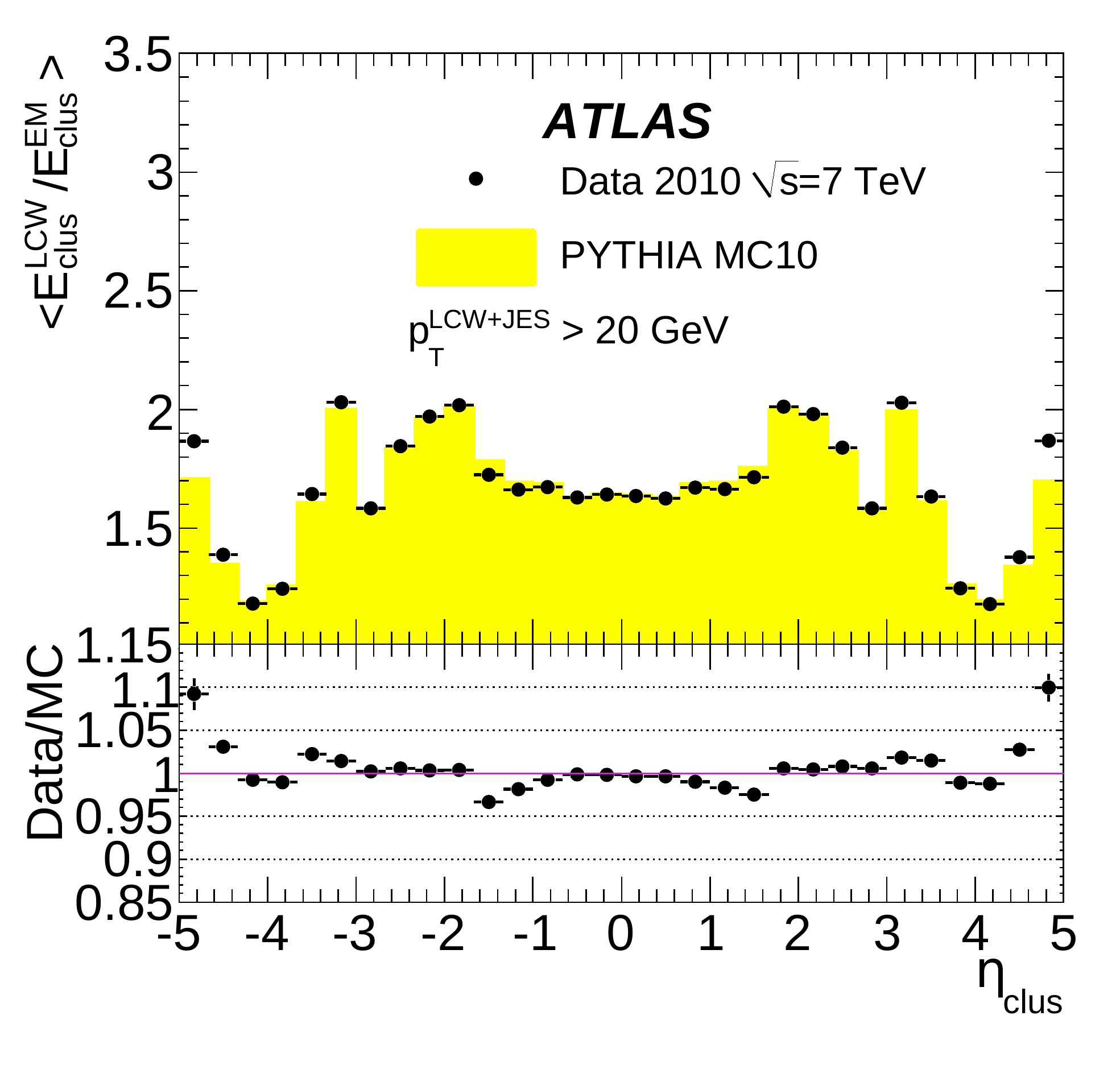}\label{fig:cluscal:eta:lcw}}
\caption[]{The average ratio of reconstructed to \EM-scale energy after each calibration step, as a function of the cluster energy \eclusem{} (\subref{fig:cluscal:e:cal}, \subref{fig:cluscal:e:cal_ooc}, and \subref{fig:cluscal:e:lcw}) for \topos{} in \antikt{} jets with $R =0.6$ and $\pT > 20$ \GeV{} and with rapidities $|y_{\text{jet}}| < 0.3$.  The corresponding average ratios as a function of \etaclus{} are shown in \subref{fig:cluscal:eta:cal}, \subref{fig:cluscal:eta:cal_ooc}, and \subref{fig:cluscal:eta:lcw}. Data recorded in 2010 is compared to the corresponding \MC{} simulations. The figures are adapted from \citRef{Aad:2011he}.}
\label{fig:cluscal}
\end{figure}

All input parameter values used in the \LCW{} calibration are derived from dedicated single-particle \MC{} simulations. 
The validity of this calibration is confirmed with data, where the cumulative effect of the hadronic calibration and the \ooc{} and \dm{} corrections on the  signal of \topos{} found in jets is analysed and compared to corresponding \MC{} simulations. 
\FigRef{fig:cluscal} summarises the quality of the \LCW{} calibration for these clusters, both as a function of the basic cluster signal \eclusem{} and the cluster direction \etaclus{} \cite{Aad:2011he}. Data are compared to \MC{} simulations after the application of the hadronic cell weights ($\ecluscal/\eclusem$ in \figMultiRefLabel~\ref{fig:cluscal}\subref{fig:cluscal:e:cal} and \ref{fig:cluscal}\subref{fig:cluscal:eta:cal}), followed by
the \ooc{} correction ($E^{\calib+\oocab}_{\clus}/\eclusem$ in \figMultiRefLabel~\ref{fig:cluscal}\subref{fig:cluscal:e:cal_ooc} and \ref{fig:cluscal}\subref{fig:cluscal:eta:cal_ooc}), and at the
\LCW{} scale after applying the \dm{} correction ($\ecluslcw/\eclusem$ in \figMultiRefLabel~\ref{fig:cluscal}\subref{fig:cluscal:e:lcw} and \ref{fig:cluscal}\subref{fig:cluscal:eta:lcw}).
The differences between data and \MC{} simulations are determined from these results as functions of \eclusem{} and \etaclus{} using the respective double-ratio 
\begin{displaymath}
	\dfrac{\left\langle\ecluscal/\eclusem\right\rangle_{\text{data}}}{\left\langle\ecluscal/\eclusem\right\rangle_{\text{\MC}}},\quad \dfrac{\left\langle E^{\calib+\oocab}_{\clus}/\eclusem\right\rangle_{\text{data}}}{\left\langle E^{\calib+\oocab}_{\clus}/\eclusem\right\rangle_{\text{\MC}}}, \quad \text{and} \quad \dfrac{\left\langle\ecluslcw/\eclusem\right\rangle_{\text{data}}}{\left\langle\ecluslcw/\eclusem\right\rangle_{\text{\MC}}}\,.
\end{displaymath} 
These double-ratios are shown in \figRef{fig:cluscal} as well, and indicate generally good agreement 
between data and \MC{} simulations. 
The particular structures shown in the \etaclus{} dependence of the magnitude of the various calibration steps indicate the cumulative effects of transition regions between calorimeters in \ATLAS, due to not only technology changes but also to changes in the \readout{} granularity. 
Especially \figRefLabel{}~\ref{fig:cluscal}\subref{fig:cluscal:eta:lcw} shows the large correction factors applied by the \LCW{} calibration in the attempt to recover signal losses introduced by (1) the transition between the central and the \EndCap{} calorimeters at $|\eta| \approx 1.45$, (2) the transition between \EndCap{} and forward calorimeters at $|\eta|\approx 3.2$, and (3) the upper limit of the \ATLAS{} calorimeter acceptance at $|\eta|\approx 4.9$.  

\renewcommand{\baselabel}{sec:perf}
\renewcommand{\thislabel}{\baselabel}
\section{Performance of the simulation of \topo{} kinematics and properties} \label{sec:perf}
The reconstruction performance of the topological cell clustering algorithm in \ATLAS{} can be evaluated in the context of
reconstructed physics objects such as jets or (isolated) single particles. In addition, features of the \topo{} signal outside these physics objects
can be studied with exclusive samples of low-multiplicity final states without jets.
These are preferably selected by muons as those 
leave only small signals in the calorimeter, nearly independent of their \pT{} (\Wlnu{} or \Zmumu{} without jets).
The \topos{} not used in reconstructing hard physics objects reflect the calorimeter sensitivity to small and dispersed energy flows generated by the \pp{} collisions in the \LHC, including \pu. The level of agreement between data and \MC{} simulations is used in all cases as a metric for the reconstruction performance.

\subsection{Single-particle response} \label{\thislabel:sp}

The calorimeter response to single isolated charged hadrons with well-measured momentum in the ID was determined using \pp{} collision data at $\sqrt{s} = 900$ \GeV{} in 2009  \cite{Aad:2012vm}. The single-hadron response at higher centre-of-mass energies was determined in 2010 at $\sqrt{s} = \unit{7}{\TeV}$ and in 2012 at $\sqrt{s} = \unit{8}{\TeV}$ \cite{Aaboud:2016hwh}.
Due to the relatively low luminosities in the 2009 and 2010 run periods, \pu{} contributions are insignificant in the corresponding data. These measurements provide important validations of the \topo{} algorithm and the calorimeter acceptance in general.   

\begin{figure}[t!] \centering
	\subfloat[$|\eta|<0.6$]{\includegraphics[width=\fighalfwidth]{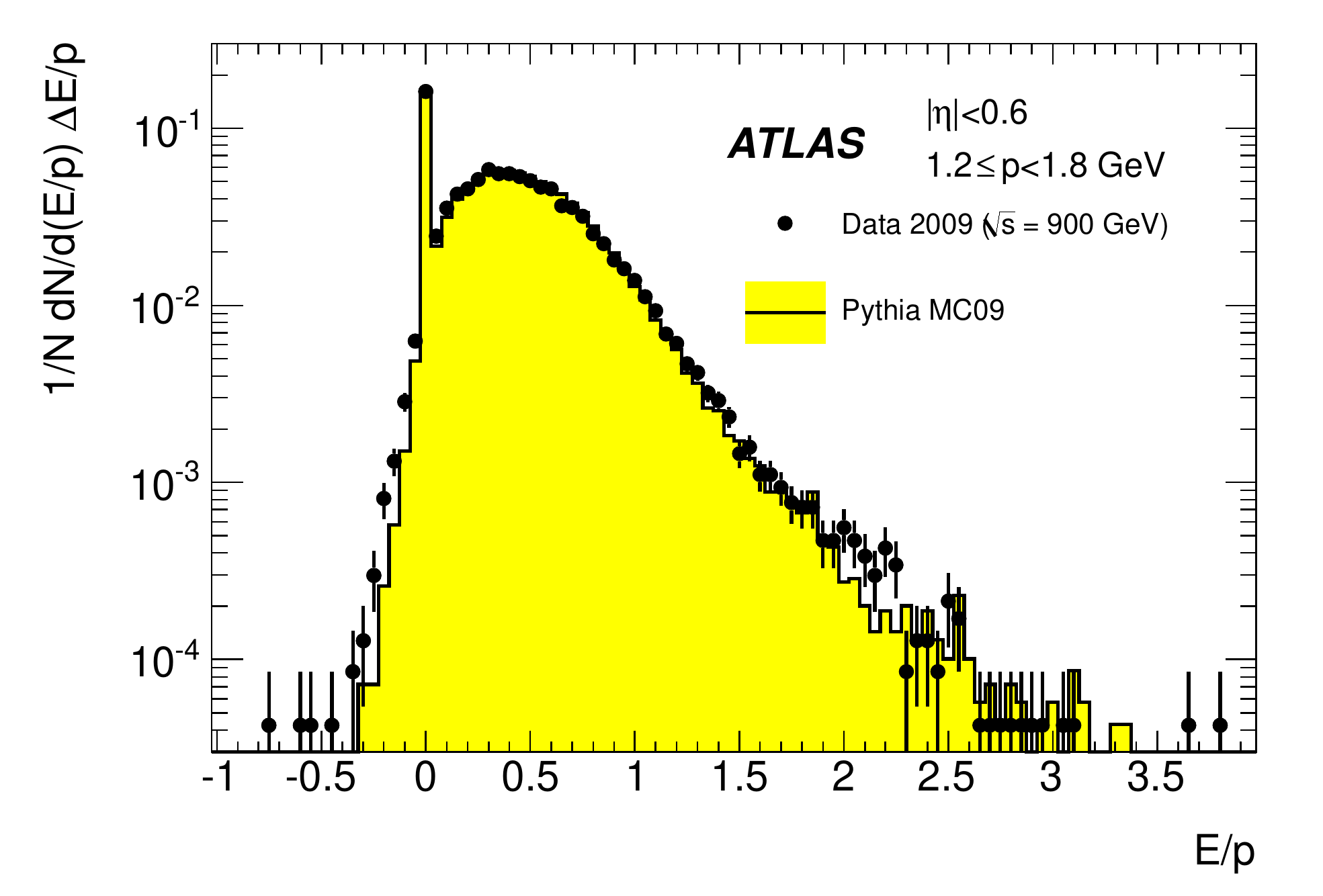}\label{fig:epdist:900:central}}
	\subfloat[$1.9<|\eta|<2.3$]{\includegraphics[width=\fighalfwidth]{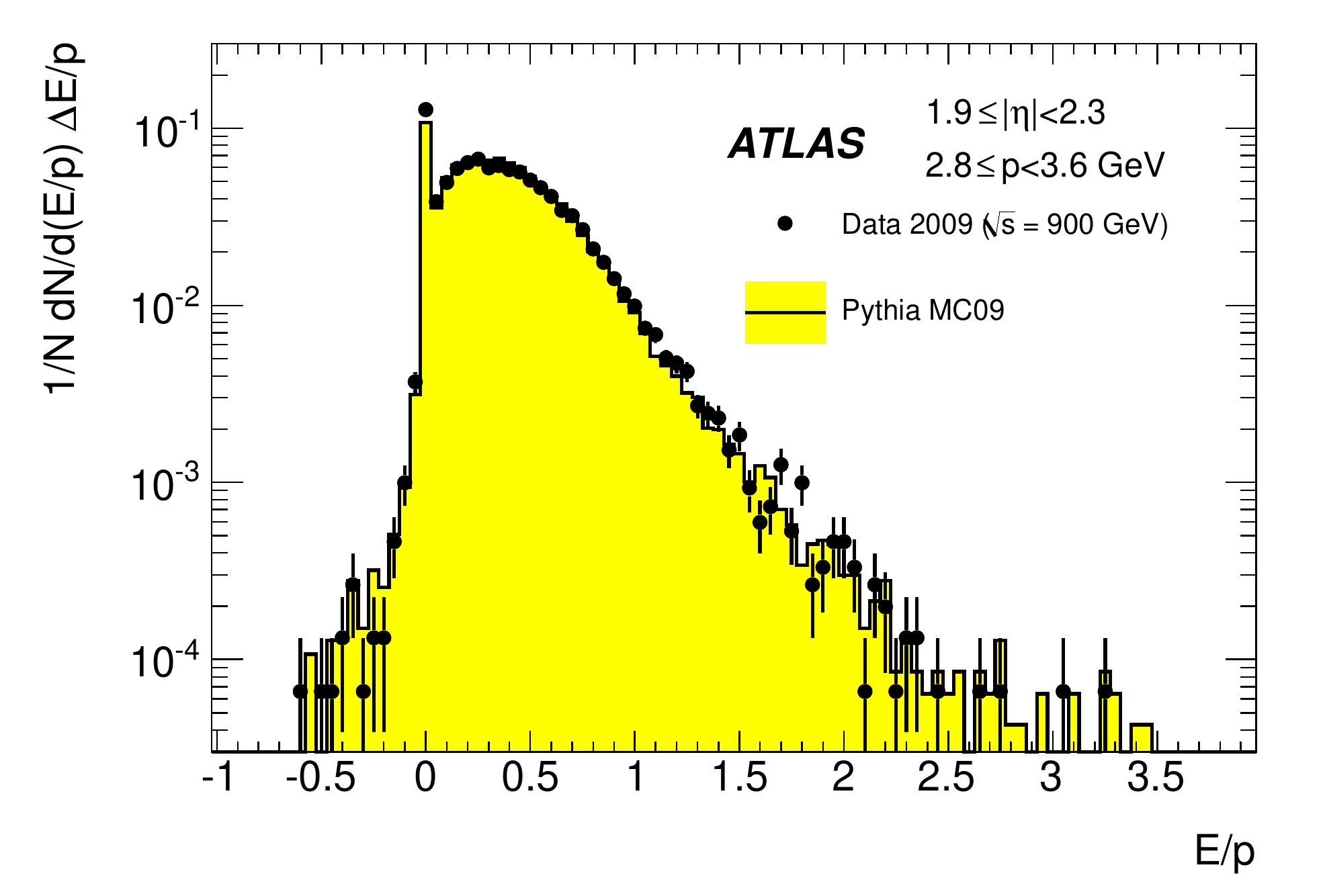}\label{fig:epdist:900:endcap}}
\caption[]{The distribution of \eoverp, the ratio of the calorimeter energy $E$ and the track momentum $p$, for \subref{fig:epdist:900:central} central tracks with $\unit{1.2}{\GeV} < p < \unit{1.8}{\GeV}$ and \subref{fig:epdist:900:endcap} forward-going tracks with $\unit{2.8}{\GeV} < p < \unit{3.6}{\GeV}$, for data and \MC{} simulations of \pp{} collisions at $\sqrt{s} = \unit{900}{\GeV}$ and no \pu{} (from \citRef{Aad:2011he}).}
\label{fig:epdist:900}
\end{figure}

The principal observable is the energy-to-momentum ratio \eoverp. The calorimeter energy $E$ is reconstructed using the \topos{} located around the direction of the track of the incoming charged particle with momentum $p$, including the ones with $\eclusem < 0$.
The effect of the axial magnetic field is taken into account by extrapolating the reconstructed tracks into the calorimeter.
The energy $E$ is then calculated by summing the \EM-scale energies from all sampling layers $s$ of \topos{} which have a barycentre $(\eta_{s},\phi_{s})$ within $\Delta R = 0.2$ of the track direction extrapolated to each $s$, as described in more detail in \citRef{Aad:2012vm}. 
The sampling layer energies are summed irrespective of their sign, i.e. $E<0$ is possible. 

The results of the measurement of \eoverp{} are shown in \figMultiRefLabel{} \ref{fig:epdist:900}\subref{fig:epdist:900:central} and \ref{fig:epdist:900}\subref{fig:epdist:900:endcap} for reconstructed isolated tracks in \pp{} collisions at $\sqrt{s} = 900$ \GeV. 
The distributions reflect the acceptance of the calorimeter for charged particles in the given momentum ranges. Entries for $\eoverp < 0$ indicate that the incoming track is matched with a \topo{} generated by significant electronic noise. The number of tracks with no matching calorimeter signal ($E = 0 \Rightarrow \eoverp = 0$) is indicative of none or only a small fraction of the particle energy reaching the calorimeter, and the signal generated by this energy fraction is not sufficiently significant to 
survive the implicit noise suppression in the \topo{} formation described in \secRef{sec:topos:formation}.

\begin{figure}[t!] \centering
	\subfloat[]{\includegraphics[width=\fighalfwidth]{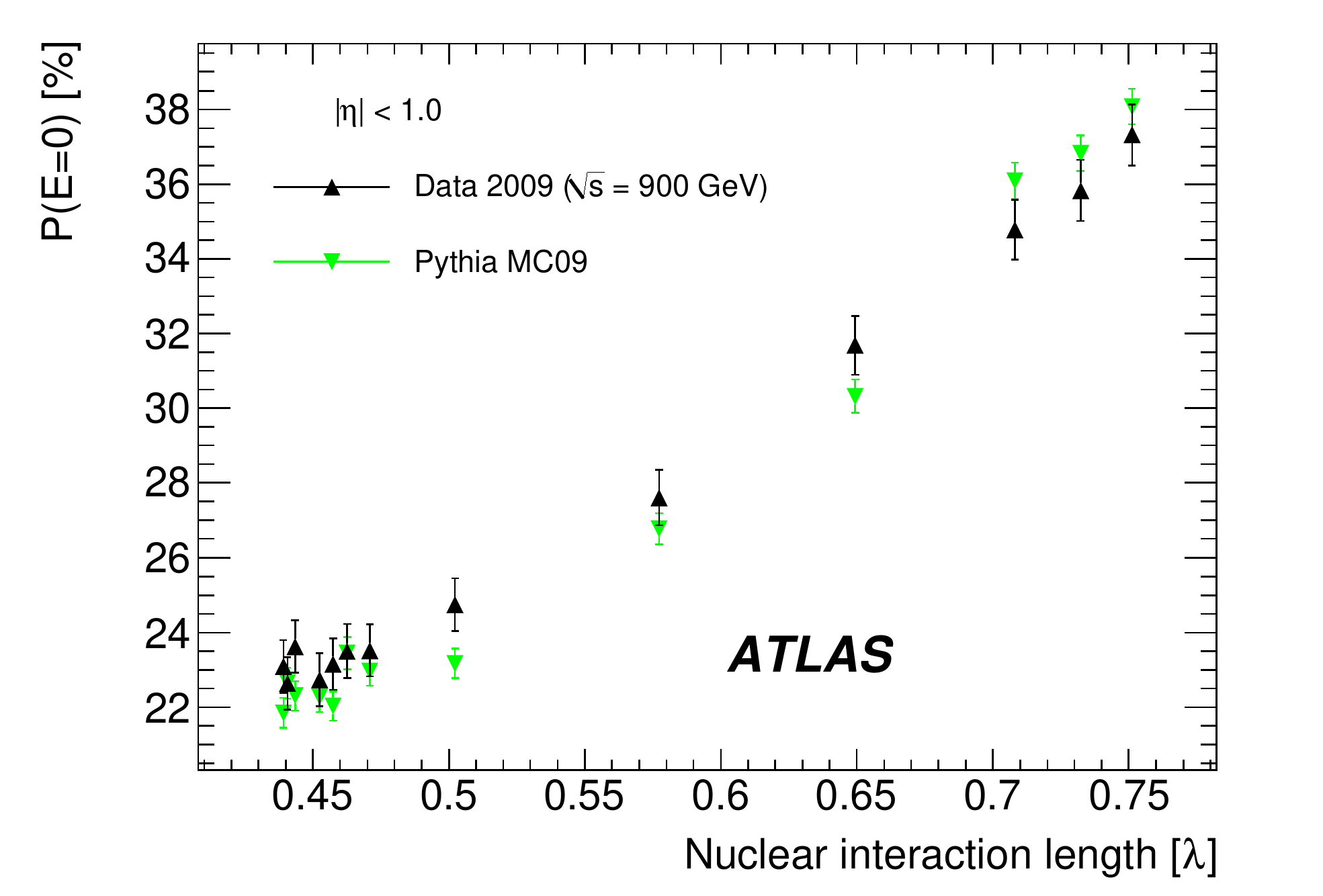}\label{fig:epave:900:lambda}}
	\subfloat[]{\includegraphics[width=\fighalfwidth]{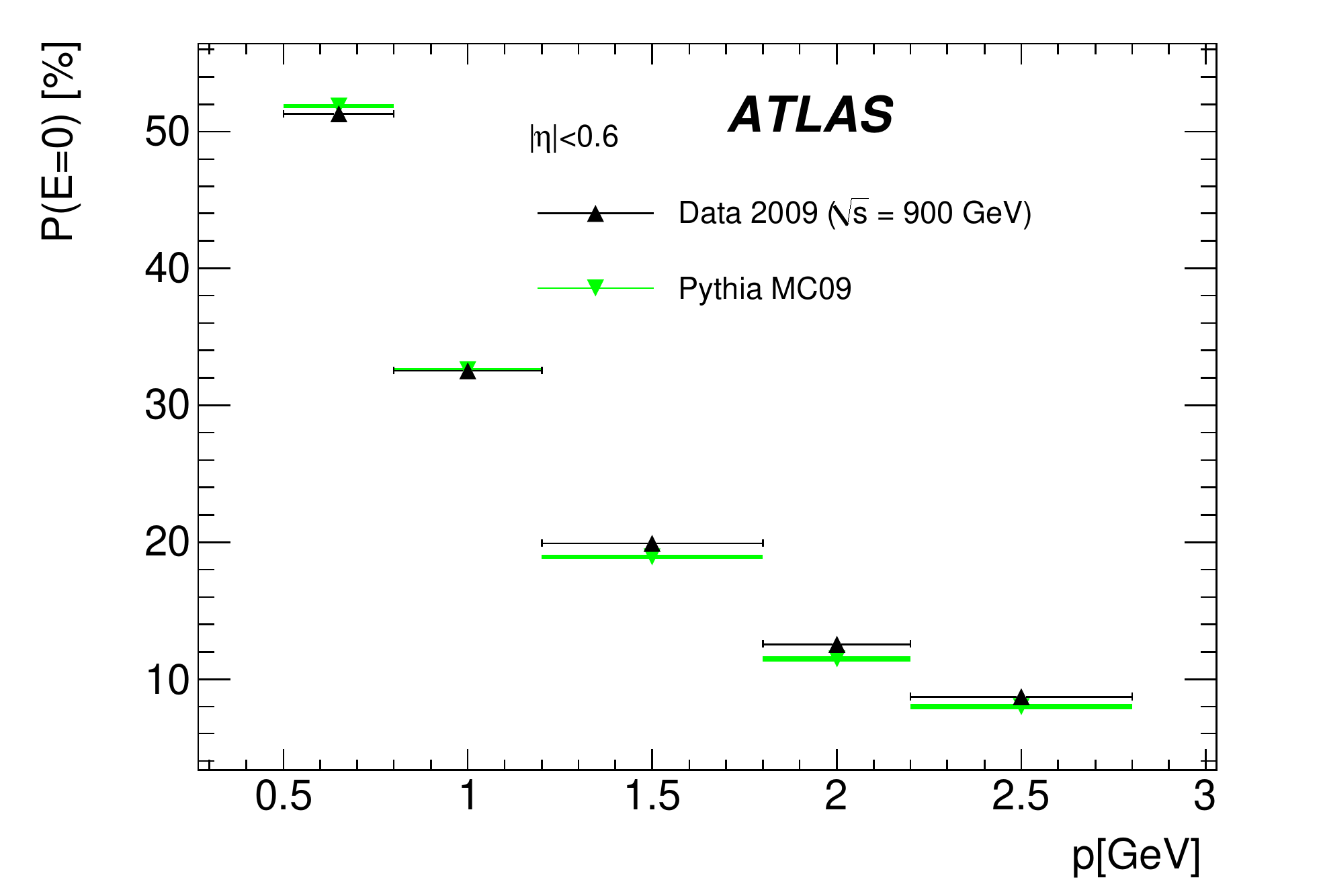}\label{fig:epave:900:p}}
	\caption[]{In \subref{fig:epave:900:lambda}, the likelihood $\mathcal{P}_{E = 0}(\ddm)$  to find no matching energy in the calorimeter ($E = 0$) for reconstructed isolated charged-particle tracks is shown as a function of the
thickness \ddm{} of the inactive material in front of the calorimeter, for data and \MC{} simulations in \pp{} collisions at $\sqrt{s} = \unit{900}{\GeV}$. The thickness of the inactive material is measured in terms of the nuclear interaction length \lamnucl. The tracks are reconstructed within $|\eta| < 1$. The likelihood to reconstruct $E = 0$ as a function of the incoming track momentum is shown for the same data and \MC{} simulations in \subref{fig:epave:900:p}, for reconstructed tracks within $|\eta| < 0.6$. Both figures are taken from \citRef{Aad:2011he}.} 
\label{fig:epave:900}
\end{figure}

The likelihood $\mathcal{P}_{E=0}(\ddm)$ to find $E = 0$ for a charged particle passing through inactive material of various thicknesses \ddm, measured in terms of the nuclear interaction length \lamnucl, is shown in  
\figRef{fig:epave:900}\subref{fig:epave:900:lambda} for isolated tracks within $|\eta| < 1.0$ in \pp{} collisions at $\sqrt{s} = \unit{900}{\GeV}$. 
The various values of \ddm{} are extracted from the detector description in the \MC{} simulation using the direction $|\eta|$ of the incoming tracks.
The data and \MC{} simulations agree well, indicating 
an appropriate 
description of the actual detector geometry in the \MC{} simulation. The likelihood to have no matching signal in the  calorimeter shows the expected increase with increasing inactive material.
 
The dependence of $\mathcal{P}_{E=0}$ 
on the track momentum is shown in \figRef{fig:epave:900}\subref{fig:epave:900:p} for
isolated tracks with $|\eta| < 0.6$. 
Good agreement between data and \MC{} simulations is observed, which together with the 
results displayed in \figMultiRefLabel~\ref{fig:epdist:900} and \ref{fig:epave:900}\subref{fig:epave:900:lambda} indicates 
a good description of the data by the QGSP\_BERT hadronic shower model used by the \MC{} simulation. 

\begin{figure}[t!] \centering
	\subfloat[$|\eta| < 0.6$]{\includegraphics[width=\fighalfwidth]{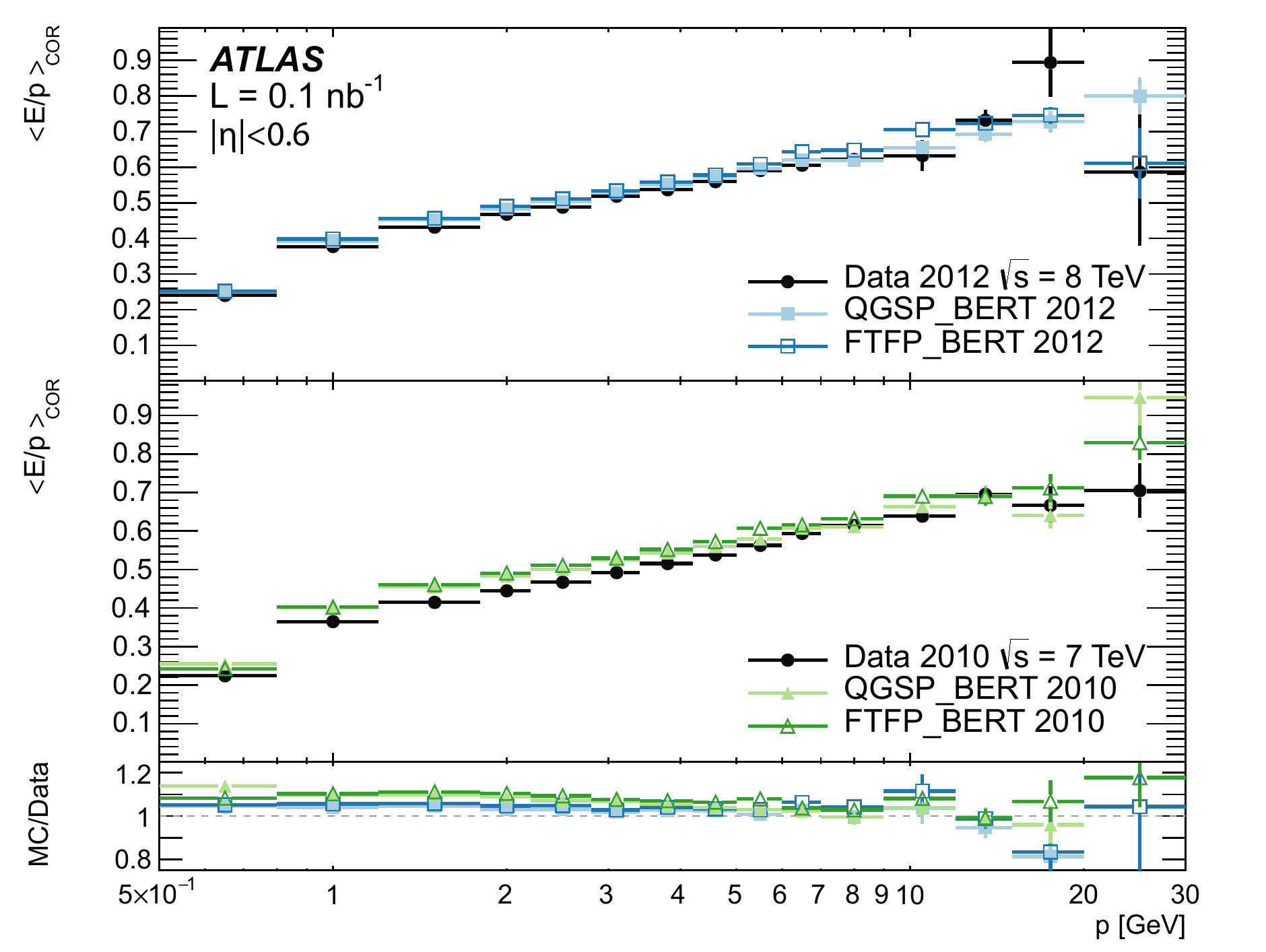}\label{fig:epall:central}}
	\subfloat[$1.9 < |\eta| < 2.3$]{\includegraphics[width=\fighalfwidth]{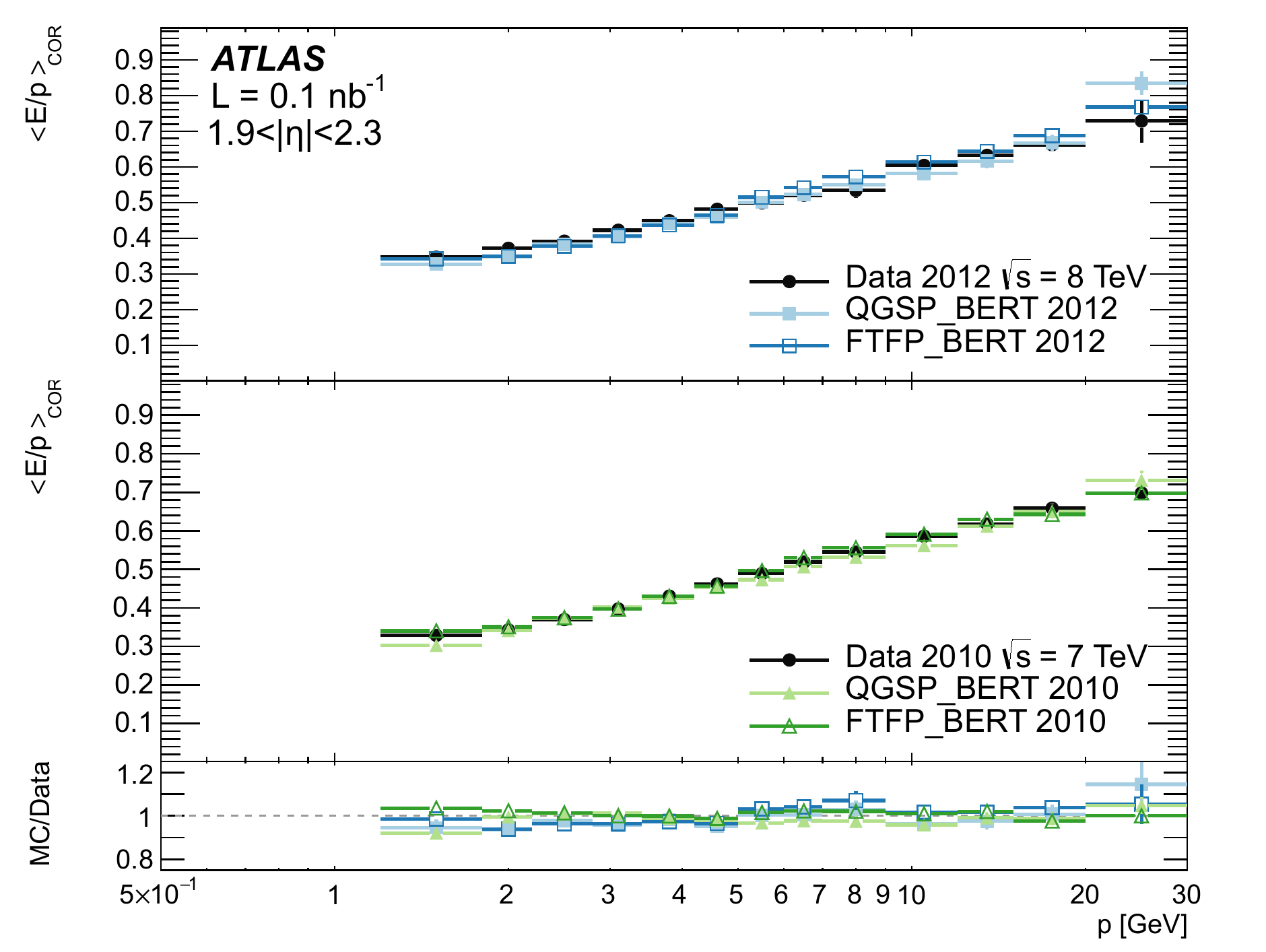}\label{fig:epall:forward}}
	\caption[]{The average $\langle\eoverp\rangle$ ratio as a function of the track momentum $p$, for \subref{fig:epall:central} tracks within $|\eta|<0.6$ and \subref{fig:epall:forward} tracks within $1.9 < |\eta| < 2.3$. Data from isolated tracks recorded in 2010 and 2012 with insignificant \pu{} are shown together with \MC{} simulations employing two different hadronic shower models. 
}
	\label{fig:epall}
\end{figure}

The dependence of \eoverp{} on the track momentum has been evaluated for two different hadronic 
shower models in \geant. In addition to the default QGSP\_BERT model introduced in \secRef{sec:atlas:mc:det}, the \fritiof{} model \cite{Andersson:1986gw,NilssonAlmqvist:1986rx} is considered together with the Bertini intra-nuclear cascade to simulate hadronic showers (FTFP\_BERT). 
The results for 2012 data from a dedicated sample with insignificant
\pu{} ($\mu \approx 0$) are presented in \figRef{fig:epall} and show good agreement between data and \MC{} simulations
without indicating a strong preference for one of the hadronic shower models. 
More results of the full systematic evaluation of the  \topo{} response to single charged hadron tracks, including for selected tracks from identified charged mesons and baryons, are available in \citRef{Aad:2012vm}.

\subsection{Effect of \pu{} on \topo{} observables} \label{\thislabel:obs}
 
\begin{figure}[tp!] \centering
        \sfcompress        
	\subfloat[]{\includegraphics[width=\figsixpanelwidth]{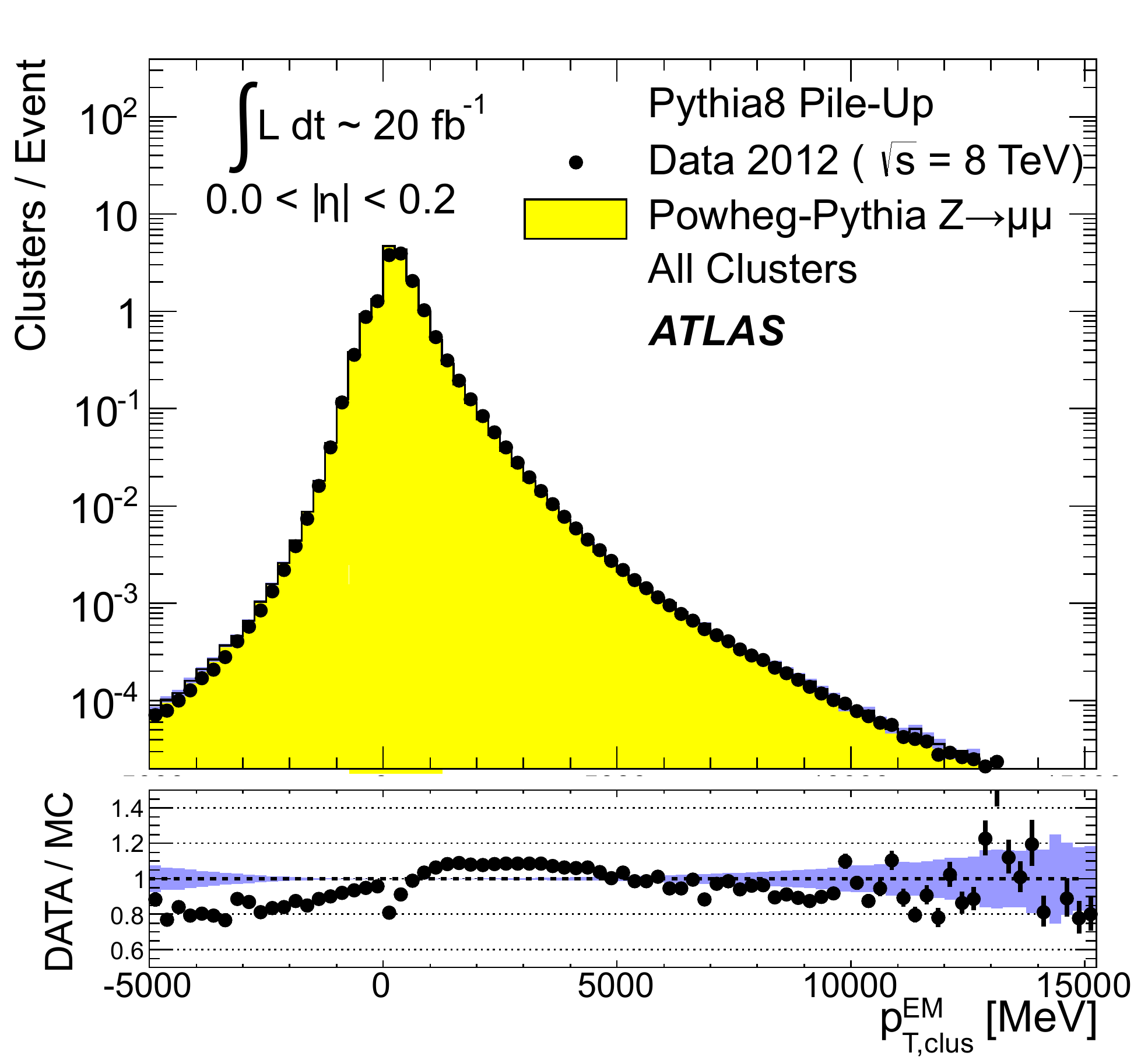}\label{fig:spectra:pt:00_02:mc}} \qquad
	\subfloat[]{\includegraphics[width=\figsixpanelwidth]{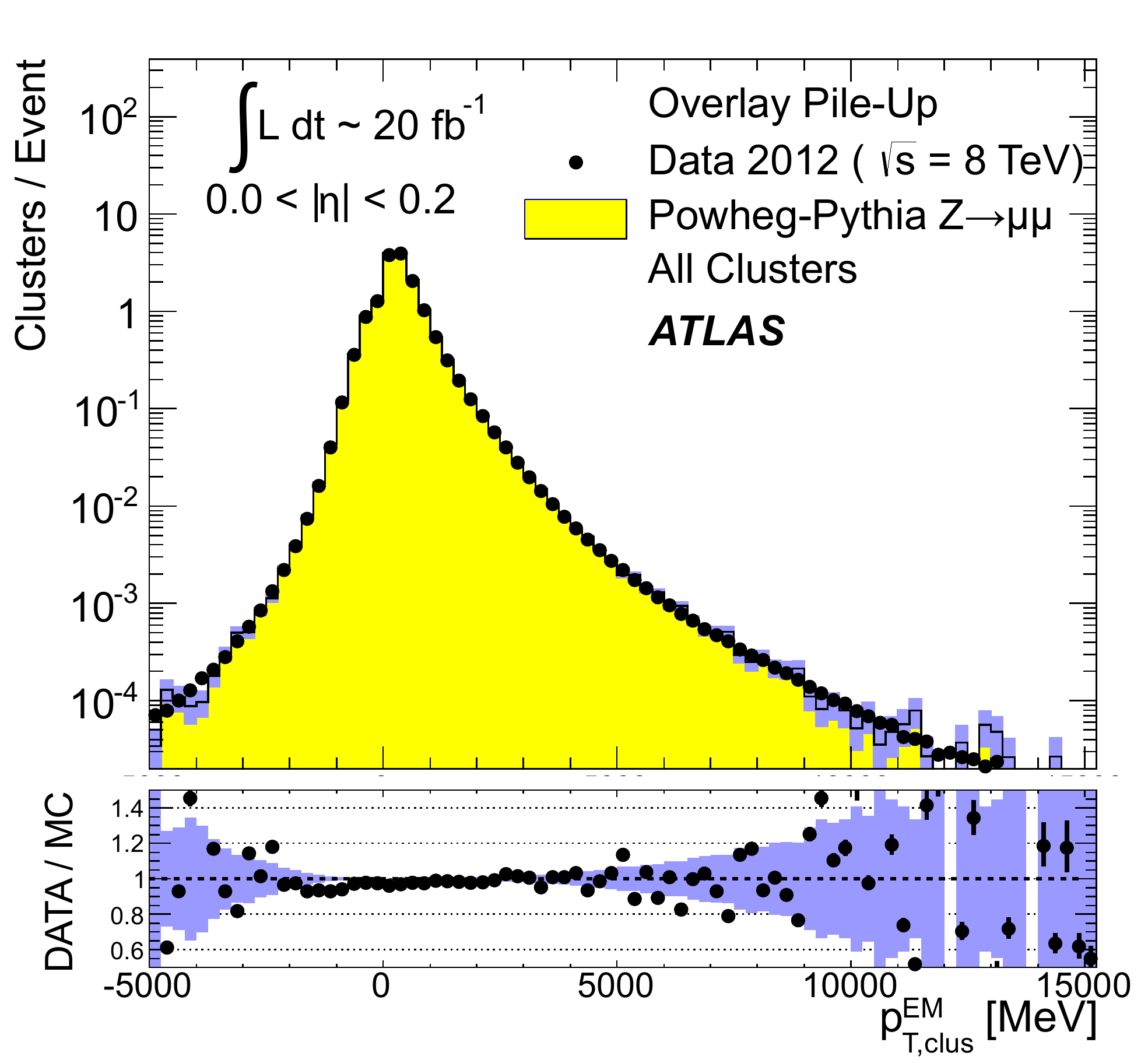}\label{fig:spectra:pt:00_02:ov}} 
        \\
	\subfloat[]{\includegraphics[width=\figsixpanelwidth]{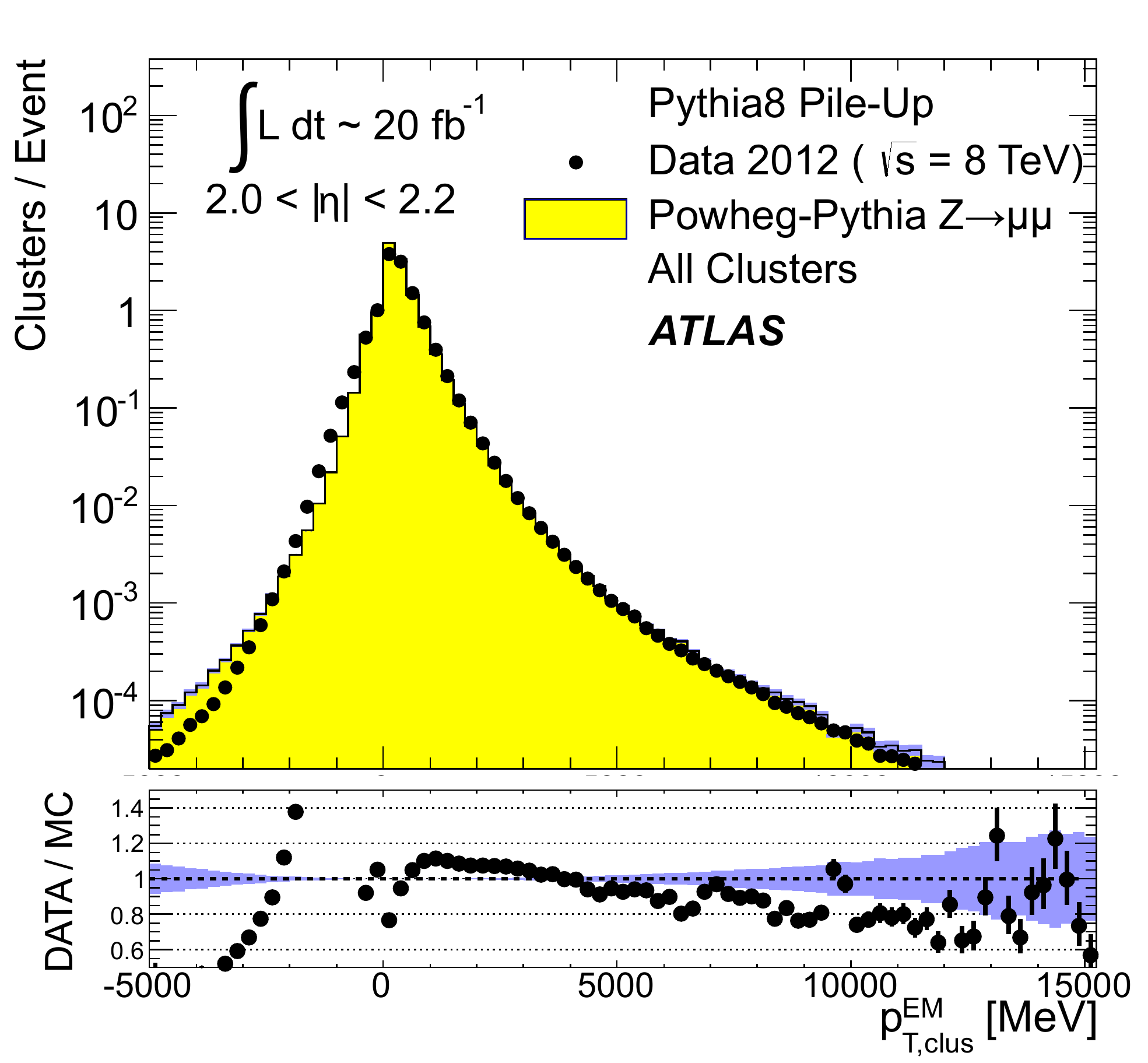}\label{fig:spectra:pt:20_22:mc}} \qquad
	\subfloat[]{\includegraphics[width=\figsixpanelwidth]{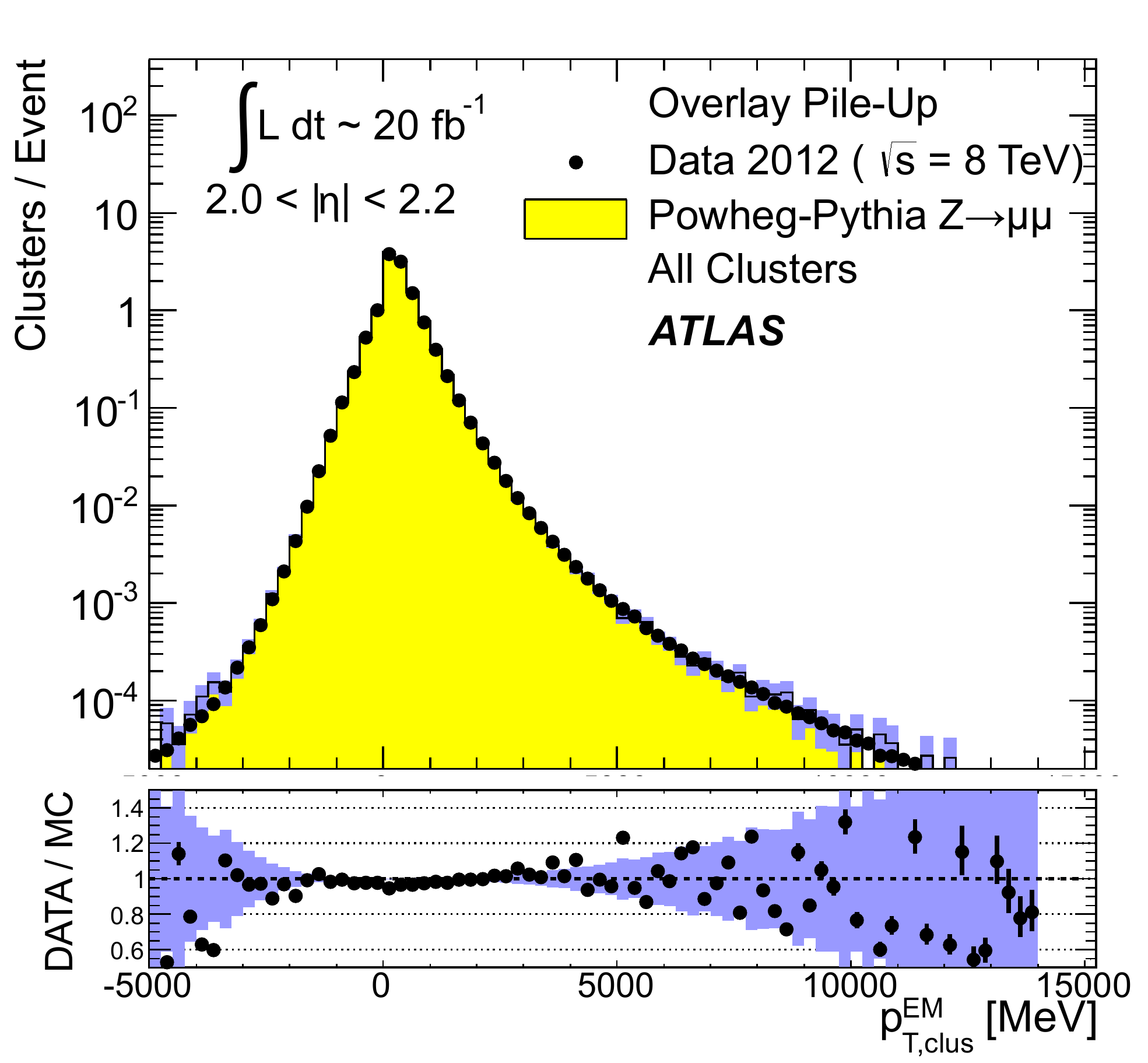}\label{fig:spectra:pt:20_22:ov}}
        \\                
	\subfloat[]{\includegraphics[width=\figsixpanelwidth]{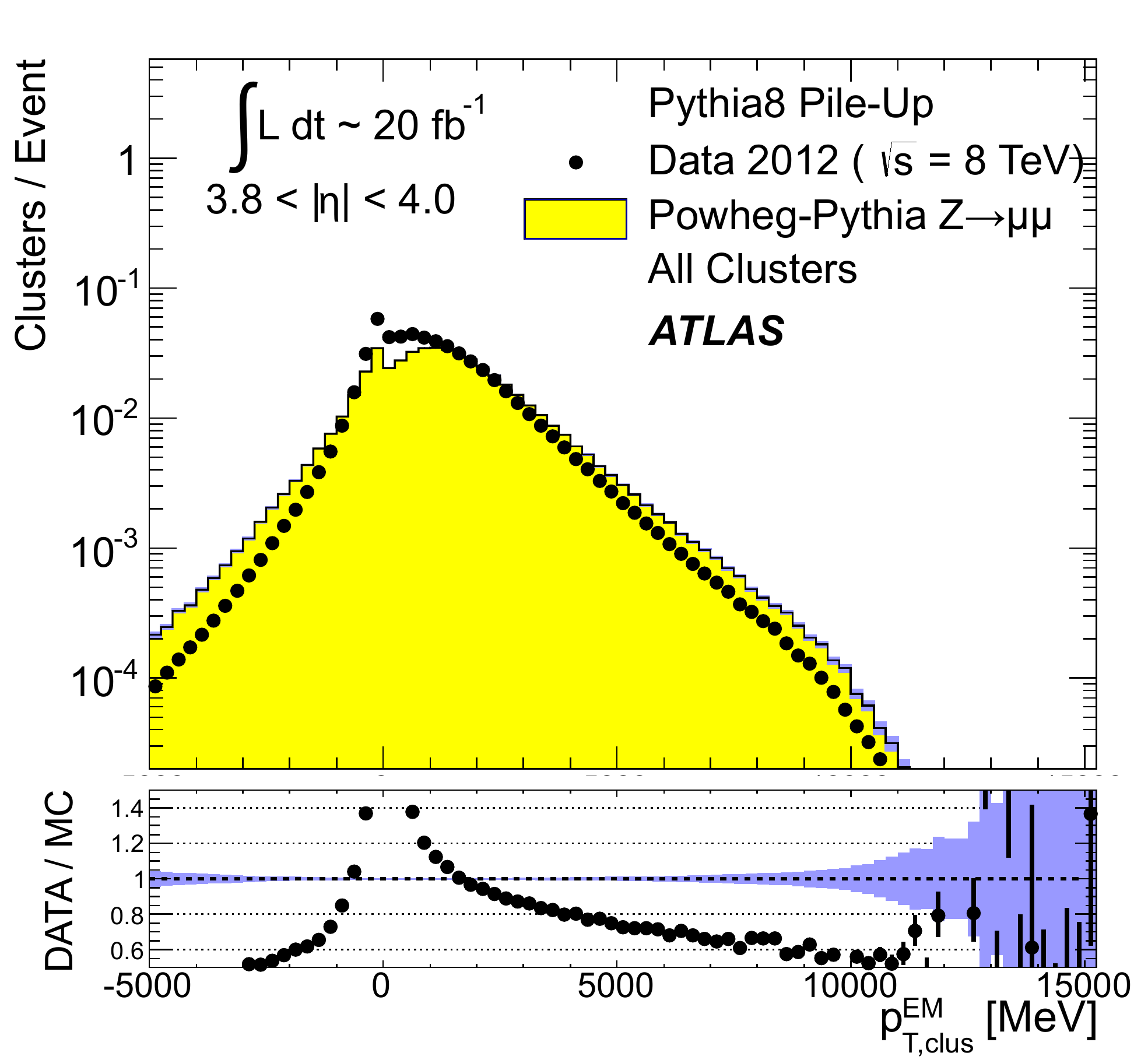}\label{fig:spectra:pt:38_40:mc}} \qquad
	\subfloat[]{\includegraphics[width=\figsixpanelwidth]{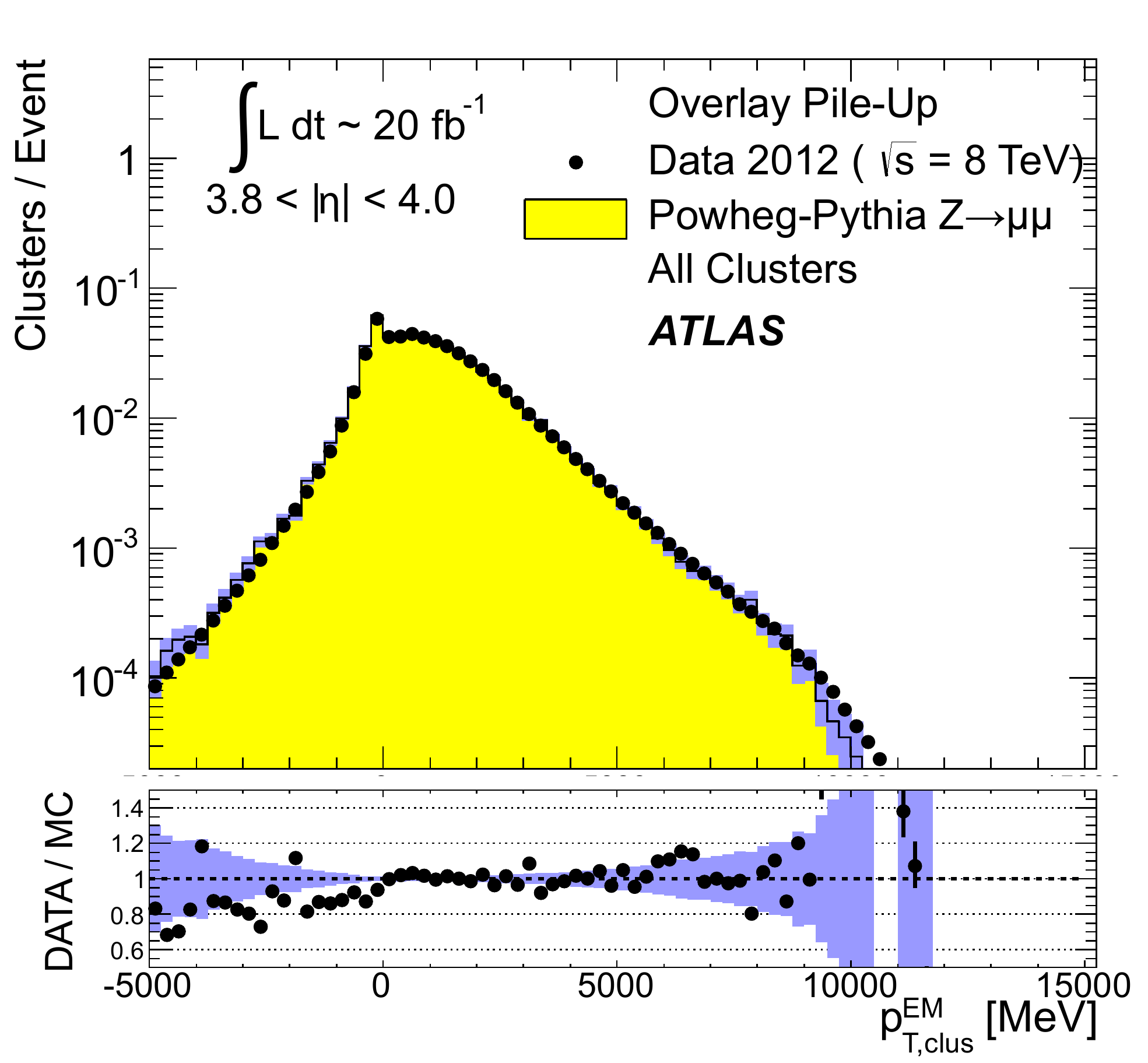}\label{fig:spectra:pt:38_40:ov}}
	\caption[]{The distribution of the transverse momentum of \topos{} reconstructed on the \EM{} scale (\ptclusem) for an inclusive \Zmumu{} event sample recorded in 2012. Data are compared to distributions from \MC{} simulations (\subref{fig:spectra:pt:00_02:mc}, \subref{fig:spectra:pt:20_22:mc}, and \subref{fig:spectra:pt:38_40:mc}) including fully simulated \pu{} and (\subref{fig:spectra:pt:00_02:ov}, \subref{fig:spectra:pt:20_22:ov}, and \subref{fig:spectra:pt:38_40:ov}) with \pu{} overlaid from data for all \topos{} within (\subref{fig:spectra:pt:00_02:mc}, \subref{fig:spectra:pt:00_02:ov}) $|\etaclus|<0.2$, (\subref{fig:spectra:pt:20_22:mc}, \subref{fig:spectra:pt:20_22:ov}) $2.0<|\etaclus|<2.2$ , (\subref{fig:spectra:pt:38_40:mc}, \subref{fig:spectra:pt:38_40:ov}) $3.8 < |\etaclus| < 4.0$. The ratio of the distribution from data to the one from \MC{} simulation is evaluated bin-by-bin and shown below the respective distribution. The shaded bands indicate the statistical uncertainties from \MC{} simulations for both the spectra and the ratios.}
\label{fig:spectra:pt}		
\end{figure}

The \topo{} reconstruction performance is affected by \intime{} and \opu. While \ipu{} is expected to usually increase the number of \topos{} 
with increasing number of reconstructed vertices (\NPV), the \opu{} leads to cluster signal and shape modifications introduced by the calorimeter signal shaping functions described in 
\secRef{sec:atlas:data:pu}. 

The high density of very significant cell signals generated inside jets in the calorimeter increases the likelihood of low-energy \pu{} signals to survive in the \topo{} formation, according to the formation rules given in \secRef{sec:topos:formation}. 
Cell signals generated by the energy flow of relatively isolated particles entering the calorimeter outside jets or (stochastic) jet-like flow structures\footnote{These can be generated by particles from different \pu{} collisions going in the same direction in a given event.}  often have less significant neighbouring cells and thus contribute less often to \topos. Consequently, the acceptance of the calorimeter for these particles, many of which are produced by \pu, is lower than for particles in or around a jet.
  
In this section the modelling of the \pu{} effects on the kinematics and moments used for the \LCW{} calibration is compared to data for \topos{} formed inside and outside jets for the conditions during 2012 running. The effect of \pu{} on jets reconstructed from \topos{} is discussed in \secRef{\thislabel:jets}, together with the stability of \topo-based observables associated with the jet and its composition.  
 

\subsubsection{Event selection} \label{\thislabel:obs:ef}

The data used for the evaluation of the \pu{} effects on \topo{} kinematics and moments are collected from \Zmumu{} events recorded in 2012. 
As indicated in \secRef{sec:atlas:det}, the corresponding sample is defined by a muon-based trigger. 
The additional event selection, applied to both data and the corresponding
\MC{} simulations, requires two muons with $\pT > \unit{25}{\GeV}$ within $|\eta| < 2.4$ and an invariant mass \twomass{\mu}{\mu} of the muon pair of $\unit{80}{\GeV} < \twomass{\mu}{\mu} < \unit{100}{\GeV}$ for the inclusive sample.  
For the analysis of an exclusive sample with softer hadronic recoil against the \Zboson{} boson transverse momentum (\pTZ), events with at least one jet reconstructed with the \antikt{} algorithm and a distance parameter $R = 0.4$ and $\pT > \unit{20}{\GeV}$ are removed. 
This sample is characterised by a final state dominated by \ipu{} signal contributions, with only a small number of \topos{} associated with the hadronic recoil.

Another exclusive sample for the analysis of \topo{} features in jets is selected by requiring at least one \antikt{} jet with $\pT > \unit{20}{\GeV}$ in the event. Like in the selection applied to collect the exclusive sample without jets, all jets are fully calibrated and corrected, including a correction for \pu{} (see \secRef{sec:atlas:jetreco}). All inclusive 
and exclusive samples are thus characterised by their stability against \pu{}.

\subsubsection{Modelling of \topo{} kinematics in events with \pu}\label{\thislabel:obs:kine}

Detailed \datatomc{} comparisons of \topo{} kinematics yield significant differences between the measured and the modelled spectra. 
The transverse momentum spectra of \topos{} reconstructed on the \EM{} scale (\ptclusem) for the final state of an inclusive \Zmumu{} sample, 
are shown in \figMultiRefLabel~\ref{fig:spectra:pt}\subref{fig:spectra:pt:00_02:mc} and \ref{fig:spectra:pt:00_02:ov} for the central,
in \figMultiRefLabel~\ref{fig:spectra:pt}\subref{fig:spectra:pt:20_22:mc} and \ref{fig:spectra:pt:20_22:ov} for the \EndCap,
and in \figMultiRefLabel~\ref{fig:spectra:pt}\subref{fig:spectra:pt:38_40:mc} and \ref{fig:spectra:pt:38_40:ov} for the
forward detector region. 
The comparison between the \ptclusem{} spectra from \MC{} simulations with fully modelled \pu{} and data in the various \etaclus{} ranges shows significant disagreements.
Possible sources 
are an imperfect detector simulation or the modelling of the
underlying soft physics processes in the \MC{} generator.

Using the data overlay method described in \secRef{sec:atlas:mc:ovly} improves the \datatomc{} comparison of the \ptclusem{} spectra significantly, especially in the low-\pT{} regime, where \pu{} is expected to have a large effect. 
This improvement can be seen in \figMultiRefLabel~\ref{fig:spectra:pt}\subref{fig:spectra:pt:00_02:ov}, \ref{fig:spectra:pt}\subref{fig:spectra:pt:20_22:ov}, and \ref{fig:spectra:pt}\subref{fig:spectra:pt:38_40:ov} for the respective \etaclus{} ranges.

\subsubsection{Transverse momentum flow in the presence of \pu} \label{\thislabel:obs:ptflow}

\begin{figure}[tp!] \centering
        \sfcompress
	\subfloat[]{\includegraphics[width=\figsixpanelwidth]{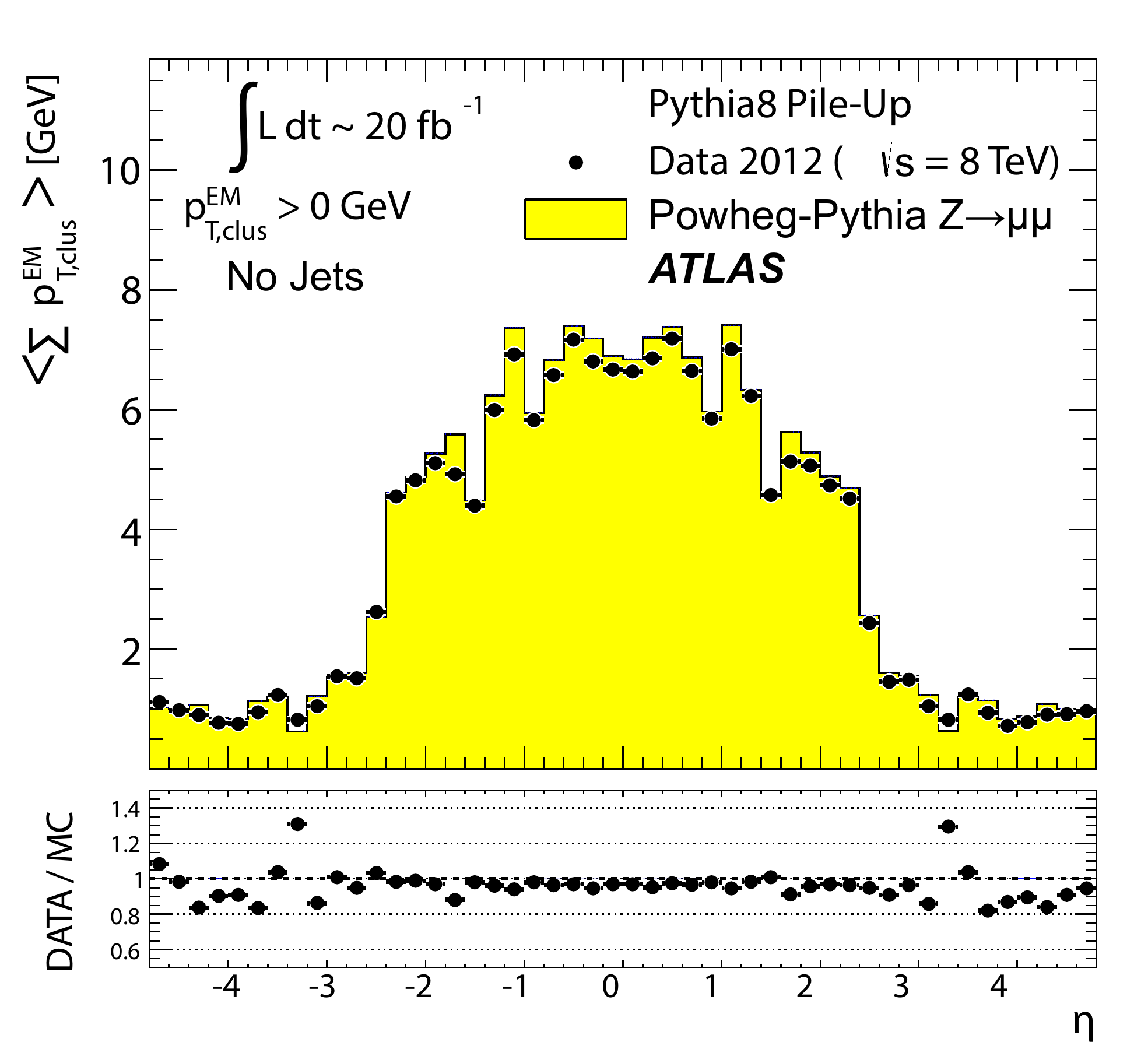}\label{fig:ptdens:all}} \qquad
	\subfloat[]{\includegraphics[width=\figsixpanelwidth]{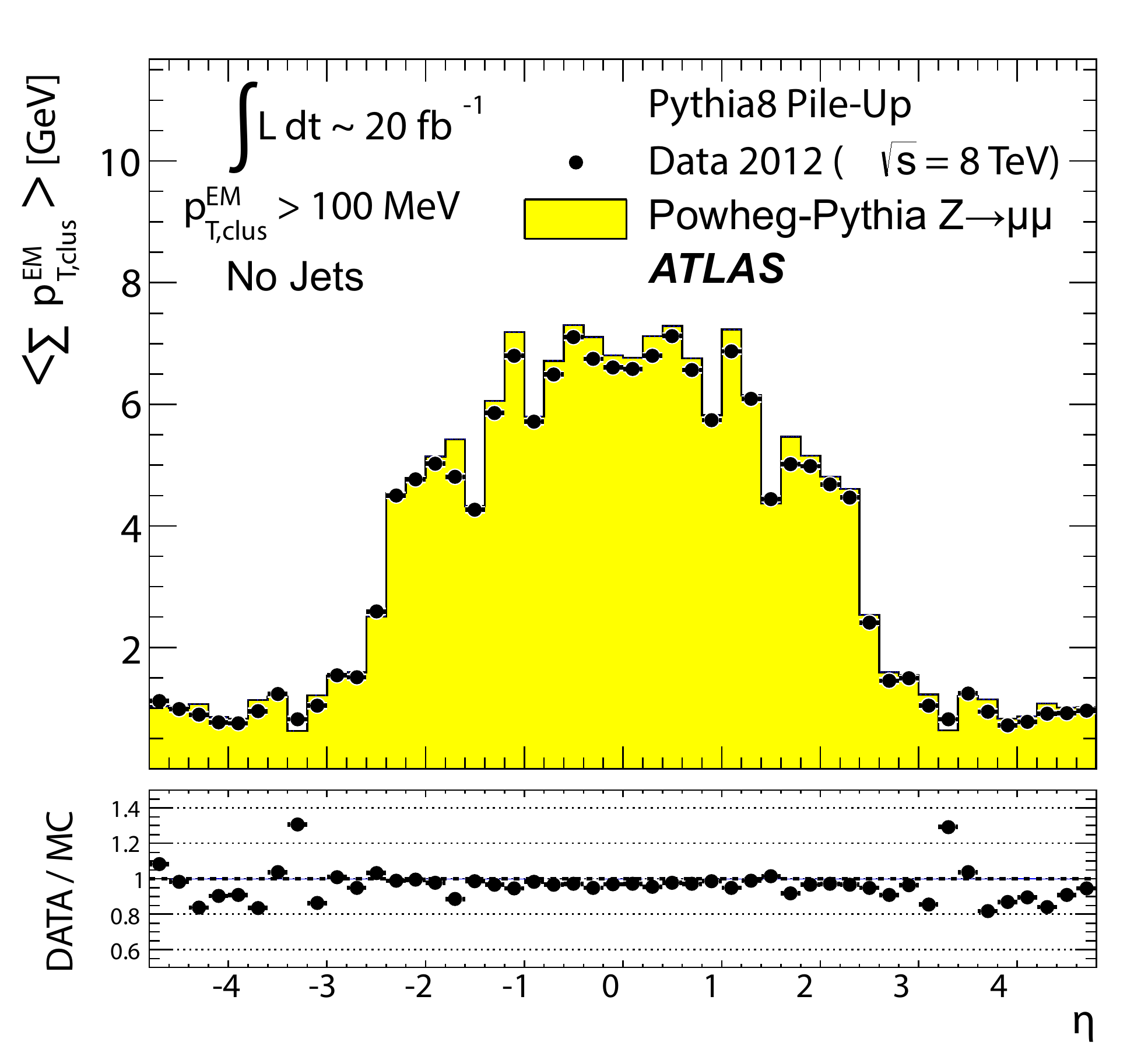}\label{fig:ptdens:100}} 
        \\
	\subfloat[]{\includegraphics[width=\figsixpanelwidth]{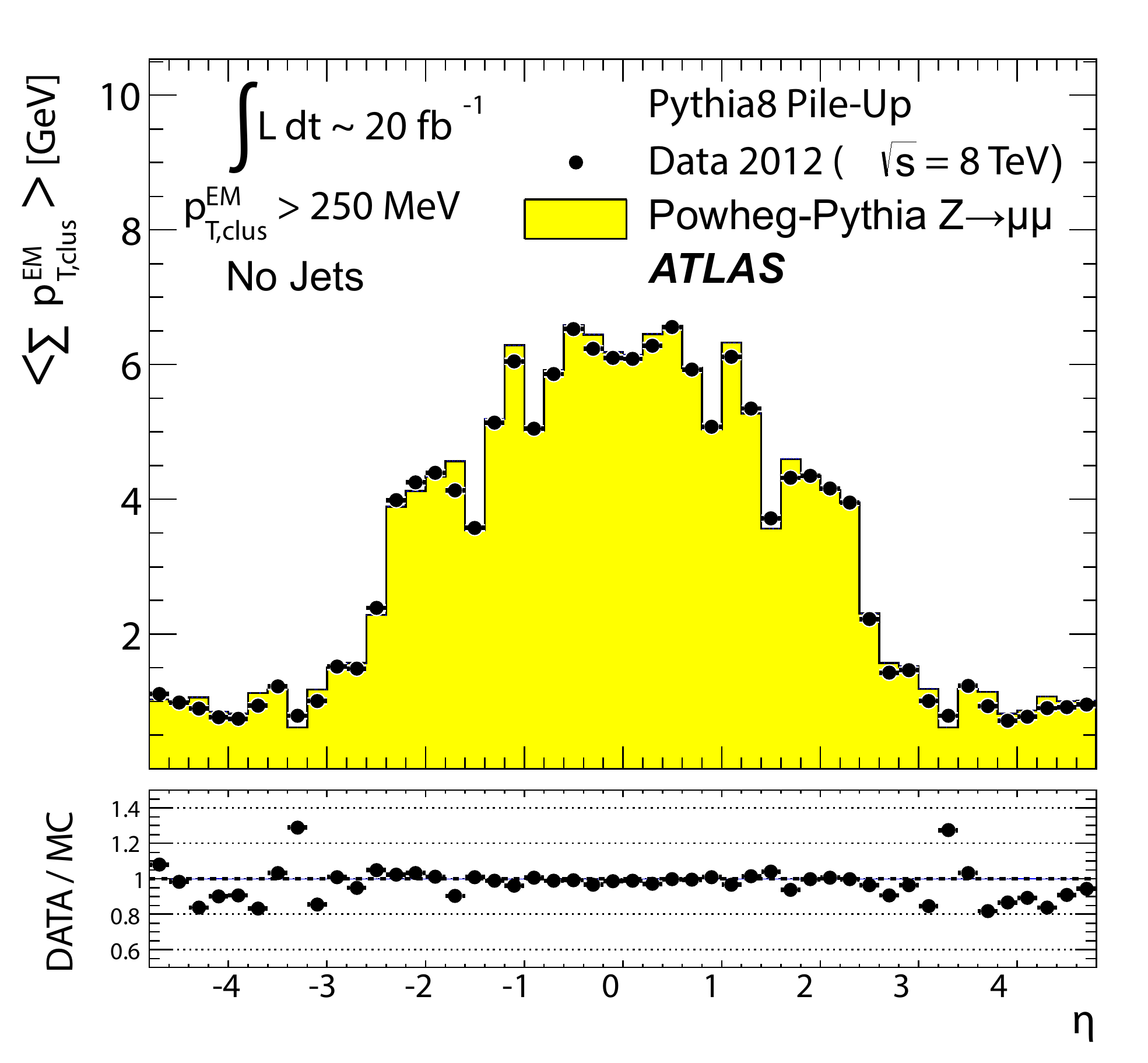}\label{fig:ptdens:250}} \qquad
	\subfloat[]{\includegraphics[width=\figsixpanelwidth]{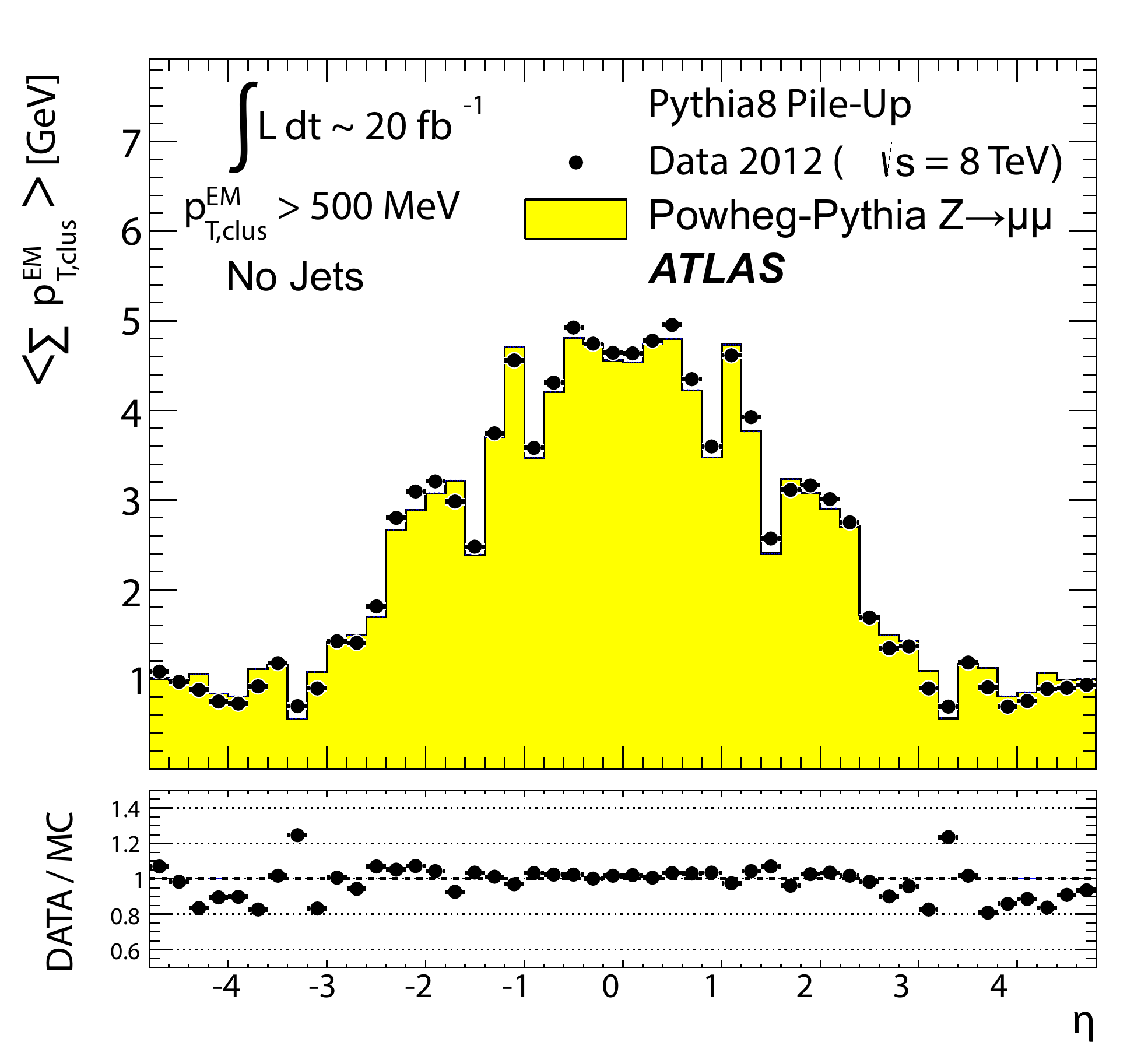}\label{fig:ptdens:500}} 
        \\
	\subfloat[]{\includegraphics[width=\figsixpanelwidth]{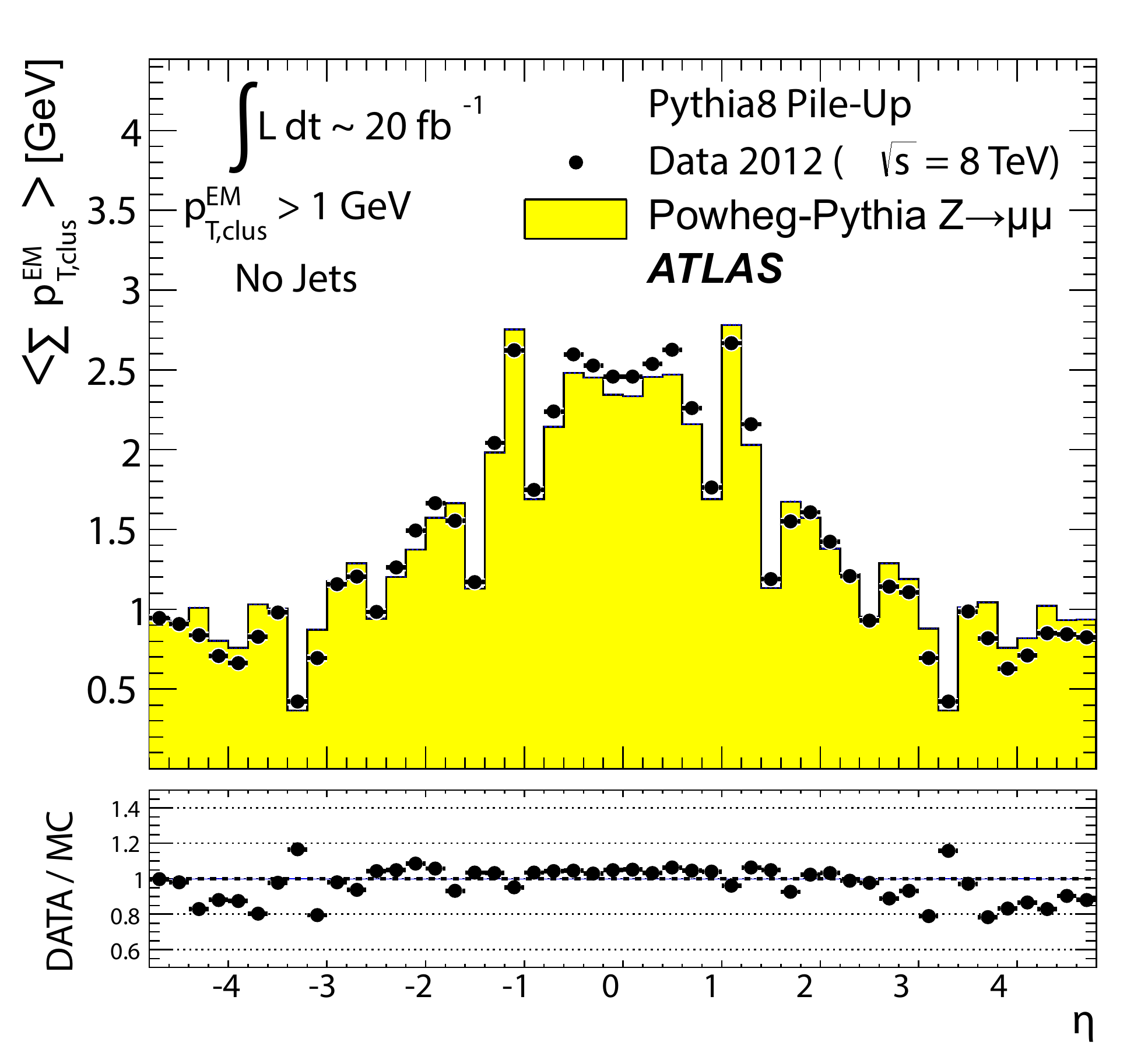}\label{fig:ptdens:1}} \qquad
	\subfloat[]{\includegraphics[width=\figsixpanelwidth]{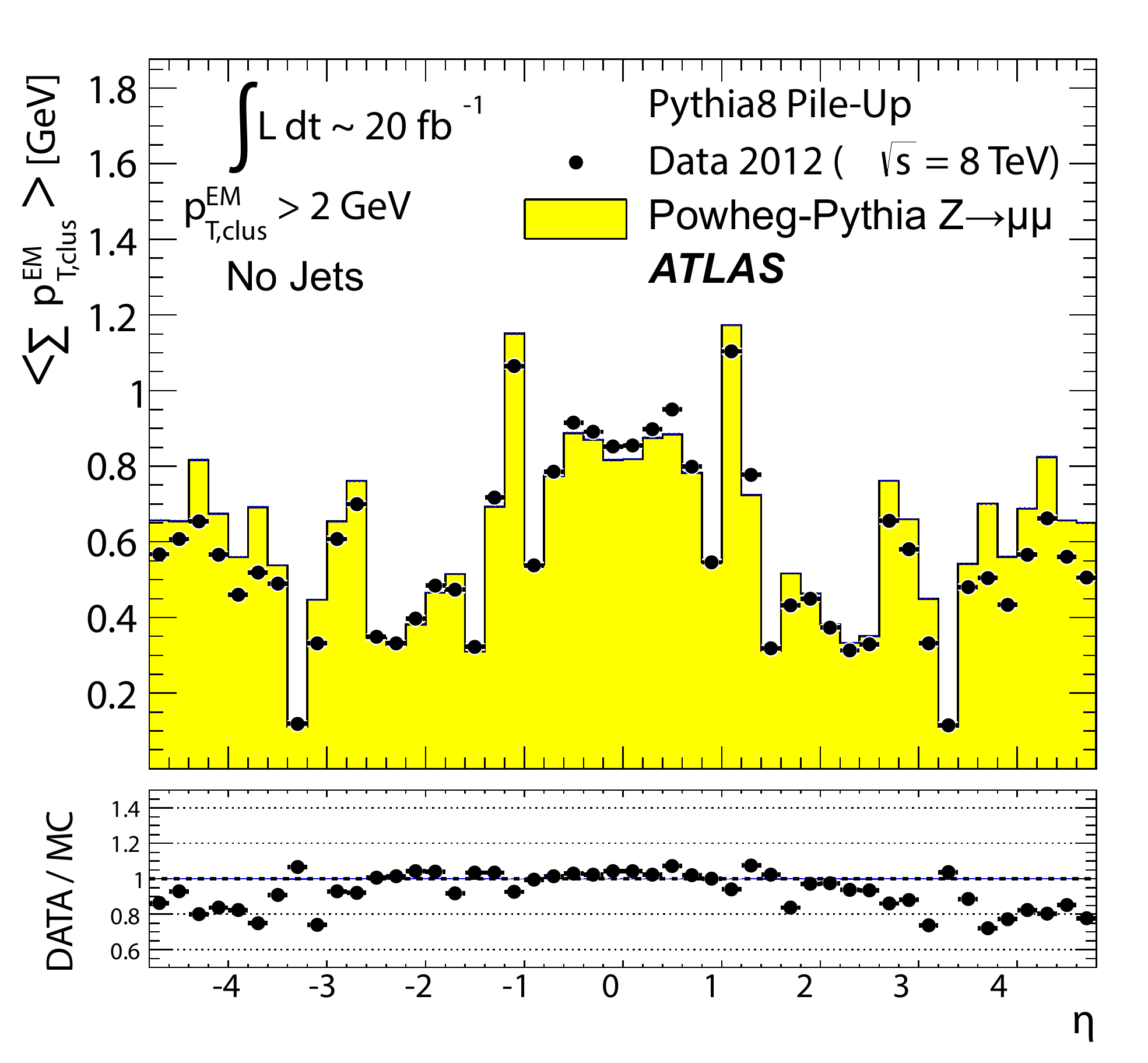}\label{fig:ptdens:2}}
	\caption[]{The average \AVE{\ptclussumem} of clusters at the \EM{} scale, calculated as function of $\eta$ using \eqRef{eq:ptsum:clus}, for clusters with \subref{fig:ptdens:all} $\ptclusem > 0$, \subref{fig:ptdens:100} $\ptclusem > \unit{100}{\MeV}$, \subref{fig:ptdens:250} $\ptclusem > \unit{250}{\MeV}$, \subref{fig:ptdens:500} $\ptclusem > \unit{500}{\MeV}$, \subref{fig:ptdens:1} $\ptclusem > \unit{1}{\GeV}$, and \subref{fig:ptdens:2} $\ptclusem > \unit{2}{\GeV}$. Results are obtained from a 2012 \Zmumu\, sample without jets with $\pT > \unit{20}{\GeV}$ in data and \MC{} simulation. The ratios of  $\AVE{\ptclussumem}(\eta)$ from data and \MC{} simulations are shown below each plot.}
	\label{fig:ptdens}
\end{figure}

The transverse momentum flow in the \Zmumu{} sample without jets with $\pT > \unit{20}{\GeV}$ is reconstructed using the exclusive selection defined in \secRef{\thislabel:obs:ef}. \Topos{} are selected by $\ptclusem > \pti{\text{min}}$, where $\pti{\text{min}} \in  \{ 0, \unit{100}{\MeV}, \unit{250}{\MeV}, \unit{500}{\MeV}, \unit{1}{\GeV}, \unit{2}{\GeV} \}$. 
The flow is measured by the average total transverse momentum \AVE{\ptclussumem}, carried by all or selected \topos{} in any given direction $\eta_{k} \leq \etaclus < \eta_{k+1}$, and averaged over a given number of events $N_{\text{evts}}$:
\begin{equation}
	\AVE{\ptclussumem}(\etaclus) = \dfrac{1}{N_{\text{evts}}} \sum_{i=1}^{N_{\text{evts}}}\left[\sum_{\{j\,|\eta_{k}<\etaclusi{j}<\eta_{k+1}\}} \ptclusemi{j}\right]_{i}  \,.
	\label{eq:ptsum:clus}
\end{equation}
Here $\eta_{k}$ denotes the lower boundary of the $k$-th $\eta$-bin used to sum the transverse momentum of the selected \topos{} in each event. 
\FigRef{fig:ptdens} shows \AVE{\ptclussumem}{} as a function of \etaclus{} for the various \topo{} selections for this \Zmumu{} data 
sample and the corresponding \MC{} simulations.

\begin{figure}[tp!] \centering
        \sfcompress
	\subfloat[]{\includegraphics[width=\figsixpanelwidth]{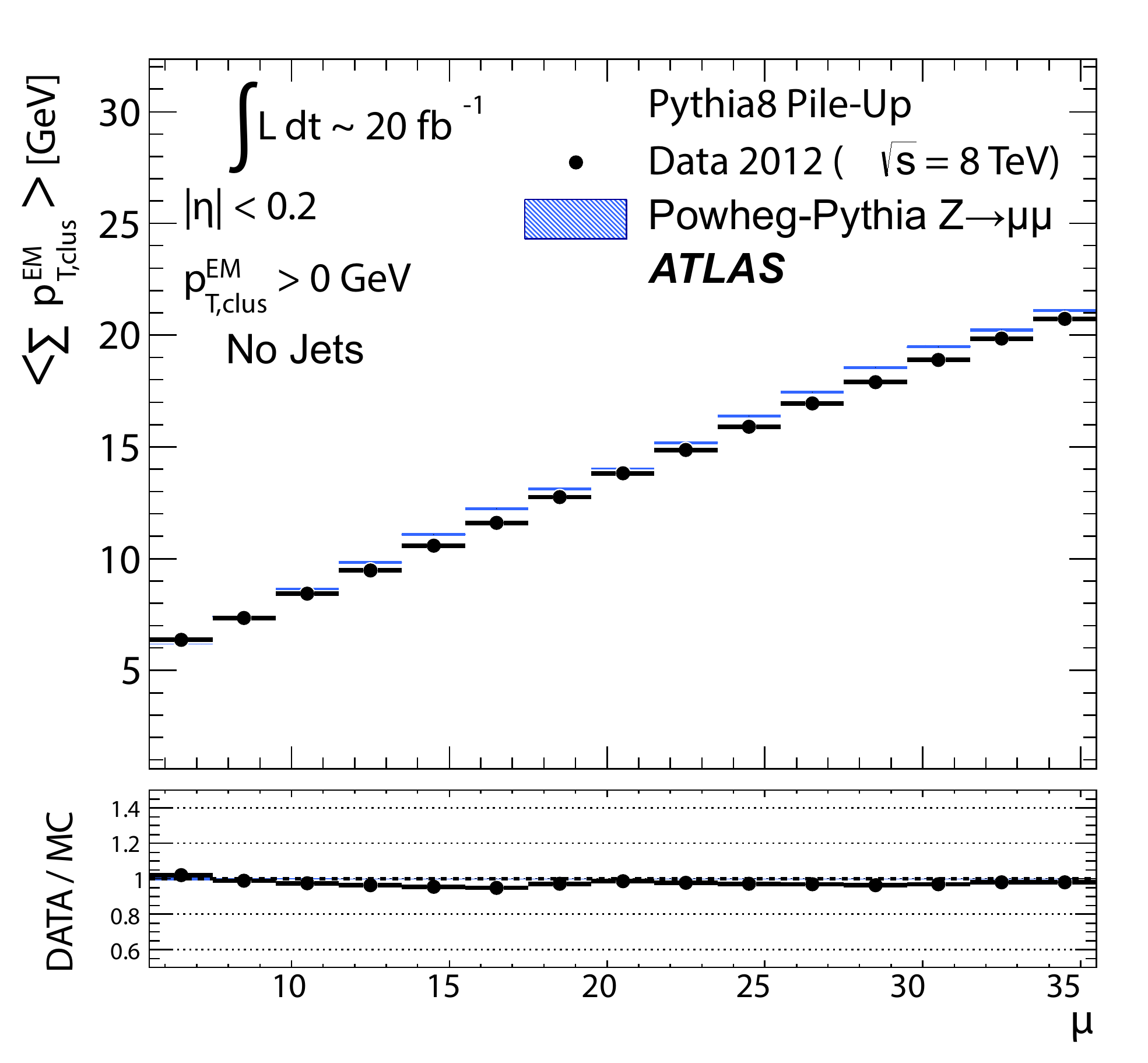}\label{fig:pu:sumpt:cent}}         \qquad 
	\subfloat[]{\includegraphics[width=\figsixpanelwidth]{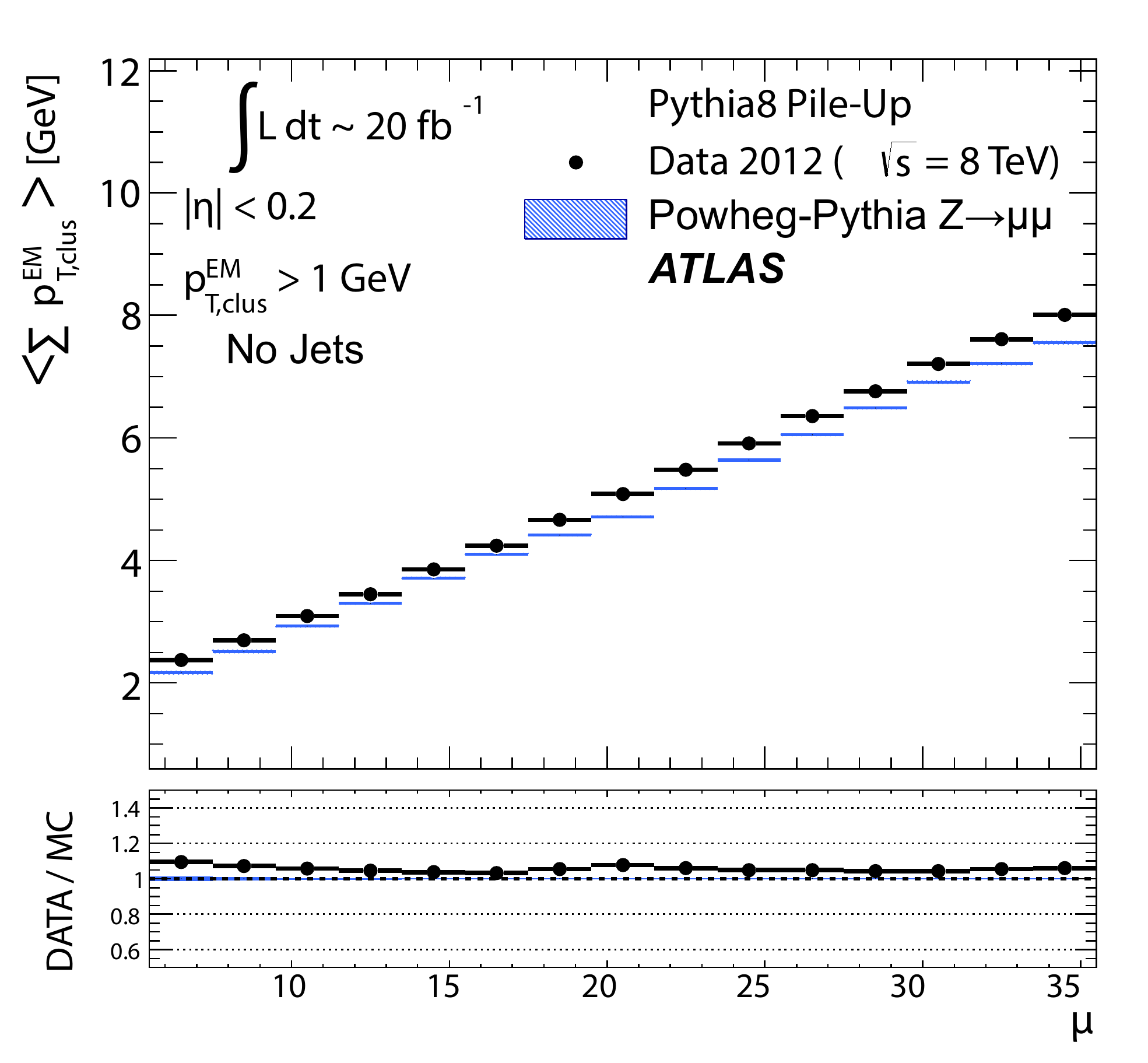}\label{fig:pu:sumptHighpT:cent}} 
        \\
	\subfloat[]{\includegraphics[width=\figsixpanelwidth]{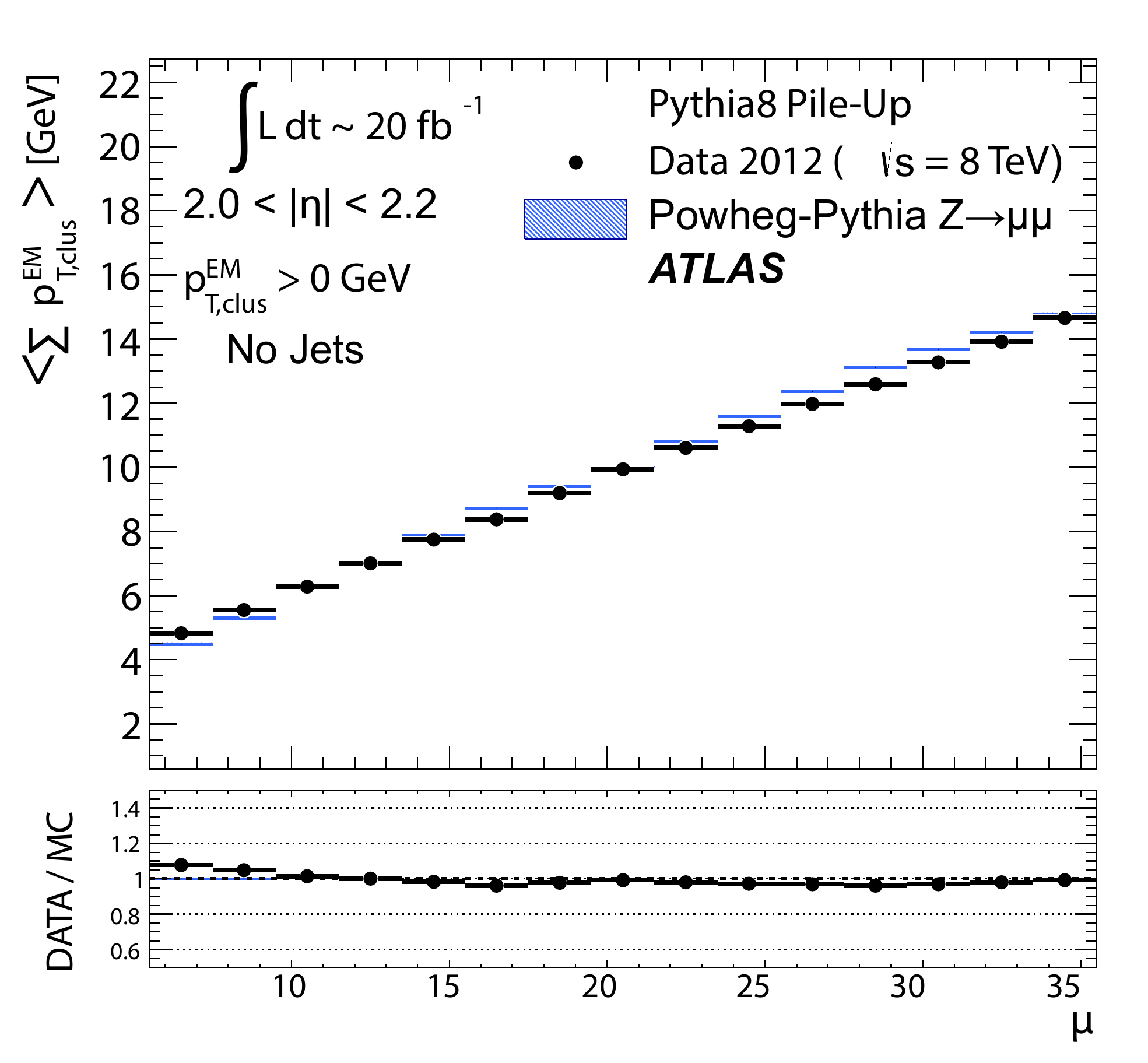}\label{fig:pu:sumpt:ec}}           \qquad
	\subfloat[]{\includegraphics[width=\figsixpanelwidth]{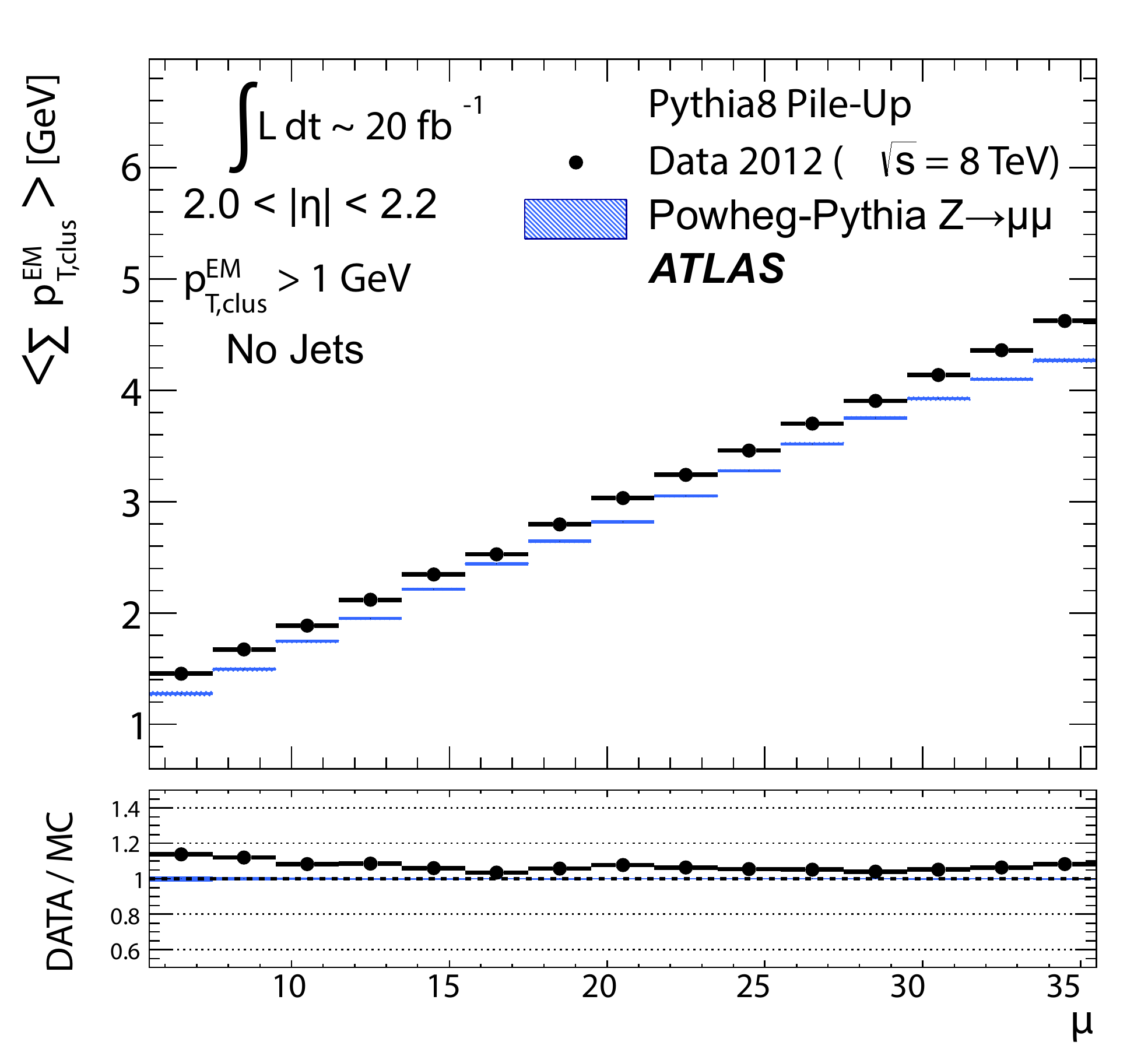}\label{fig:pu:sumptHighpT:ec}}
	\\
	\subfloat[]{\includegraphics[width=\figsixpanelwidth]{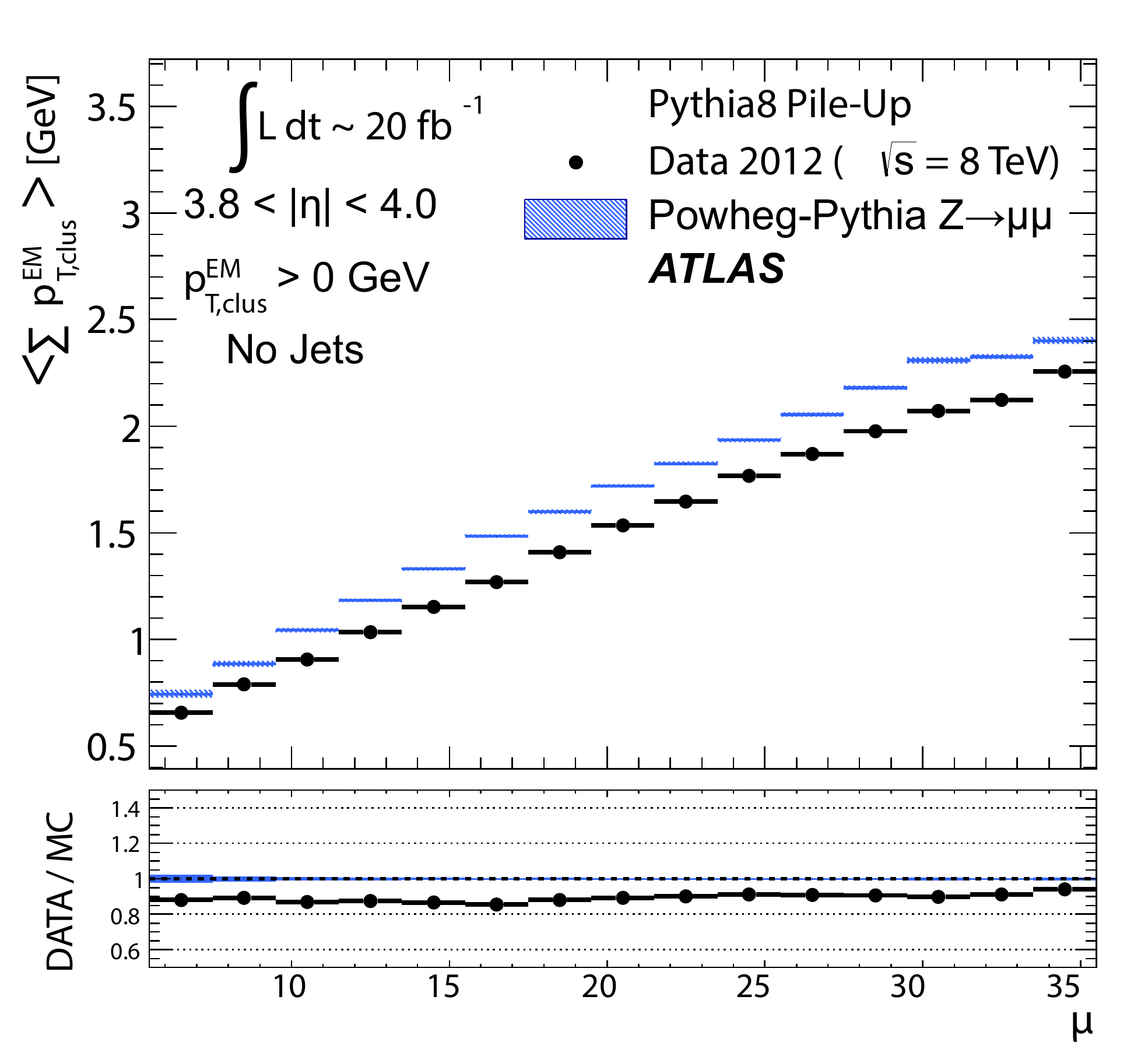}\label{fig:pu:sumpt:fwd}}          \qquad 
	\subfloat[]{\includegraphics[width=\figsixpanelwidth]{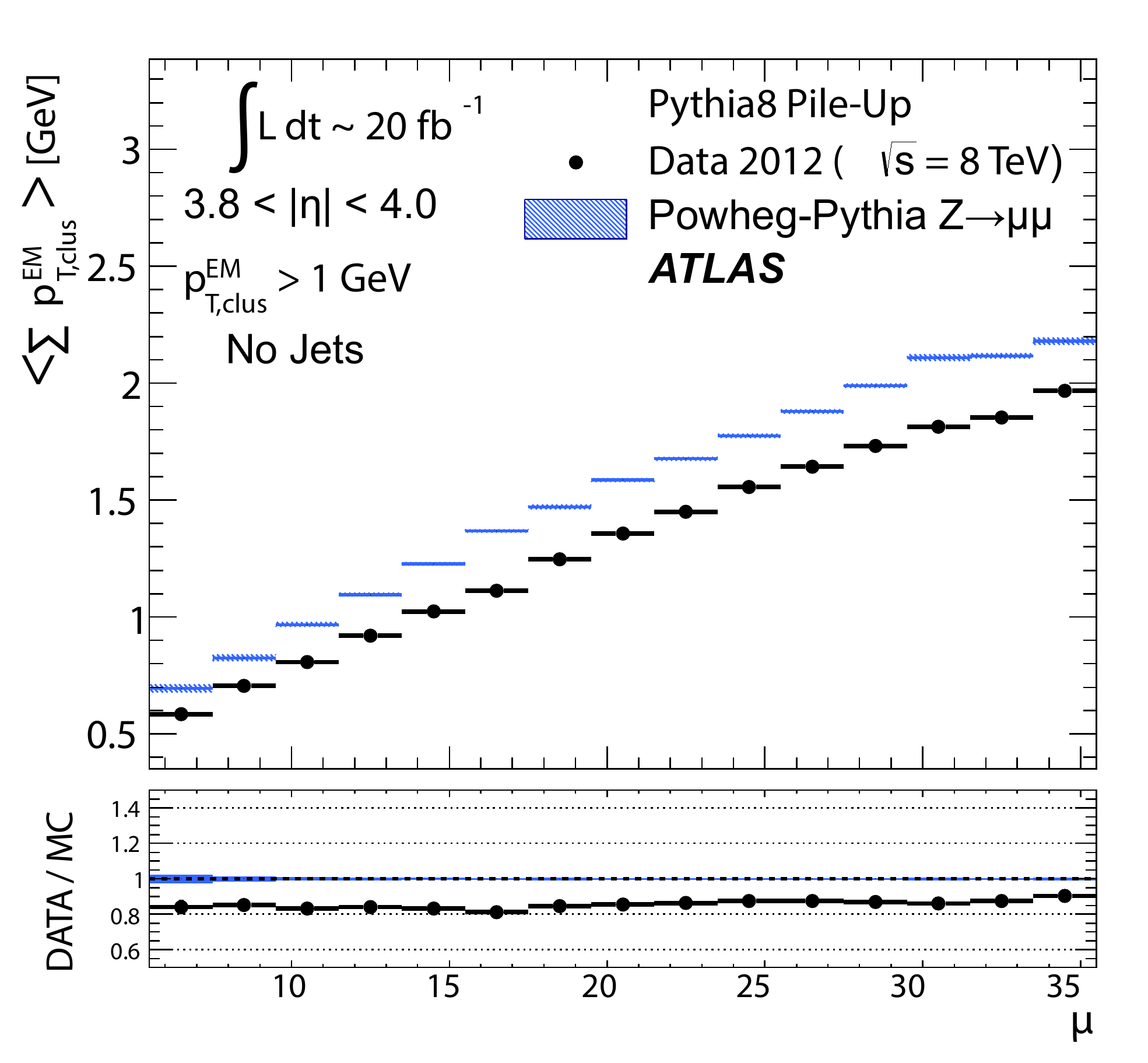}\label{fig:pu:sumptHighpT:fwd}} 
	\caption[]{The average transverse momentum flow $\AVE{\ptclussumem}$ evaluated as function of the \pu{} activity measured by the number of \pp{} interactions per bunch crossing $\mu$, in several calorimeter regions. In \subref{fig:pu:sumpt:cent}, \subref{fig:pu:sumpt:ec}, and \subref{fig:pu:sumpt:fwd}, $\AVE{\ptclussumem}(\mu)$ is shown in the central ($|\eta| < 0.2$), \EndCap{} ($2.0<|\eta|<2.2$), and the forward ($3.8<|\eta|<4.0$) region, respectively, using \topos{} with $\ptclusem > 0$. The corresponding results using \topos{} with $\ptclusem > \unit{1}{\GeV}$ are presented in \subref{fig:pu:sumptHighpT:cent}, \subref{fig:pu:sumptHighpT:ec}, and \subref{fig:pu:sumptHighpT:fwd}. Results are obtained from a 2012 \Zmumu{} sample without jets with $\pT > \unit{20}{\GeV}$ in data and \MC{} simulations. The narrow shaded bands around the results for \MC{} simulations indicate statistical uncertainties, both for $\AVE{\ptclussumem}(\mu)$ and the corresponding data-to-\MC{} simulation ratios shown below each plot.}
	\label{fig:pu:sumpt}
\end{figure}

The \pu{} dependence of the average transverse momentum flow in various detector regions, as expressed by $\AVE{\ptclussumem}(\mu)$, 
is shown in \figRef{fig:pu:sumpt} for an inclusive ($ptclusem > 0$) and a exclusive ($ptclusem > \unit{1}{\GeV}$) \topo{} selection. 
The \MC{} simulations predict the flow in the detector regions $|\eta| < 0.2$ and $2.0<|\eta|<2.2$ well, in particular for the more \pu-sensitive cluster selection shown in \figMultiRefLabel{}~\ref{fig:pu:sumpt}\subref{fig:pu:sumpt:cent} and \ref{fig:pu:sumpt}\subref{fig:pu:sumpt:ec}.
%
Larger deviations are observed for these two regions with the exclusive selection in \figMultiRefLabel~\ref{fig:pu:sumpt}\subref{fig:pu:sumptHighpT:cent} and \ref{fig:pu:sumpt}\subref{fig:pu:sumptHighpT:ec}.
In the forward region, \MC{} simulations predict higher \pT-flow for both \topo{} selections, as can be seen in \figMultiRefLabel~\ref{fig:pu:sumpt}\subref{fig:pu:sumpt:fwd} and \ref{fig:pu:sumpt}\subref{fig:pu:sumptHighpT:fwd}. 
The slope of the $\AVE{\ptclussumem}(\mu)$ dependence in this region is very similar for data and \MC{} simulations. 

The observations in \figMultiRef{fig:spectra:pt}{to}{fig:pu:sumpt} indicate that in the case of the fully simulated \pu{} the simulation of the \topo{} response to the underlying transverse energy flow outside jets suffers from \MC{} simulation deficiencies. The use of overlaid \pu{} from data, while not demonstrated here in all details, promises significant improvements for the modelling of the soft-event signals. 

\subsubsection{\Topo{} multiplicity in the presence of \pu} \label{\thislabel:obs:occ}

\begin{figure}[tp!] \centering
        \sfcompress
	\subfloat[all \topos]{\includegraphics[width=\figsixpanelwidth]{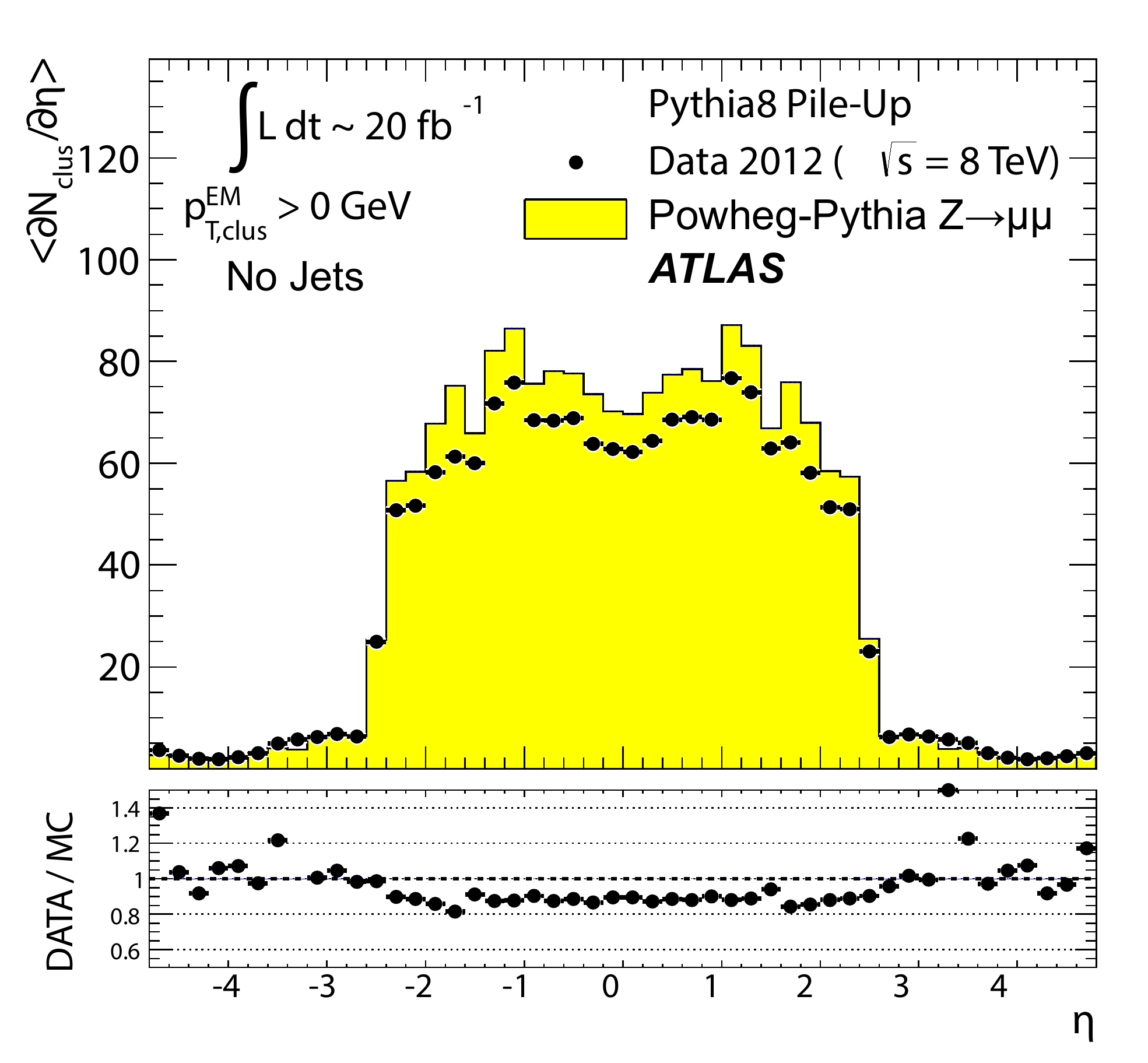}\label{fig:ndens:all}}                          \qquad
	\subfloat[$\ptclus > \unit{100}{\MeV}$]{\includegraphics[width=\figsixpanelwidth]{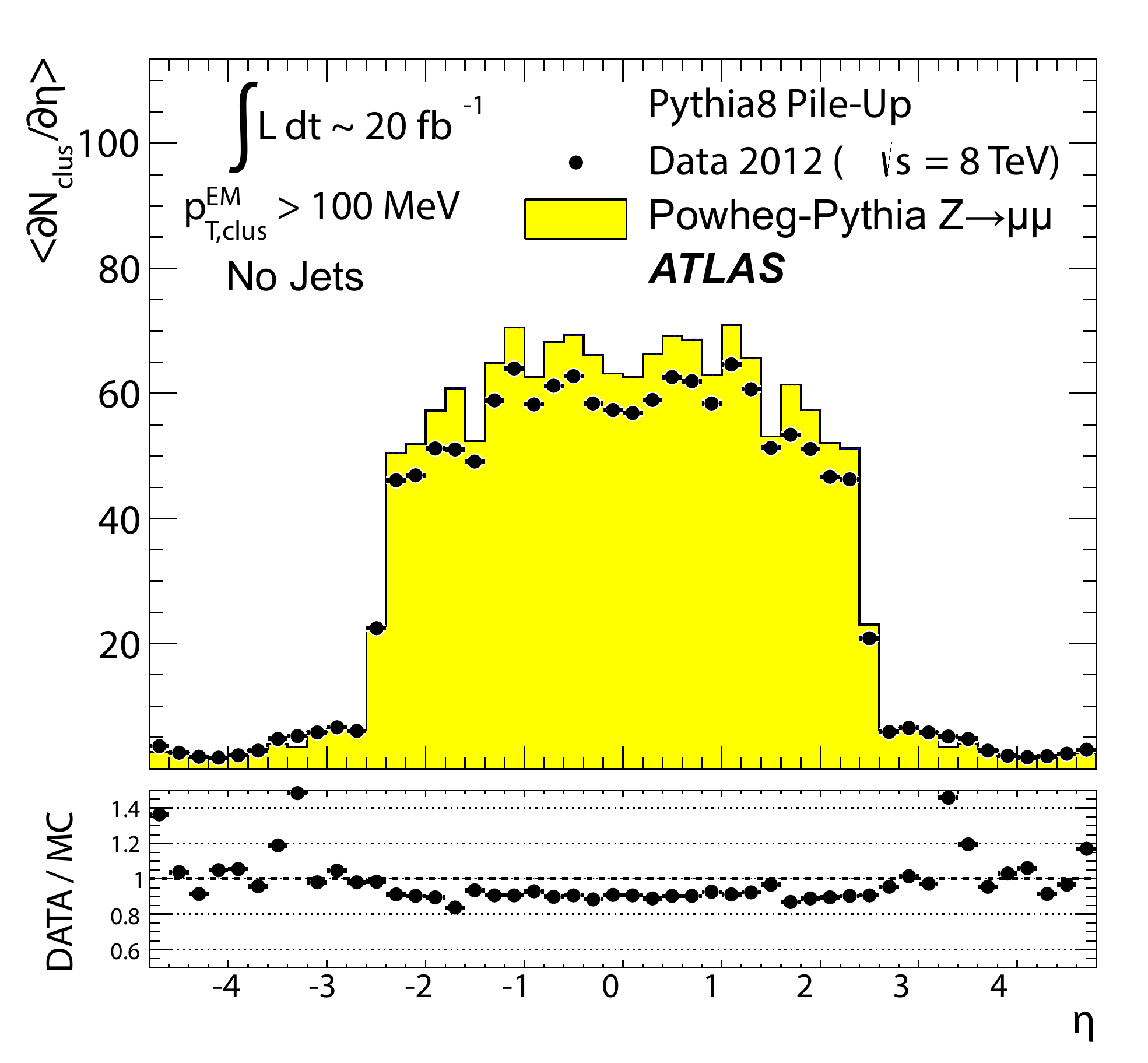}\label{fig:ndens:100}}
        \\      
	\subfloat[$\ptclus > \unit{250}{\MeV}$]{\includegraphics[width=\figsixpanelwidth]{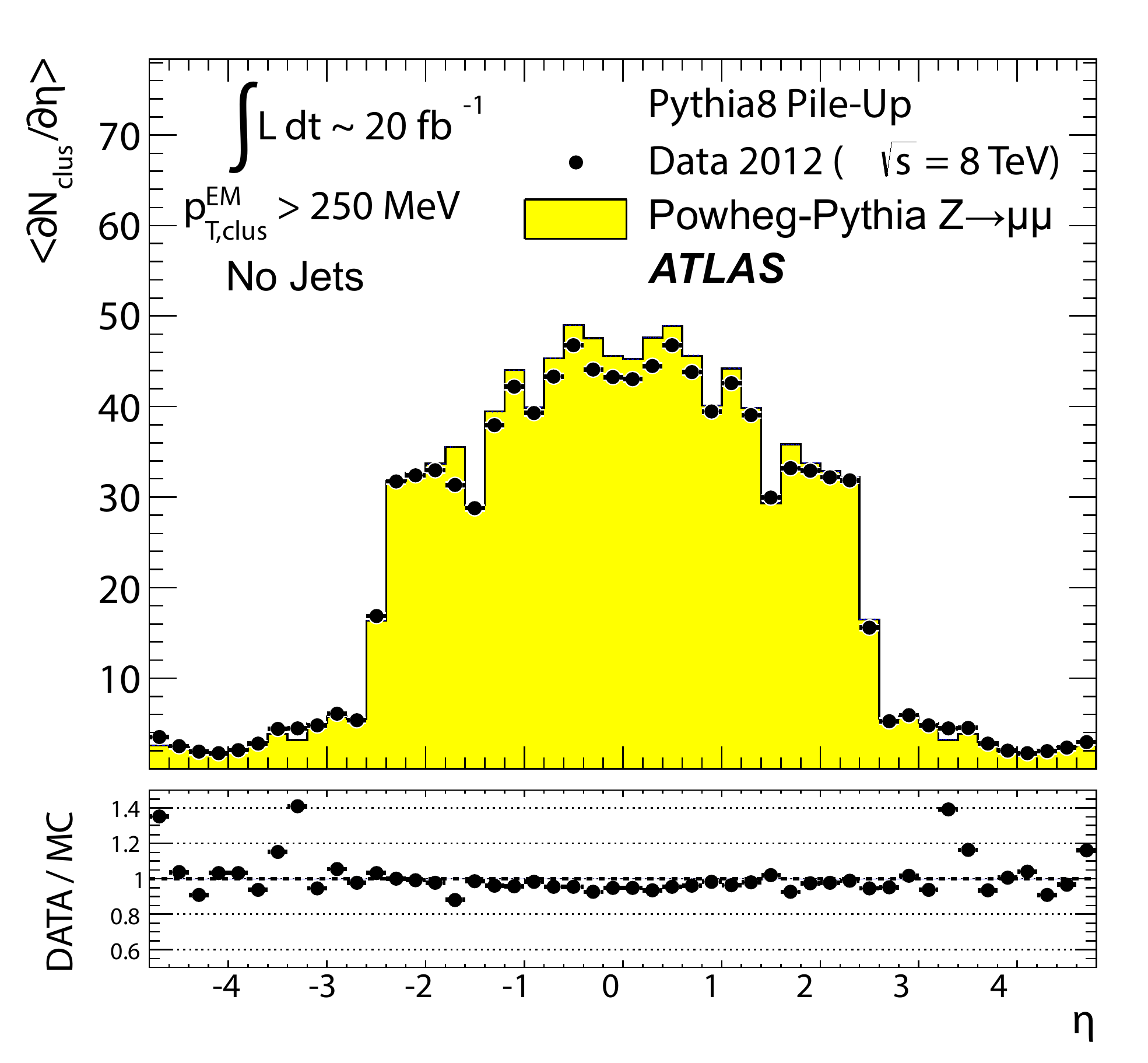}\label{fig:ndens:250}} \qquad
	\subfloat[$\ptclus > \unit{500}{\MeV}$]{\includegraphics[width=\figsixpanelwidth]{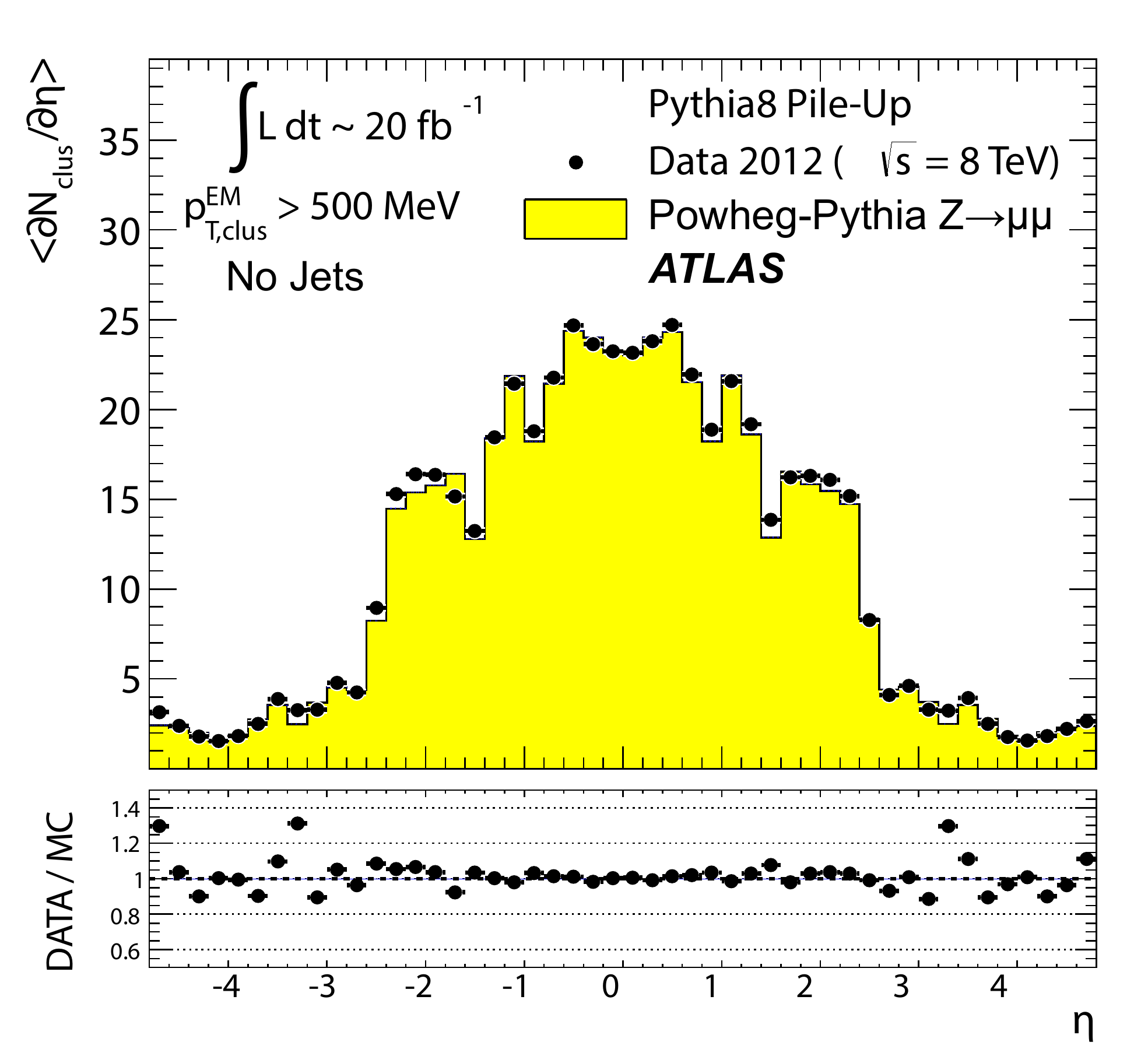}\label{fig:ndens:500}}
        \\
	\subfloat[$\ptclus > \unit{1}{\GeV}$]{\includegraphics[width=\figsixpanelwidth]{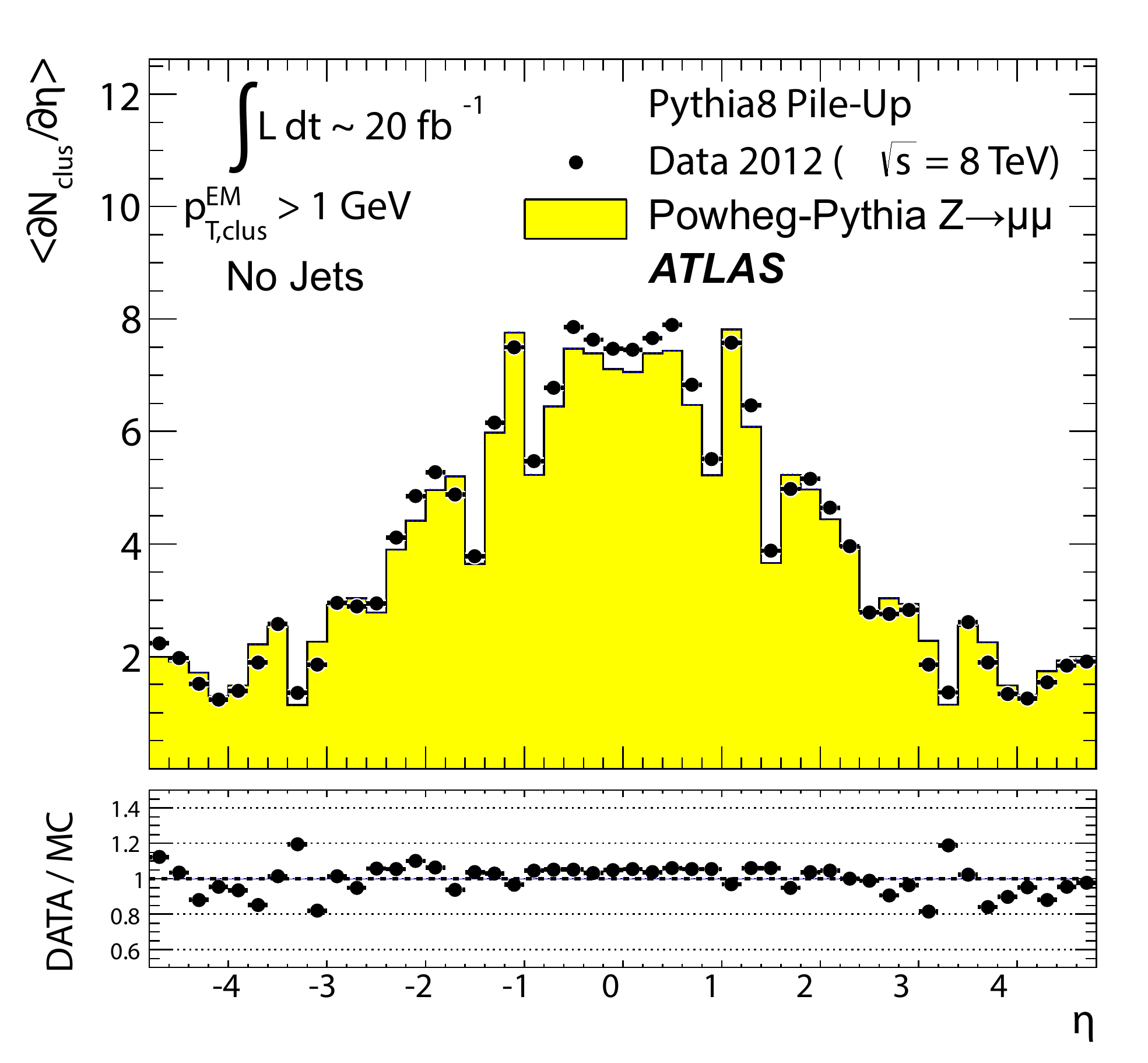}\label{fig:ndens:1}}       \qquad
	\subfloat[$\ptclus > \unit{2}{\GeV}$]{\includegraphics[width=\figsixpanelwidth]{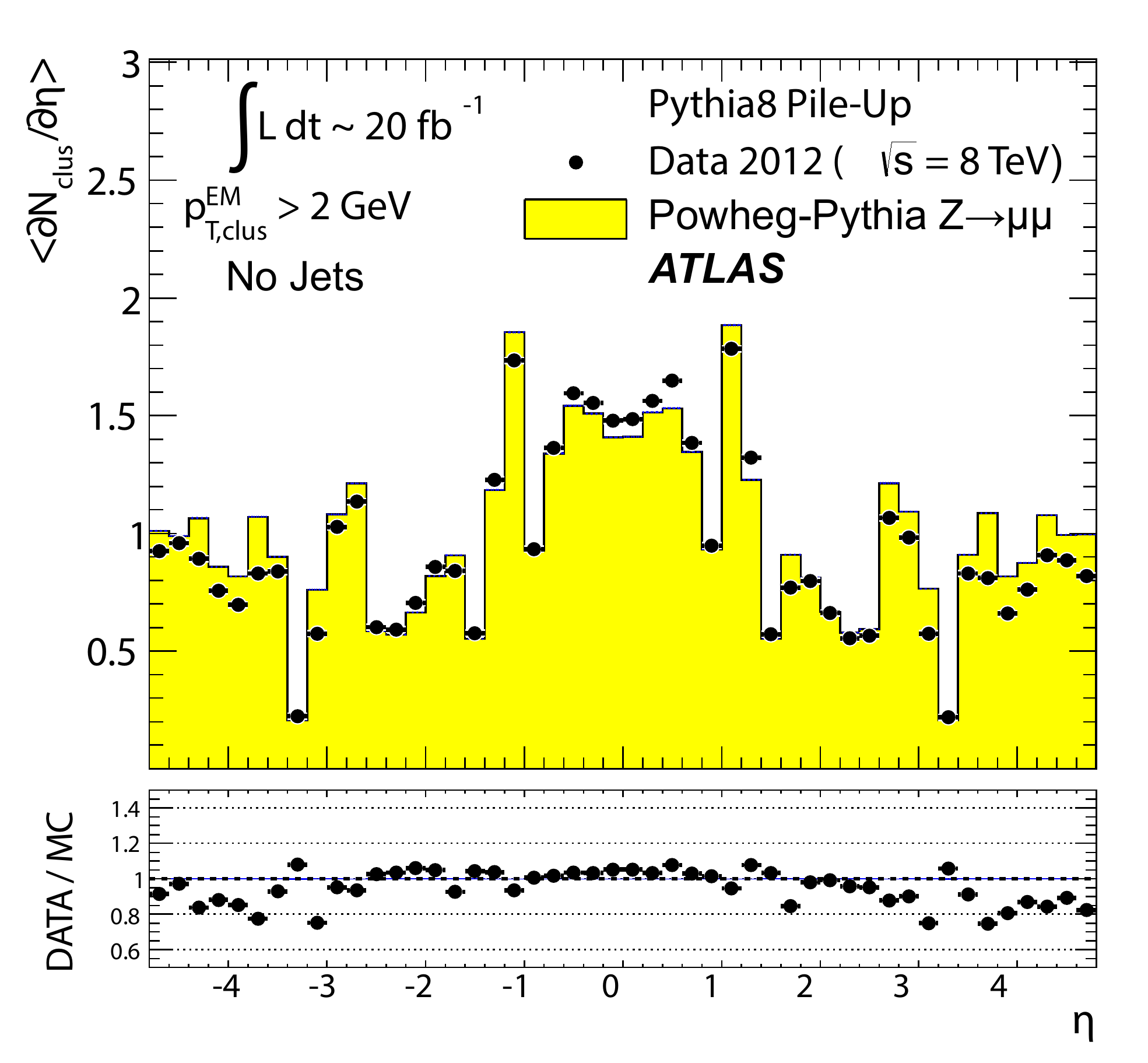}\label{fig:ndens:2}}
	\caption[]{Average \topo{} number density $\langle\partial N/\partial\eta\rangle$ as a function of \etaclus, for clusters with $\ptclusem > \pti{\text{min}}$, for various \pti{\text{min}}{} values. Results are obtained from a 2012 \Zmumu{} sample without jets with $\pT > \unit{20}{\GeV}$ in data and \MC{} simulations. The corresponding data-to-\MC{} simulation ratios are shown below each figure.}
	\label{fig:ndens}
\end{figure}

\begin{figure}[t!]\centering
	\includegraphics[width=\fighalfwidth]{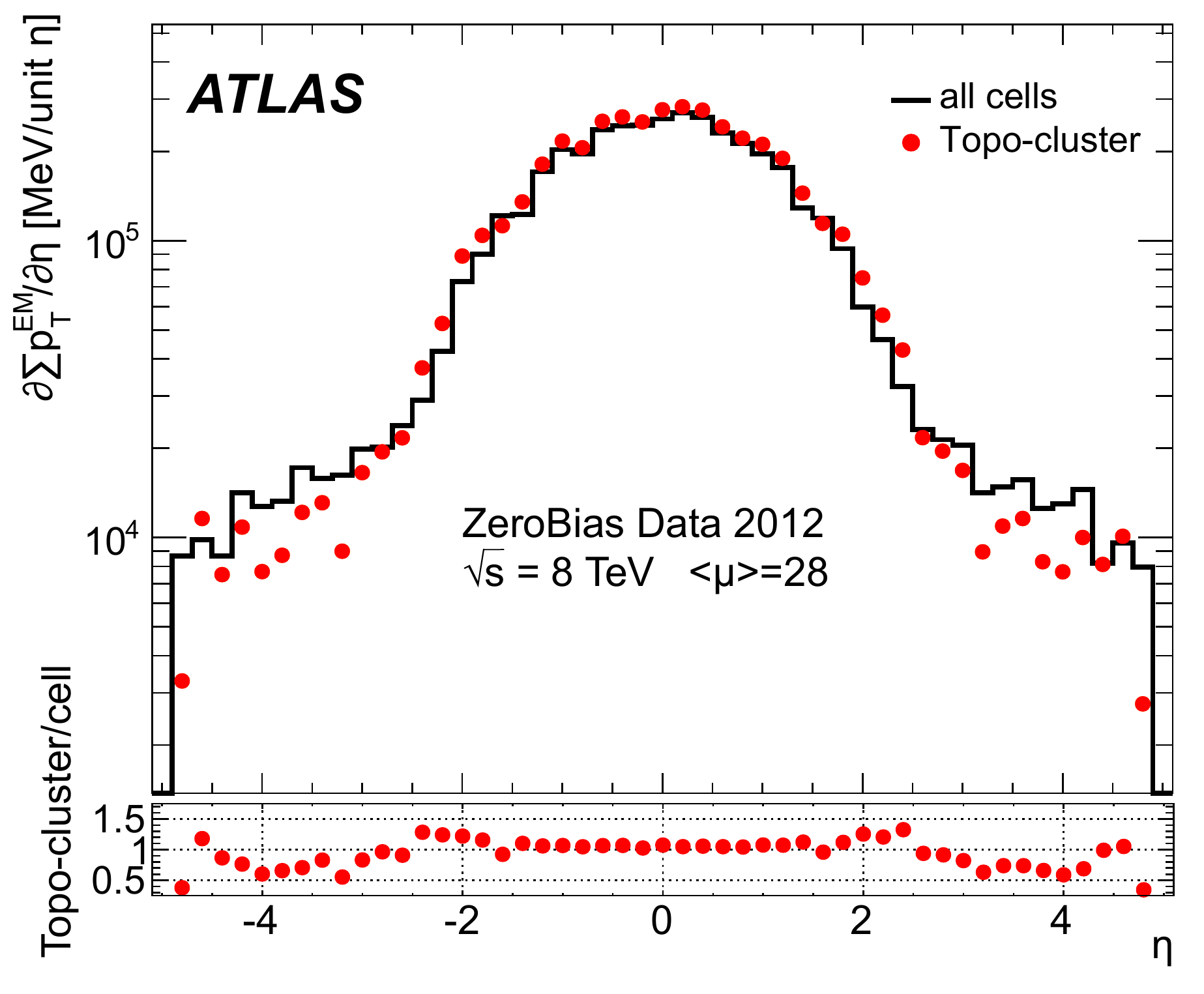}
	\caption[]{The reconstructed average transverse momentum flow on \EM{} scale, measured with \topos{} in bins of $\eta$ using $\AVE{\ptclussumem}(\eta)$ in \eqRef{eq:ptsum:clus} and with all calorimeter cells in the same $\eta$-bins using $\AVE{\ptcellsumem}(\eta)$ given in \eqRef{eq:ptsum:cell}, in 2012 \MB{} data.}
	\label{fig:cell_clus}
\end{figure}  

\begin{figure}[tp!] \centering
        \sfcompress
	\subfloat[]{\includegraphics[width=\figsixpanelwidth]{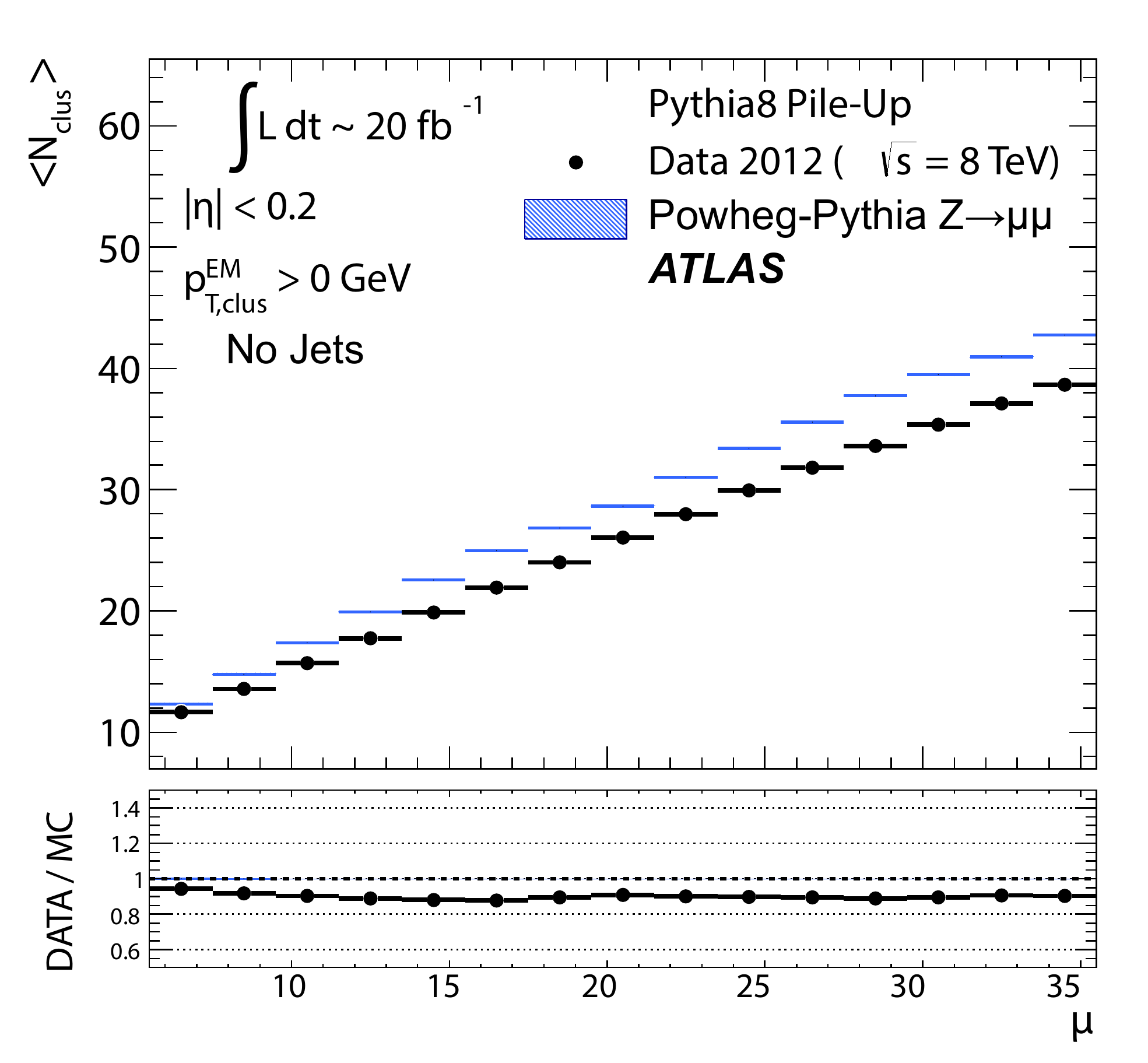}\label{fig:pu:nclus:cent}}          \qquad
	\subfloat[]{\includegraphics[width=\figsixpanelwidth]{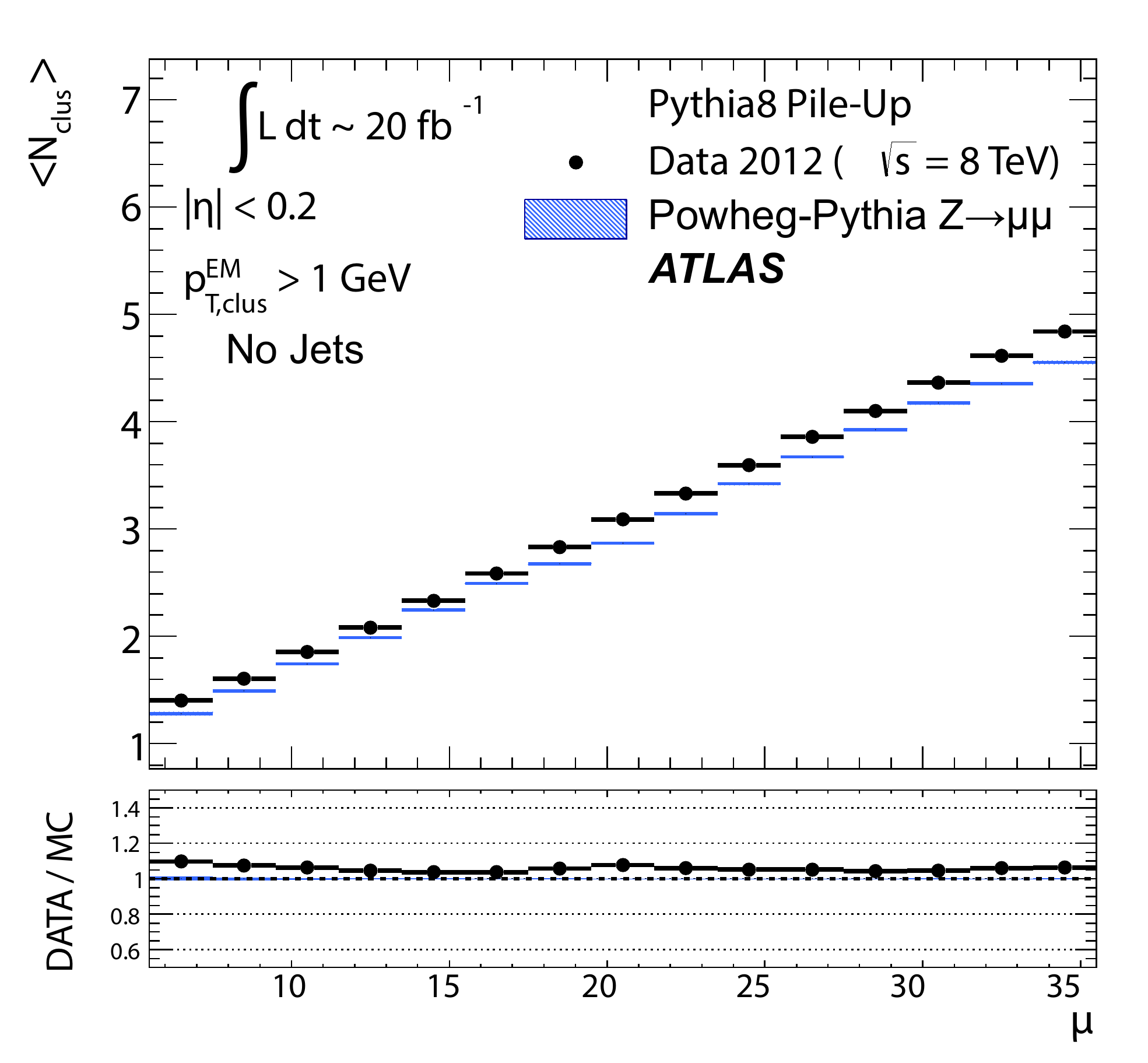}\label{fig:pu:nclusHighpT:cent}} 
	\\
	\subfloat[]{\includegraphics[width=\figsixpanelwidth]{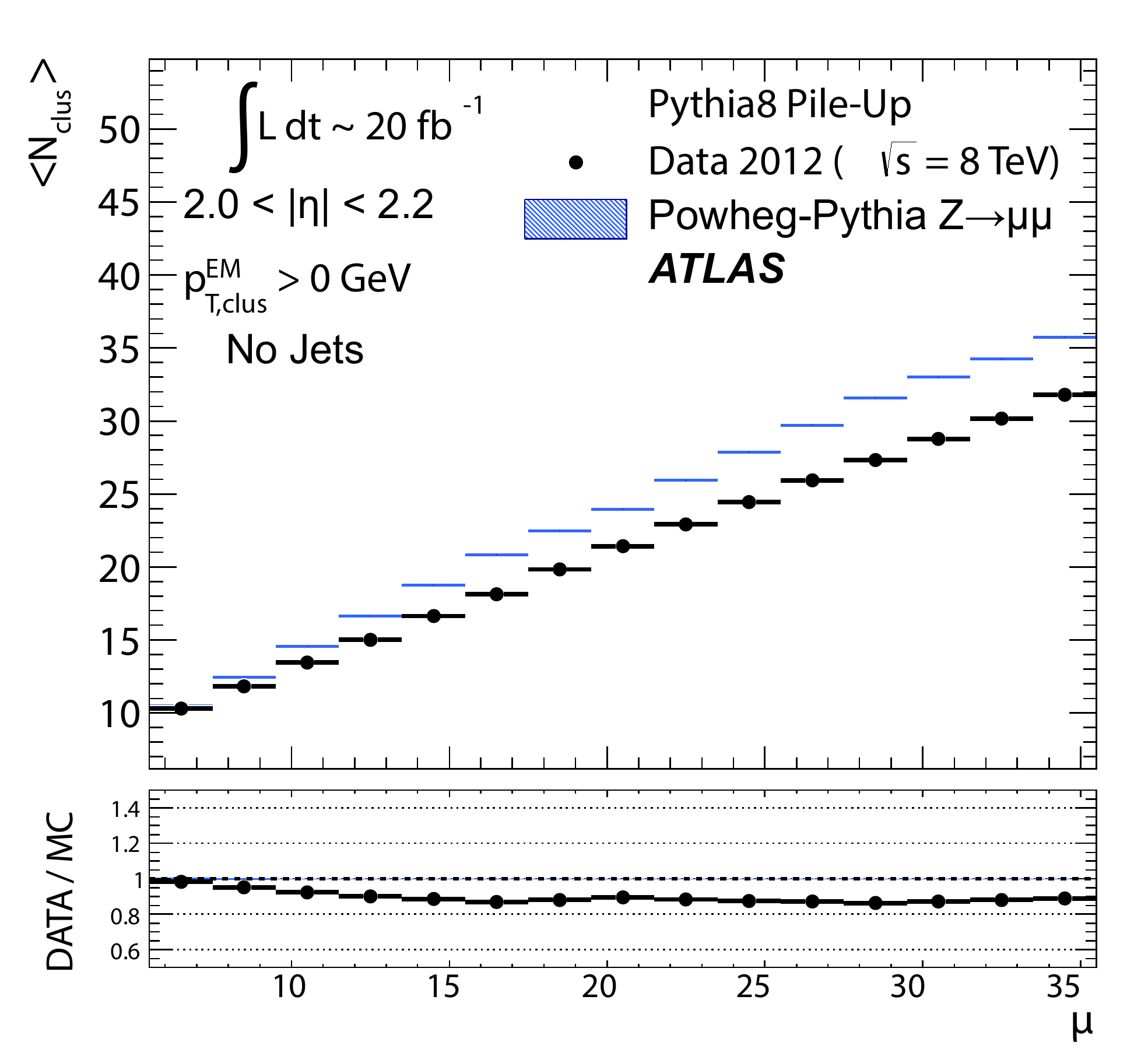}\label{fig:pu:nclus:ec}}            \qquad
	\subfloat[]{\includegraphics[width=\figsixpanelwidth]{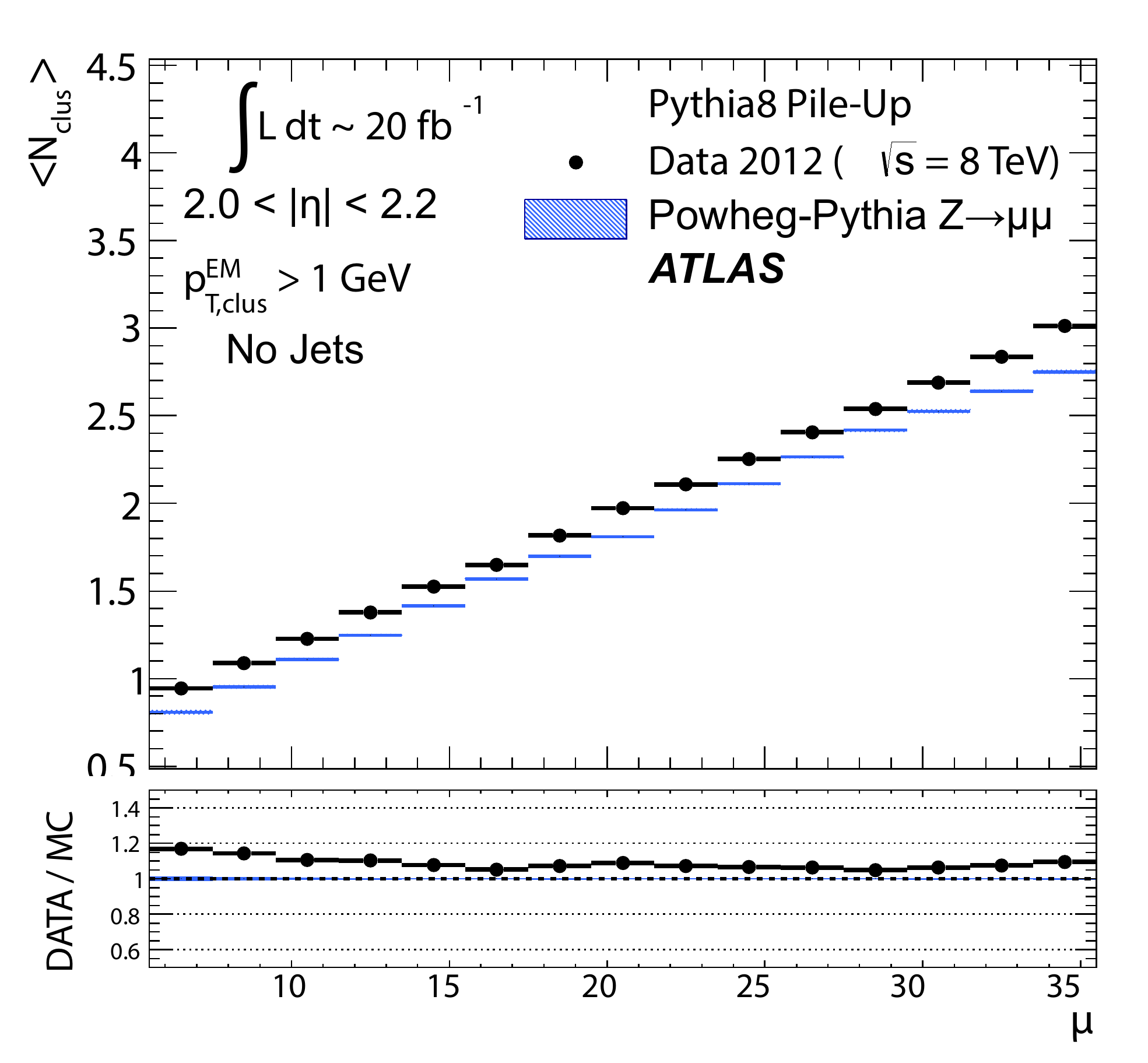}\label{fig:pu:nclusHighpT:ec}}
	\\
	\subfloat[]{\includegraphics[width=\figsixpanelwidth]{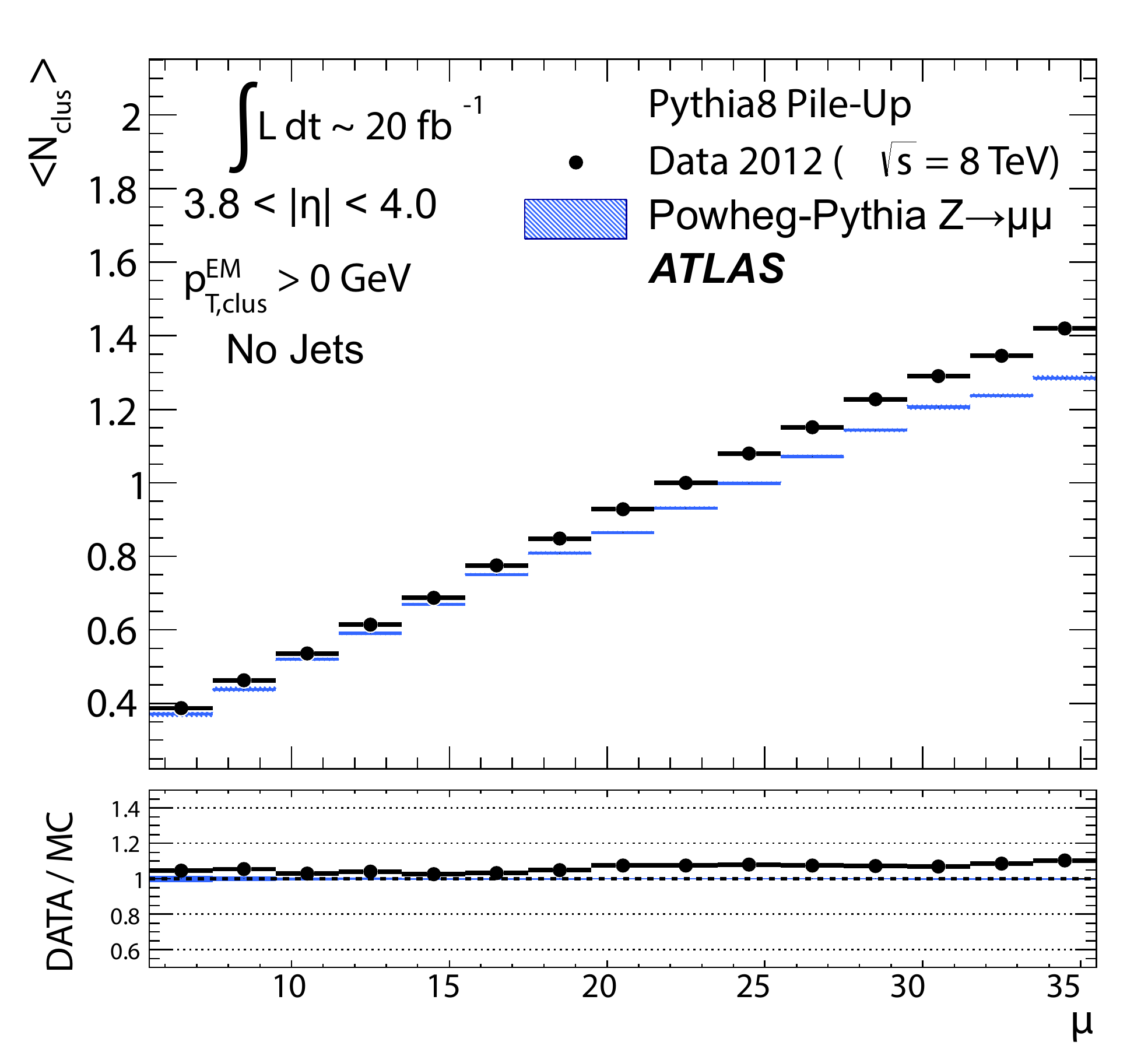}\label{fig:pu:nclus:fwd}}           \qquad	
	\subfloat[]{\includegraphics[width=\figsixpanelwidth]{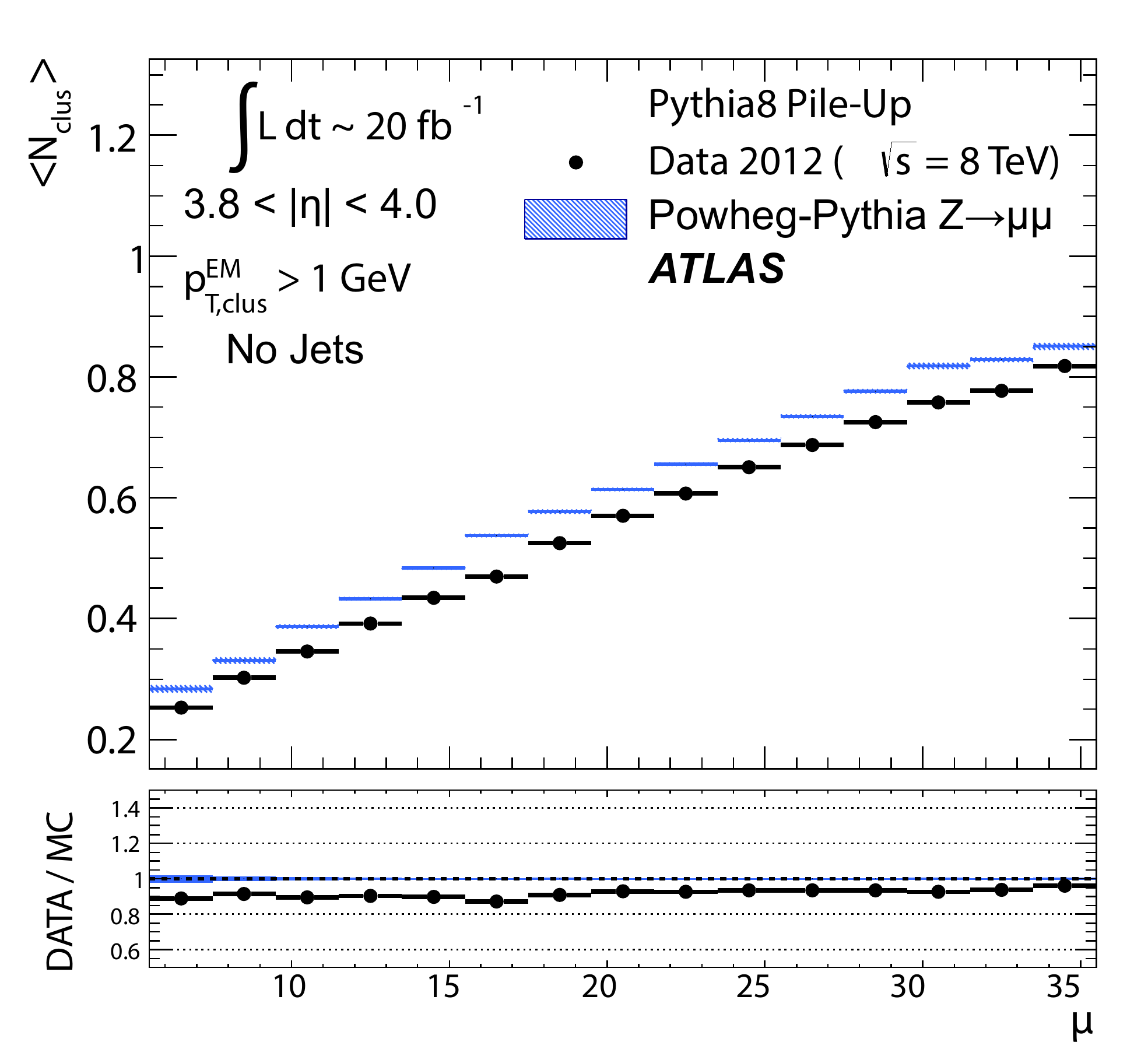}\label{fig:pu:nclusHighpT:fwd}} 
	\caption[]{The dependence of the average number	of positive-energy \topos{} on the \pu{} activity measured by the number of \pp{} collisions per bunch crossings $\mu$ in several regions of the detector is shown in (\subref{fig:pu:nclus:cent}, \subref{fig:pu:nclusHighpT:cent}) for $|\eta| < 0.2$, in (\subref{fig:pu:nclus:ec}, \subref{fig:pu:nclusHighpT:ec}) for $2.0<|\eta|<2.2$, and in (\subref{fig:pu:nclus:fwd}, \subref{fig:pu:nclusHighpT:fwd}) for $3.8<|\eta|<4.0$. Plots \subref{fig:pu:nclus:cent}, \subref{fig:pu:nclus:ec} and \subref{fig:pu:nclus:fwd} show the results for counting all clusters with $\ptclusem > 0$, while \subref{fig:pu:nclusHighpT:cent}, \subref{fig:pu:nclus:ec} and \subref{fig:pu:nclusHighpT:fwd} show the results for only counting clusters with $\ptclusem > \unit{1}{\GeV}$. The corresponding ratio of data to \MC{} simulations is shown below each plot. 
All results are obtained from a 2012 \Zmumu\, sample without jets with $\pT > \unit{20}{\GeV}$ in data and \MC{} simulations.  The narrow shaded bands indicate the statistical uncertainties associated with the results from \MC{} simulations for the mean values and the ratios.}
	\label{fig:pu:nclus}
\end{figure}

The calorimeter signal occupancy in the exclusive \Zmumu{} sample is determined using selected \topos{} with $\ptclusem > \pti{\text{min}}$ and the $\pti{\text{min}}$ values used in \secRef{\thislabel:obs:ptflow}. The relevant ob\-ser\-va\-ble is the cluster number density, which is given by the number of \topos{} per unit $\eta$  ($\partial\Nclus/\partial\eta$).
\FigRef{fig:ndens} shows the average $\AVE{\partial\Nclus/\partial\eta}(\etaclus)$ for these \topo{} selections, for data and \MC{} simulations with fully simulated \pu. The shape observed especially for the less restrictive selections with $\pti{\text{min}} \leq \unit{500}{\MeV}$ in \figMultiRefLabel~\ref{fig:ndens}\subref{fig:ndens:all} to \ref{fig:ndens}\subref{fig:ndens:500}, reflects the variations of the calorimeter segmentation and the effect of sub-detector transition regions on the \topo{} formation across the full \ATLAS{} acceptance $|\etaclus| < 4.9$. 
Generally, \MC{} simulations describe the \pT-flow better than the number of clusters. This is expected as the description of the summed \pT-flow is constrained with more weight in the numerical fits for the \ATLAS{} tunes than the particle number density.

The \topo{} number density changes rapidly at $|\etaclus|=2.5$. This is a consequence of the reduction of the calorimeter cell granularity by about a factor of four in terms of pseudorapidity and azimuth ($\Delta\eta\times\Delta\phi$), which reduces the number of potential \topo{} seeds. The granularity change also introduces more signal overlap between individual particles in any given cell and thus less spatial resolution for the reconstruction of the corresponding energy flow due to this merging of particle signals. In addition, the larger cells increase the noise thresholds, as shown in \figMultiRefLabel~\ref{fig:noise}\subref{fig:noise:2011} and
\ref{fig:noise}\subref{fig:noise:2012}, which changes the calorimeter sensitivity.  
This change of sensitivity can be evaluated by comparing \AVE{\ptclussumem}{} with the corresponding quantity
\begin{equation}
\AVE{\ptcellsumem}(\etacell) = \dfrac{1}{N_{\text{evts}}}\sum_{i=1}^{N_{\text{evts}}}\left[\sum_{\{j\,|\eta_{k}<\etacelli{j}<\eta_{k+1}\}} \ptcellemi{j}\right]_{i}\,,
\label{eq:ptsum:cell}
\end{equation}
reconstructed from all calorimeter cell signals in each $\eta$ bin, similar to \eqRef{eq:ptsum:clus} for clusters. The cell-based \pT-flow
expressed by $\AVE{\ptcellsumem}(\etacell)$ is unbiased with respect to noise suppression as none is applied. Consequently, it is subject to larger fluctuations. \FigRef{fig:cell_clus} shows this measurement for a 2012 \MB{} data sample with \pu{} close to the nominal $\mu = 30$ used for the noise thresholds (see \secRef{sec:atlas:data:noise}). 
It indicates signal losses due to clustering up to about \unit{50}{\%}{} for $2.5 < |\eta| < 4.5$, and some signal increase due to suppression of cells with $E < 0$, in particular in the \EndCap{} region $1.5 < |\eta| < 2.5$. All \topos{} and calorimeter cell signals are accepted for this study.

The geometry effect yields the steep drop in \topo{} number density at this boundary. Raising the transverse momentum threshold for accepted \topos{} increasingly mitigates the geometrical and noise effects on the cluster number density. 
The \datatomc{} comparison shows larger deficiencies for more inclusive \topo{} selections, which capture more signals from \pu. 
It improves as the $\pti{\text{min}}$ threshold increases, when the selections are dominated by clusters that are generated by harder emissions than those due to \pu. 

The dependence of the average number of \topos{} in a given calorimeter region on the \pu{} activity, expressed in terms of $\mu$, is shown for cluster with $\ptclusem > 0$ and $\ptclusem > \unit{1}{\GeV}$ in 
\figRef{fig:pu:nclus}. 
Applying the (inclusive) $\ptclusem > 0$ selection yields more \topos{} in \MC{} simulations than in data in the selected central ($|\eta|<0.2$) and \EndCap{} ($2.0<|\eta|<2.2$) regions, with the difference rising with increasing $\mu$ in \figMultiRefLabel{}~\ref{fig:pu:nclus}\subref{fig:pu:nclus:cent} and \ref{fig:pu:nclus}\subref{fig:pu:nclus:ec}.
In the forward region the number of \topos{} in \MC{} simulations is closer to the number in data for low $\mu$, but tends to be lower than data at higher $\mu$, as seen in \figRefLabel{}~\ref{fig:pu:nclus}\subref{fig:pu:nclus:fwd}.

These qualitative differences between the observations for the central and \EndCap{} regions and the forward region can arise from the modelling of soft physics, which is tuned with reconstructed charged tracks in the detector region $|\eta| < 2.5$ but is not experimentally constrained in the forward region. 
In addition, imperfections in the description of the inactive material in front 
of the calorimeter in the detector simulation can change the acceptance for low-energy particles significantly in different ways in the various $\eta$-regions.   
Also, mismodelling in the simulation of the (mostly hadronic) lateral
and longitudinal shower spreads in the calorimeters, as e.g. documented in \citMultiRef{Adragna:2010zz,Abat:2010zz}, can lead to different \topo{} splitting
behaviour in the different calorimeter regions. In particular the increased signal overlap between particles in the forward region is suspected to introduce
a higher sensitivity of the cluster splitting to the detector simulation.

As can be seen in \figMultiRefLabel~\ref{fig:pu:nclus}\subref{fig:pu:nclusHighpT:cent}, \ref{fig:pu:nclus}\subref{fig:pu:nclusHighpT:cent} and \ref{fig:pu:nclus}\subref{fig:pu:nclusHighpT:fwd}, counting only \topos{} with $\ptclusem > \unit{1}{\GeV}$ introduces a more similar slope in the cluster number density as a function of $\mu$. 
The qualitative behaviour of $\AVE{\partial\Nclus/\partial\eta}(\etaclus)$ in the various detector regions is different than for the more inclusive \topo{} selection, with \MC{} simulation predicting fewer clusters in the central and \EndCap{} regions shown in \figMultiRefLabel~\ref{fig:pu:nclus}\subref{fig:pu:nclusHighpT:cent} and
\ref{fig:pu:nclus}\subref{fig:pu:nclusHighpT:ec}. 
In the forward region, data shows overall fewer clusters than \MC{} simulation, as can be seen in  \figRef{fig:pu:nclus}\subref{fig:pu:nclusHighpT:fwd}, with larger differences at any given $\mu$, but a very similar number of additional clusters per additional \pp{} interaction.

\subsubsection{Modelling of the \topo{} depth location in the presence of \pu} \label{\thislabel:obs:moms}

\begin{figure}[tp!] \centering
        \sfcompress         
	\subfloat[]{\includegraphics[width=\figsixpanelwidth]{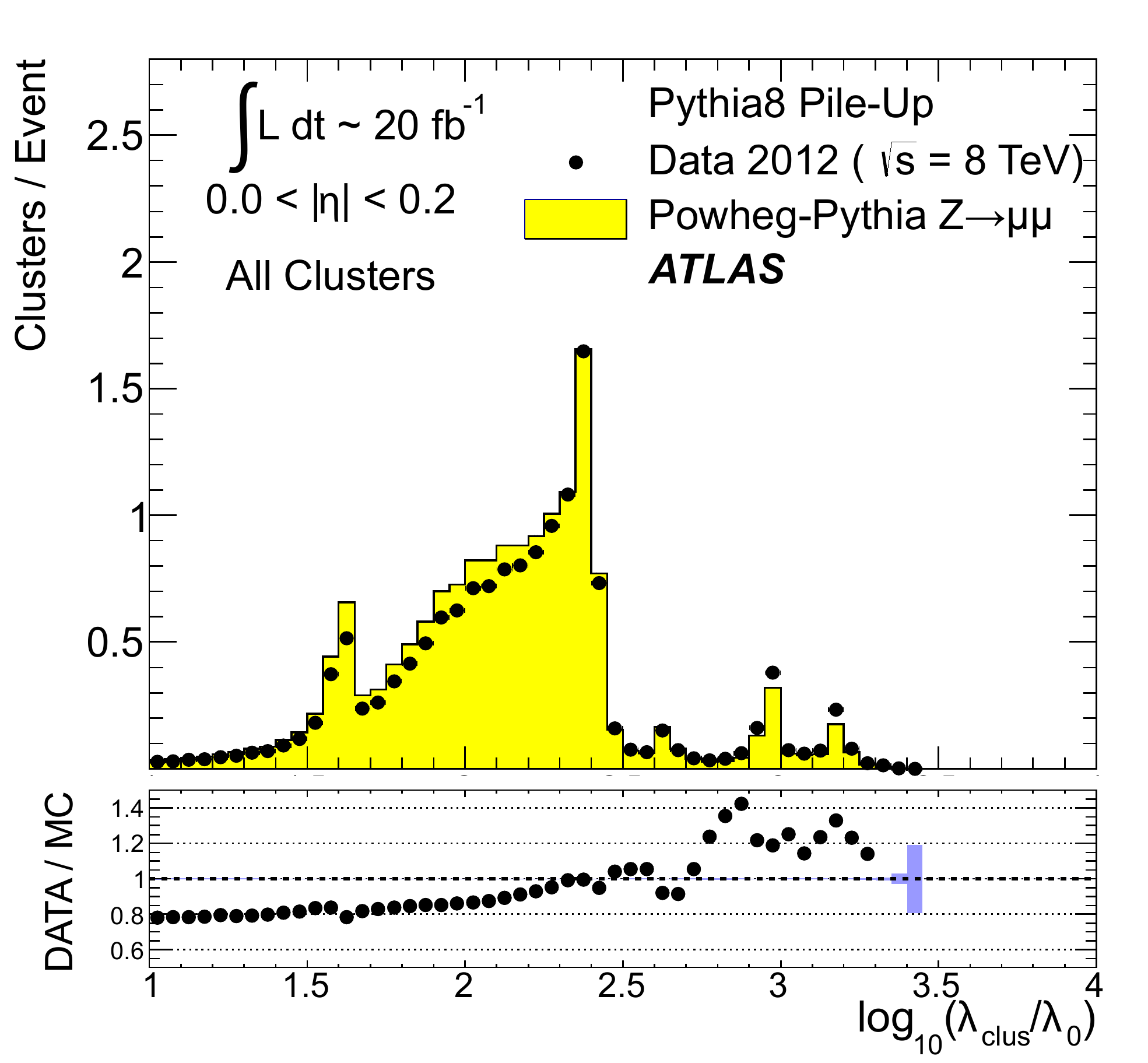}\label{fig:spectra:lam:00_02:mc}}      \qquad
	\subfloat[]{\includegraphics[width=\figsixpanelwidth]{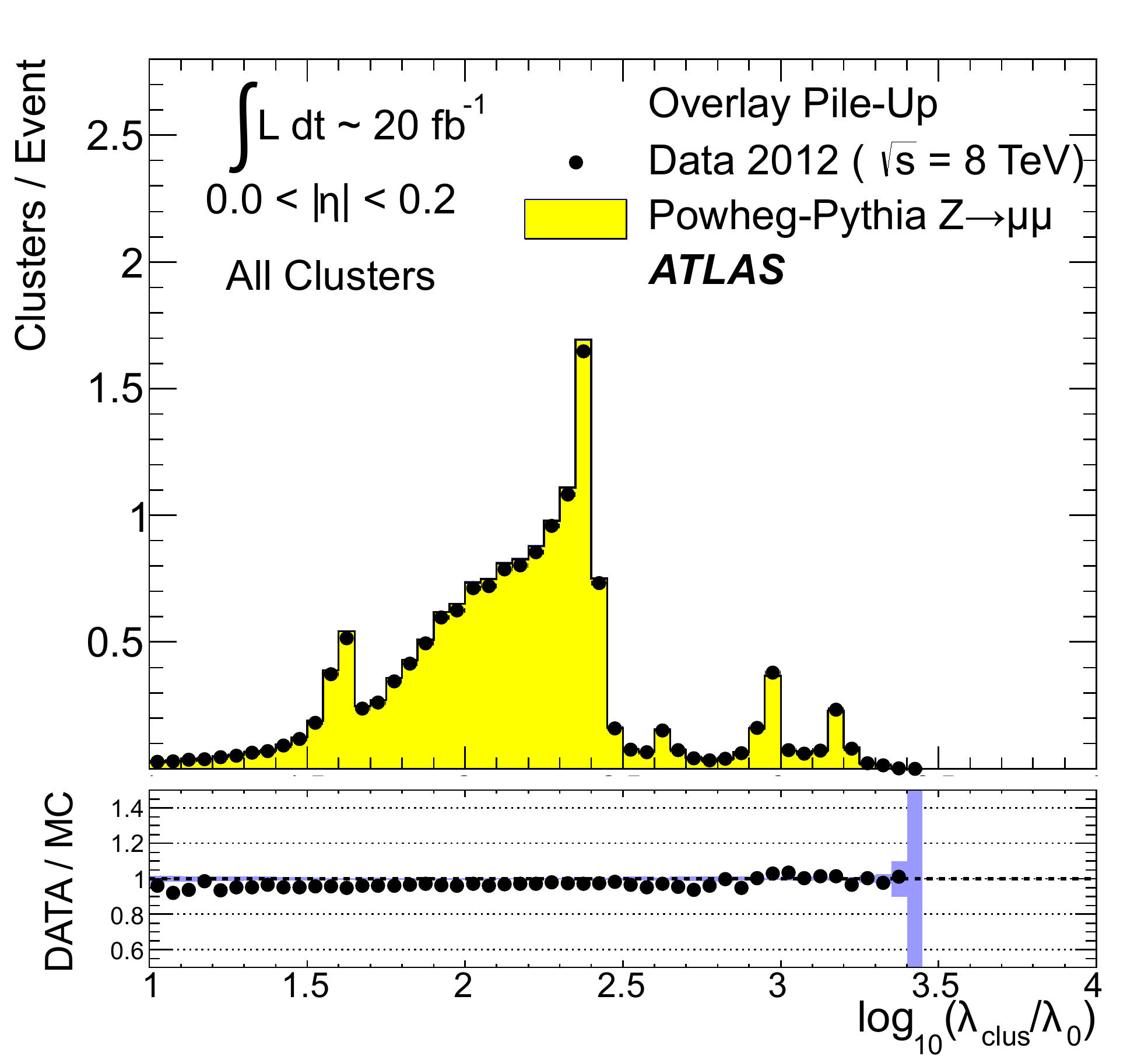}\label{fig:spectra:lam:00_02:ov}}
	\\
	\subfloat[]{\includegraphics[width=\figsixpanelwidth]{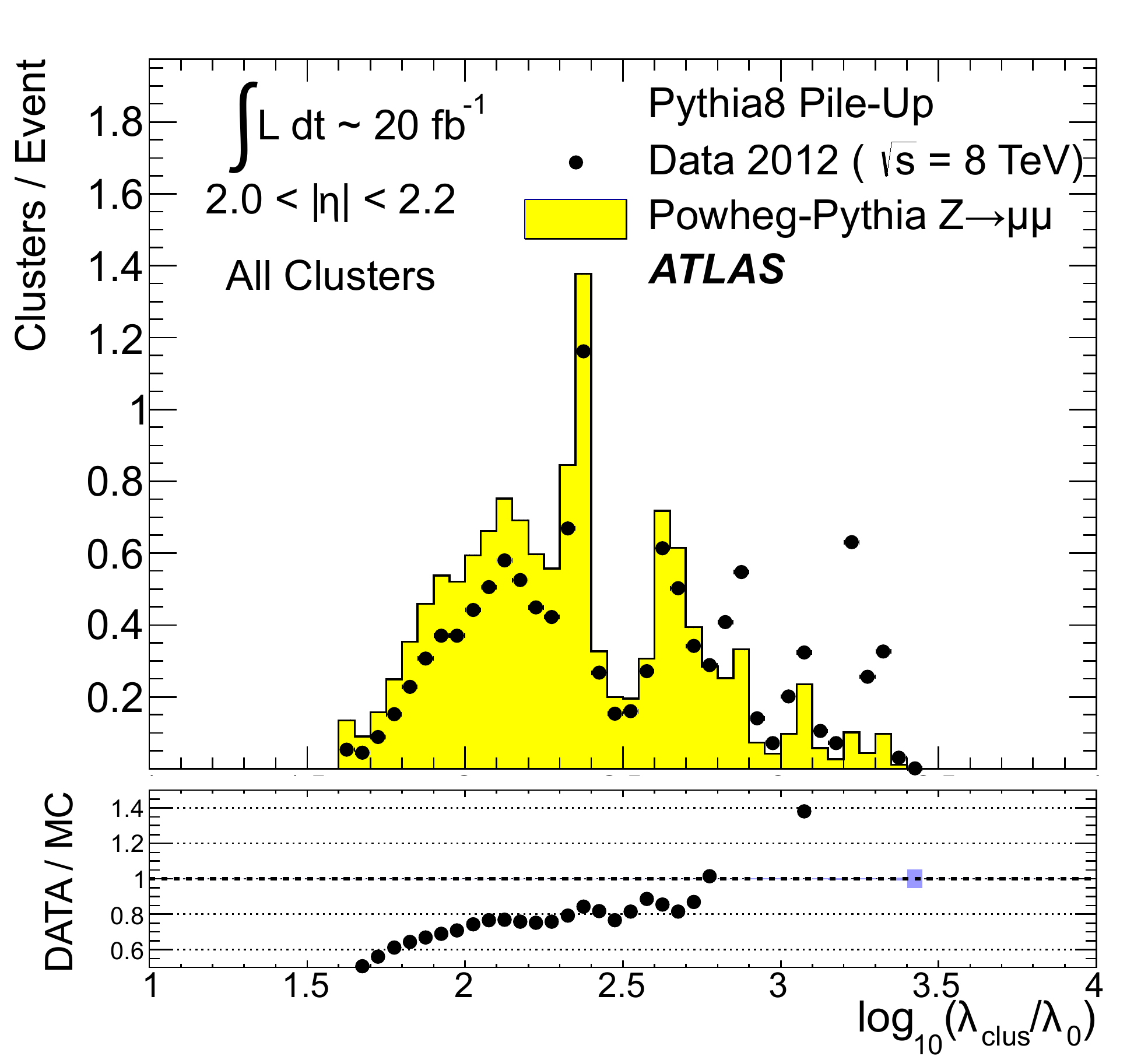}\label{fig:spectra:lam:20_22:mc}}      \qquad
	\subfloat[]{\includegraphics[width=\figsixpanelwidth]{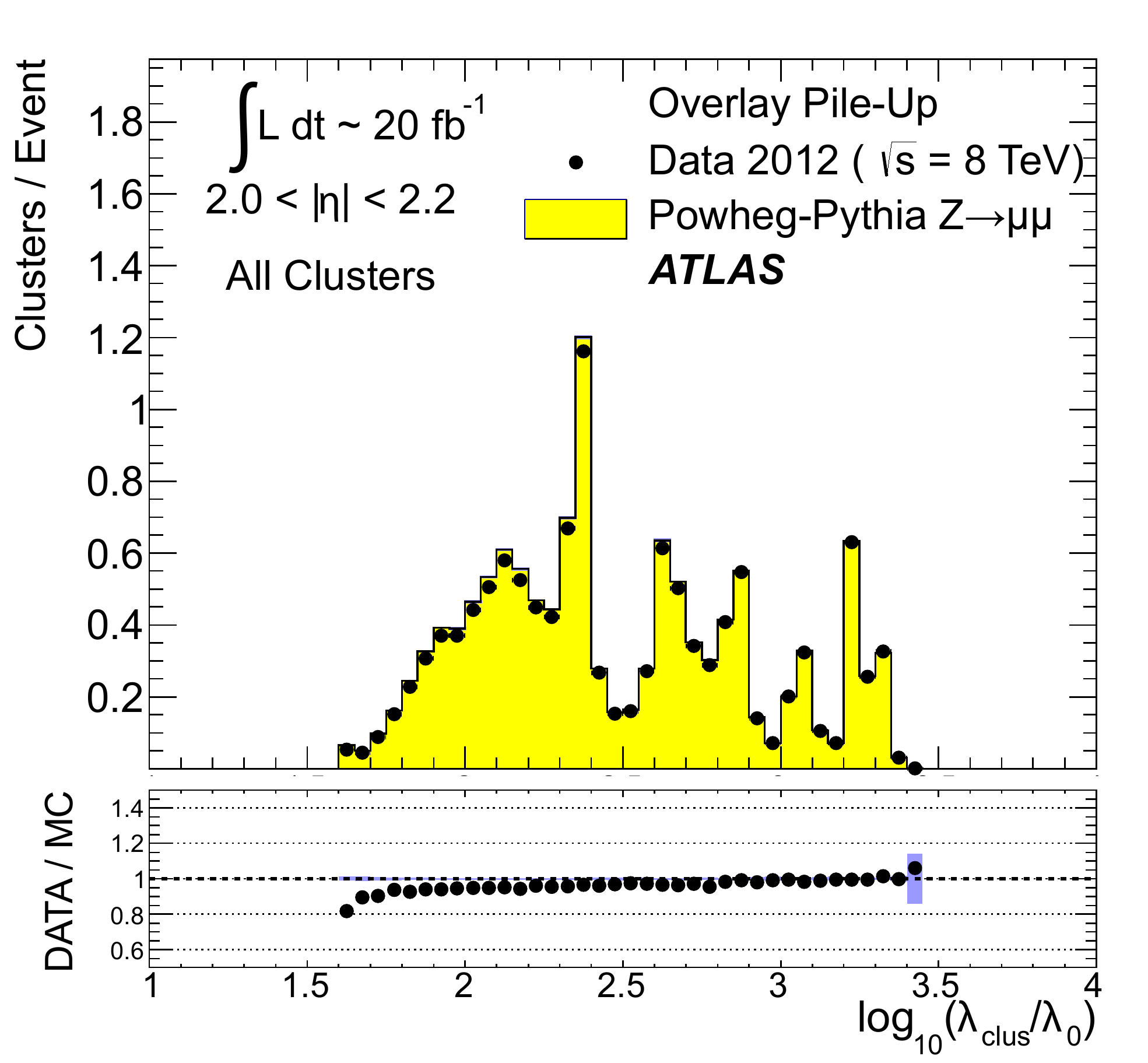}\label{fig:spectra:lam:20_22:ov}}
	\\
	\subfloat[]{\includegraphics[width=\figsixpanelwidth]{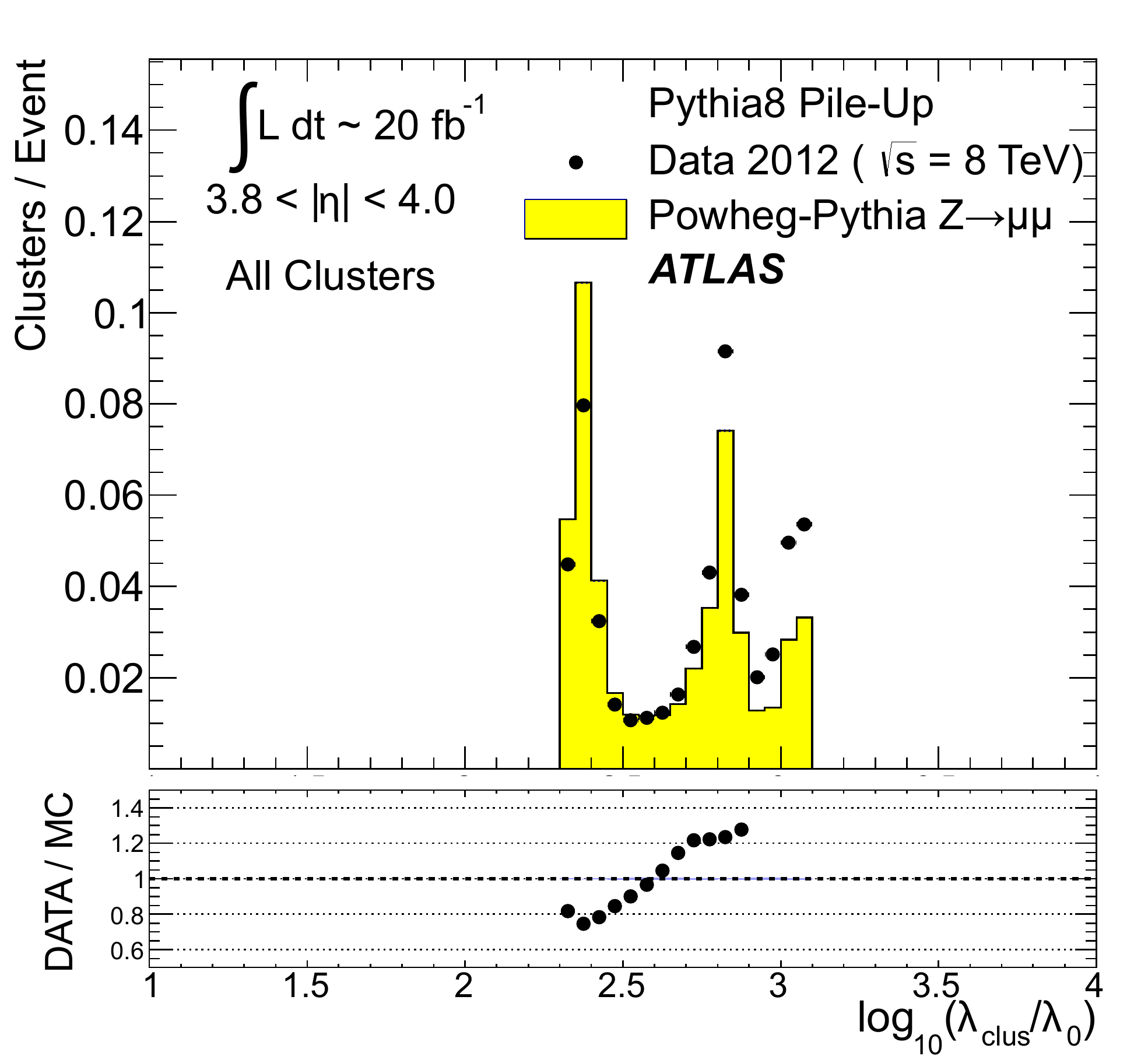}\label{fig:spectra:lam:38_40:mc}}      \qquad	
	\subfloat[]{\includegraphics[width=\figsixpanelwidth]{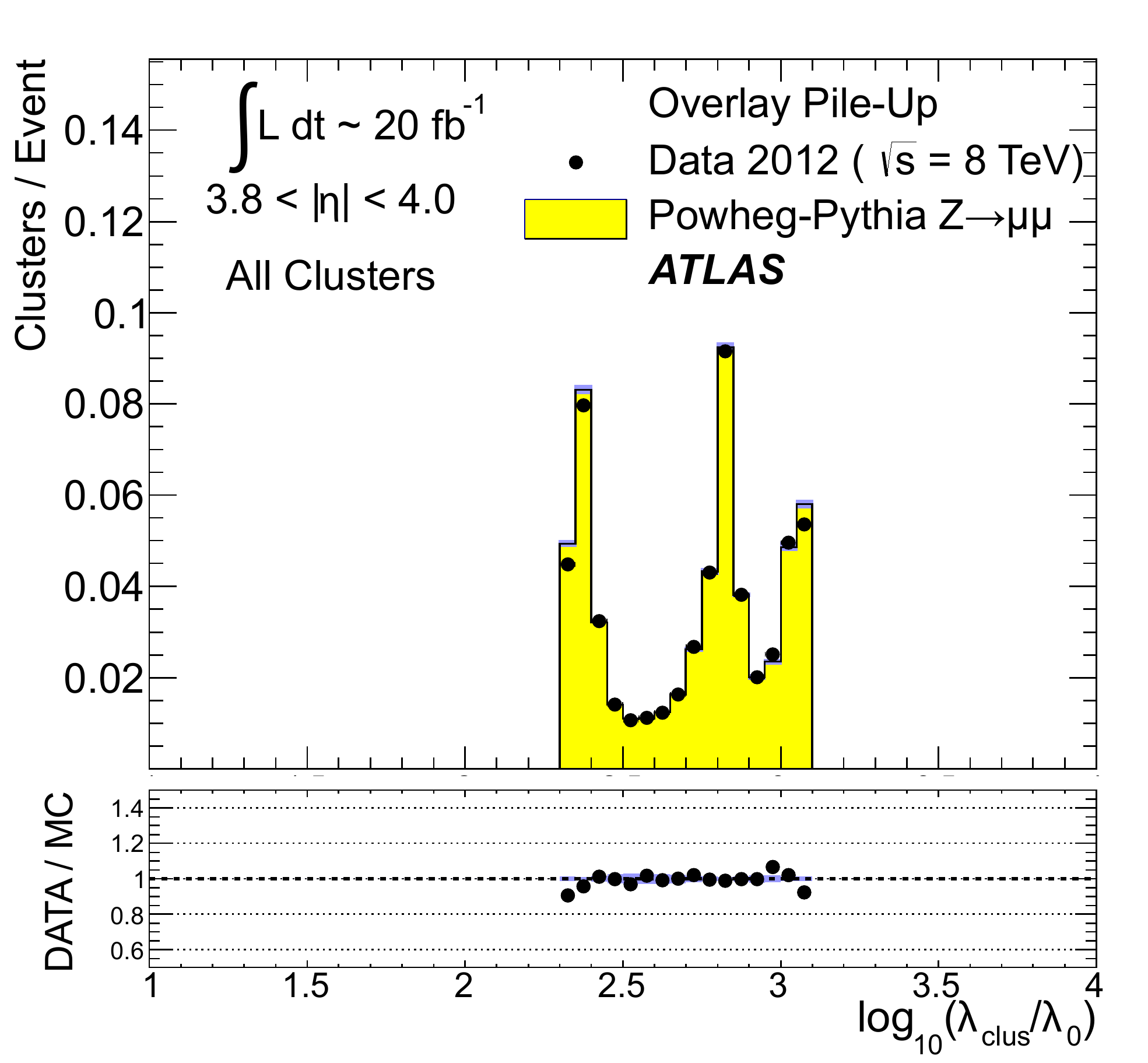}\label{fig:spectra:lam:38_40:ov}}
	\caption[]{The distribution of the \topo{} depth location, measured in terms of $\log_{10}(\lamctr/\lambda_{0})$, for clusters in various bins of \etaclus{} for an inclusive \Zmumu{} event sample recorded in 2012. Data is compared to distributions from \MC{} simulations including fully simulated \pu{} for all \topos{} within  \subref{fig:spectra:lam:00_02:mc} $|\etaclus| < 0.2$, \subref{fig:spectra:lam:20_22:mc} $2.0 < |\etaclus| < 2.2$, and \subref{fig:spectra:lam:38_40:mc} $3.8 < |\etaclus| < 4.0$. The corresponding distributions for \MC{} simulations with \pu{} from data overlaid are depicted in \subref{fig:spectra:lam:00_02:ov}, \subref{fig:spectra:lam:20_22:ov}, and \subref{fig:spectra:lam:38_40:ov}. The ratios of the distributions for data and \MC{} simulations are shown below the respective distributions. The shaded bands indicate the statistical uncertainties for \MC{} simulations.} 
\label{fig:spectra:lam}		
\end{figure}

\begin{figure}[tp!] \centering
        \sfcompress
	\subfloat[]{\includegraphics[width=\figsixpanelwidth]{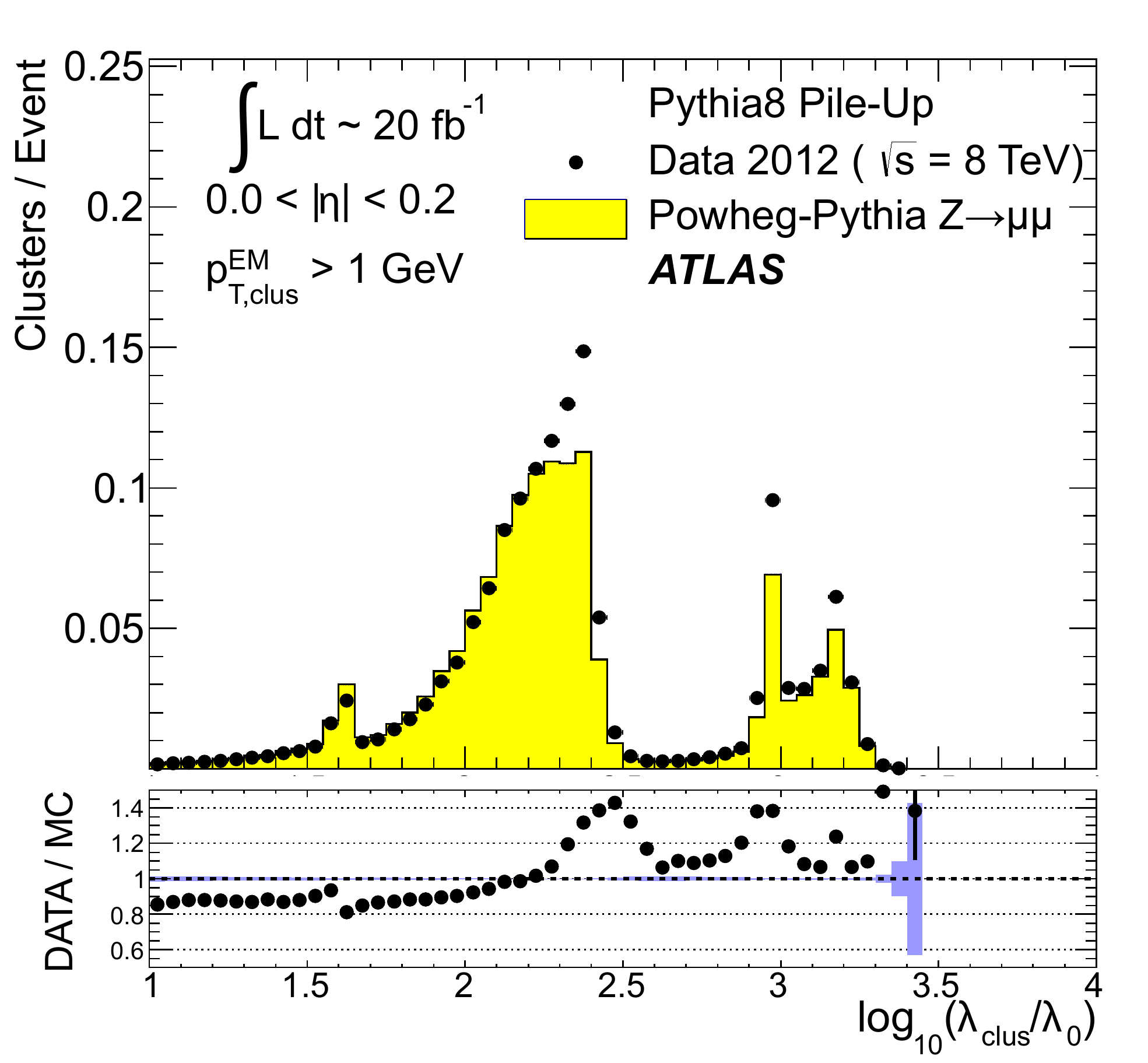}\label{fig:spectra:lam:ptexcl:1:mc}}	\qquad
	\subfloat[]{\includegraphics[width=\figsixpanelwidth]{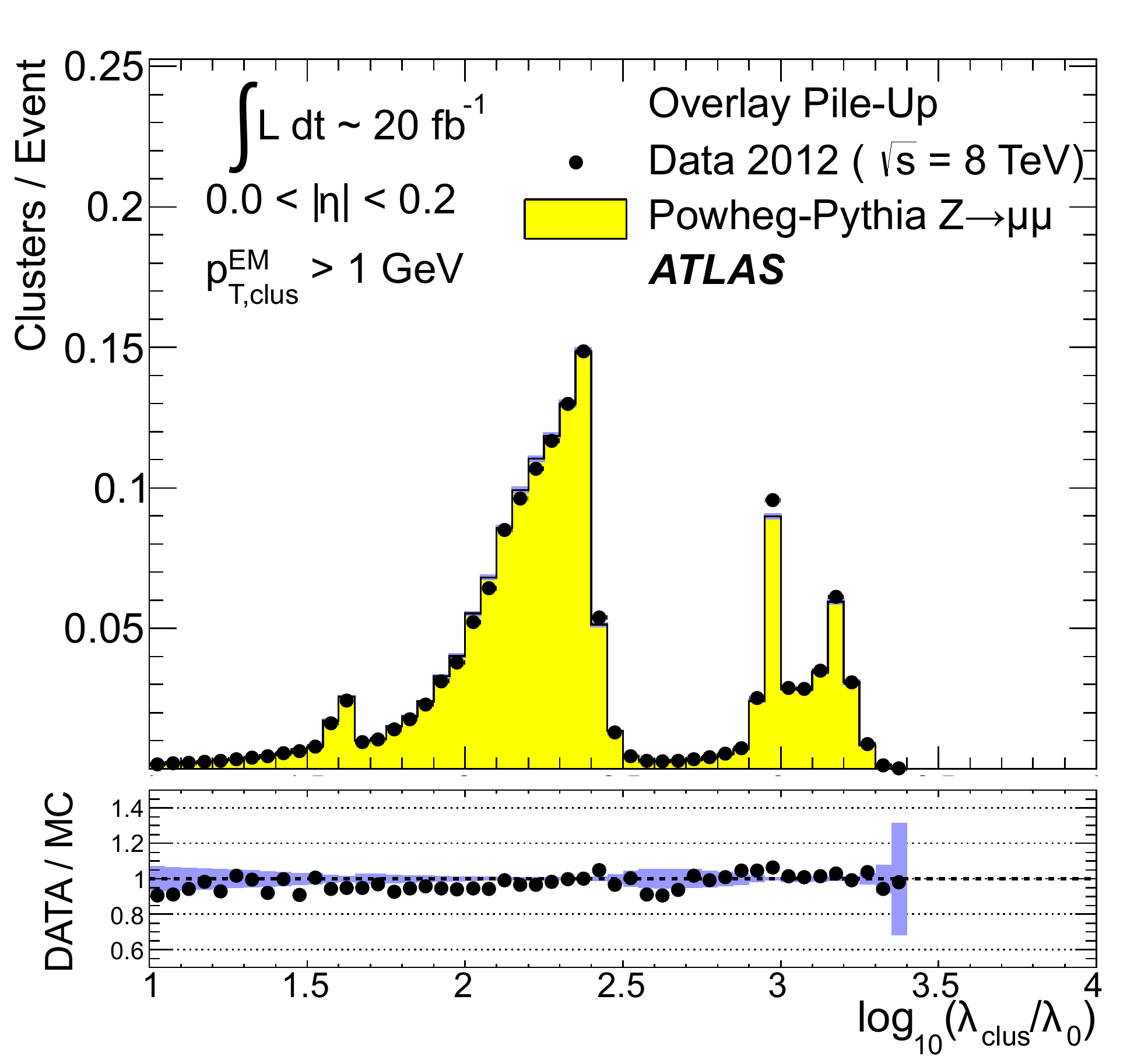}\label{fig:spectra:lam:ptexcl:1:ov}}
	\\
	\subfloat[]{\includegraphics[width=\figsixpanelwidth]{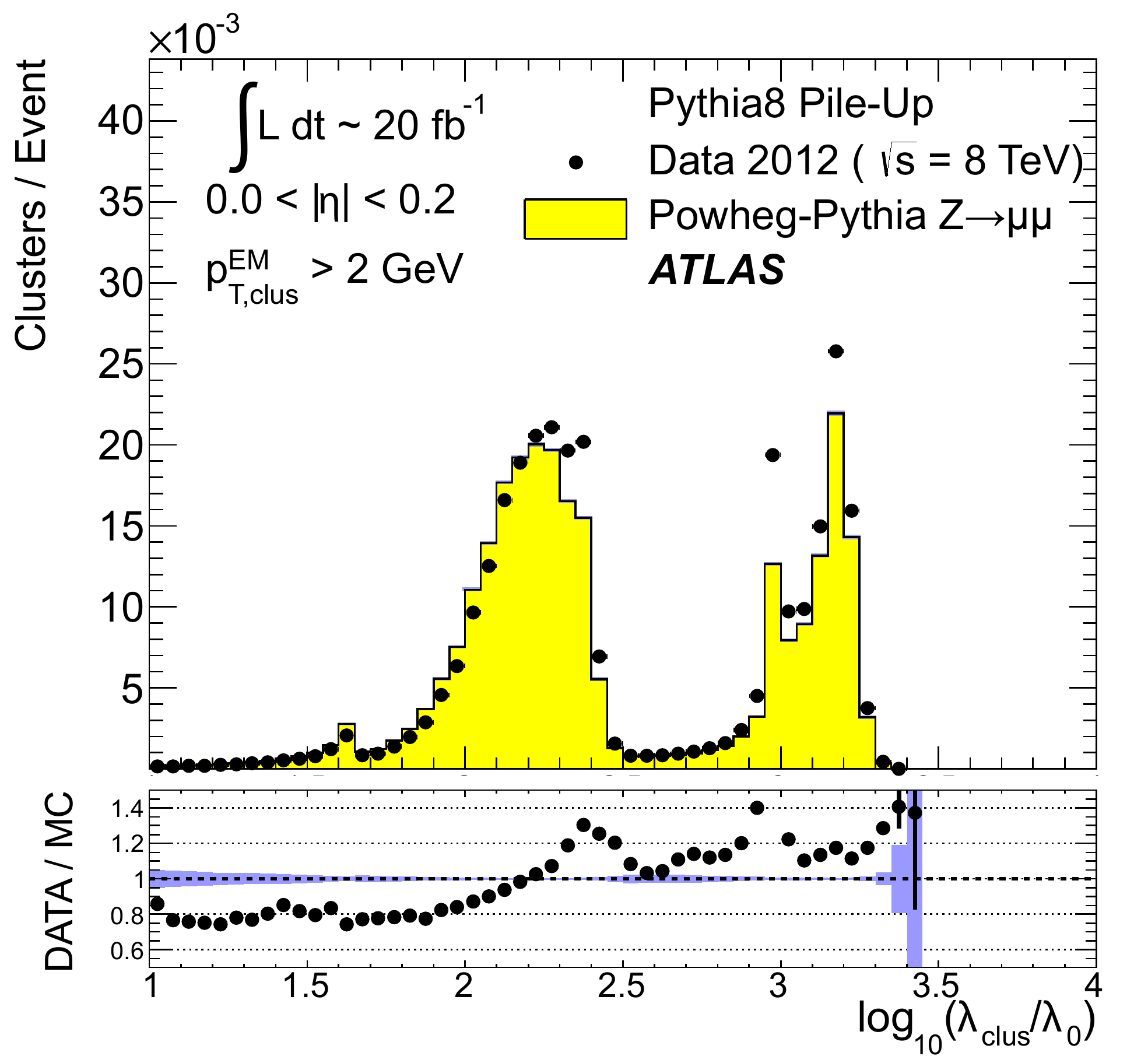}\label{fig:spectra:lam:ptexcl:2:mc}}      \qquad
	\subfloat[]{\includegraphics[width=\figsixpanelwidth]{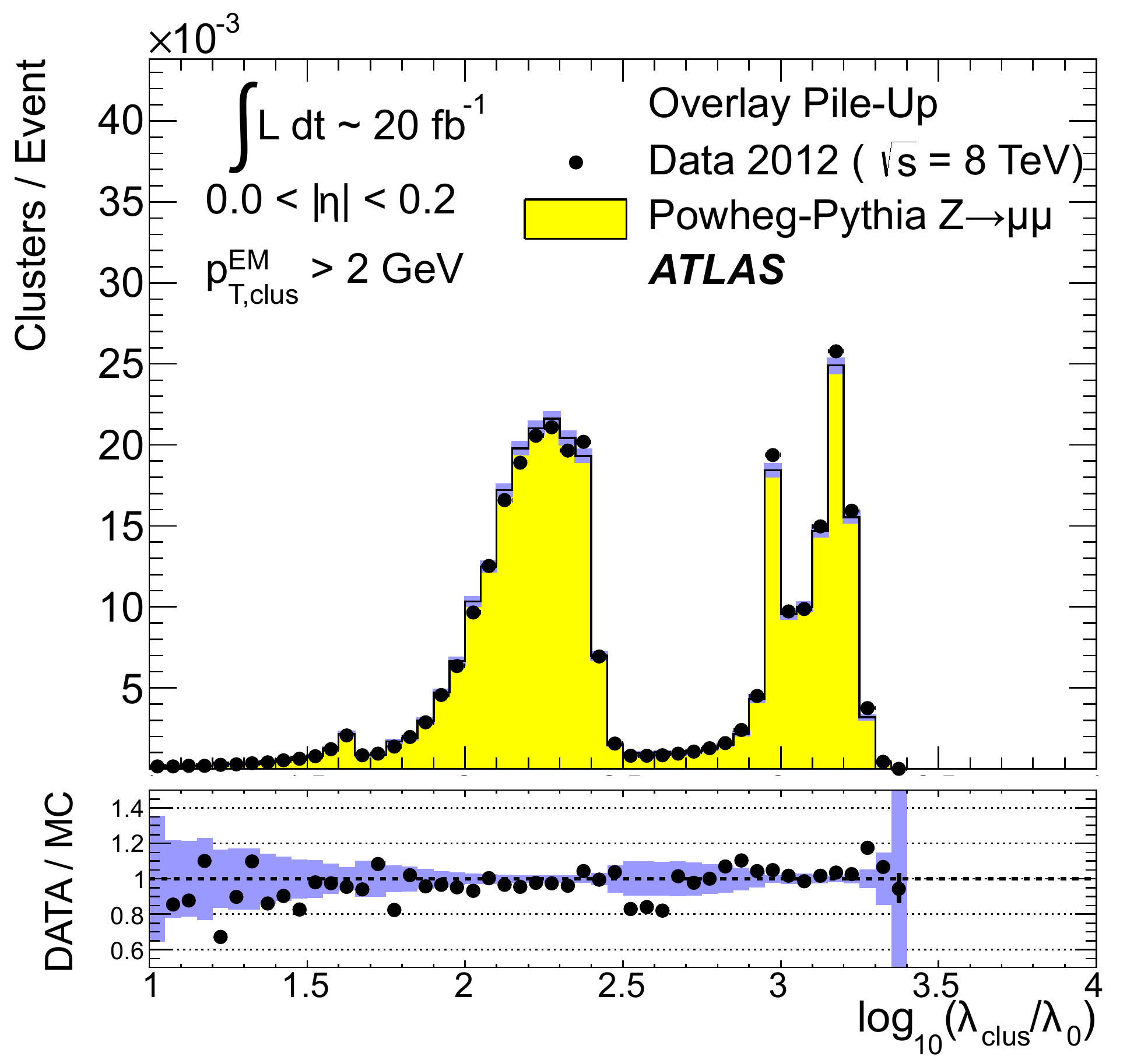}\label{fig:spectra:lam:ptexcl:2:ov}}
        \\
	\subfloat[]{\includegraphics[width=\figsixpanelwidth]{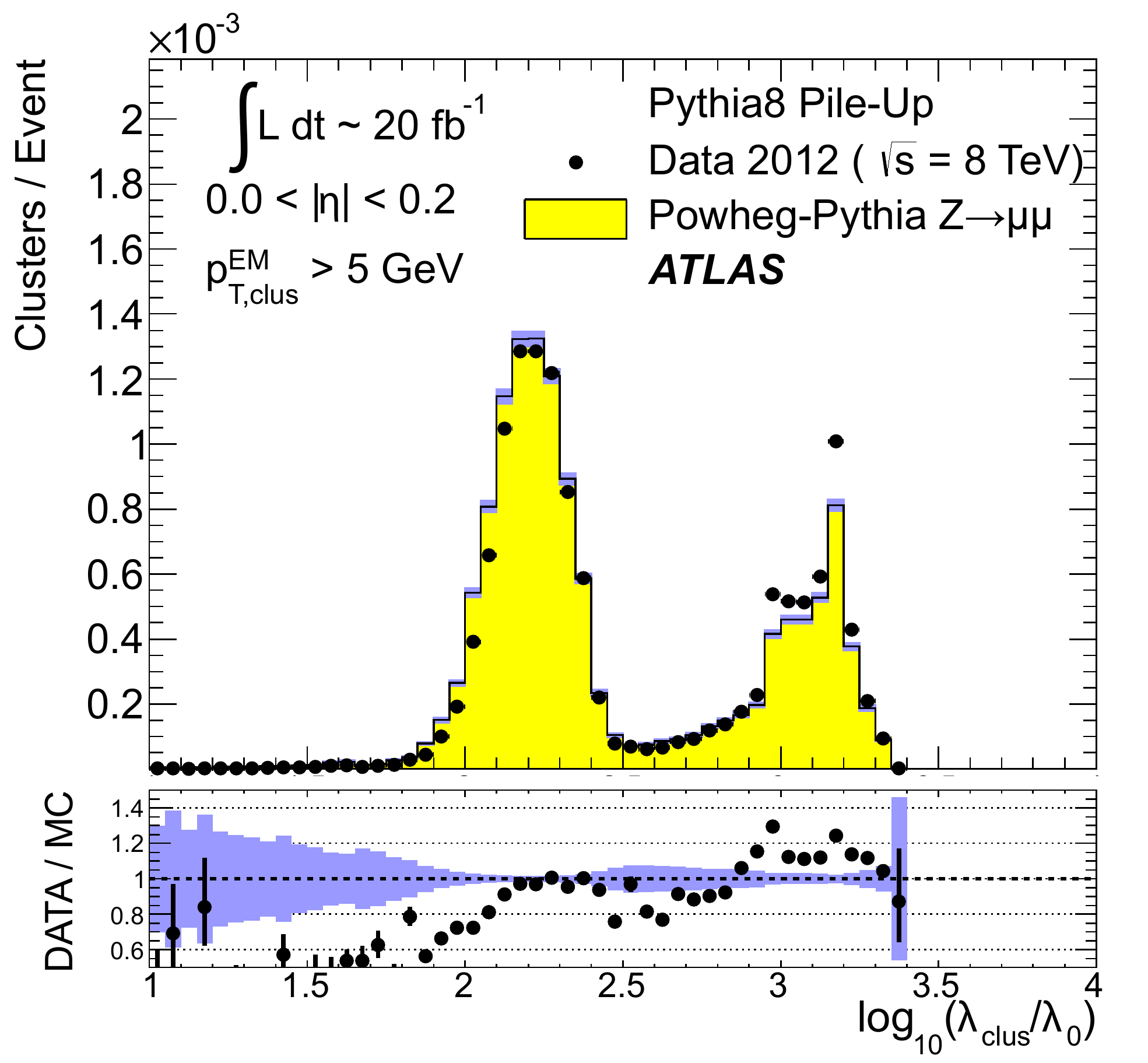}\label{fig:spectra:lam:ptexcl:5:mc}}      \qquad	
	\subfloat[]{\includegraphics[width=\figsixpanelwidth]{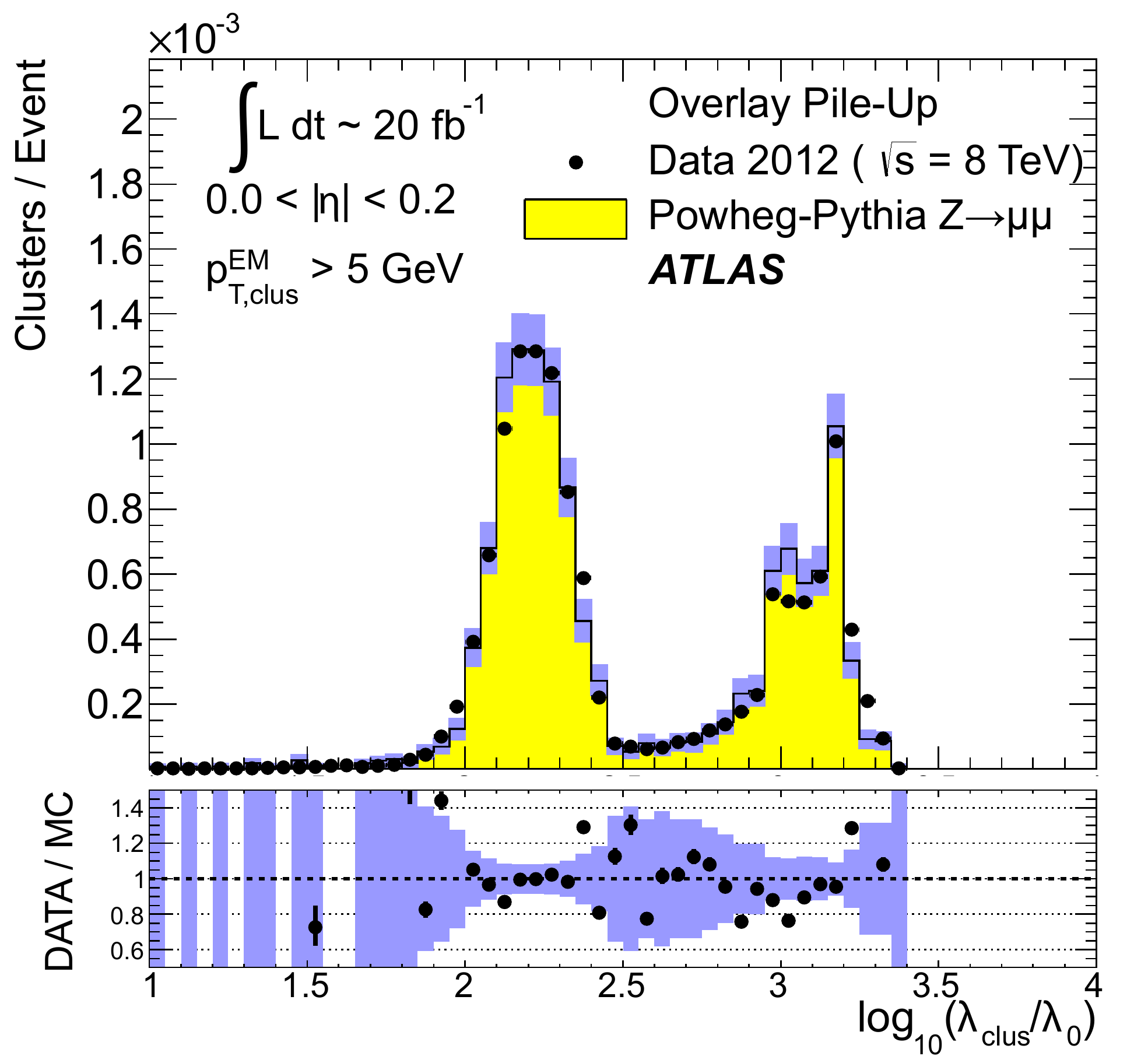}\label{fig:spectra:lam:ptexcl:5:ov}}
	\caption[]{The distribution of the \topo{} depth location, measured in terms of $\log_{10}(\lamctr/\lambda_{0})$, for selected \topos{} within $|\etaclus| < 0.2$ and with a transverse momentum \ptclusem, evaluated on the \EM{} scale, larger than various thresholds. Results are shown for an inclusive \Zmumu{} event sample recorded in 2012. Data are compared to distributions from \MC{} simulations including fully simulated \pu{} for all \topos{} with  \subref{fig:spectra:lam:ptexcl:1:mc} $\ptclusem > \unit{1}{\GeV}$, \subref{fig:spectra:lam:ptexcl:2:mc} $\ptclusem > \unit{2}{\GeV}$, and \subref{fig:spectra:lam:ptexcl:5:mc} $\ptclusem > \unit{5}{\GeV}$.
The corresponding distributions for \MC{} simulations with \pu{} from data overlaid are depicted in \subref{fig:spectra:lam:ptexcl:1:ov}, \subref{fig:spectra:lam:ptexcl:2:ov}, and \subref{fig:spectra:lam:ptexcl:5:ov}. 
The shaded bands indicate the statistical uncertainties for the distributions obtained from \MC{} simulations and the corresponding uncertainties in the ratio plots.}
\label{fig:spectra:lam:ptexcl}		
\end{figure}

\PU{} is expected to affect cluster moments as well as the overall cluster kinematics. Its diffuse energy emission can not only produce additional \topos, but also change the \cog, the barycentre, and other cluster shapes. 
In some cases, \pu{} can actually increase the cluster splitting, as additional local signal maxima can be inserted into a \topo{} by \pu. In addition, the increased cell noise can produce additional signal minima in groups of previously connected cells in the \topo. 
This last effect can be more important for \topos{} in jets and is discussed in \secRef{\thislabel:jets}. The \topo{} depth location \lamctr{} discussed here serves as an example for the quality of modelling cluster moments in the presence of \pu. Other moments are investigated in the context of jets.

The modelling of \lamctr{} in the calorimeter is compared to data in \figRef{fig:spectra:lam} for the inclusive \Zmumu{} sample in the same bins of  \etaclus{} used for the study of \pTclus{} in
\figRef{fig:spectra:pt}. 
The fully simulated events with \pu{} from the minimum-bias simulations show significant differences from the data, while the \MC{} simulations overlaid with \pu{} from data show good agreement with respect to all features of these distributions. 
The complex structure of the distributions reflects the longitudinal calorimeter segmentation in the various regions defined by \etaclus. 
For example, the forward direction $3.8 < |\etaclus| < 4.0$ is covered by the \LArFCAL, which has three coarse and deep longitudinal segments (approximately $\unit{2.5/3.5/3.5}{\lamnucl}$). 
This structure generates \topos{} preferably in the depth centre of each module, as can be seen in \figMultiRefLabel~\ref{fig:spectra:lam}\subref{fig:spectra:lam:38_40:mc} and \ref{fig:spectra:lam}\subref{fig:spectra:lam:38_40:ov}. 
These distributions are dominated by low-energy clusters associated with \pu{} interactions such that the improvement seen by using data overlay is expected.

Similarly to the studies of the kinematic and flow properties of \topos{} discussed in \secMultiRef{\thislabel:obs:kine}{and}{\thislabel:obs:ptflow}, more exclusive \topo{} selections are also investigated. \FigRef{fig:spectra:lam:ptexcl} shows \datatomc{} comparisons of the \lamctr{} distributions for clusters within $|\etaclus| < 0.2$ for $\ptclusem > \pti{\text{min}}$ with $\pti{\text{min}} \in \unit{\{ 1, 2, 5 \}}{\GeV}$, for \MC{} simulations with fully simulated \pu{} and for \MC{} simulations with \pu{} from data overlaid. The \MC{} simulation with overlaid \pu{} 
agrees better with data than the one with fully simulated \pu{}, particularly in the case of the least restrictive $\pti{\text{min}} = \unit{1}{\GeV}$ \topo{} selection.

\subsection{\Topos{} in jets} \label{\thislabel:jets}

Jets are important in many analyses at the \LHC. 
Therefore, the performance of the simulation of their constituents is important, in particular for analyses employing jet substructure techniques or relying on the jet mass.
In order to study the \topo{} features in jets and the jet \topo{}
composition, exclusive jet samples are extracted from data and \MC{}
simulation using the \Zmumu{} and jet selection described in \secRef{\thislabel:obs:ef}. As the jets are globally corrected for \pu{} \cite{Aad:2014bia}, they form a stable kinematic reference for the evaluation of \pu{} effects on the \topos{} used to reconstruct them. Jets include only \topos{} with $E > 0$, as required by the kinematic recombination.

The full evaluation of the reconstruction performance for jets formed with \topos{} on both \EM{} and \LCW{} scale is presented in \citMultiRef{Aad:2011he,Aad:2014bia}. 
The evaluation of the jet energy resolution can be found in \citRef{Aad:2012ag}.

\subsubsection{Jet energy scale and \topo-based response in \pu}\label{\thislabel:resp}

\begin{figure}[t!] \centering
	\subfloat[]{\includegraphics[width=\fighalfwidth]{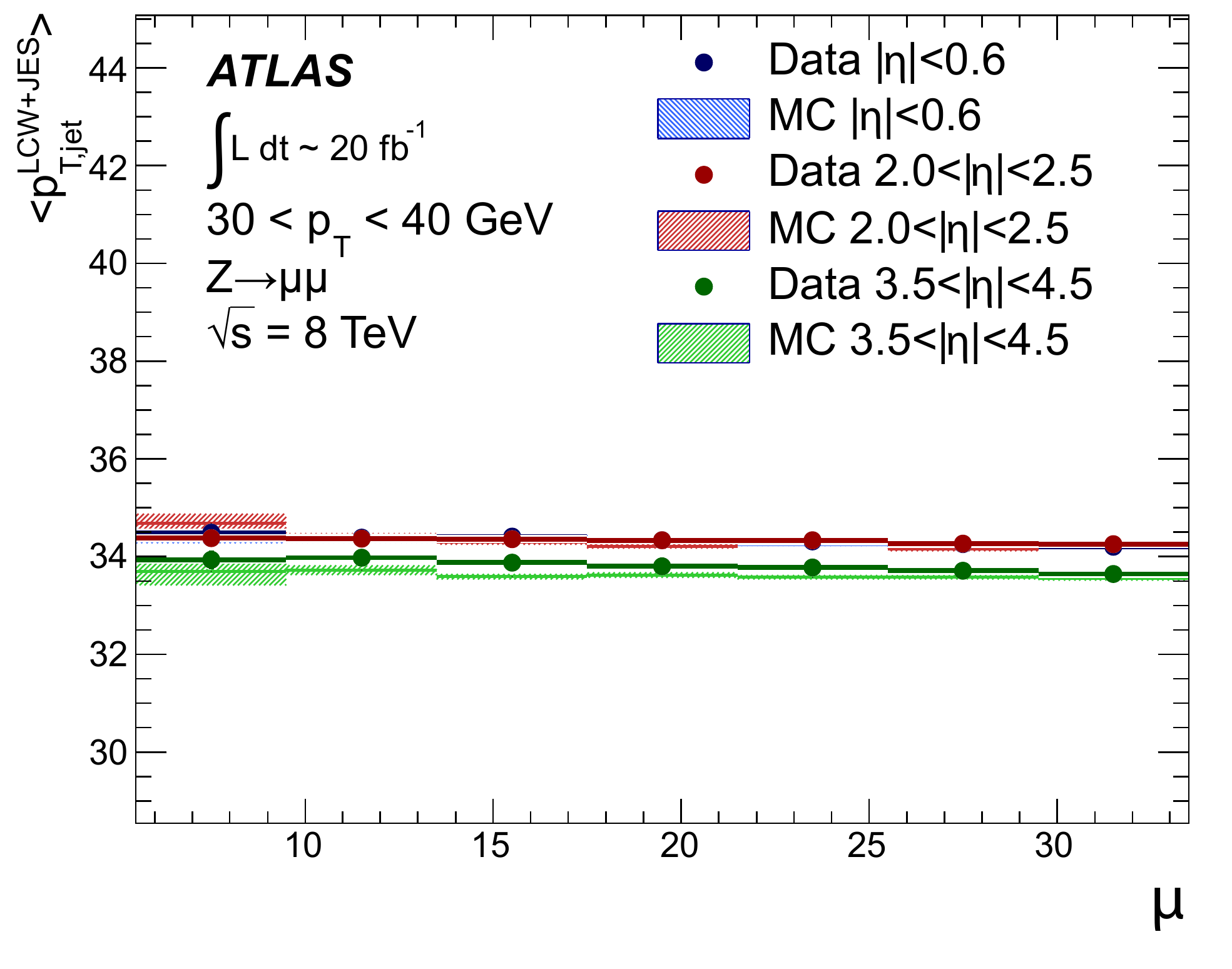}\label{fig:jet:pt:1}} 
	\subfloat[]{\includegraphics[width=\fighalfwidth]{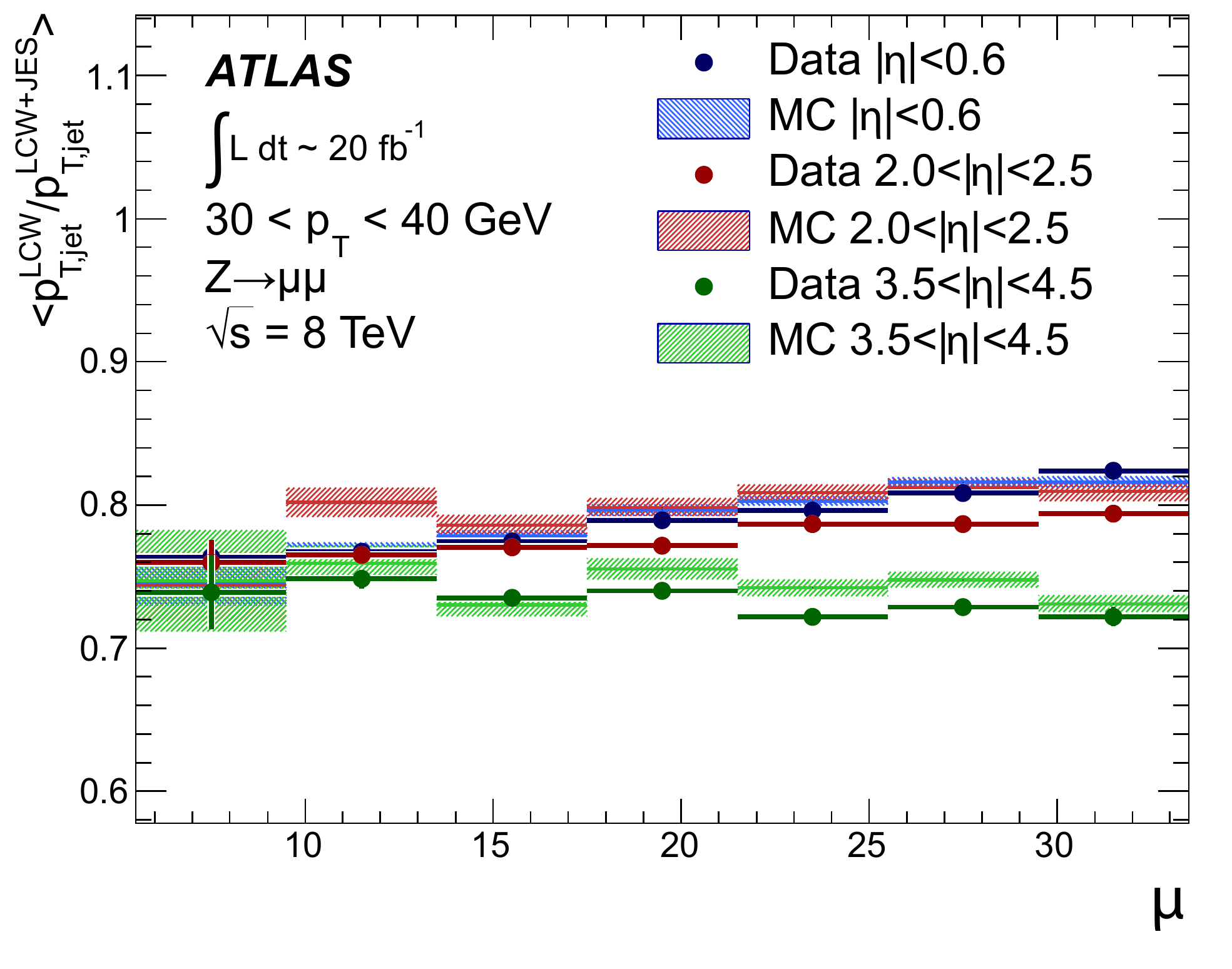}\label{fig:jet:pt:2}} 
	\caption[]{In \subref{fig:jet:pt:1}, the fully calibrated and corrected jet \pT{} response measured by \ptjetlcwjes{} is shown as a function of the \pu{} activity measured by $\mu$, in three different detector regions for \Zmumu{} events with one \antikt{} jet with $R = 0.4$ with $\unit{30}{\GeV} < \ptjetlcwjes < \unit{40}{\GeV}$, for 2012 data and \MC{} simulations with fully simulated \pu.  The $\mu$ dependence of the uncorrected jet \pT{} response is shown in \subref{fig:jet:pt:2}. It is measured in terms of its ratio to the fully calbrated jet response, $\ptjetlcw/\ptjetlcwjes(\mu)$, for the same events and in the same detector regions.
The shaded bands shown for the results from \MC{} simulations indicate statistical uncertainties.}
	\label{fig:jet:pt}
\end{figure}

As mentioned above, the fully calibrated four-momentum \Pjet{} of jets reconstructed from \topos{} is corrected for \pu{} effects. Therefore, the corresponding transverse momentum \ptjet{} provides a stable signal for event selections and the kinematics of the true particle flow. The basic jet four-momentum is reconstructed on two different scales, the \EM{} scale and the \LCW{} scale using locally calibrated \topos{} with $E > 0$:
\begin{eqnarray}
	\Pjetem & = & \sum_{i=1}^{\Nclusjet} \Pclusem  \label{eq:pjet:em} \\
	\Pjetlcw & = & \sum_{i=1}^{\Nclusjet} \Pcluslcw \label{eq:pjet:lcw}
\end{eqnarray}
The sum runs over the number \Nclusjet{} of \topos{} in a given jet.
Both \Pjetem{} and \Pjetlcw{} are not corrected further. The corresponding \pT{} responses \ptjetem{} and \ptjetlcw{} are therefore affected by \pu. A full jet energy scale (\JES) calibration is applied to both scales, yielding \Pjetemjes{} and \Pjetlcwjes, respectively. This \JES{} calibration includes \pu{} corrections, response calibration, direction corrections and refinements from \emph{in situ} transverse momentum balances, similar to those outlined for 2011 data in \citRef{Aad:2014bia}. The respective fully calibrated transverse momentum is then \ptjetemjes{} and \ptjetlcwjes. 

\begin{figure}[tp!] \centering
        \sfcompress         
	\subfloat[]{\includegraphics[width=\figsixpanelwidth]{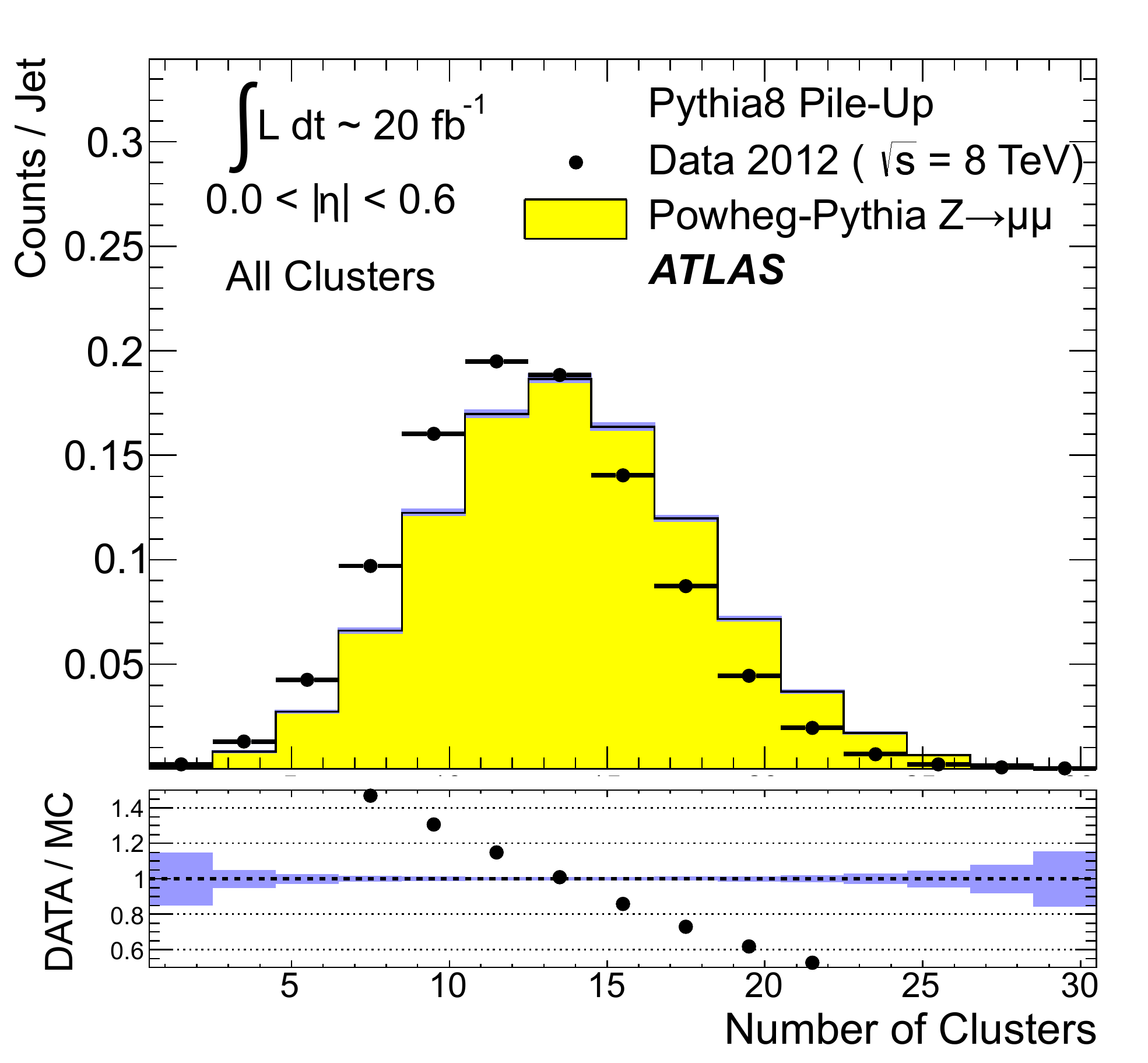}\label{fig:jets:nClus:1:mc}}      \qquad
	\subfloat[]{\includegraphics[width=\figsixpanelwidth]{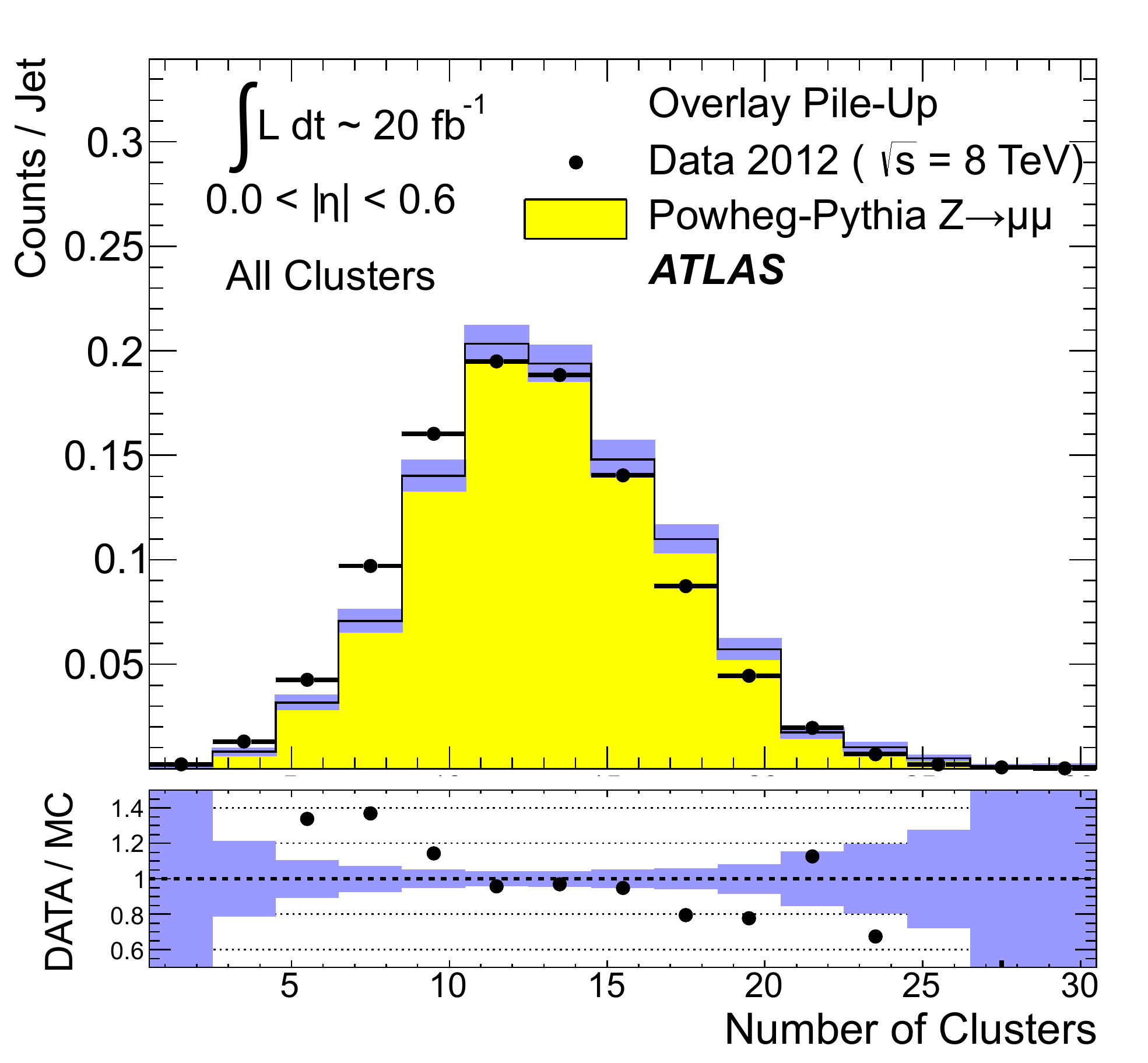}\label{fig:jets:nClus:1:ov}}
	\\
	\subfloat[]{\includegraphics[width=\figsixpanelwidth]{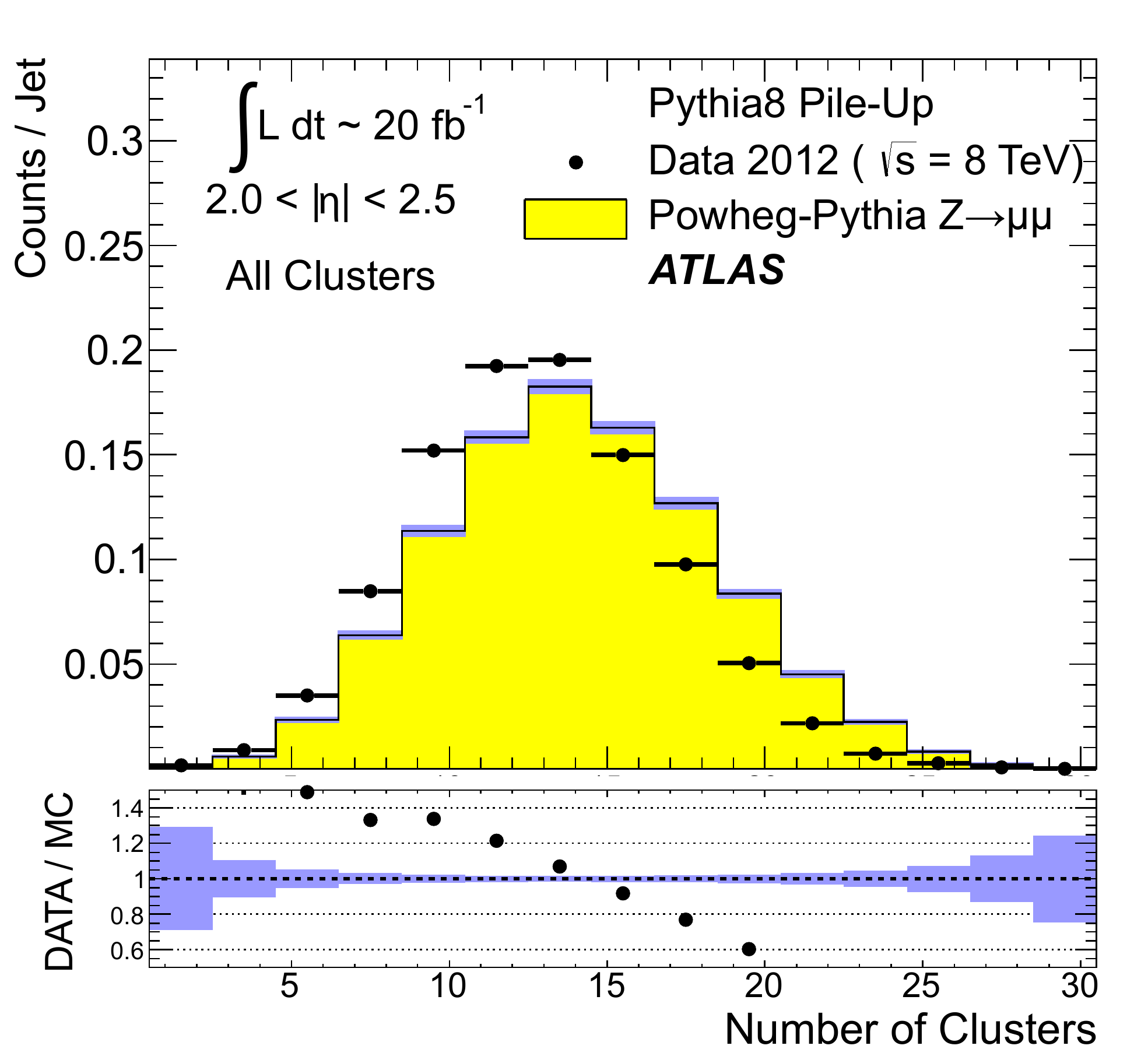}\label{fig:jets:nClus:2:mc}}      \qquad
	\subfloat[]{\includegraphics[width=\figsixpanelwidth]{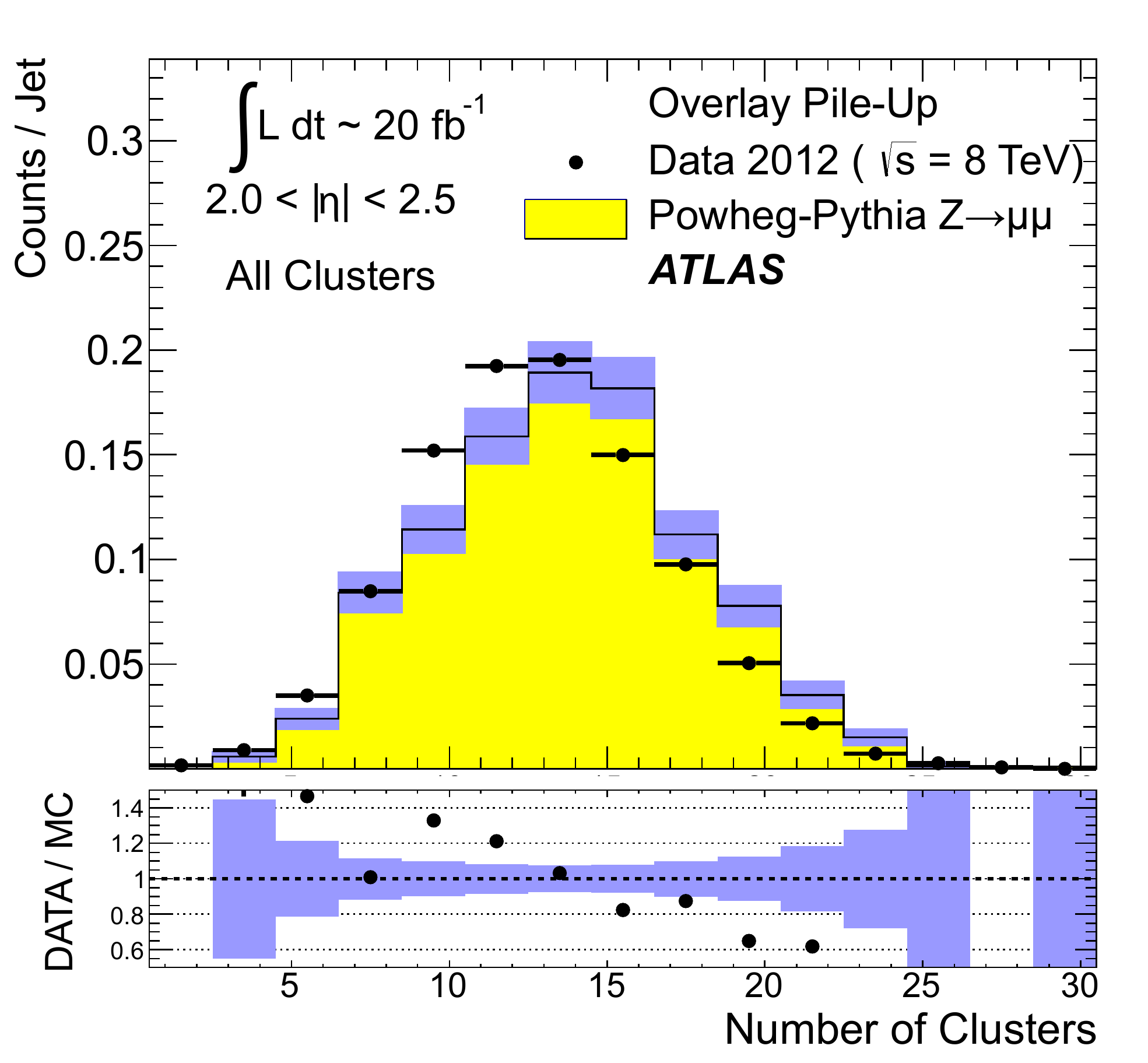}\label{fig:jets:nClus:2:ov}}
	\\
	\subfloat[]{\includegraphics[width=\figsixpanelwidth]{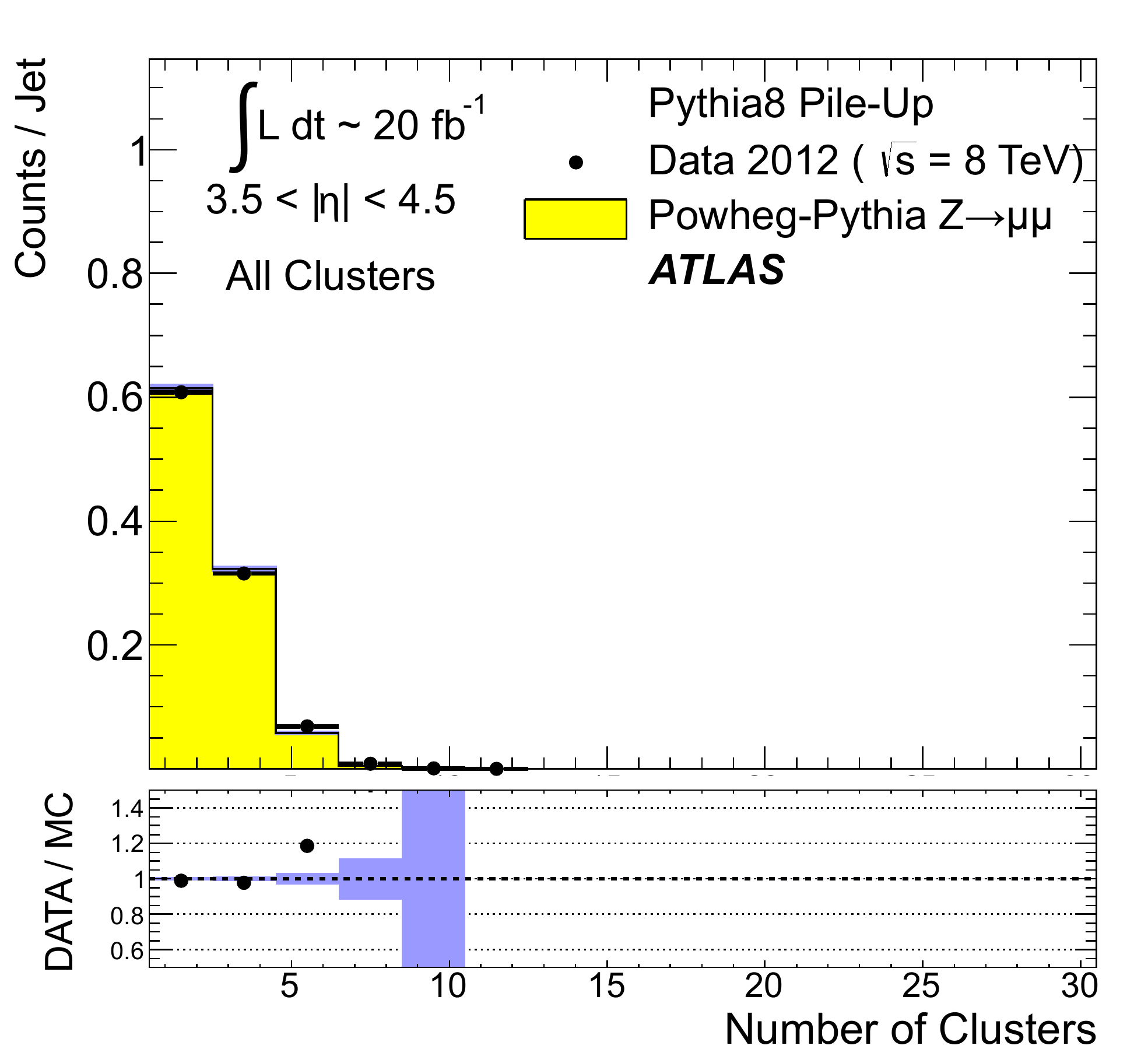}\label{fig:jets:nClus:3:mc}}      \qquad	
	\subfloat[]{\includegraphics[width=\figsixpanelwidth]{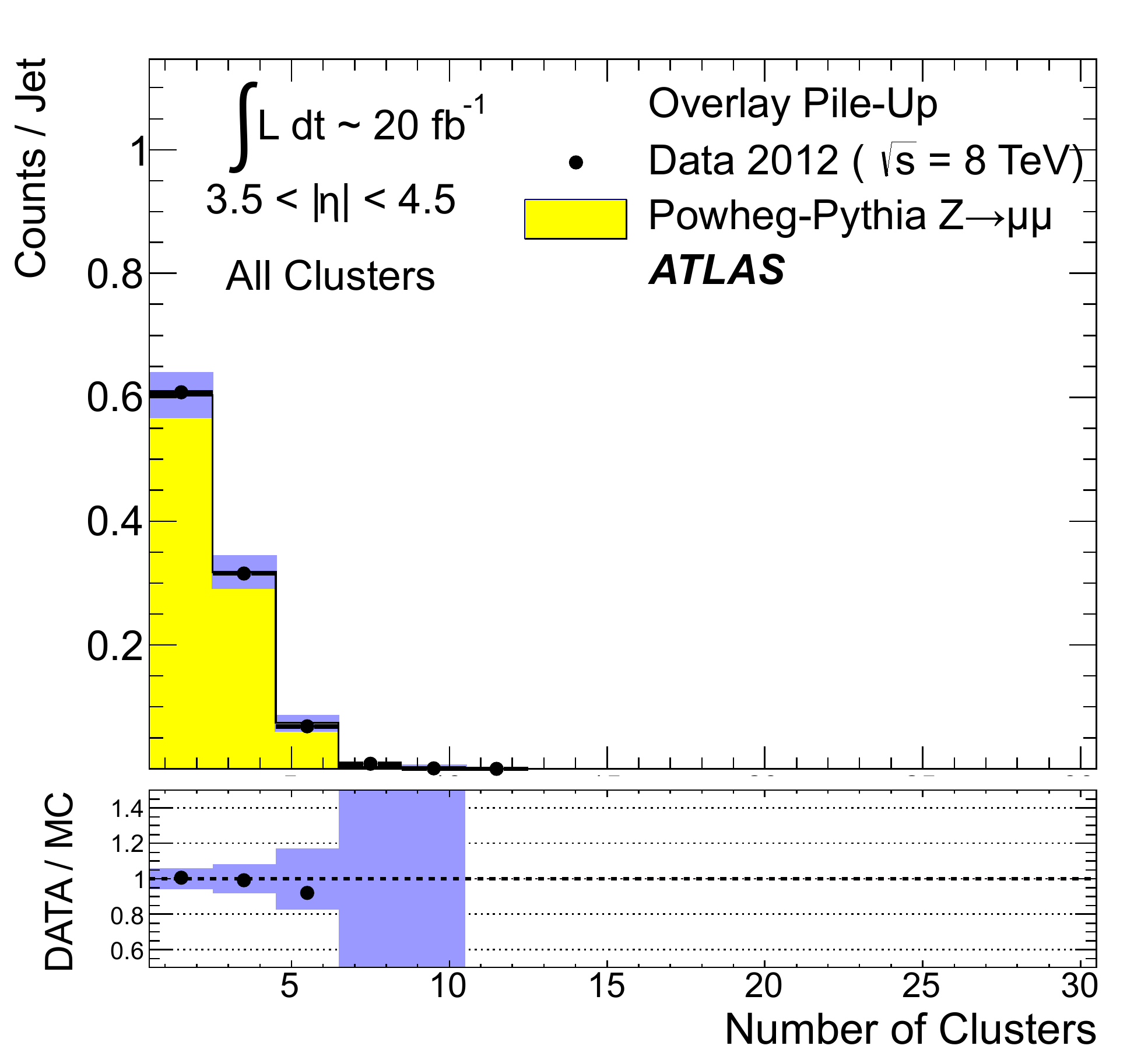}\label{fig:jets:nClus:3:ov}}
	\caption[]{The distribution of the number of \topos{} inside \antikt{} jets formed with $R = 0.4$ in the (\subref{fig:jets:nClus:1:mc}, \subref{fig:jets:nClus:1:ov}) central ($|\eta|<0.6$), the (\subref{fig:jets:nClus:2:mc}, \subref{fig:jets:nClus:2:ov}) \EndCap{} ($2.0<|\eta|<2.5$), and the (\subref{fig:jets:nClus:3:mc}, \subref{fig:jets:nClus:3:ov}) forward detector region ($3.5<|\eta|<4.5$) of \ATLAS. Shown are results from the analysis of \Zmumu{} events with at least one jet with $\unit{30}{\GeV} < \pT < \unit{40}{\GeV}$ in 2012
data and \MC{} simulations with fully simulated \pu{} in \subref{fig:jets:nClus:1:mc}, \subref{fig:jets:nClus:2:mc} and \subref{fig:jets:nClus:3:mc}, and with \pu{} from data overlaid in 
\subref{fig:jets:nClus:1:ov}, \subref{fig:jets:nClus:2:ov} and \subref{fig:jets:nClus:3:ov}. The ratios of results for data and \MC{} simulations are shown below the distributions.
The shaded bands show the statistical uncertainties for the distributions obtained from \MC{} simulations and the corresponding uncertainty bands in the ratio plots.}
\label{fig:jets:nClus}		
\end{figure}

\FigRef{fig:jet:pt} shows the \pu{} dependence of the fully calibrated \ptjetlcwjes{} and the uncorrected \ptjetlcw{} on the \pu{} activity in the event,
measured by $\mu$. Results are obtained from a \Zmumu{} sample of events with one jet with $\unit{30}{\GeV} < \ptjetlcwjes < \unit{40}{\GeV}$ in data and \MC{} simulations. 
While \figRef{fig:jet:pt}\subref{fig:jet:pt:1} shows the stability of the corrected jet \pT{} scale, \figRef{fig:jet:pt}\subref{fig:jet:pt:2} indicates
the different sensitivities of the uncorrected response to \pu{} in the various detector regions. 
The different shapes seen in this figure are mostly related to the calorimeter granularity and the specific shaping functions in the different \LAr{} calorimeters. 
While the general expectation that every \pu{} interaction adds energy 
to the jet is indicated in the rise of $\ptjetlcw/\ptjetlcwjes$ with increasing $\mu$, the dependence of this ratio on $\mu$ is less
pronounced for jets with $3.5 < |\etajet| < 4.5$ in the \FCAL{} calorimeter. This observation is related to the much coarser calorimeter geometry in this region, in addition to the different (faster) shaping function in the \FCAL, yielding a better average online \ipu{} suppression by the \opu{} signal history in 2012 running conditions (\unit{50}{\ns} bunch crossings).

\subsubsection{\Topo{} multiplicity in jets} \label{\thislabel:multi}

\FigRef{fig:jets:nClus} shows the distributions of the number of \topos{} (\Nclusjet) 
in central, \EndCap{} and forward jets. Distributions are shown using fully simulated \pu{} and using data overlay. The discrepancies
between \MC{} simulations and data, while slightly reduced in the
simulations employing the \pu{} overlaid from data, generally
persist.

\begin{figure}[tp!] \centering
        \sfcompress
	\subfloat[]{\includegraphics[width=\figsixpanelwidth]{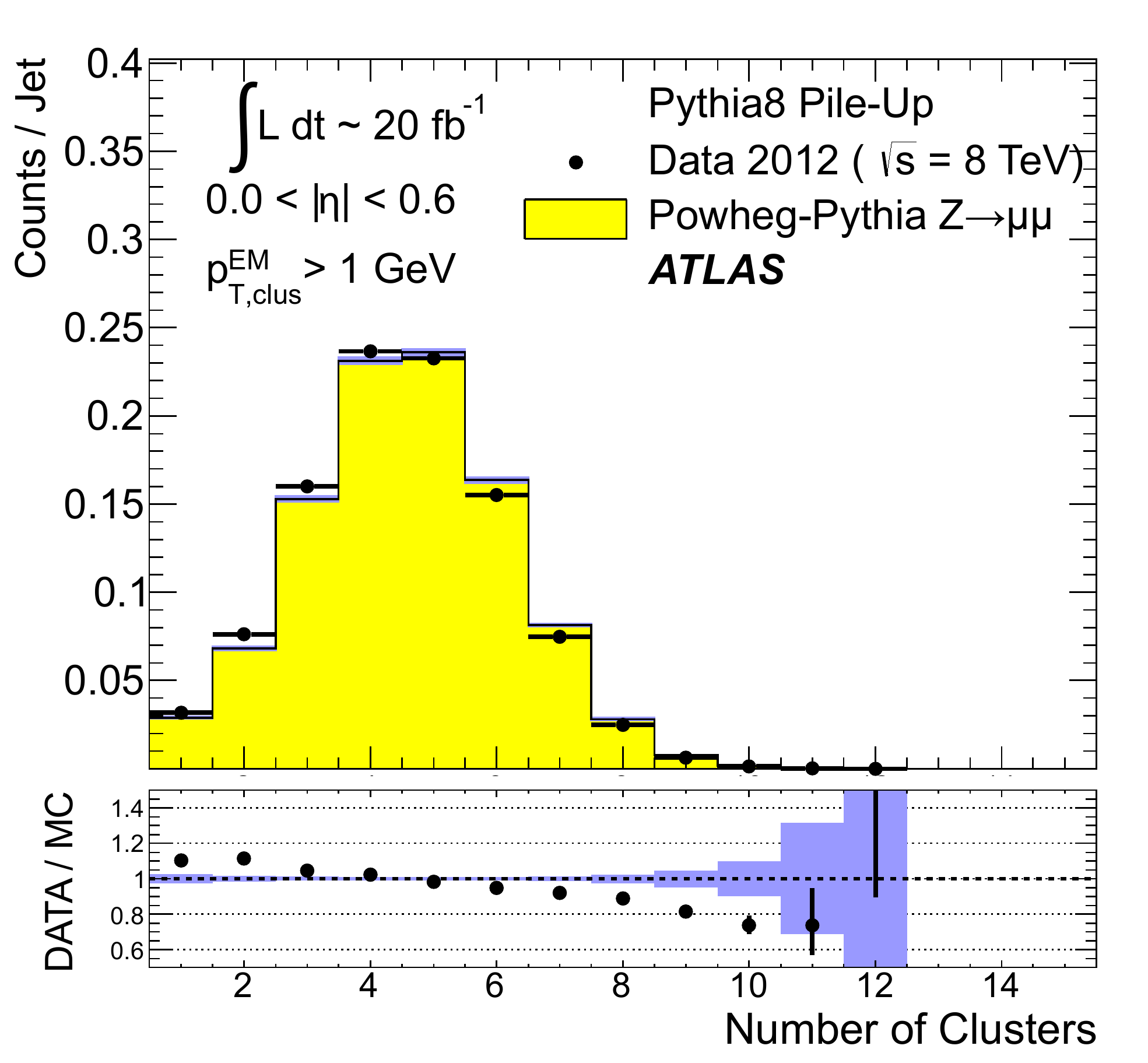}\label{fig:jets:nClus1GeV:1:mc}}      \qquad
	\subfloat[]{\includegraphics[width=\figsixpanelwidth]{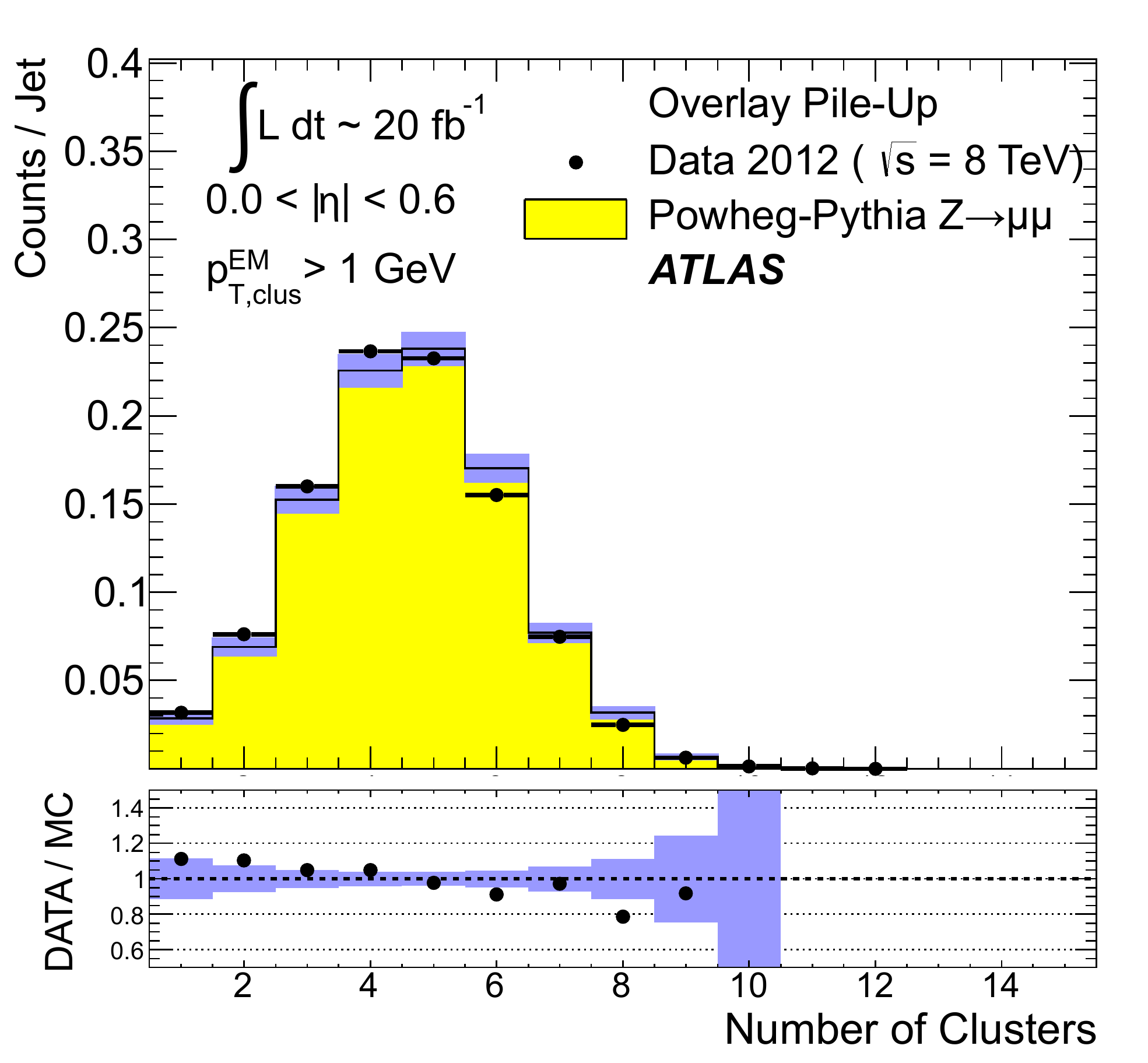}\label{fig:jets:nClus1GeV:1:ov}}
	\\
	\subfloat[]{\includegraphics[width=\figsixpanelwidth]{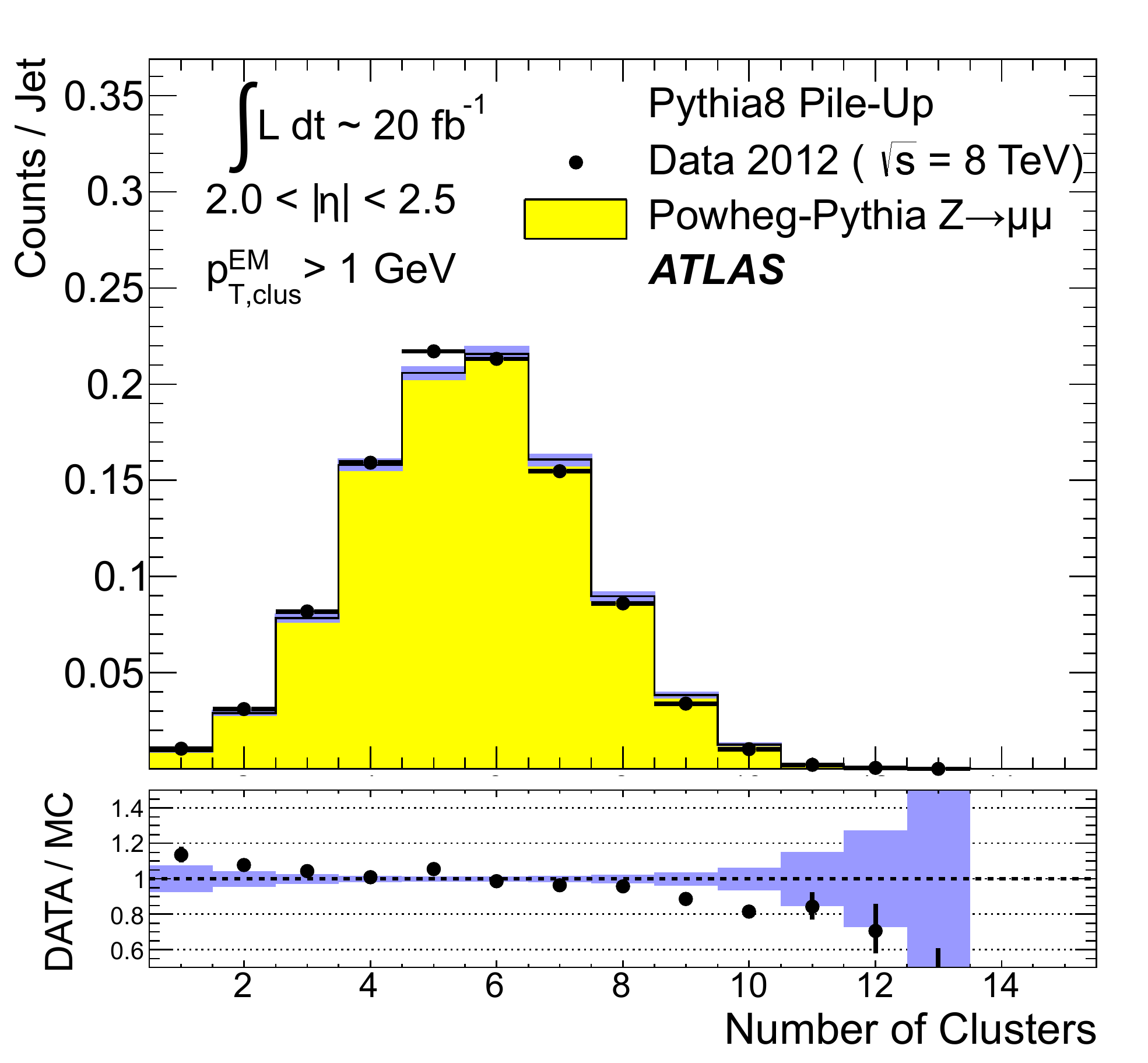}\label{fig:jets:nClus1GeV:2:mc}}      \qquad
	\subfloat[]{\includegraphics[width=\figsixpanelwidth]{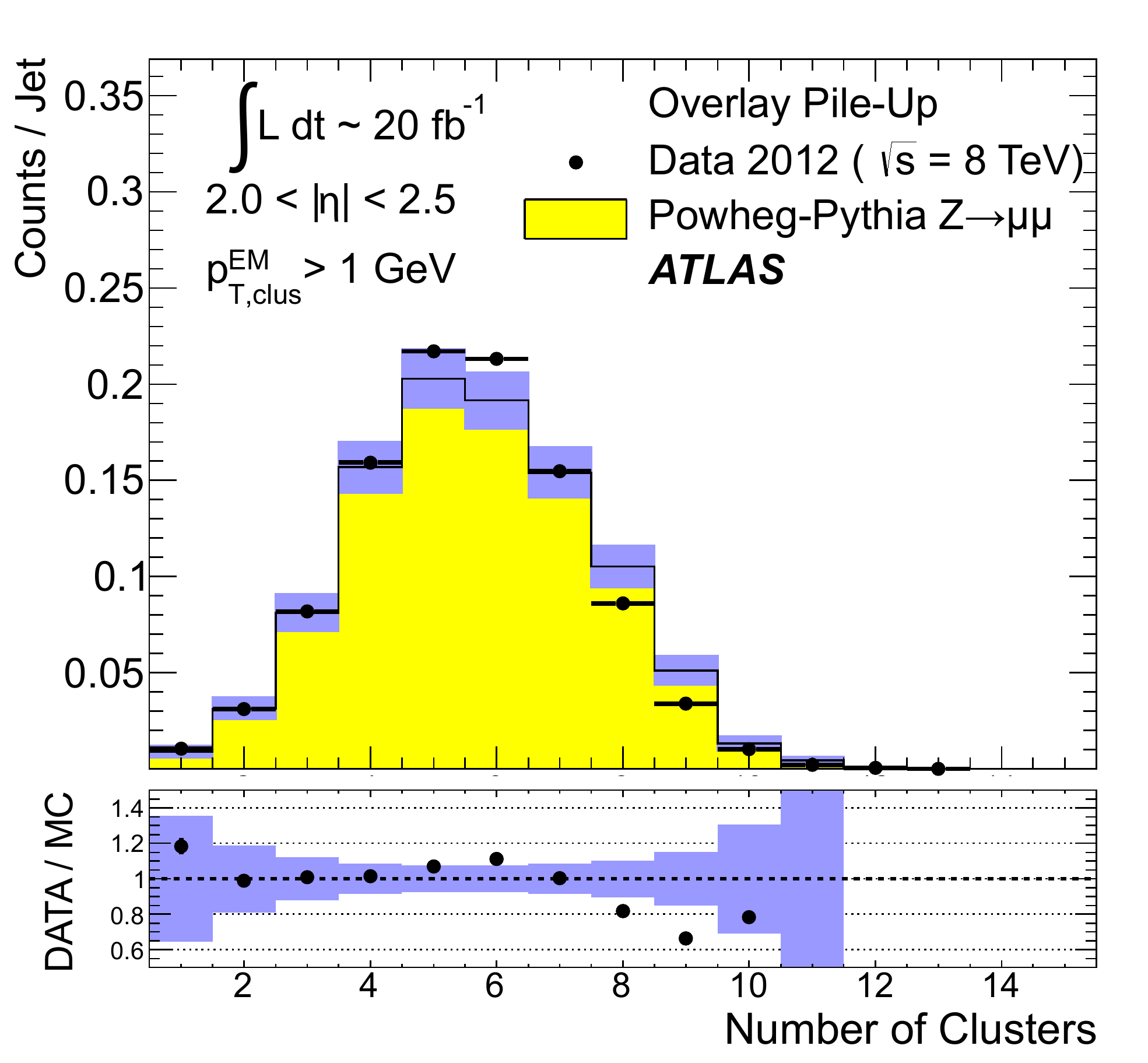}\label{fig:jets:nClus1GeV:2:ov}}
        \\      
	\subfloat[]{\includegraphics[width=\figsixpanelwidth]{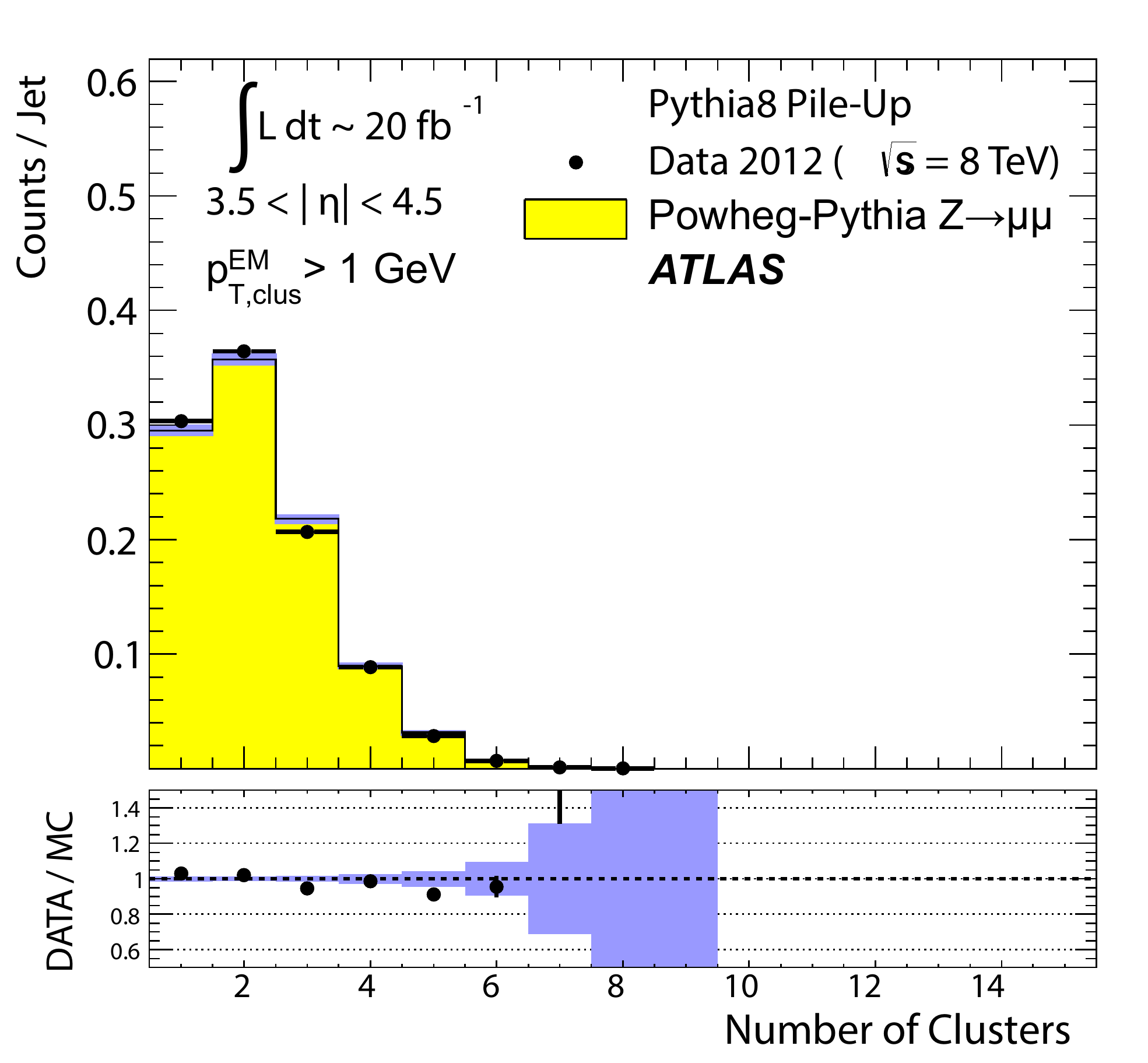}\label{fig:jets:nClus1GeV:3:mc}}      \qquad	
	\subfloat[]{\includegraphics[width=\figsixpanelwidth]{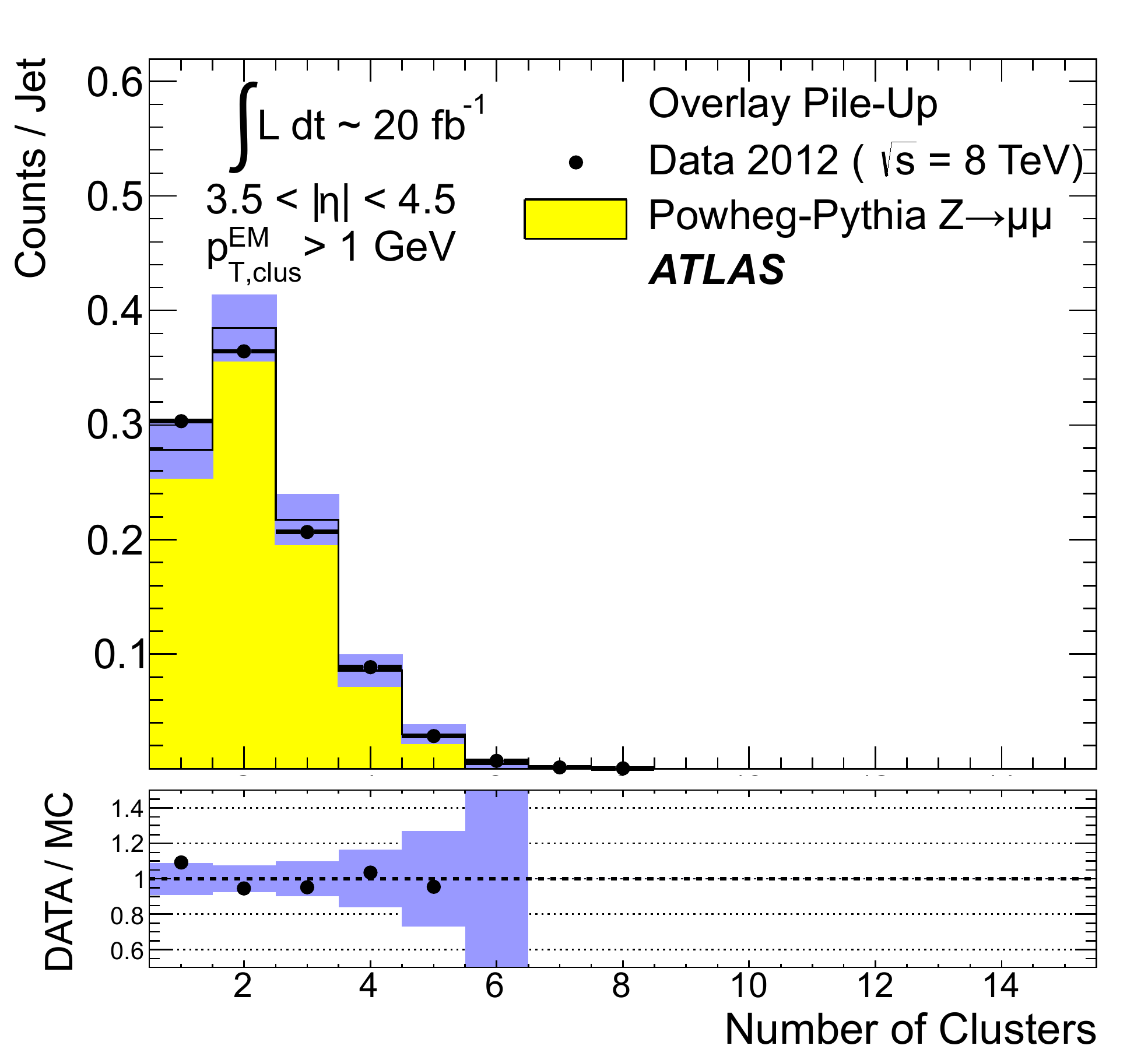}\label{fig:jets:nClus1GeV:3:ov}}
	\caption[]{The distribution of the number of \topos{} with $\ptclusem > \unit{1}{\GeV}$ inside \antikt{} jets  with $R = 0.4$ in the (\subref{fig:jets:nClus1GeV:1:mc}, \subref{fig:jets:nClus1GeV:1:ov}) central ($|\eta|<0.6$), the (\subref{fig:jets:nClus1GeV:2:mc}, \subref{fig:jets:nClus1GeV:2:ov}) \EndCap{} ($2.0<|\eta|<2.5$), and the (\subref{fig:jets:nClus1GeV:3:mc}, \subref{fig:jets:nClus1GeV:3:ov}) forward detector region ($3.5<|\eta|<4.5$) of \ATLAS. Shown are results from the analysis of \Zmumu{} events with at least one jet with $\unit{30}{\GeV} < \pT < \unit{40}{\GeV}$ in 2012 data and \MC{} simulations with fully simulated \pu{} in \subref{fig:jets:nClus1GeV:1:mc}, \subref{fig:jets:nClus1GeV:2:mc} and \subref{fig:jets:nClus1GeV:3:mc}, and with \pu{} from data overlaid in \subref{fig:jets:nClus1GeV:1:ov}, \subref{fig:jets:nClus1GeV:2:ov} and \subref{fig:jets:nClus1GeV:3:ov}. 
The data-to-\MC{} simulation ratios are shown below the respective distributions. 
The shaded bands indicate statistical uncertainties for the distributions from \MC{} simulations and the corresponding statistical uncertainty bands for the ratios.} 
\label{fig:jets:nClus1GeV}		
\end{figure}

The \datatomc{} comparisons of the cluster multiplicity distributions counting only \topos{} with $\ptclusem > \unit{1}{\GeV}$ for the same \Zmumu{} data and \MC{} simulations are shown in \figRef{fig:jets:nClus1GeV}. This comparison is significantly improved with respect to \figRef{fig:jets:nClus}, indicating that the number of low-energy \topos{} in jets is poorly simulated. 
The comparison of data to \MC{} simulations with fully simulated \pu{} and with \pu{} overlaid from data for the more inclusive cluster multiplicities in \figRef{fig:jets:nClus} indicates that \pu{} is likely not the main source for the deficiencies in the \MC{} simulation, as the comparison does not improve significantly when \pu{} is taken from the data. 
This observation, together with the insensitivity of the \datatomc{} comparison of the multiplicity of harder \topos{} to the choice of \pu{} modelling in \MC{} simulations shown in \figRef{fig:jets:nClus1GeV}, suggests that the deficiencies in the simulation of the low-energy \topo{} multiplicity arise from imperfections in the detector model, response or tuning of the parton shower and other sources of soft emissions, including multiple parton interactions in the underlying event, rather than from the modelling of \pu{} or electronic noise. Further investigations concerning the distribution of the \topo{} location in jets confirm this interpretation and are presented in \secRef{\thislabel:depth}. 

\begin{figure}[t!]\centering
	\subfloat[all \topos{} in jet]{\includegraphics[width=\fighalfwidth]{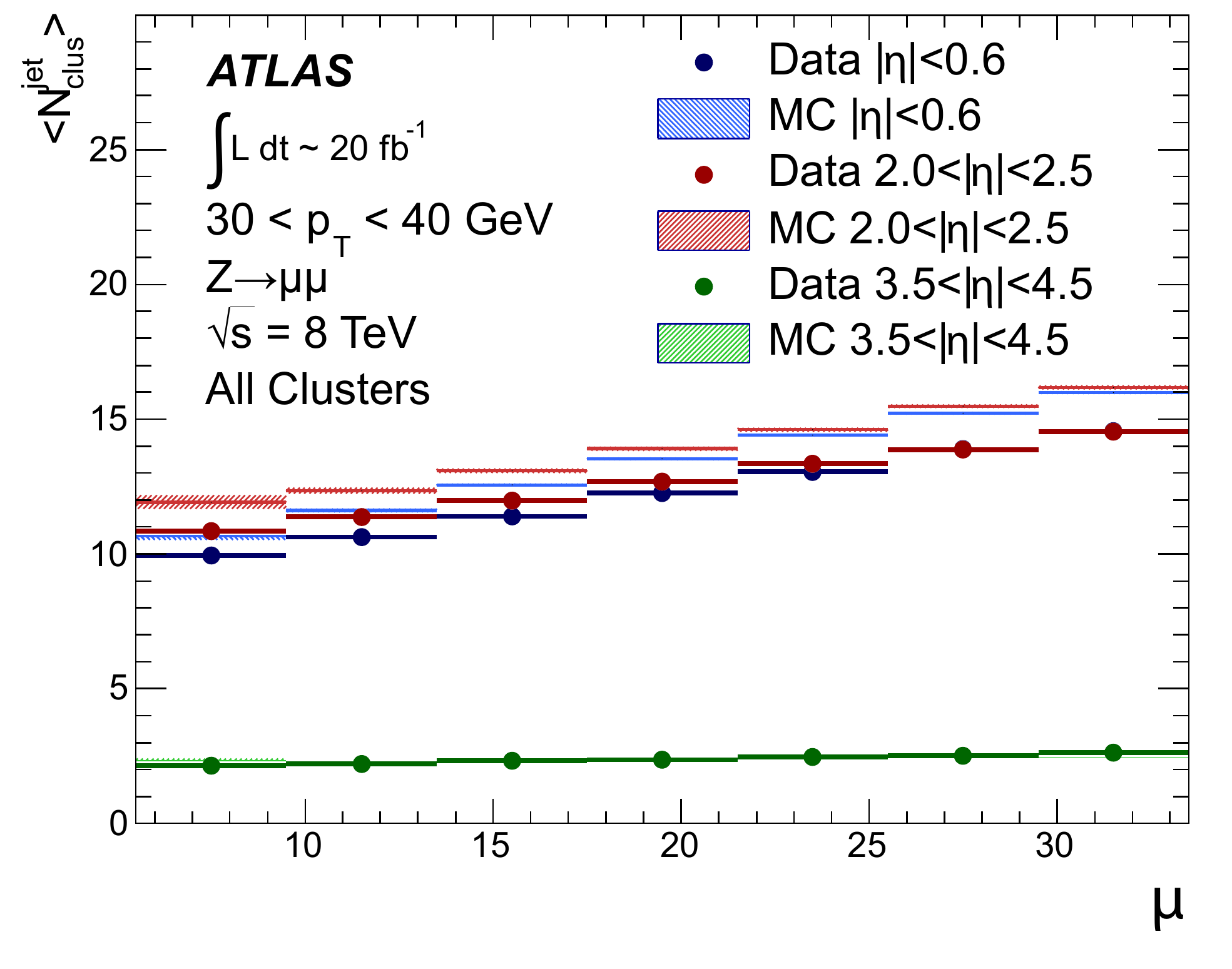}\label{fig:jet:nClus:pu:1}}
	\subfloat[all \topos{} in jet core]{\includegraphics[width=\fighalfwidth]{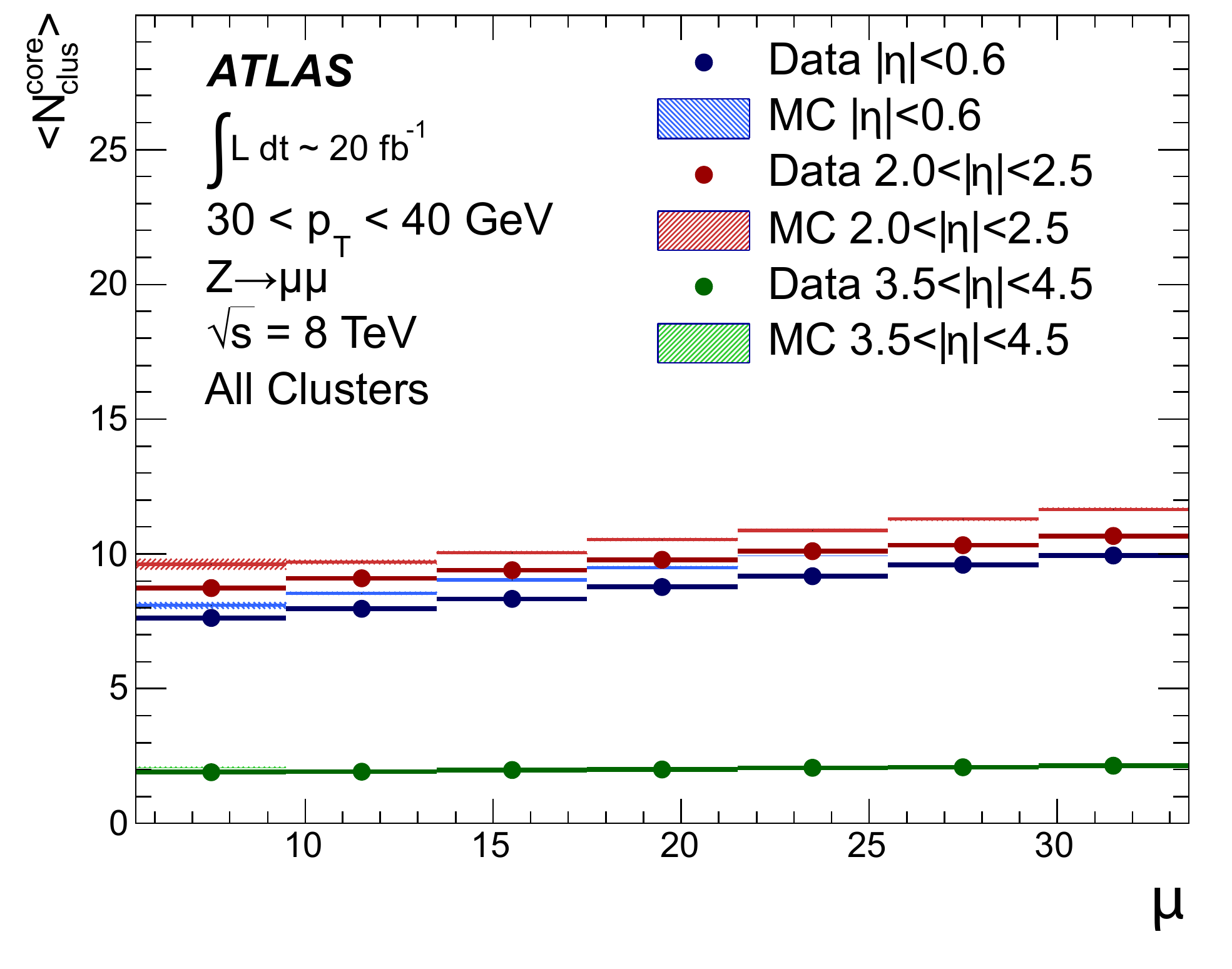}\label{fig:jet:nClus:pu:2}} 
	\\
	\subfloat[\topos{} with $\ptclusem > \unit{2}{\GeV}$]{\includegraphics[width=\fighalfwidth]{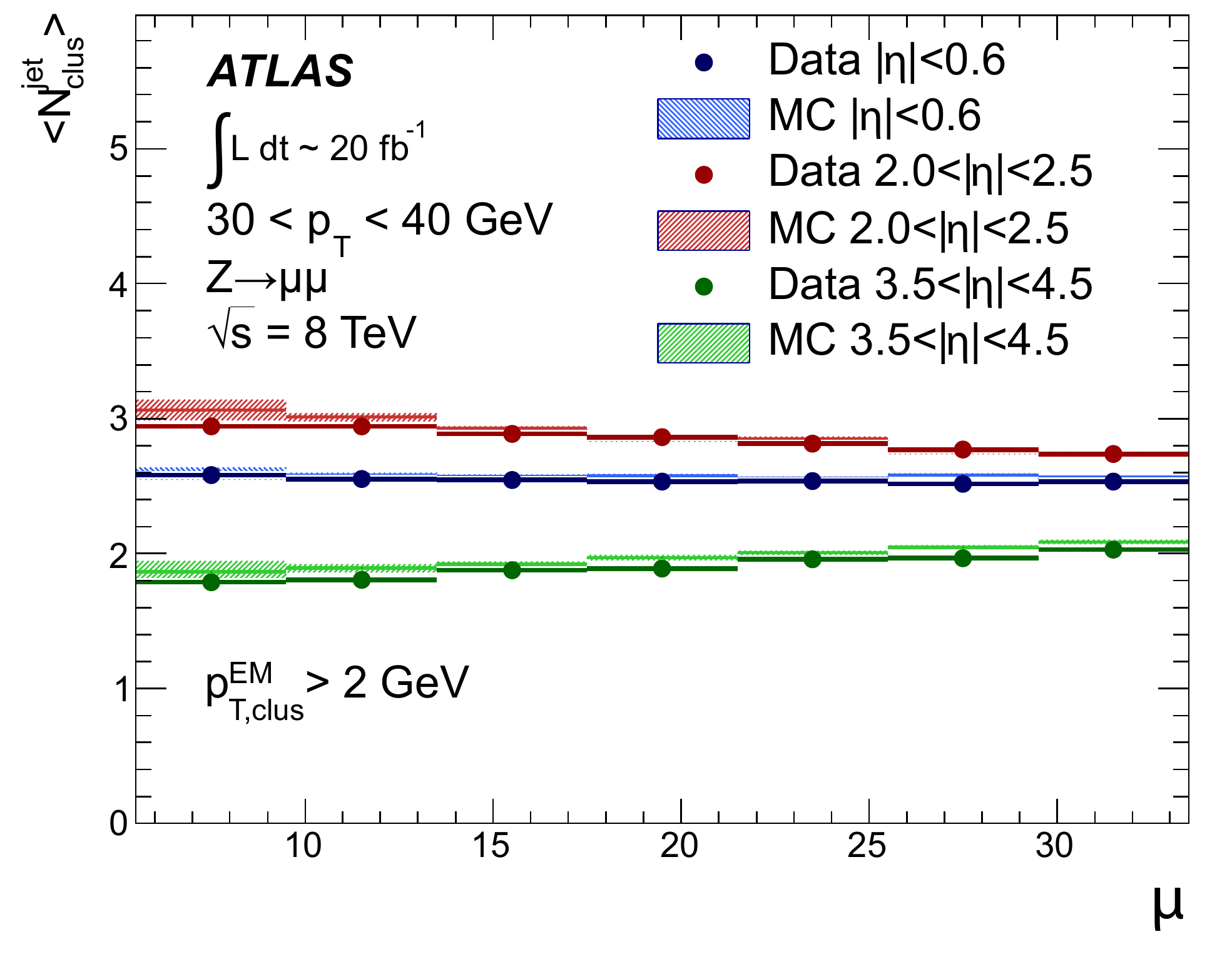}\label{fig:jet:nClus:pu:3}}
	\subfloat[\topos{} with $\ptclusem > \unit{2}{\GeV}$ in jet core]{\includegraphics[width=\fighalfwidth]{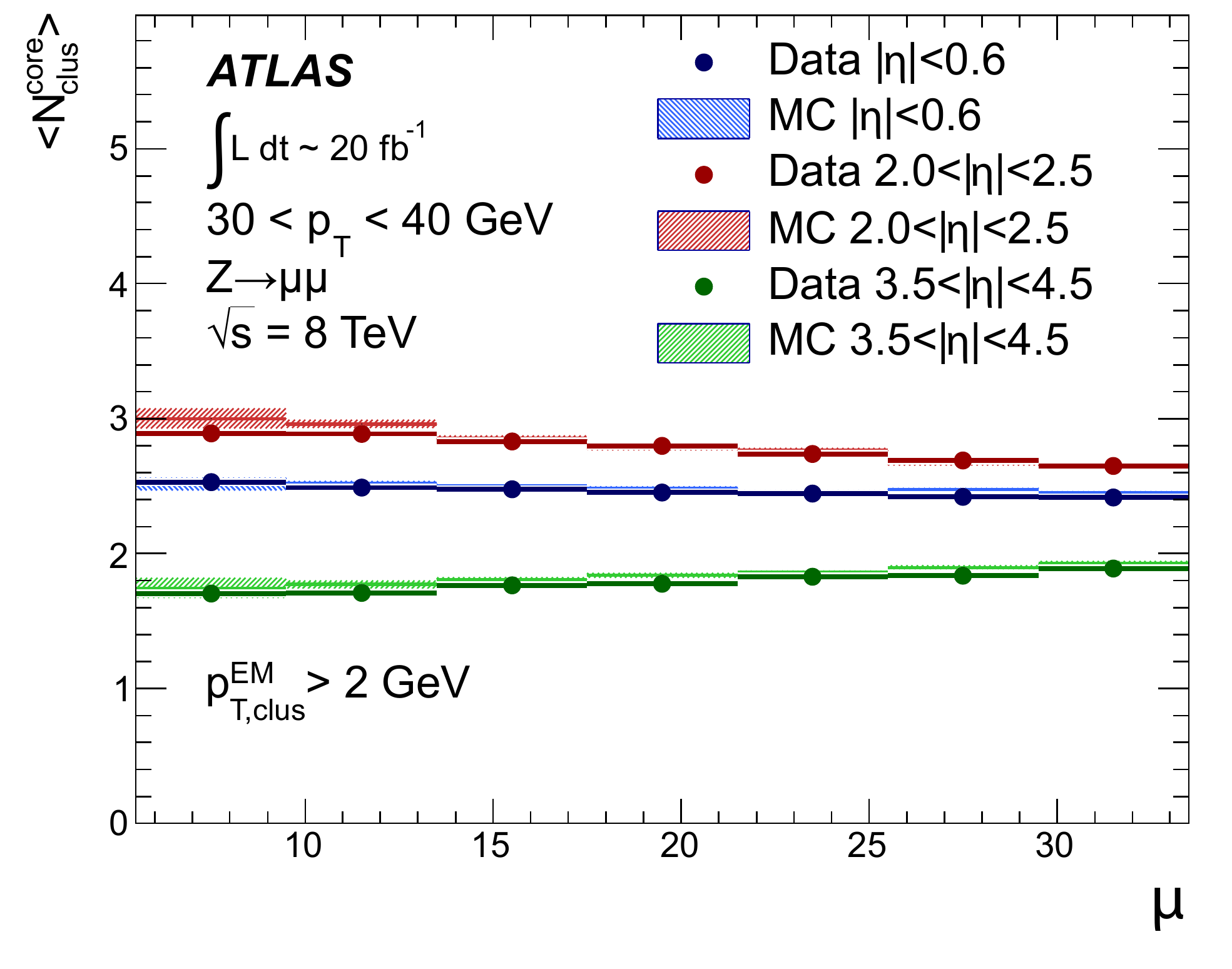}\label{fig:jet:nClus:pu:4}}
	\caption[]{The average number of \topos{} \AVE{\Nclusjet}{} in \antikt{} jets reconstructed with $R = 0.4$ within $\unit{30}{\GeV} < \ptjetlcwjes < \unit{40}{\GeV}$ as a function of $\mu$, in \Zmumu{} events in 2012 data and \MC{} simulations \subref{fig:jet:nClus:pu:1}. The \pu{} dependence of the average number of \topos{}  \AVE{\Ncluscore}{} in the core of the jet, defined by the distance to jet axis $\Delta R < 0.3$, is shown in \subref{fig:jet:nClus:pu:2}. Selecting \topos{} by $\ptclusem > \unit{2}{\GeV}$ inside jets and in the core of the jet yields the $\mu$ dependencies shown in \subref{fig:jet:nClus:pu:3} and \subref{fig:jet:nClus:pu:4}.
The shaded bands shown for the results obtained from \MC{} simulations indicate statistical uncertainties. }
	\label{fig:jet:nClus:pu}
\end{figure}

The dependence of the number of clusters \Nclusjet{} forming the \antikt{} jets of size $R = 0.4$ and with $\unit{30}{\GeV} < \ptjetlcwjes < \unit{40}{\GeV}$ as a function of the \pu{} activity, measured by $\mu$, is shown in \figRef{fig:jet:nClus:pu}. As indicated in \figRefLabel~\ref{fig:jet:nClus:pu}\subref{fig:jet:nClus:pu:1}, \Nclusjet{} rises approximately linearly with increasing $\mu$ in the central and \EndCap{} detector regions. 
The gradient of this rise is much smaller in the forward region, where the coarser \readout{} geometry and the signal shaping effects already discussed  in \secRef{\thislabel:resp} in the context of \figRefLabel~\ref{fig:jet:pt}\subref{fig:jet:pt:2} lead to merging and suppression of \pu{} signals. 
\FigRefLabel~\ref{fig:jet:nClus:pu}\subref{fig:jet:nClus:pu:1} also confirms the already mentioned  deficiencies in the \MC{} simulation of the absolute values of the most inclusive $\AVE{\Nclusjet}$ in any given $\mu$ range, except for the forward detector region. The slope of $\AVE{\Nclusjet}(\mu)$, on the other hand, compares well with data. 

The number of \topos{} in the core of the jet (\Ncluscore) is defined by counting the clusters at distances $\Delta R < 0.3$ around the jet axis. 
\FigRef{fig:jet:nClus:pu}\subref{fig:jet:nClus:pu:2} shows a residual dependence of the average \AVE{\Ncluscore}{} on $\mu$ in the central and \EndCap{} regions, with significant differences between data and the predictions from \MC{} simulations. 
The figure shows good \datatomc{} agreement for  \AVE{\Ncluscore}{}  in the forward region. 
Comparing $\AVE{\Nclusjet}(\mu)$ in \figRef{fig:jet:nClus:pu}\subref{fig:jet:nClus:pu:1} with $\AVE{\Ncluscore}(\mu)$ in \figRef{fig:jet:nClus:pu}\subref{fig:jet:nClus:pu:2} shows a steeper slope for  $\AVE{\Nclusjet}(\mu)$ than for $\AVE{\Ncluscore}(\mu)$ in the central and \EndCap{} calorimeter regions. 
\PU{} interactions add more \topos{} at the margin of the jet than in the core. 
For forward jets, \AVE{\Nclusjet}{} rises only slightly with increasing $\mu$, while \AVE{\Ncluscore} shows no observable dependency on \pu.
  
Calculating \Nclusjet{} and \Ncluscore{} with only considering \topos{} with $\ptclusem > \unit{2}{\GeV}$ yields the result for the \pu{} dependence
of \AVE{\Nclusjet}{} and \AVE{\Ncluscore}{} displayed in \figRef{fig:jet:nClus:pu}\subref{fig:jet:nClus:pu:3} and \figRef{fig:jet:nClus:pu}\subref{fig:jet:nClus:pu:4}, respectively.  
While both \AVE{\Nclusjet}{} and \AVE{\Ncluscore} are nearly independent of $\mu$  in the central detector region, they show more complex dependencies on the \pu{} activity in the \EndCap{} region. 
The loss of hard \topos{} in both the overall jet and in its core with increasing $\mu$ reflects additional cluster splitting induced by the diffuse energy flow from \pu{} in the \EndCap{} calorimeters. 
The observations in both the central and the \EndCap{} regions are well described by \MC{} simulations. 

A good quality of the \MC{} predictions is also achieved when comparing the number of hard \topos{} above the \ptclusem{} threshold in forward jets. 
This number shows only a small increase with increasing $\mu$, as shown in \figRef{fig:jet:nClus:pu}\subref{fig:jet:nClus:pu:3}. 
This is due to the fact that the cluster splitting behaviour does not change with increasing \pu{} in the coarse granularity of the \LArFCAL. In this module, the residual signal contribution from \pu{} shifts a small number of additional clusters above the \unit{2}{\GeV}{} threshold, yielding an increase of about \unit{10}{\%}{} for both \AVE{\Nclusjet} and \AVE{\Ncluscore} for $\mu < 10$ to $\mu > 30$. 
A comparison of $\AVE{\Nclusjet}(\mu)$ with and without the $\ptclusem > \unit{2}{\GeV}$ selection shows that the cut occasionally removes a \topo{} from a forward jet such that \AVE{\Nclusjet}{} is reduced by not more than \unit{15}{\%}{} for any given $\mu$. 
The selection affects $\AVE{\Ncluscore}(\mu)$ in 
a different way. 
While $\AVE{\Ncluscore}(\mu) \approx \const$ without the cut, the average number of \topos{} in the jet core passing the \ptclusem{} selection is smaller by  approximately \unit{15}{\%}{} in the region of lower \pu{} activity, where $\AVE{\Nclusjet}(\mu < 10) \approx  \AVE{\Ncluscore}(\mu < 10)$ both with and without the selection. It is only about \unit{5}{\%}{} smaller for higher \pu, where $\AVE{\Nclusjet}(\mu > 30) > \AVE{\Ncluscore}(\mu > 30)$ independent of the cut, as can be seen by comparing \figMultiRefLabel~\ref{fig:jet:nClus:pu}\subref{fig:jet:nClus:pu:2} with \ref{fig:jet:nClus:pu}\subref{fig:jet:nClus:pu:4} for forward jets.

\subsubsection{\Topo{} location in jets} \label{\thislabel:depth}

\begin{figure}[tp!] \centering
        \sfcompress
	\subfloat[]{\includegraphics[width=\figsixpanelwidth]{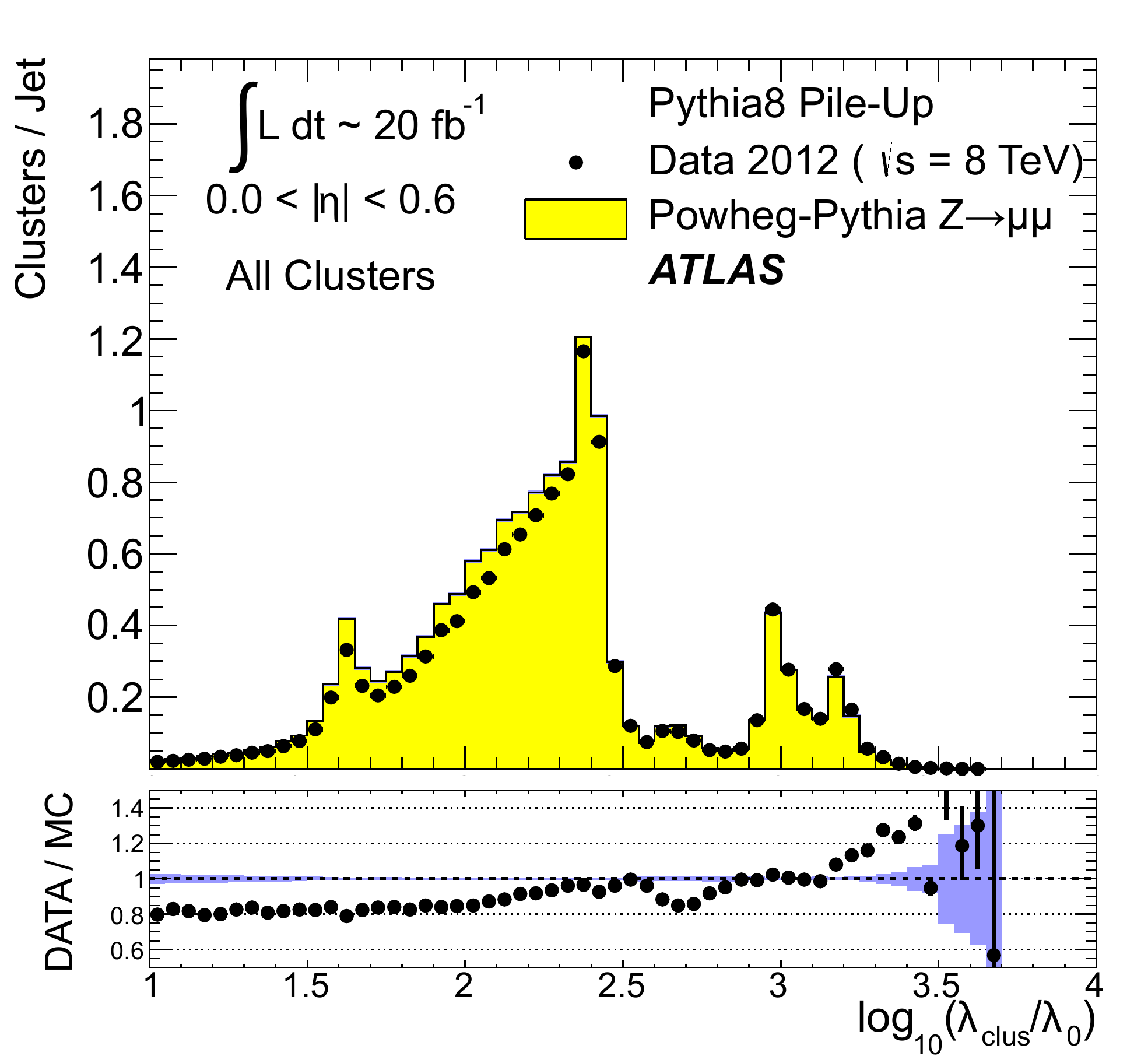}\label{fig:jet:logL:1:mc}} \qquad
	\subfloat[]{\includegraphics[width=\figsixpanelwidth]{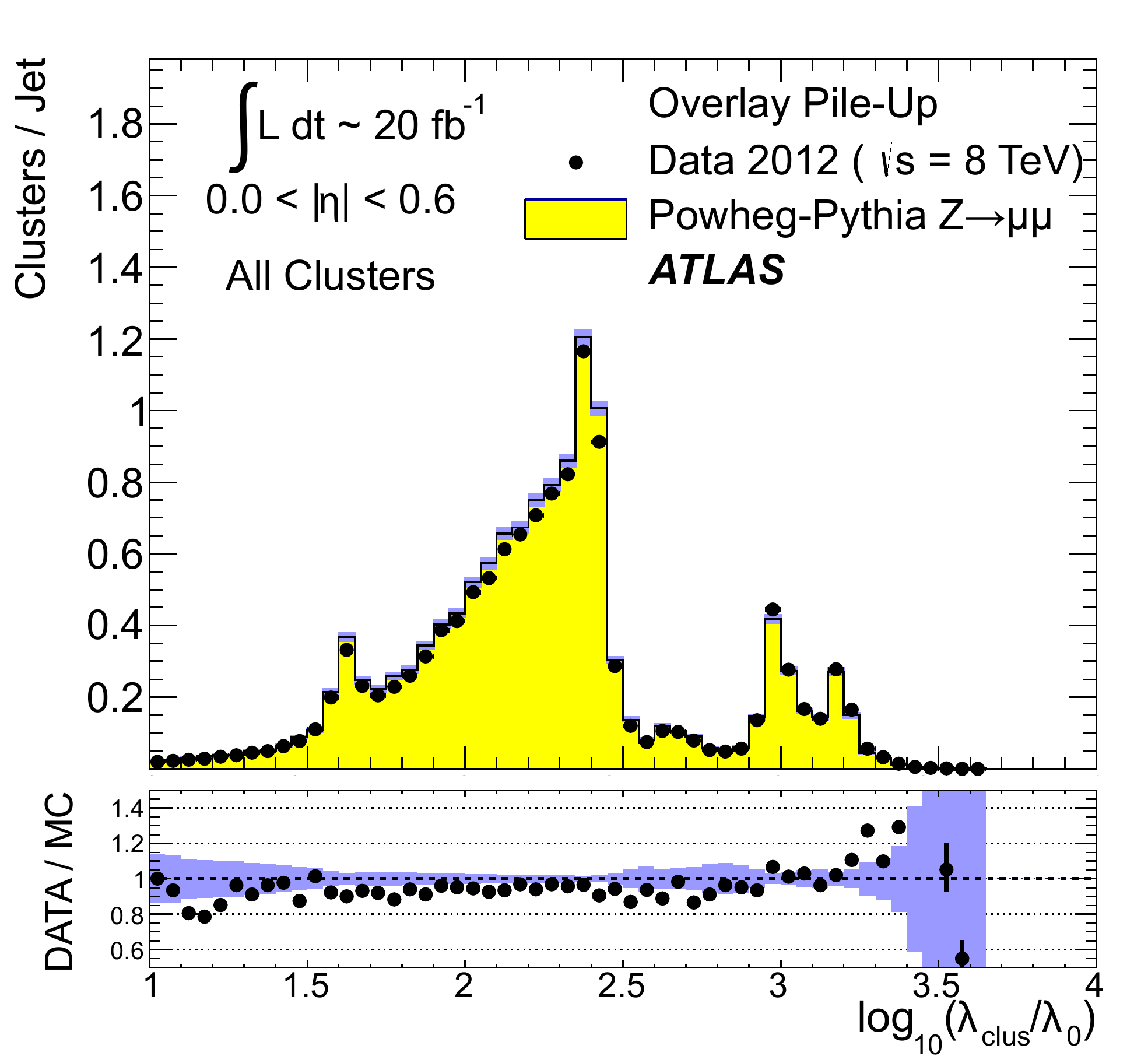}\label{fig:jet:logL:1:ov}}
        \\      
	\subfloat[]{\includegraphics[width=\figsixpanelwidth]{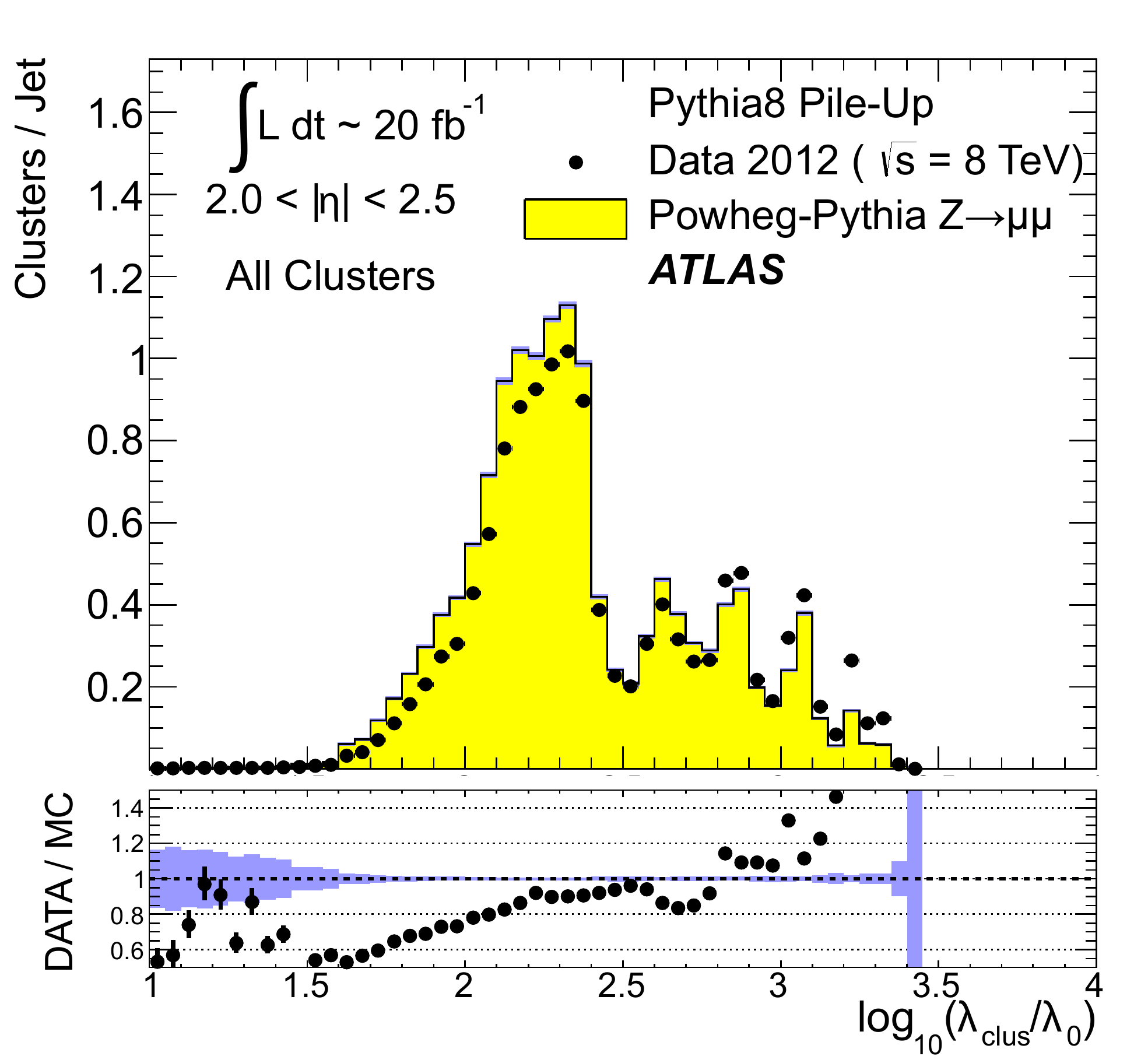}\label{fig:jet:logL:2:mc}} \qquad
	\subfloat[]{\includegraphics[width=\figsixpanelwidth]{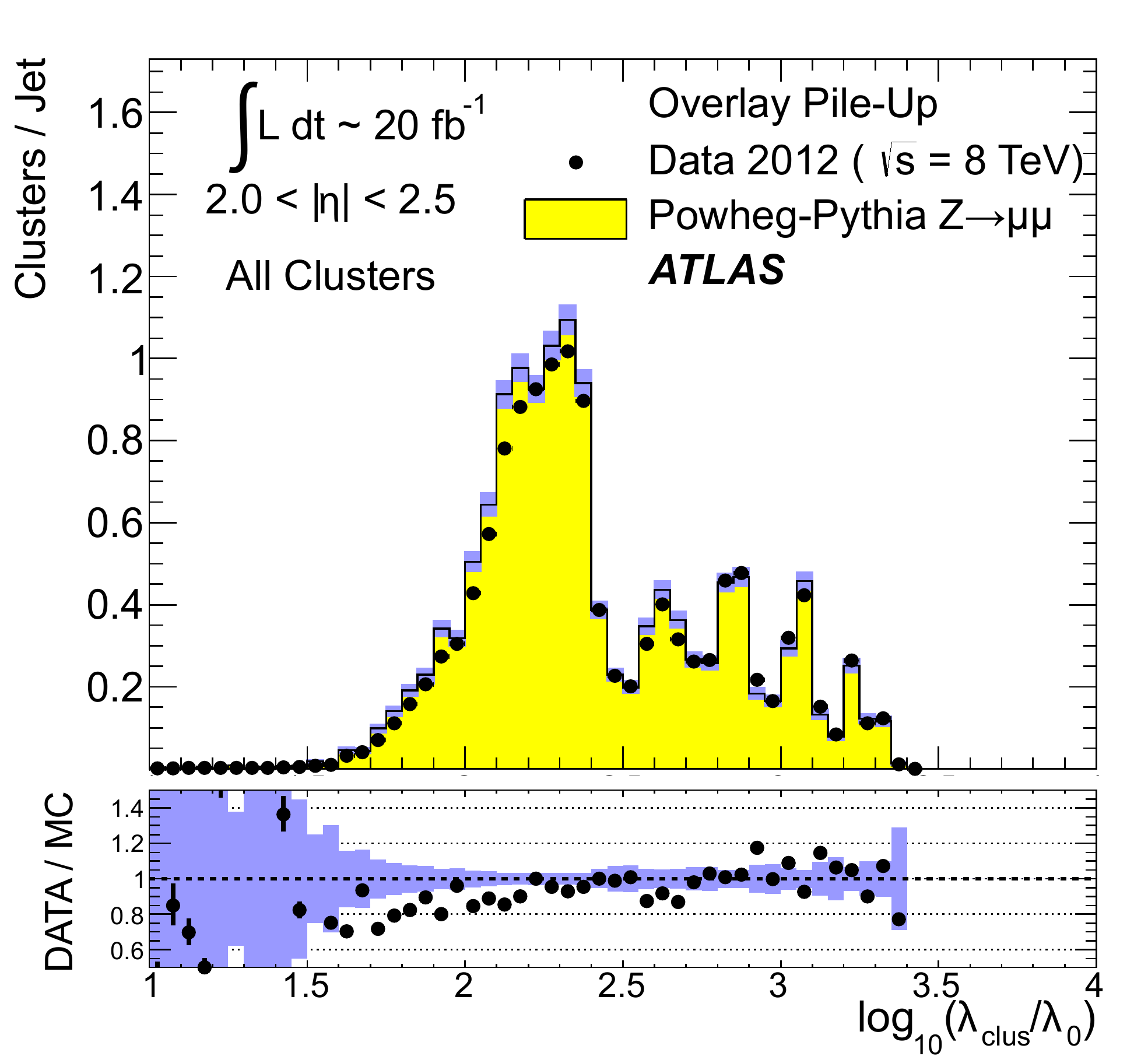}\label{fig:jet:logL:2:ov}}
	\\
	\subfloat[]{\includegraphics[width=\figsixpanelwidth]{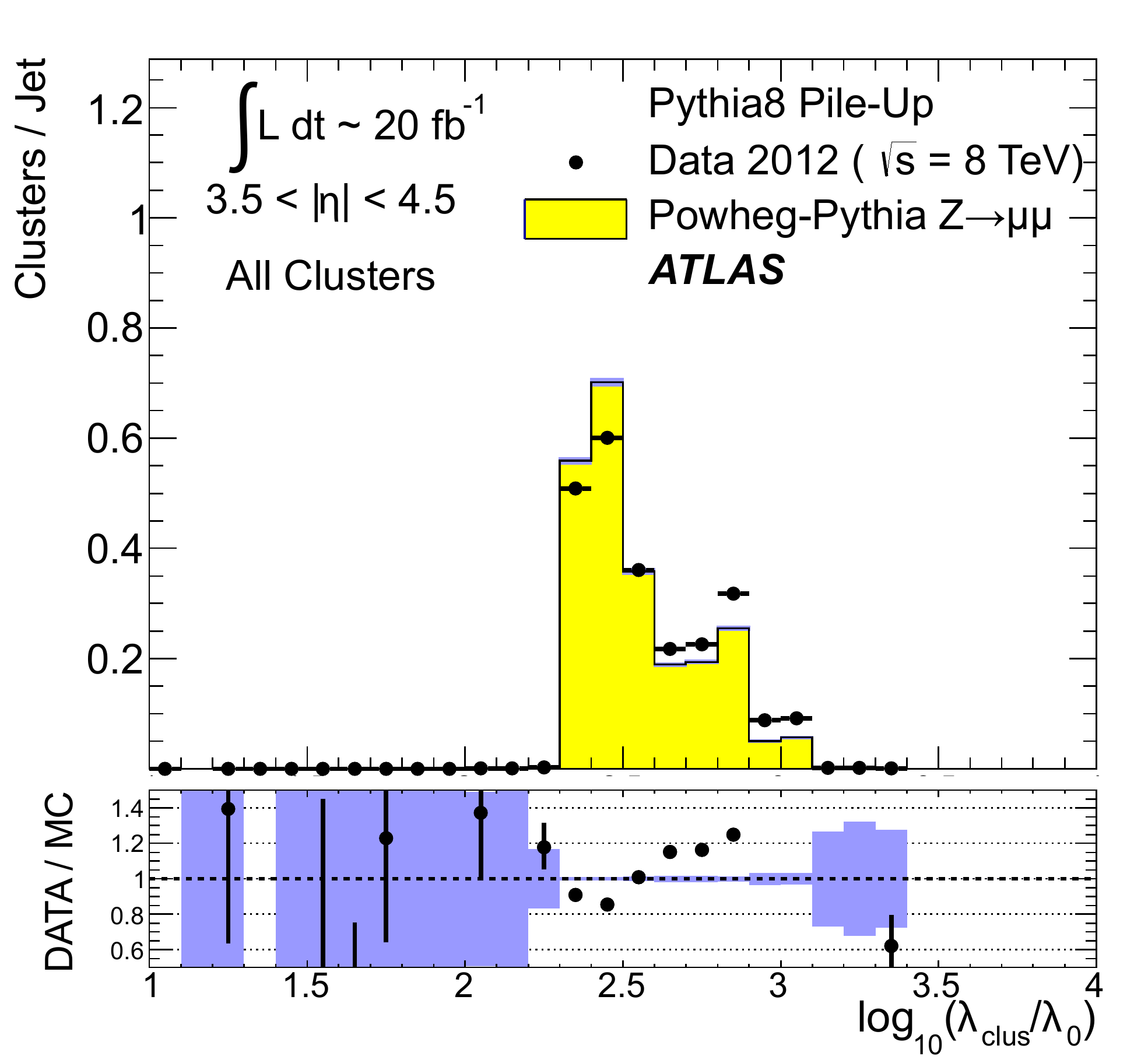}\label{fig:jet:logL:3:mc}} \qquad
	\subfloat[]{\includegraphics[width=\figsixpanelwidth]{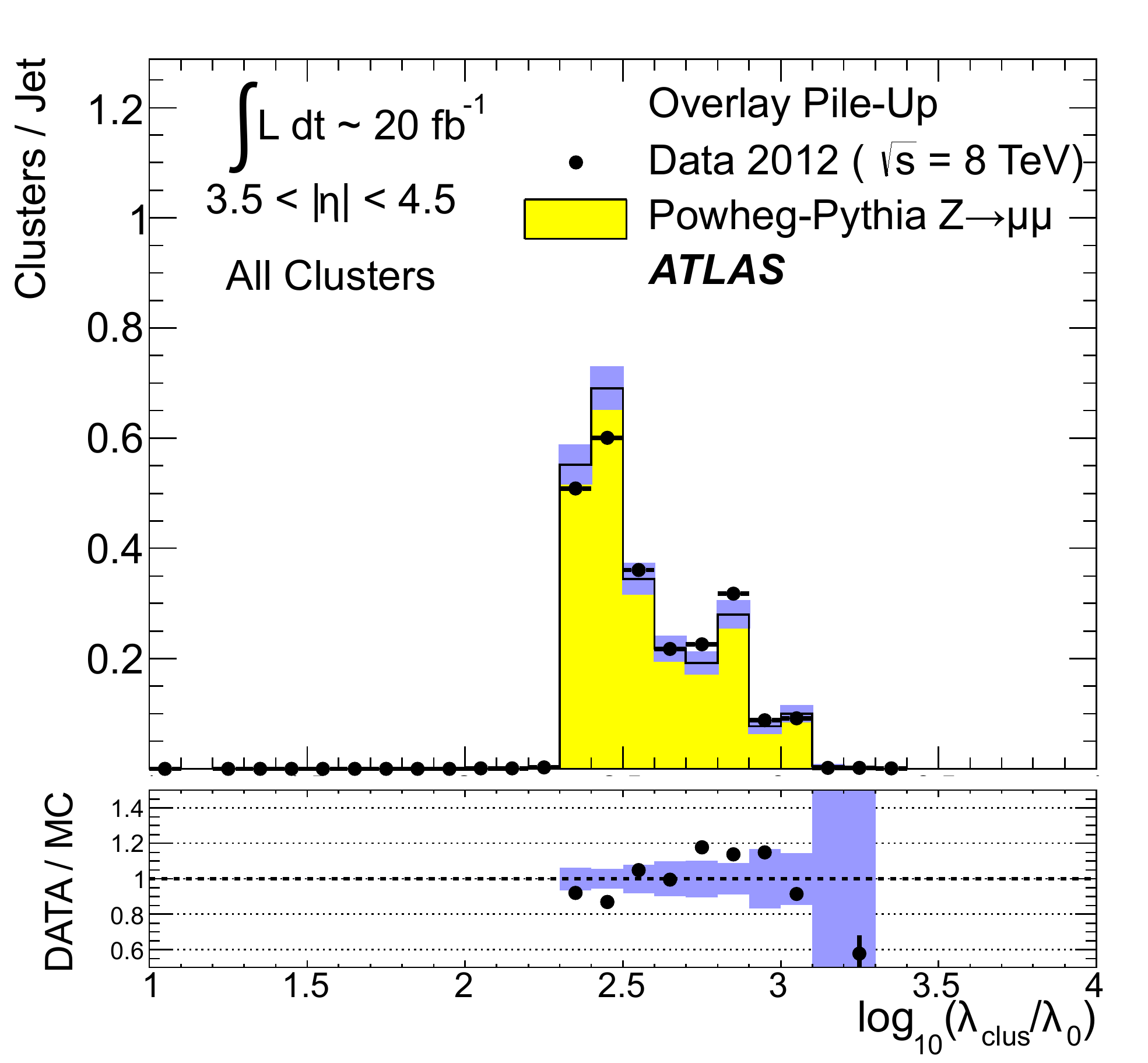}\label{fig:jet:logL:3:ov}}
	\caption[]{The distribution of the depth location, measured in terms of $\log_{10}(\lamctr/\lambda_{0})$ with $\lambda_{0} = 1\ \mm$, of all \topos{} in jets reconstructed with the \antikt{} algorithm with $R = 0.4$ and with $\unit{30}{\GeV} < \ptjetlcwjes < \unit{40}{\GeV}$ in \Zmumu{} events in 2012 data and \MC{} simulations with (\subref{fig:jet:logL:1:mc}, \subref{fig:jet:logL:2:mc}, and \subref{fig:jet:logL:3:mc}) fully simulated \pu{} and with (\subref{fig:jet:logL:1:ov}, \subref{fig:jet:logL:2:ov}, and \subref{fig:jet:logL:3:ov}) overlaid \pu{} from data. Distributions are shown for jets in the (\subref{fig:jet:logL:1:mc}, \subref{fig:jet:logL:1:ov}) central ($|\eta|<0.6$), the (\subref{fig:jet:logL:2:mc}, \subref{fig:jet:logL:2:ov}) \EndCap{} ($2.0<|\eta|<2.5$), and the (\subref{fig:jet:logL:3:mc}, \subref{fig:jet:logL:3:ov}) forward detector region ($3.5<|\eta|<4.5$). The bin-by-bin ratios of the distributions from data and \MC{} simulations are shown below the plots. The shaded bands indicate statistical uncertainties for the distributions from \MC{} simulations and the corresponding uncertainty bands in the ratio plots.} 
\label{fig:jet:logL}		
\end{figure}

\begin{figure}[tp!] \centering
        \sfcompress
	\subfloat[]{\includegraphics[width=\figsixpanelwidth]{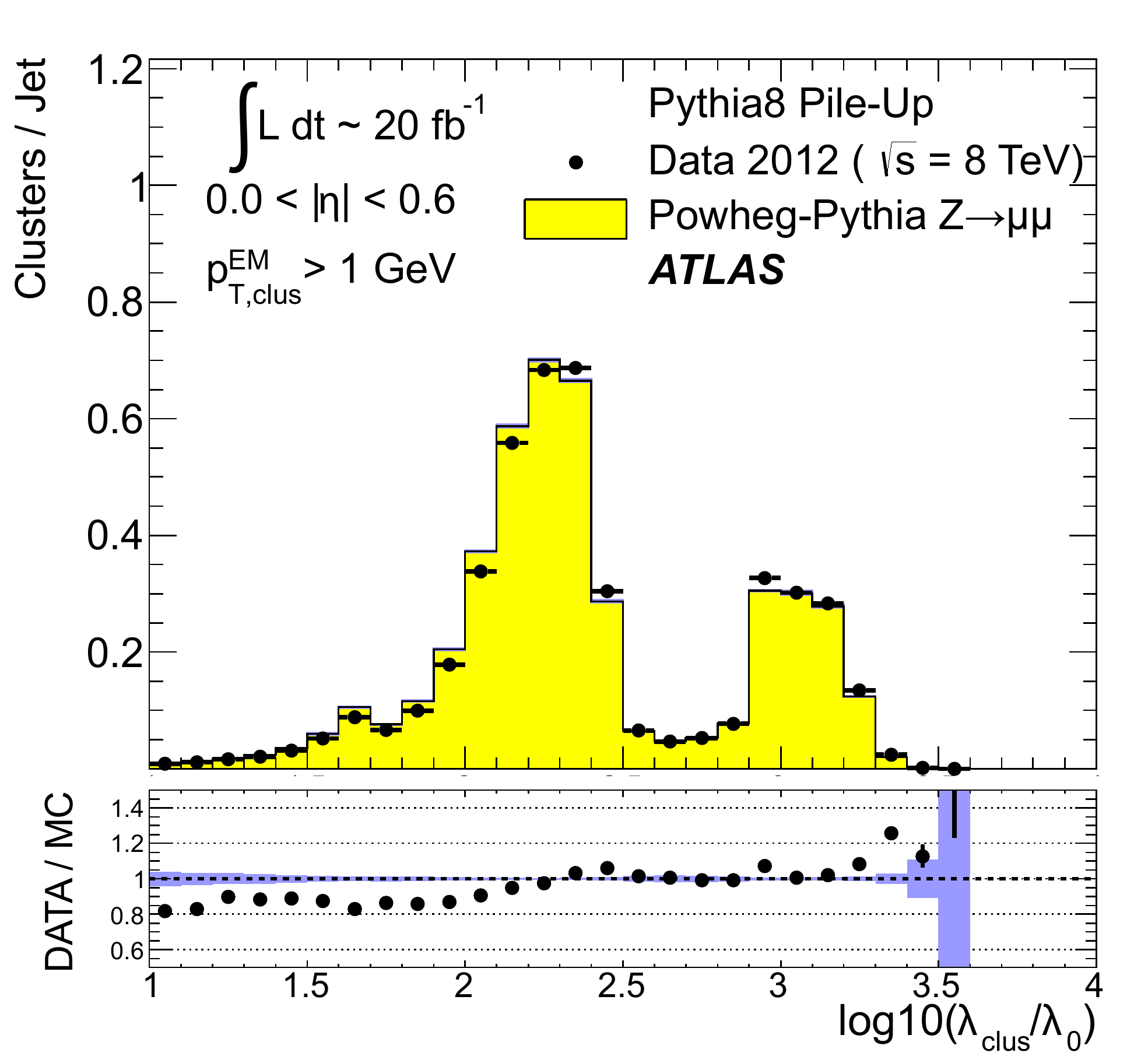}\label{fig:jet:logL1GeV:1:mc}}     \qquad 
	\subfloat[]{\includegraphics[width=\figsixpanelwidth]{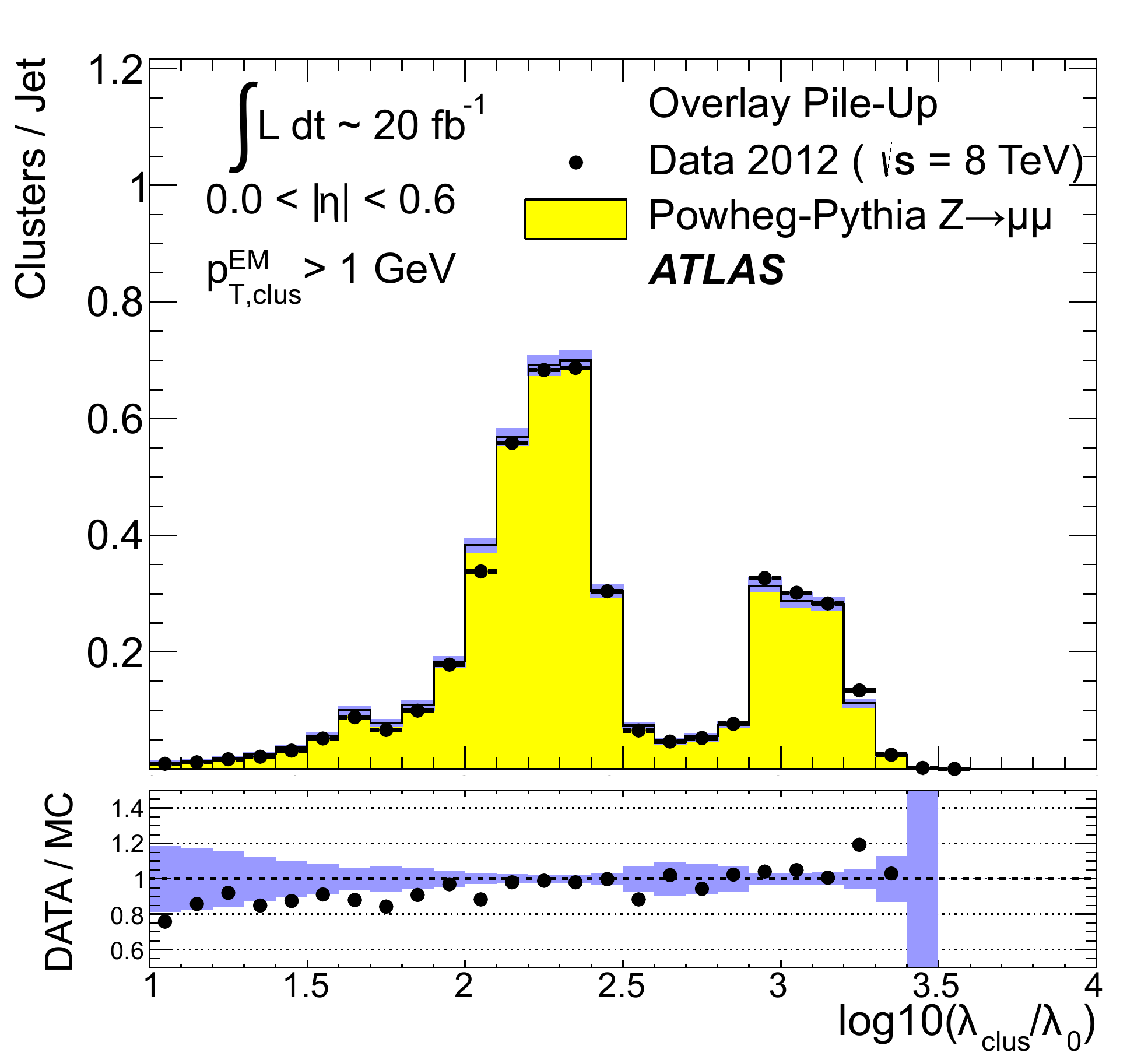}\label{fig:jet:logL1GeV:1:ov}}
	\\
	\subfloat[]{\includegraphics[width=\figsixpanelwidth]{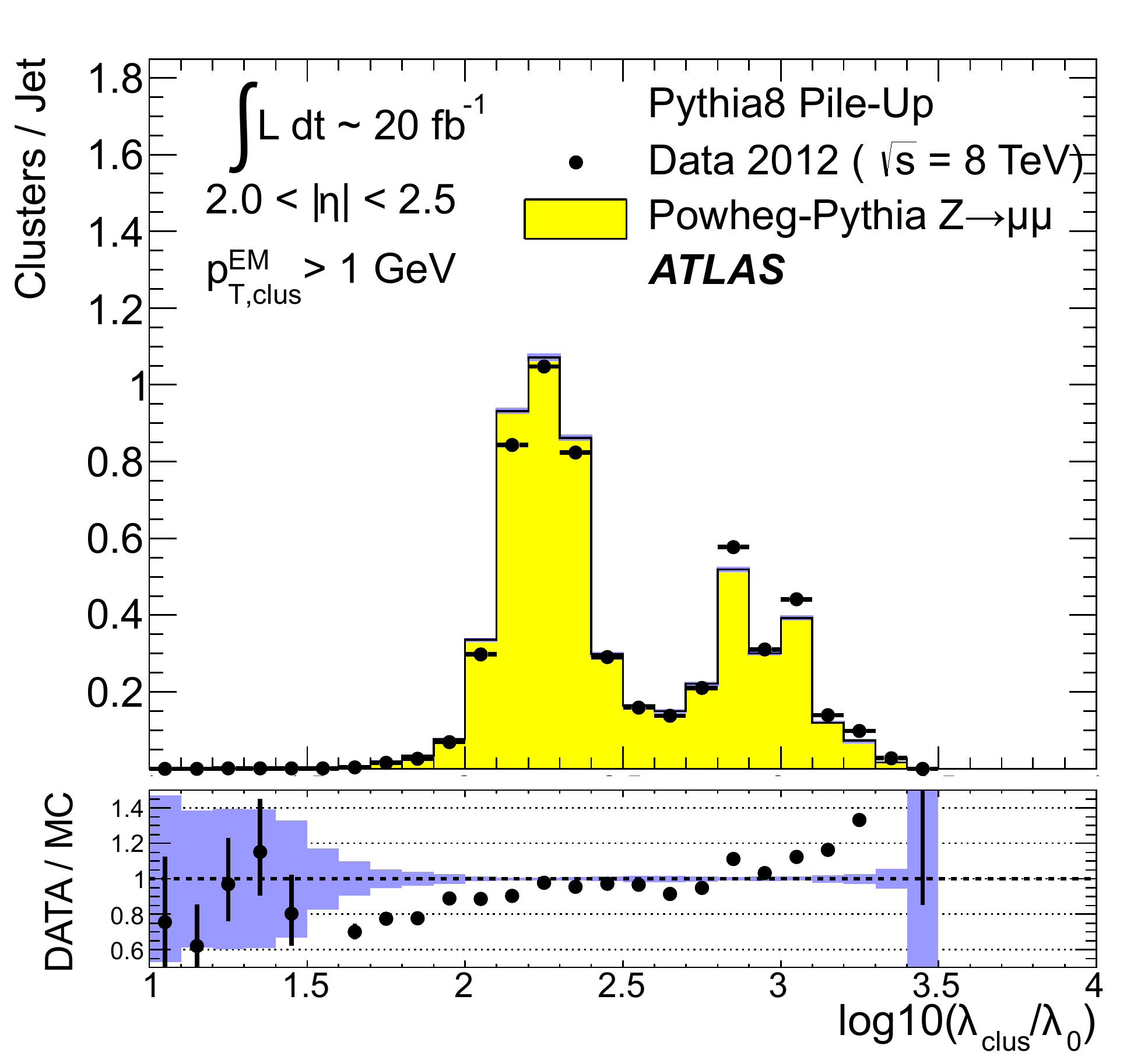}\label{fig:jet:logL1GeV:2:mc}}     \qquad 
	\subfloat[]{\includegraphics[width=\figsixpanelwidth]{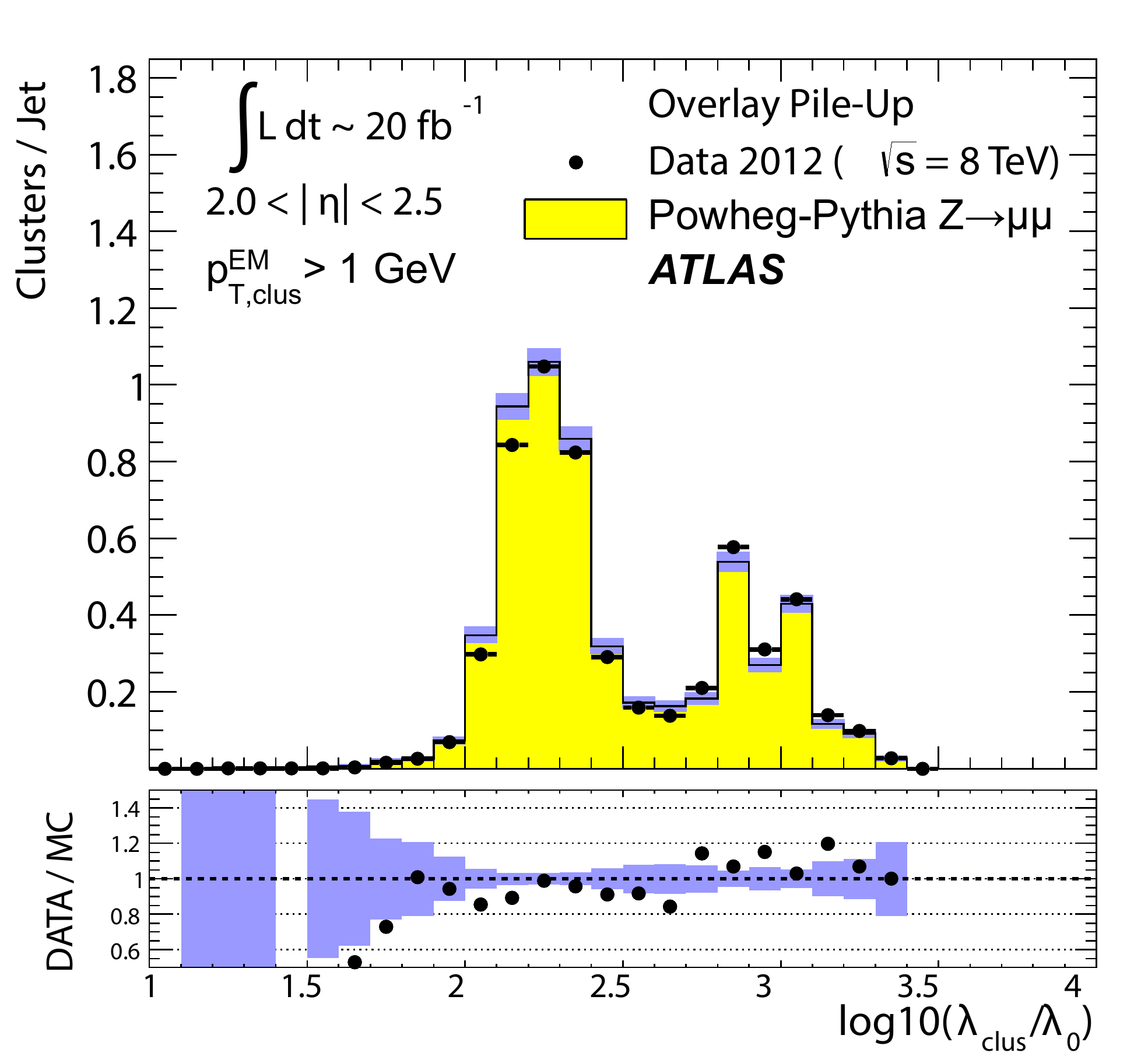}\label{fig:jet:logL1GeV:2:ov}}
	\\
	\subfloat[]{\includegraphics[width=\figsixpanelwidth]{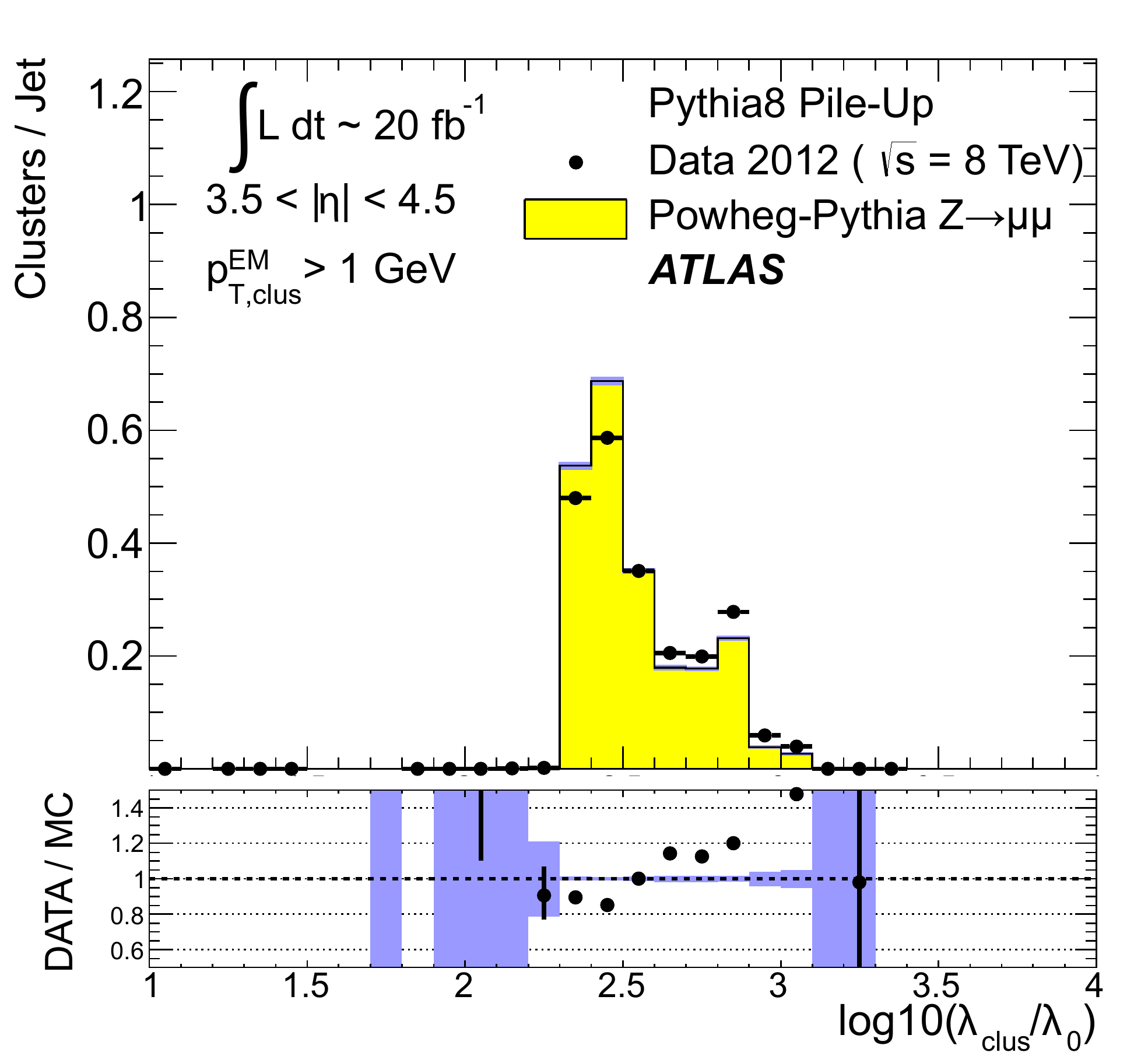}\label{fig:jet:logL1GeV:3:mc}}     \qquad 
	\subfloat[]{\includegraphics[width=\figsixpanelwidth]{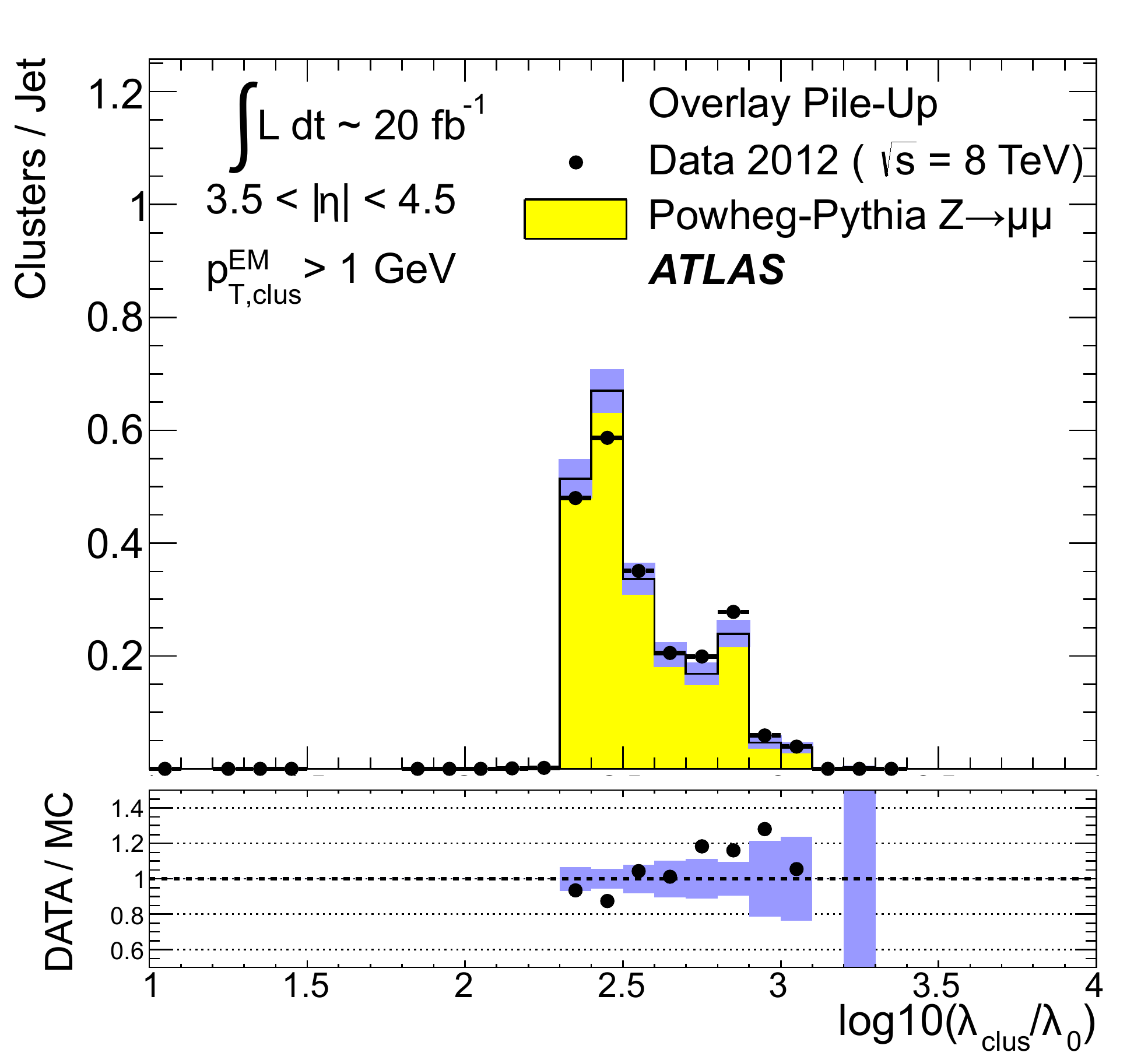}\label{fig:jet:logL1GeV:3:ov}}
	\caption[]{The distribution of the depth location, measured in terms of $\log_{10}(\lamctr/\lambda_{0})$ with $\lambda_{0} = \unit{1}{\mm}$, of \topos{} with $\ptclusem > \unit{1}{\GeV}$ in jets reconstructed with the \antikt{} algorithm with $R = 0.4$ and with $\unit{30}{\GeV} < \ptjetlcwjes < \unit{40}{\GeV}$ in \Zmumu{} events in 2012 data and \MC{} simulations with (\subref{fig:jet:logL1GeV:1:mc}, \subref{fig:jet:logL1GeV:2:mc}, and \subref{fig:jet:logL1GeV:3:mc}) fully simulated \pu{} and with (\subref{fig:jet:logL1GeV:1:ov}, \subref{fig:jet:logL1GeV:2:ov}, and \subref{fig:jet:logL1GeV:3:ov}) overlaid \pu{} from data. Distributions are shown for jets in the (\subref{fig:jet:logL1GeV:1:mc}, \subref{fig:jet:logL1GeV:1:ov}) central ($|\eta|<0.6$),  the (\subref{fig:jet:logL1GeV:2:mc}, \subref{fig:jet:logL1GeV:2:ov}) \EndCap{} ($2.0<|\eta|<2.5$), and the (\subref{fig:jet:logL1GeV:3:mc}, \subref{fig:jet:logL:3:ov}) forward detector region ($3.5<|\eta|<4.5$). The data-to-\MC{} simulation ratios are shown below the distributions. The shaded bands shown for the distributions obtained from \MC{} simulations indicate statistical uncertainties and the corresponding uncertainty bands in the ratio plots.} 
\label{fig:jet:logL1GeV}		
\end{figure}

The distribution of the depth location of all \topos{} inside \antikt{} jets reconstructed with $R = 0.4$ and with $\unit{30}{\GeV} < \ptjetlcwjes < \unit{40}{\GeV}$ in \Zmumu{} events in 2012 data and \MC{} simulations is shown in \figRef{fig:jet:logL}. 
Like for the depth distribution of \topos{} in the inclusive \Zmumu{} sample presented in \figRef{fig:spectra:lam}, the \MC{} simulations with overlaid \pu{} data show better agreement with data than the ones with fully simulated \pu. The differences in the jet context are significantly smaller than observed for the inclusive selection.

Applying a $\ptclusem > \unit{1}{\GeV}$ cut to the \topos{} in the jets results in the depth distributions shown in \figRef{fig:jet:logL1GeV}.
This selection also shows better \datatomc{} agreement for the sample with fully simulated \pu, an indicator consistent with the  better simulation of harder signals observed in e.g.~\figRef{fig:spectra:lam:ptexcl}. 
A noticeable difference from the depth distributions obtained from the inclusive sample in \figRef{fig:spectra:lam:ptexcl}\subref{fig:spectra:lam:ptexcl:1:mc} is that for \topos{} in jets the \datatomc{} agreement in the case of the fully simulated \pu{} is already better for the $\ptclusem > \unit{1}{\GeV}$ selection, as can be seen in \figRef{fig:jet:logL1GeV}\subref{fig:jet:logL1GeV:1:mc}.
In addition, comparing \figMultiRef{fig:jet:logL}{and}{fig:jet:logL1GeV} shows that the $\ptclusem > \unit{1}{\GeV}$ selection predominantly removes \topos{} at small depth \lamctr, as the 
distributions are depopulated more for smaller values of \lamctr{} than for larger ones. 
This means that mostly \topos{} generated by soft particles with little penetration depth into the calorimeters, including those consistent with \pu, are removed. The \datatomc{} comparisons are thus less sensitive to \pu{} modelling issues, and therefore show better agreement.

\subsubsection{Calibration and signal features of the leading \topo} \label{\thislabel:hardest:signal}

\begin{figure}[t!]\centering
	\subfloat[]{\includegraphics[width=\fighalfwidth]{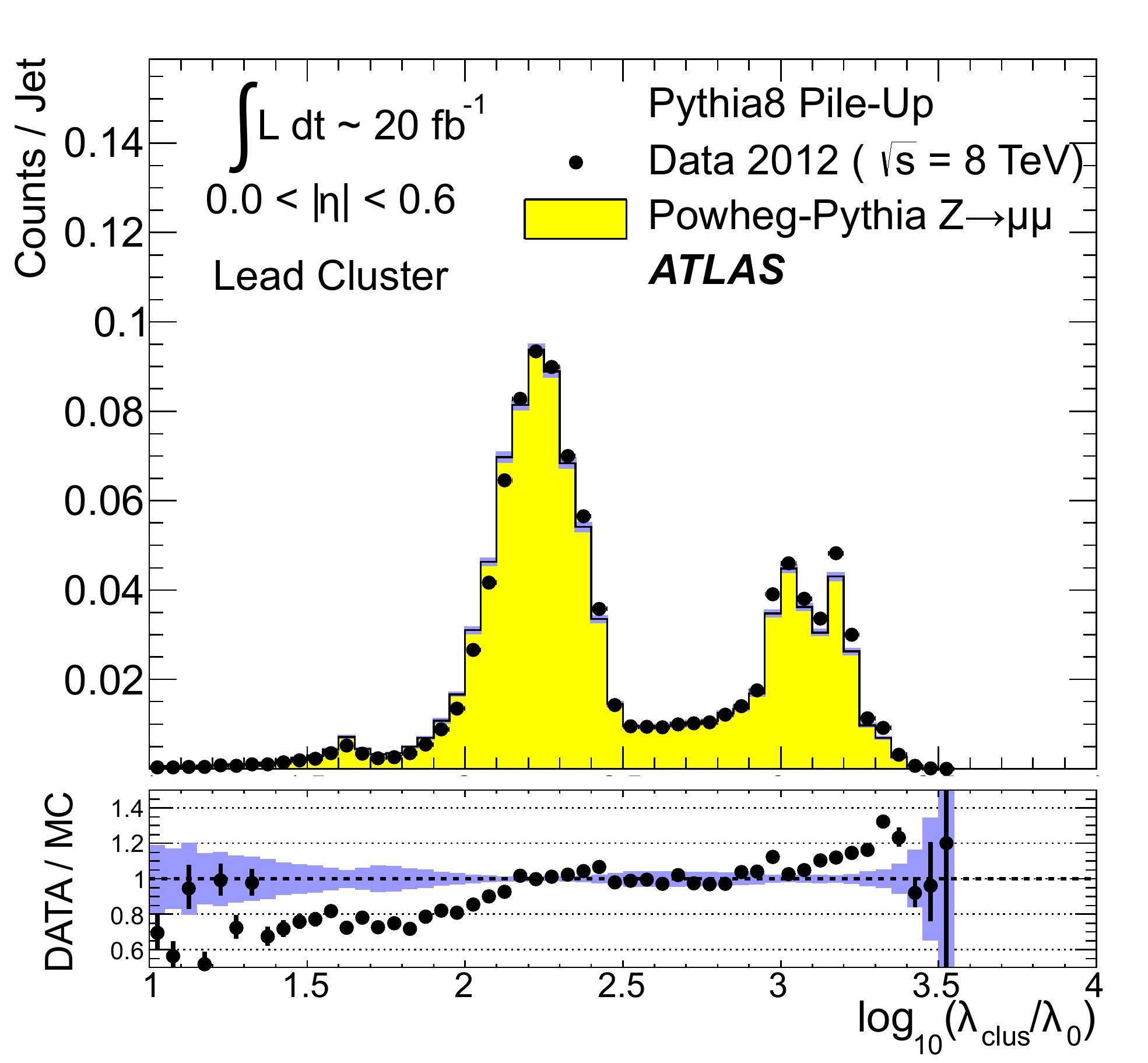}\label{fig:jet:hardest:lam:1:mc}}
	\subfloat[]{\includegraphics[width=\fighalfwidth]{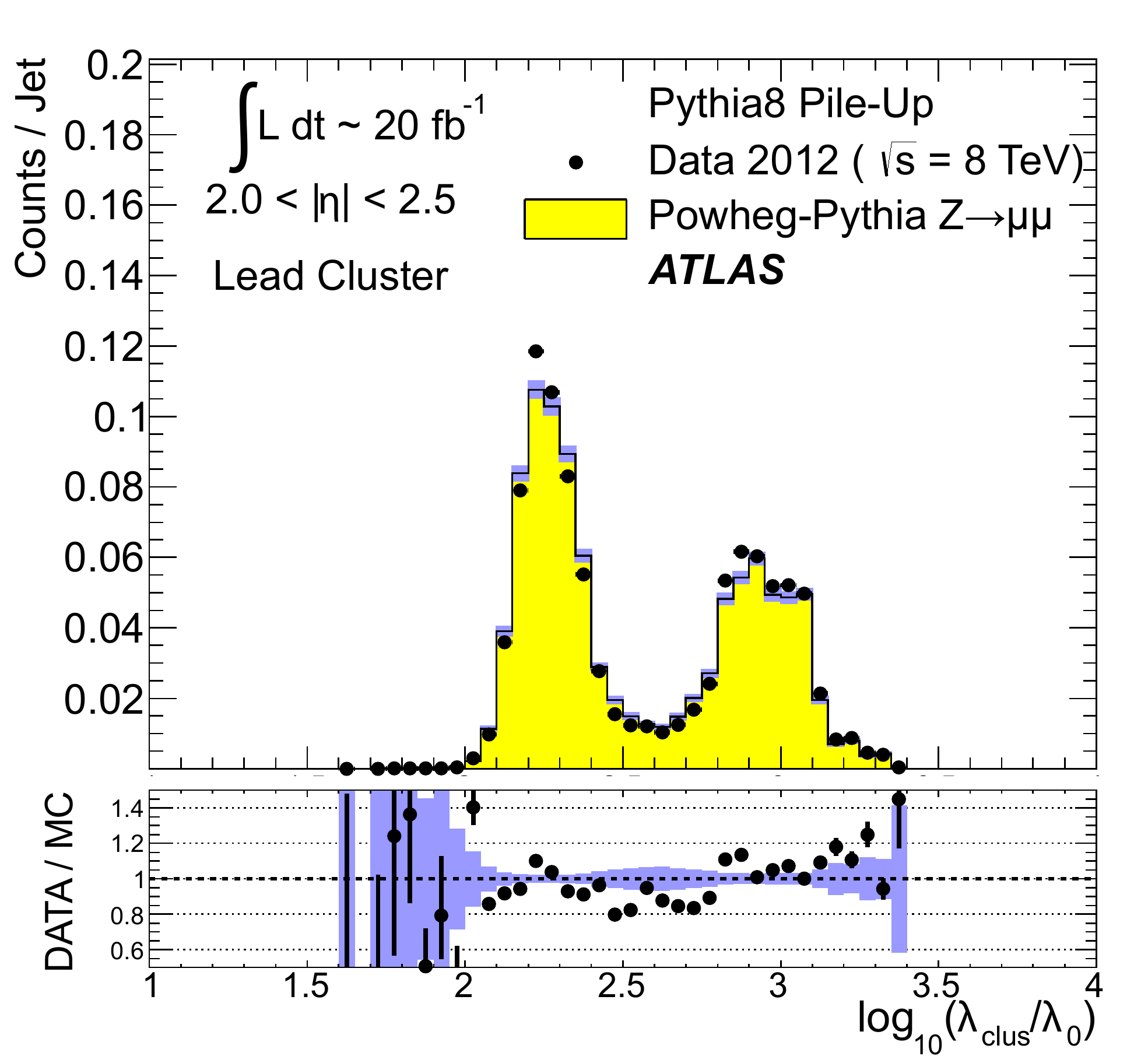}\label{fig:jet:hardest:lam:2:mc}}
        \\      
	\subfloat[]{\includegraphics[width=\fighalfwidth]{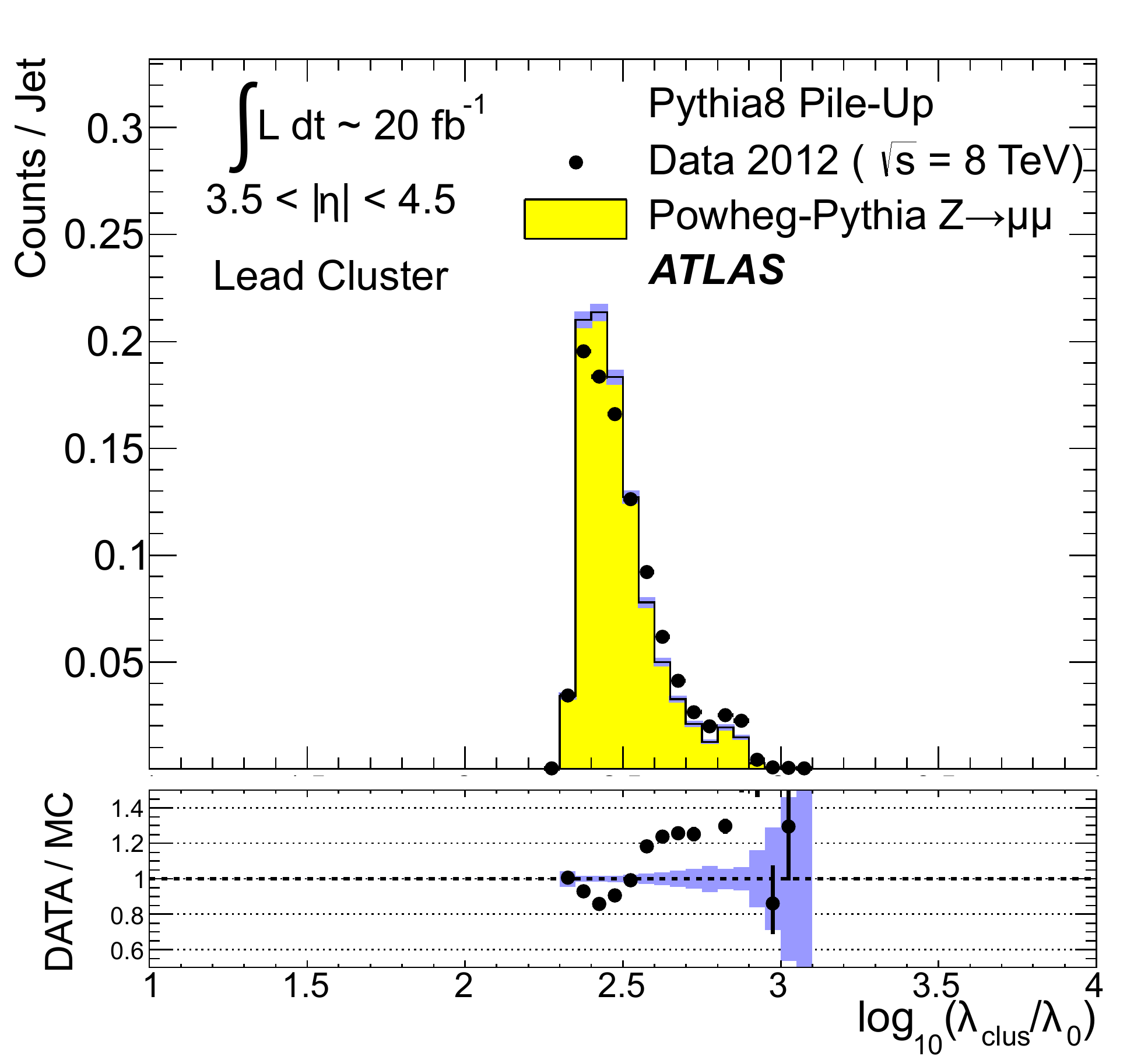}\label{fig:jet:hardest:lam:3:mc}}
	\caption[]{The distribution of the leading \topo{} depth location measure $\log_{10}(\lamctr/\lambda_{0})$ in fully calibrated jets reconstructed with the \antikt{} algorithm with $R = 0.4$ and $\unit{30}{\GeV} < \ptjetlcwjes < \unit{40}{\GeV}$ in regions of \subref{fig:jet:hardest:lam:1:mc} the central ($|\etajet|<0.6$), \subref{fig:jet:hardest:lam:2:mc} the \EndCap{} ($2.0<|\etajet|<2.5$), and the \subref{fig:jet:hardest:lam:3:mc} forward ($3.5<|\etajet|<4.5$) calorimeters in \ATLAS. Data is compared to \MC{} simulations with fully simulated \pu{} for \Zmumu{} events recorded in 2012. The ratio of the distribution from data to the one from \MC{} simulations is shown below each plot. The shaded bands show statistical uncertainties for the distributions from \MC{} simulations and the corresponding uncertainty bands in the ratio plots. The reference scale for \lamctr{} is $\lambda_{0} = \unit{1}{\mm}$.}
	\label{fig:jet:hardest:lam}
\end{figure}

\begin{figure}[t!]\centering
	\subfloat[]{\includegraphics[width=\fighalfwidth]{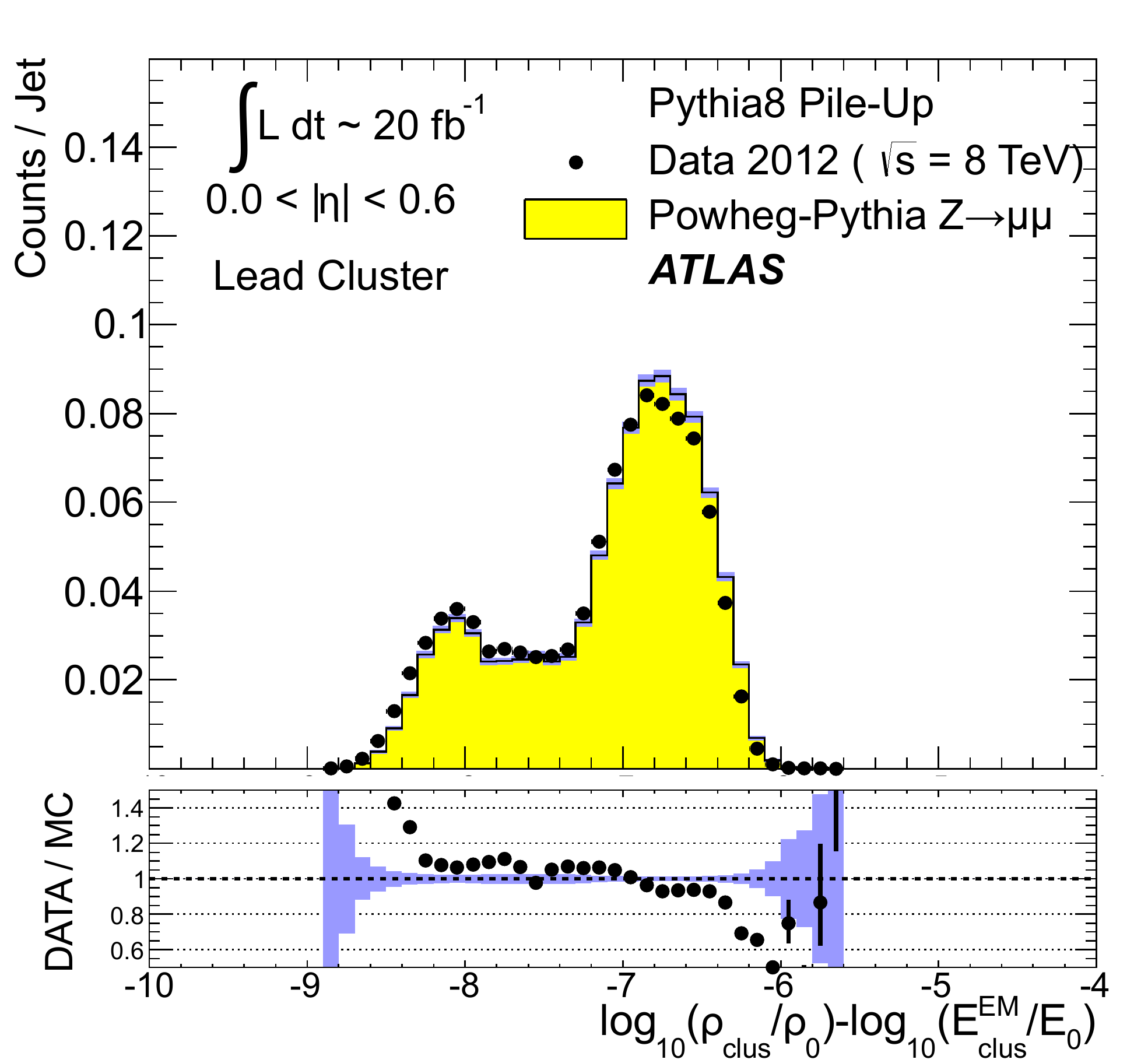}\label{fig:jet:hardest:firstEden:1:mc}} 
	\subfloat[]{\includegraphics[width=\fighalfwidth]{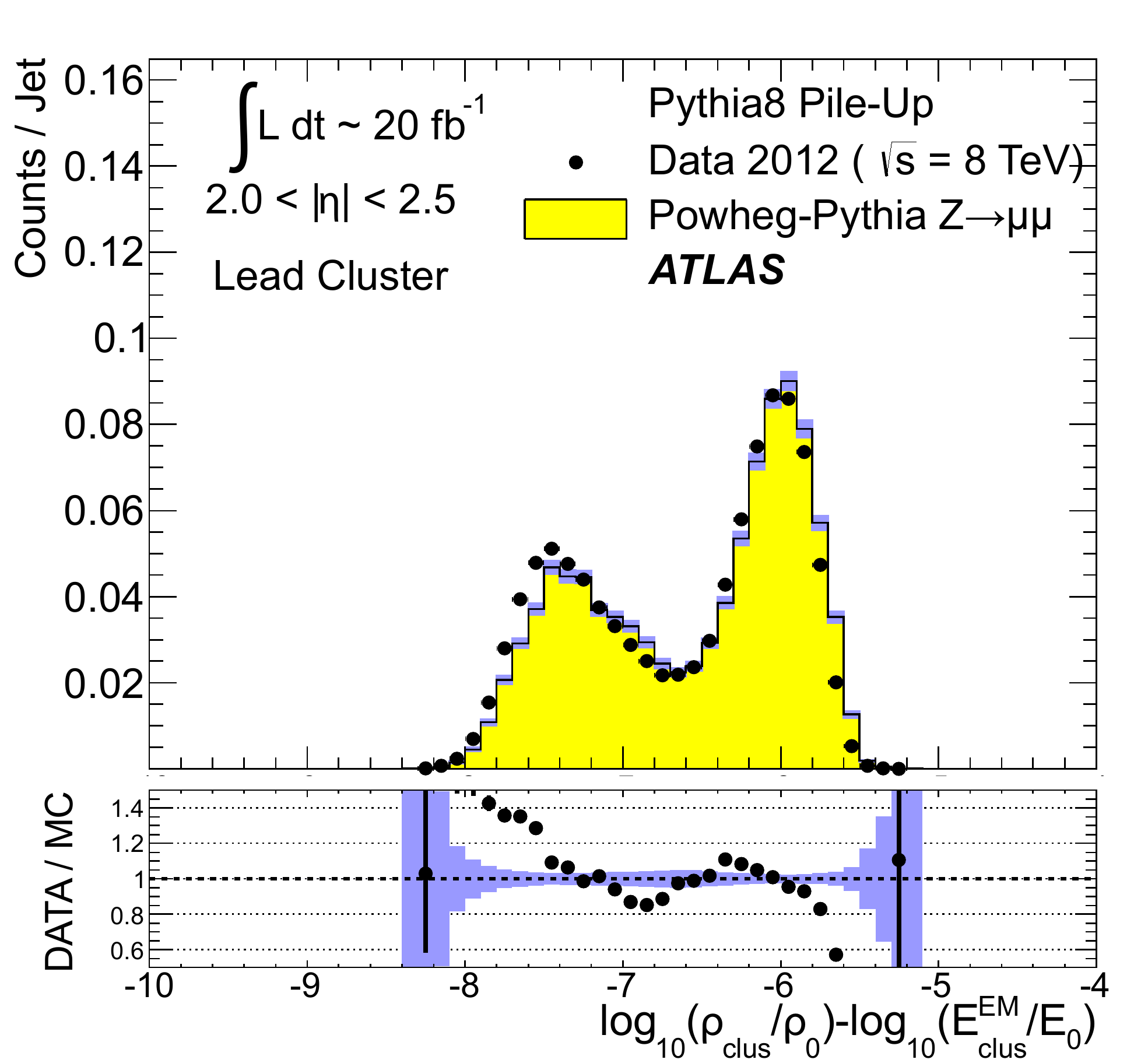}\label{fig:jet:hardest:firstEden:2:mc}}
	\\
	\subfloat[]{\includegraphics[width=\fighalfwidth]{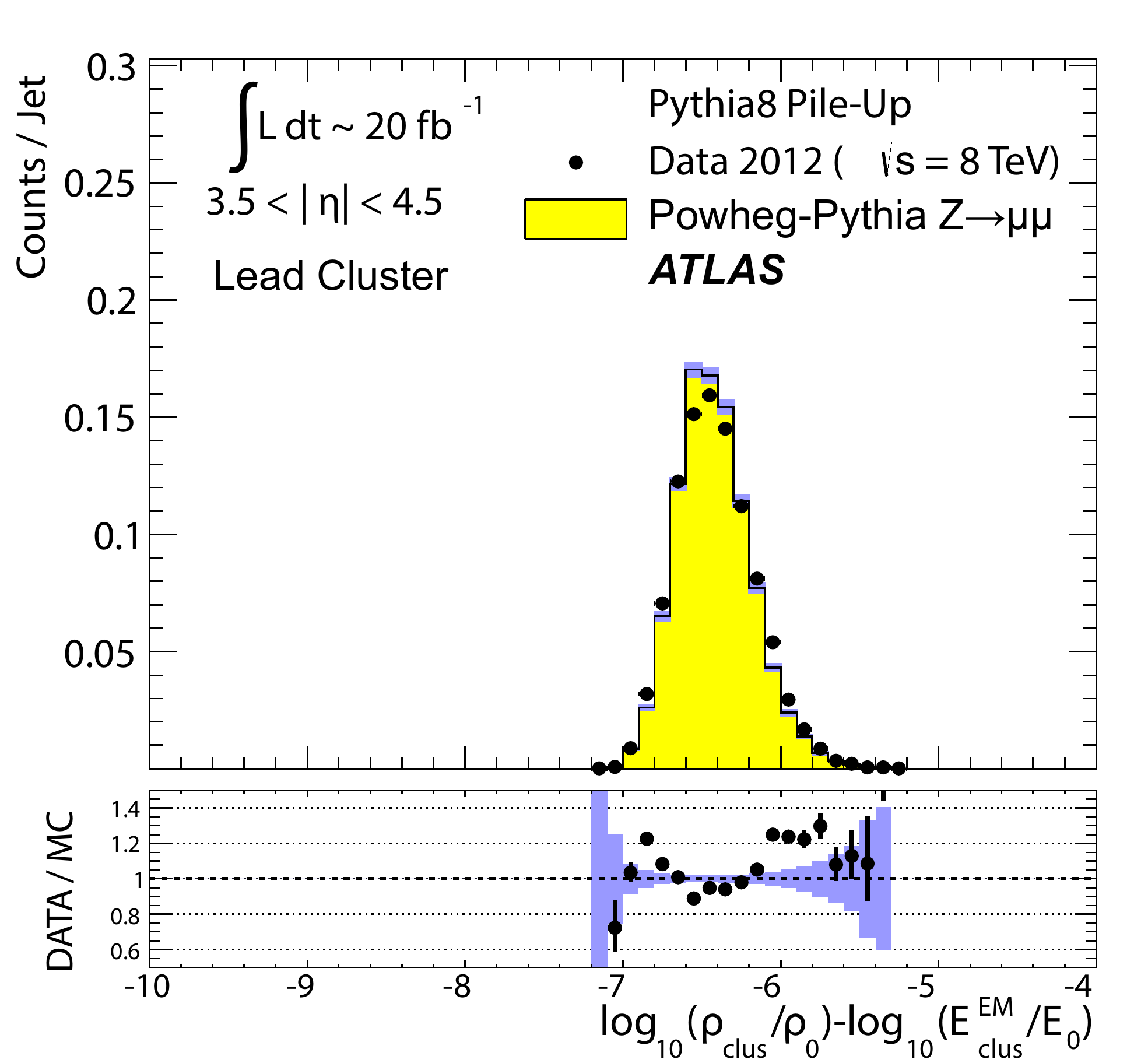}\label{fig:jet:hardest:firstEden:3:mc}} 
	\caption[]{The distribution of the leading \topo{} signal density measure $\log_{10}(\rhoclus/\rho_{0})-\log_{10}(\eclusem/E_{0})$ in fully calibrated jets reconstructed with the \antikt{} algorithm with $R = 0.4$ and $\unit{30}{\GeV} < \ptjetlcwjes < \unit{40}{\GeV}$ in regions of \subref{fig:jet:hardest:firstEden:1:mc} the central ($|\etajet|<0.6$), \subref{fig:jet:hardest:firstEden:2:mc} the \EndCap{} ($2.0<|\etajet|<2.5$), and \subref{fig:jet:hardest:firstEden:3:mc} the forward ($3.5<|\etajet|<4.5$) calorimeters in \ATLAS. Data is compared to \MC{} simulations with fully simulated \pu{} for \Zmumu{} events recorded in 2012. The ratio of the distribution from data to the one from \MC{} simulations is shown below each plot. The shaded bands show statistical uncertainties for the distributions from \MC{} simulations and the corresponding uncertainty bands in the ratio plots. The reference scale for \rhoclus{} is $\rho_{0} = \unit{1}{\MeV/\mm^{3}}$, and for the energy $E_{0} = \unit{1}{\MeV}$.}
	\label{fig:jet:hardest:firstEden}
\end{figure}

\begin{figure}[t!]\centering
	\subfloat[]{\includegraphics[width=\fighalfwidth]{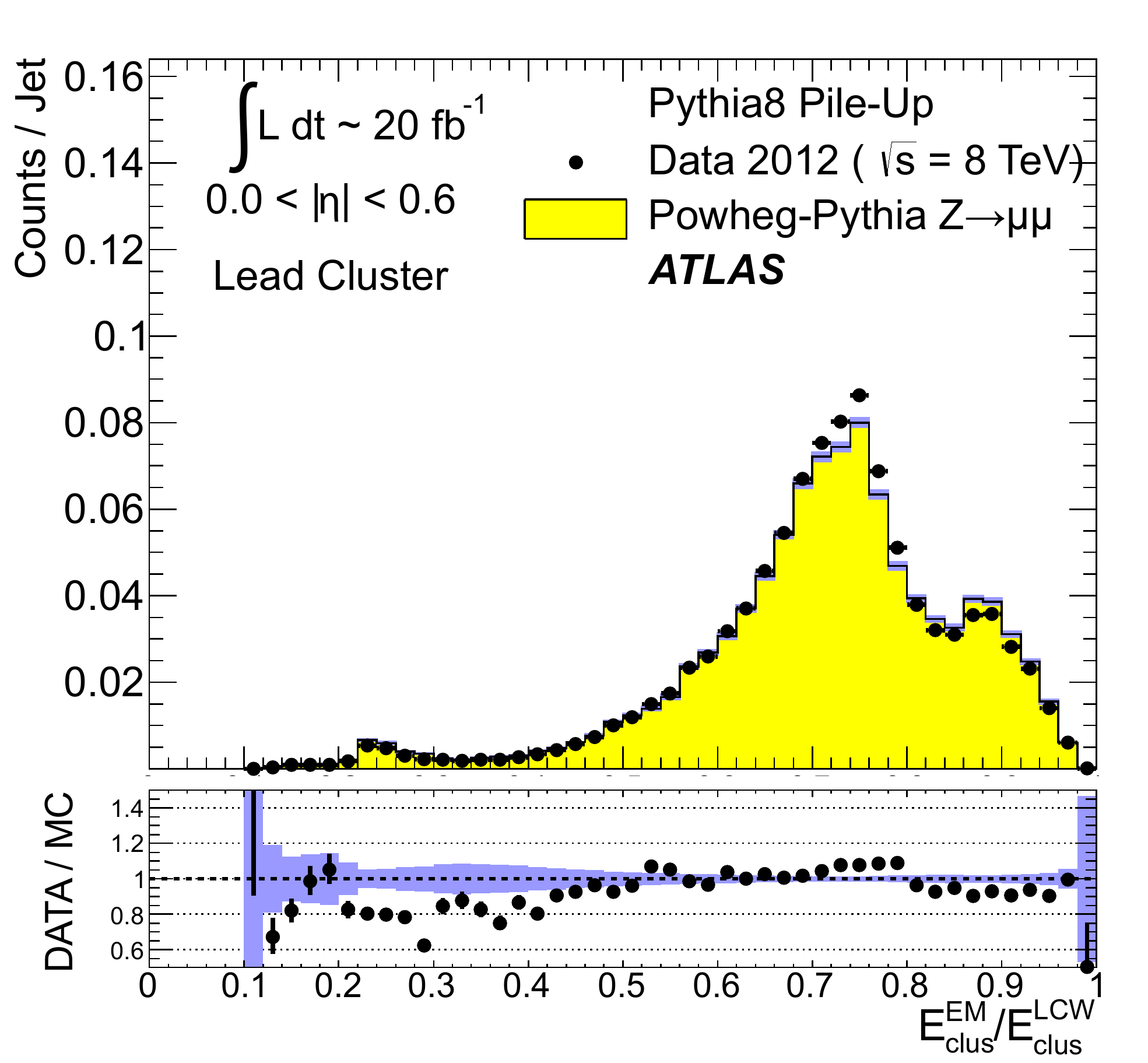}\label{fig:jet:hardest:LC:1:mc}} 
	\subfloat[]{\includegraphics[width=\fighalfwidth]{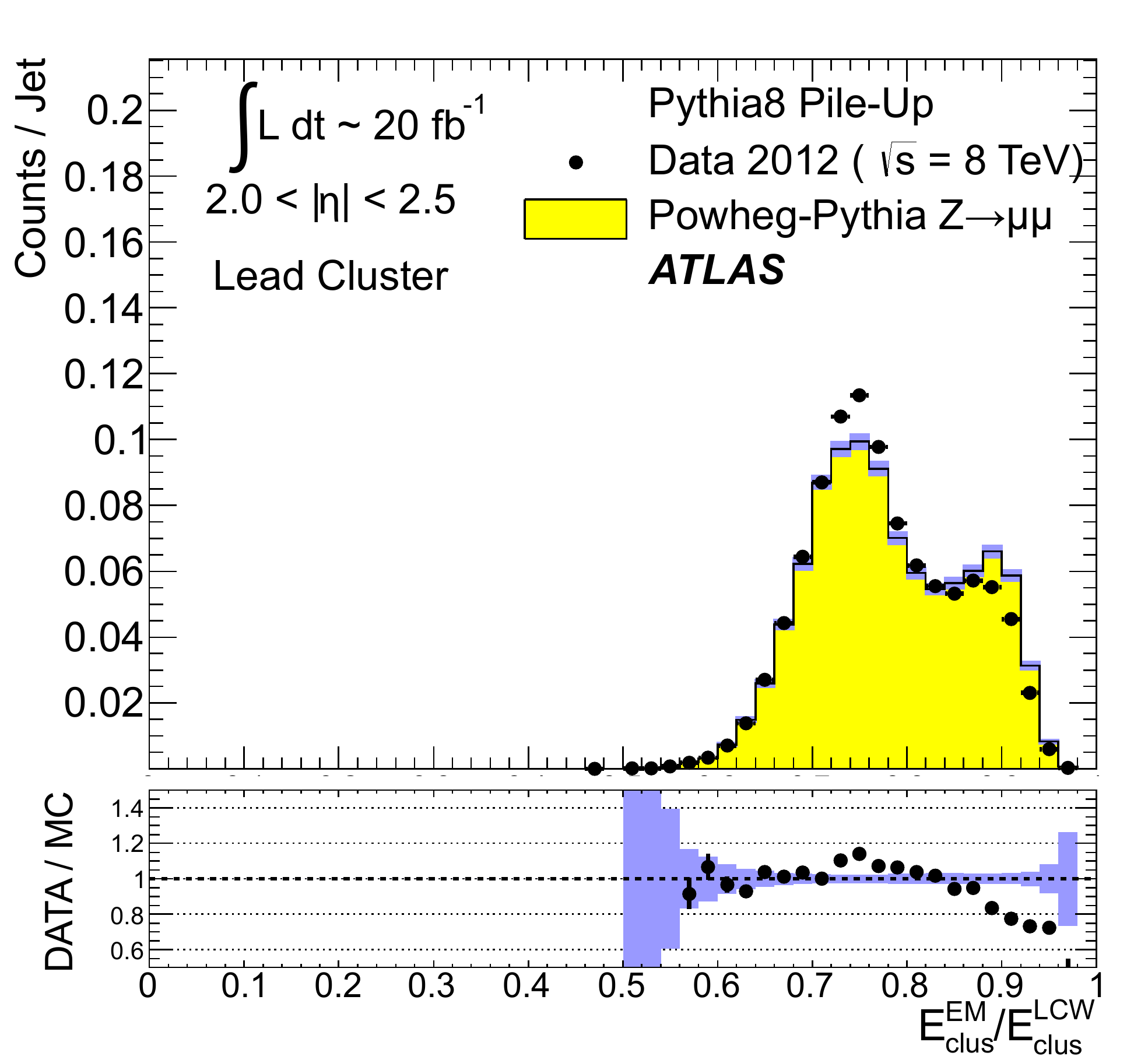}\label{fig:jet:hardest:LC:2:mc}}
	\\
	\subfloat[]{\includegraphics[width=\fighalfwidth]{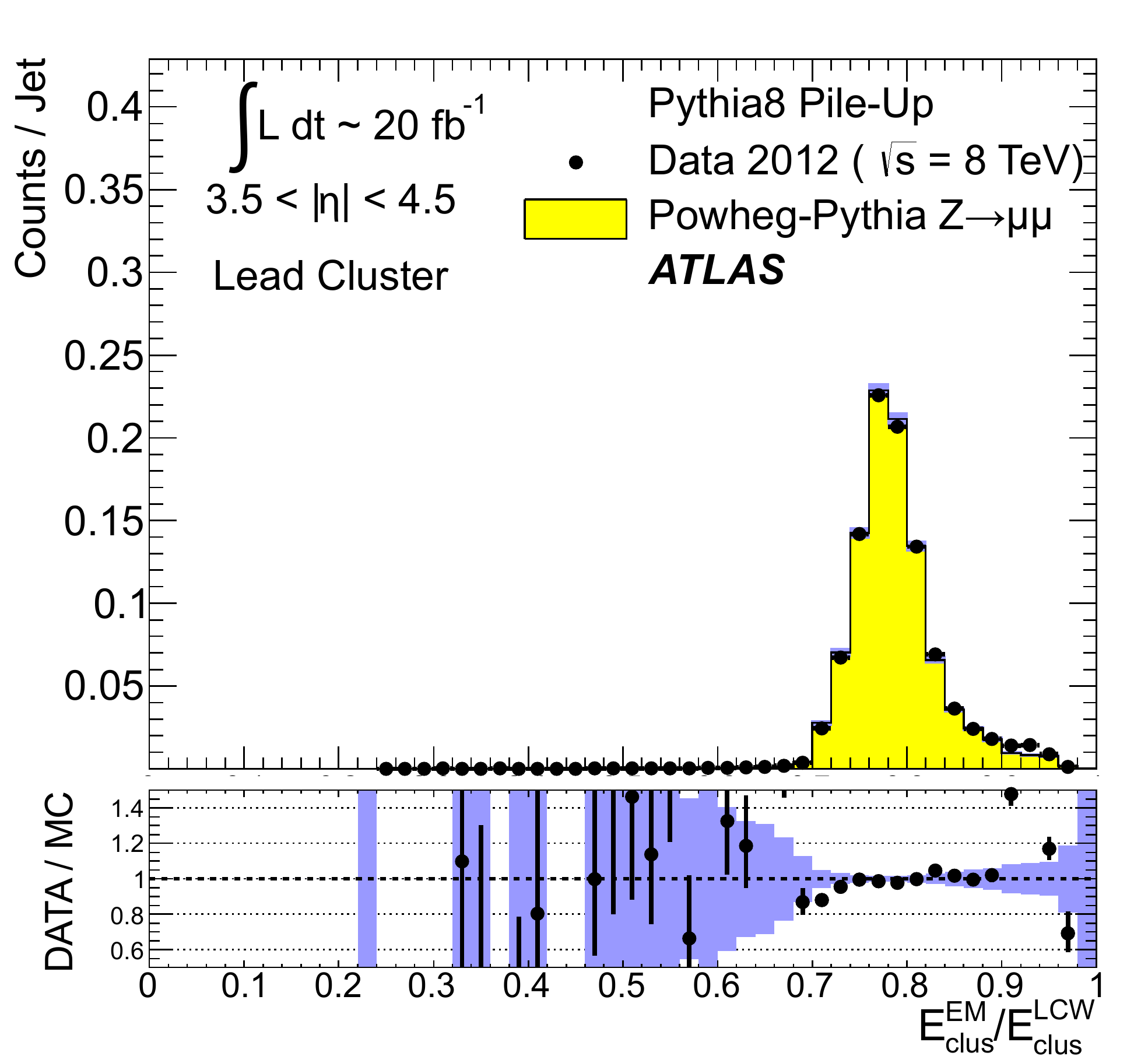}\label{fig:jet:hardest:LC:3:mc}} 
	\caption[]{The distribution of the ratio of the cluster signal reconstructed on \EM{} scale \eclusem{} to the fully calibrated signal \ecluslcw{} for the leading \topo{} in fully calibrated jets reconstructed with the \antikt{} algorithm with $R = 0.4$ and $\unit{30}{\GeV} < \ptjetlcwjes < \unit{40}{\GeV}$ in regions of \subref{fig:jet:hardest:LC:1:mc} the central ($|\etajet|<0.6$), \subref{fig:jet:hardest:LC:2:mc} the \EndCap{} ($2.0<|\etajet|<2.5$), and \subref{fig:jet:hardest:LC:3:mc} the forward ($3.5<|\etajet|<4.5$) calorimeters in \ATLAS. Data is compared to \MC{} simulations with fully simulated \pu{} for \Zmumu{} events recorded in 2012. The ratio of the distribution from data to the one from \MC{} simulations is shown below each plot. The shaded bands show statistical uncertainties for the distributions from \MC{} simulations and the corresponding uncertainty bands in the ratio plots. The reference scale for \rhoclus{} is $\rho_{0} = \unit{1}{\MeV/\mm^{3}}$, and for the energy $E_{0} = \unit{1}{\MeV}$.}
	\label{fig:jet:hardest:LC}
\end{figure}

The leading \topo{} in a jet is defined as the one with the highest \ptclusem. Its moments and its signal contribution to the jet provide a good reference for the dependence of important 
\topo{} calibration inputs on \pu. 
The leading cluster is found in the \antikt{} jets reconstructed with $R = 0.4$ and with $\unit{30}{\GeV} < \ptjetlcwjes < \unit{40}{\GeV}$ in the 2012 \Zmumu{} sample in data and \MC{} simulations with full \pu{} simulation. 
The distributions of the \topo{} moments relevant to the \LCW{} calibration for the leading cluster in the jet are shown in \figMultiRef{fig:jet:hardest:lam}{and}{fig:jet:hardest:firstEden}. 
The distribution of the overall \LCW{} calibration weight described in \secRef{sec:lcw:full} is shown in \figRef{fig:jet:hardest:LC}. 

The distribution of the depth location of the leading \topo, already discussed for all and selected \topos{} in the inclusive \Zmumu{} sample in \secRef{sec:perf:obs:moms} and the \Zmumu{} sample with jets  in \secRef{\thislabel:depth}, is shown in \figMultiRefLabel~\ref{fig:jet:hardest:lam}\subref{fig:jet:hardest:lam:1:mc}, \ref{fig:jet:hardest:lam}\subref{fig:jet:hardest:lam:2:mc}, and \ref{fig:jet:hardest:lam}\subref{fig:jet:hardest:lam:3:mc}
for jets reconstructed in the central, \EndCap, and the forward detector region, respectively. 
As expected from the previous observations, \MC{} simulations agree reasonably well with data. 
It is also observed that the leading cluster in the central and \EndCap{} detector regions is most often located either in the electromagnetic or in the hadronic calorimeters, and rarely between the modules. 
In the forward region, the hardest cluster is most often located in the first \FCAL{} module.   

The signal density \rhoclus{} of \topos{} is defined in \secRef{sec:moments:signal:density}. 
\FigRef{fig:jet:hardest:firstEden} shows the \rhoclus{} distributions for the leading \topo{} in the jet. 
The complex structures of these distributions are  well modelled. 
Their shape in the central and \EndCap{} regions is driven by the jet fragmentation. 
Jets with a leading photon, or two nearby photons from a neutral pion decay, can produce the leading \topo{} with a high signal density,  reflecting the single or the two largely overlapping compact electromagnetic shower(s) reconstructed in the highly granular electromagnetic calorimeters.
Jets with a leading hadron that reaches the detector typically produce less dense \topo{} signals in the corresponding hadronic shower.  
For these jets an additional  geometric effect is introduced, as the leading \topo{} is more likely located in the hadronic calorimeters in \ATLAS.\footnote{In the case of a leading (stable) hadron in the jet, the leading \topo{} may still arise from a photon, as the selection of this cluster is performed on the \EM{} scale. This introduces a bias due to $e/\pi >1$, which is nevertheless well modelled in \MC{} simulations, according to \figRef{fig:jet:hardest:firstEden}.}
The typically larger cell sizes in these detectors introduce lower density signals even for compact showers. 

The forward detector region has a coarser longitudinal segmentation, with the first module \LArFCALN{0}{} closest to the collision vertex being about \unit{30}{X_{0}}{} and \unit{2.5}{\lamnucl}{} deep \cite{Armitage:2007zz}. 
Consequently, most leading \topos{} in jets going in this direction are located in \LArFCALN{0}, as can be seen in the \lamctr{} distribution in \figRef{fig:jet:hardest:lam}\subref{fig:jet:hardest:lam:3:mc}. 
The \rhoclus{} distribution in \figRef{fig:jet:hardest:firstEden}\subref{fig:jet:hardest:firstEden:3:mc} therefore does not show the features seen in \figMultiRefLabel~\ref{fig:jet:hardest:firstEden}\subref{fig:jet:hardest:firstEden:1:mc} and \ref{fig:jet:hardest:firstEden}\subref{fig:jet:hardest:firstEden:2:mc}, because the calorimeter \readout{} granularity changes smoothly within this module. 
The hard transitions between calorimeter modules with very different granularity affecting the \rhoclus{} distributions in the central and \EndCap{} regions are absent. 

\begin{figure}[t!] \centering
	\subfloat[$|\eta|<0.6$]{\includegraphics[width=\fighalfwidth]{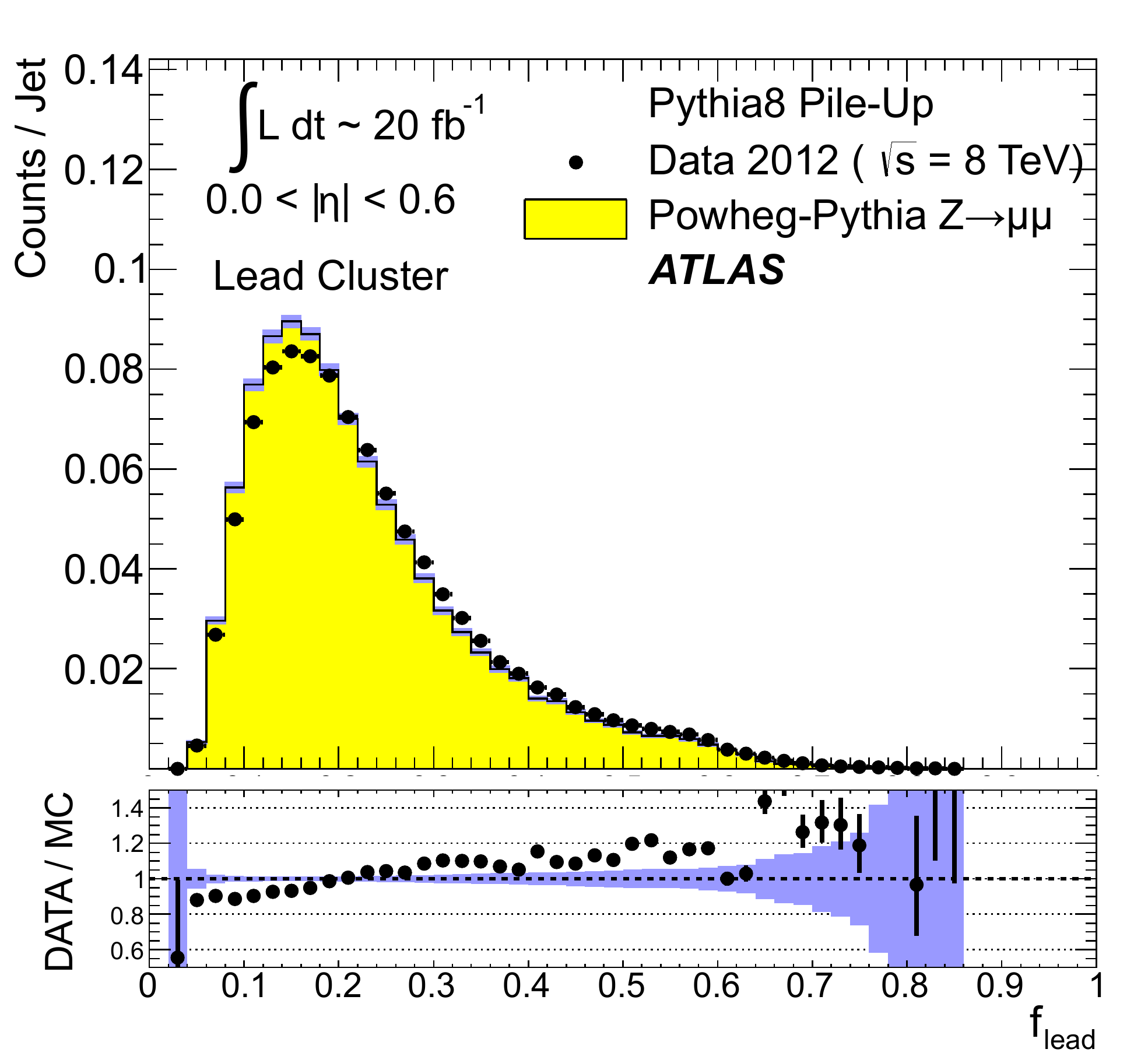}\label{fig:jet:lead:fHard:1:mc}} 
	\subfloat[$2.0<|\eta|<2.5$]{\includegraphics[width=\fighalfwidth]{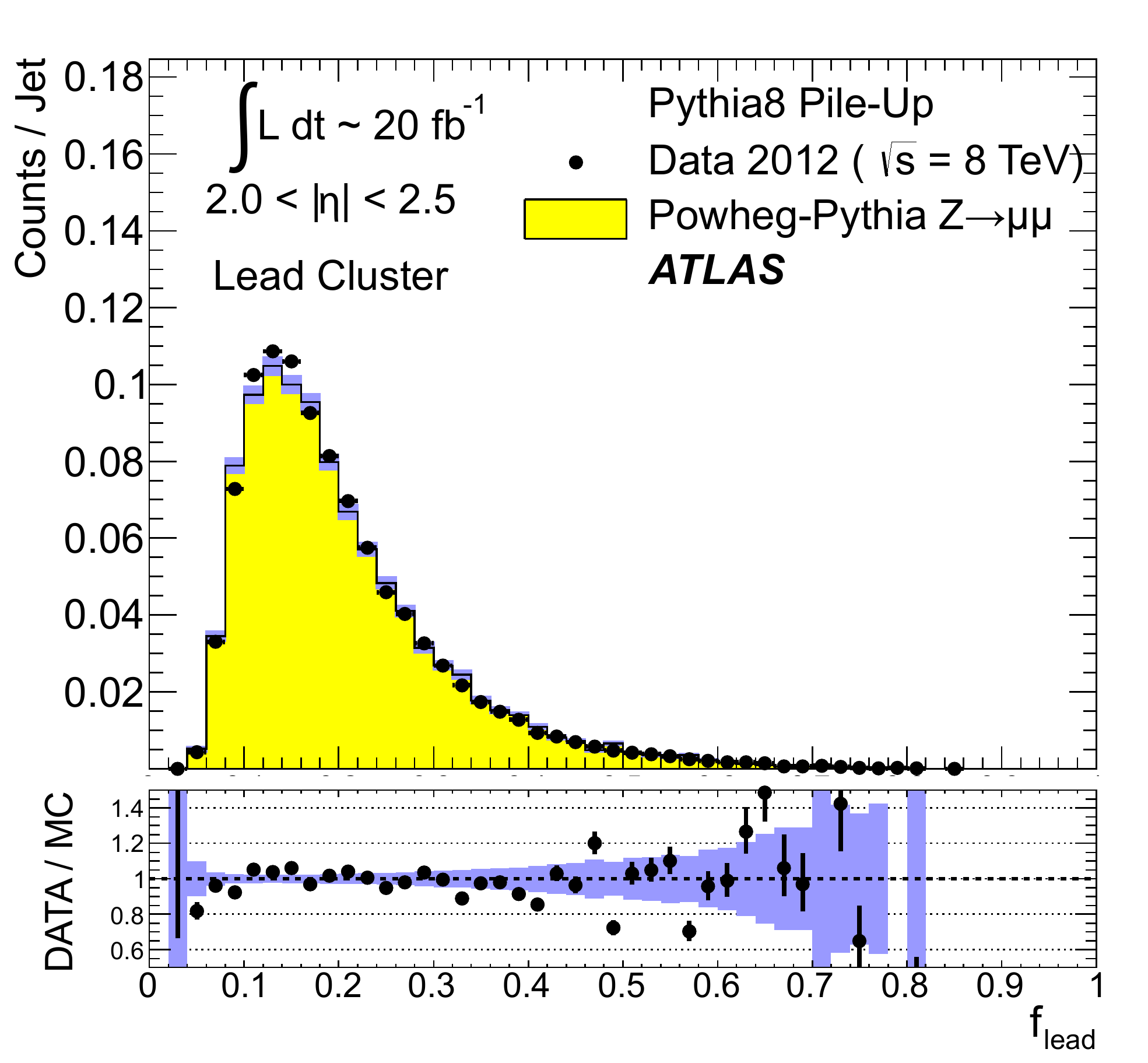}\label{fig:jet:lead:fHard:2:mc}}
	\\
	\subfloat[$3.5<|\eta|<4.5$]{\includegraphics[width=\fighalfwidth]{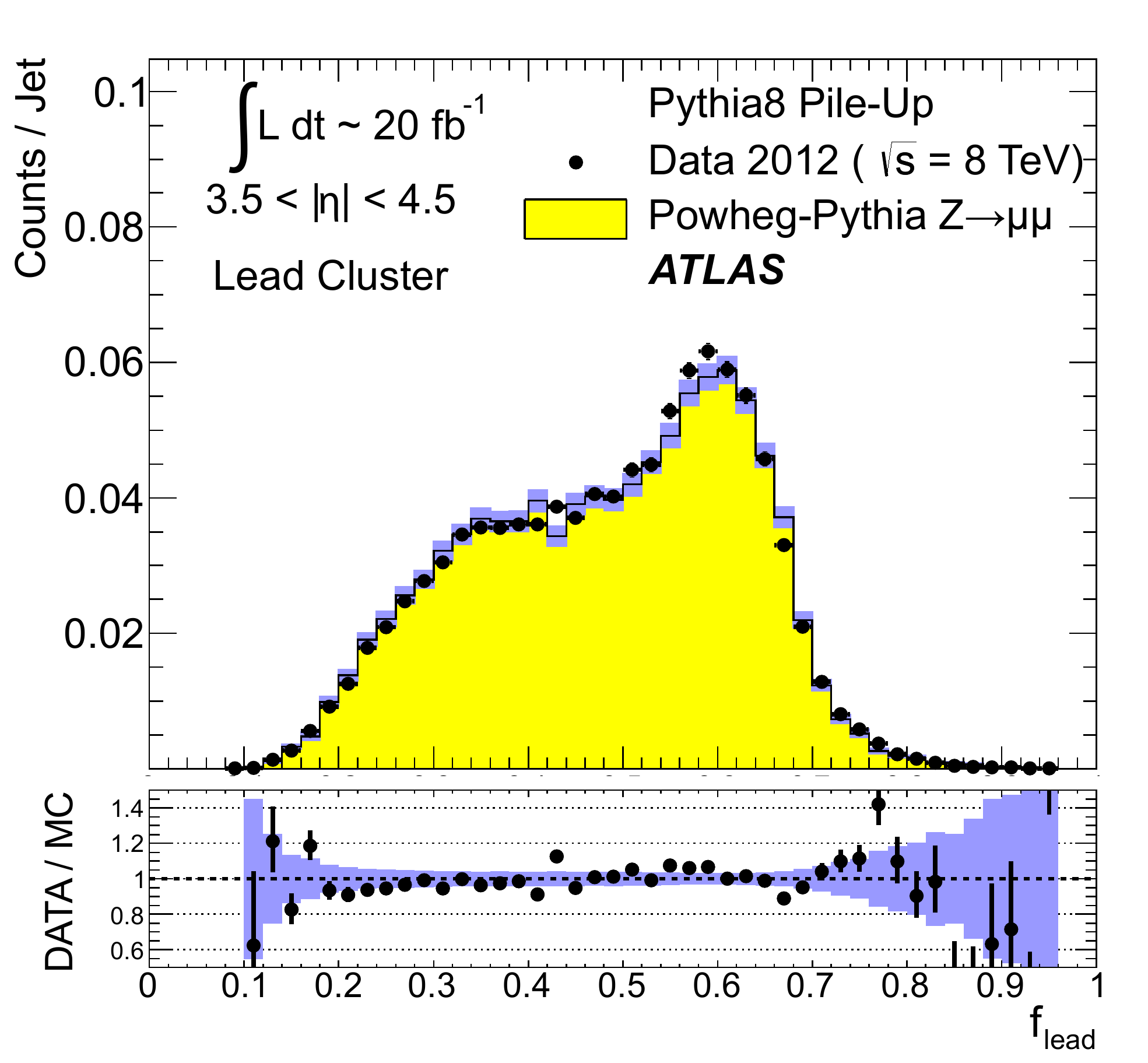}\label{fig:jet:lead:fHard:3:mc}} 
	\qquad	
	\caption[]{The distribution of the signal fraction \fhardest{} carried by the leading \topo{} in jets, as defined in \eqRef{eq:fhardest}, in \subref{fig:jet:lead:fHard:1:mc} the central, \subref{fig:jet:lead:fHard:2:mc} the \EndCap, and \subref{fig:jet:lead:fHard:3:mc} the forward detector region. The jets are reconstructed using the \antikt{} algorithm with $R = 0.4$ and with $\unit{30}{\GeV} < \ptjetlcwjes < \unit{40}{\GeV}$ in \Zmumu{} events in 2012 data and \MC{} simulations with fully simulated \pu. The data-to-\MC{} simulation ratios are shown below the distributions. The shaded bands shown for the distributions obtained from \MC{} simulations indicate statistical uncertainties and the corresponding uncertainty bands in the ratio plots.}
	\label{fig:jet:lead}		
\end{figure}

The overall effect of the \LCW{} calibration described in \secRef{sec:lcw} on the signal scale of the leading \topo{} can be measured by the ratio of the basic \EM{} scale signal \eclusem{} to the
fully calibrated cluster signal \ecluslcw. 
The distribution of this ratio is shown for the three detector regions in \figRef{fig:jet:hardest:LC}\subref{fig:jet:hardest:LC:1:mc}.
These distributions are inclusive with respect to the \topo{} classification described in \secRef{sec:lcw:classification}.  
The shapes observed in the central and \EndCap{} detector regions reflect this classification of the leading \topo. 
The rightmost peak is mostly produced by \topo{} that are generated by electromagnetic showers and predominantly calibrated as such.
In this case the calibration corrections consist of relatively small \ooc{} and \dm{} corrections only, as outlined in \secRef{sec:lcw}. 
As a consequence, $\eclusem/\ecluslcw$ is closer to unity.  
\Topos{} classified as hadronic receive much larger corrections, and are more likely to populate the lower side of the $\eclusem/\ecluslcw$ spectrum. 

The $\eclusem/\ecluslcw$ distribution in the forward detector region shown in \figRef{fig:jet:hardest:LC}\subref{fig:jet:hardest:LC:3:mc} does not display these shapes. 
This is due to a lack of classification power in the coarse geometry of the \FCAL. 
Here most \topos{} are classified as hadronic and receive relatively large corrections.
The populated ranges of $\eclusem/\ecluslcw$ in \figMultiRefLabel~\ref{fig:jet:hardest:LC}\subref{fig:jet:hardest:LC:1:mc} and \ref{fig:jet:hardest:LC}\subref{fig:jet:hardest:LC:2:mc} indicate that the magnitude of the total correction scaling the basic cluster signal \eclusem{} up to the locally calibrated signal \ecluslcw{} reaches considerably higher values in the central region than in the \EndCap{} detector regions. 
This reflects the fact that the incoming particle energies are higher at larger $|\eta|$ for a given range in jet \pT. 
Therefore, the calorimeter response to hadrons relative to the response to electrons and photons ($e/\pi$) rises with increasing $|\eta|$, and reduces the amount of correction needed. 
This is effect is initially expected to be observed when comparing the \EndCap{} with the forward region displayed in \figRef{fig:jet:hardest:LC}\subref{fig:jet:hardest:LC:3:mc} as well, yet in the \LArFCAL{} the \ooc{} and \dm{} corrections are larger than the hadronic calibration addressing $e/\pi >1$ and thus dominate the overall \LCW{} calibration. 

The signal fraction carried by the leading \topo{} in the jet is calculated relative to the fully corrected and calibrated \ptjetlcwjes, which provides a stable signal reference in the presence of \pu{} (see \figRefLabel~\ref{fig:jet:pt}\subref{fig:jet:pt:1}),
\begin{equation}
	\fhardest = \dfrac{\ptclusemi{\text{lead}}}{\ptjetlcwjes}\,.
	\label{eq:fhardest}
\end{equation}
This means that \fhardest{} is expected to satisfy  $0 < \fhardest < 1$. 
\FigRef{fig:jet:lead} shows the distribution of \fhardest{} in the three detector regions. 
The \fhardest{} distributions in the central region shown in \figRefLabel~\ref{fig:jet:lead}\subref{fig:jet:lead:fHard:1:mc} and the \EndCap{} region shown in \figRefLabel~\ref{fig:jet:lead}\subref{fig:jet:lead:fHard:2:mc} display very similar features and indicate the most probable value\footnote{The particular choice of normalisation in the definition of \fhardest{} in \eqRef{eq:fhardest} means that even for jets with only one \topo{} \fhardestmop{} is expected to be smaller than unity.} is $\fhardestmop \approx \unit{\text{12--15}}{\%}$.  
The distribution of \fhardest{} in the forward detector region shown in \figRefLabel~\ref{fig:jet:lead}\subref{fig:jet:lead:fHard:3:mc} displays a significantly different shape introduced by the relatively low \topo{} multiplicity in jets in this region, as shown in  \figMultiRefLabel~\ref{fig:jets:nClus}\subref{fig:jets:nClus:3:mc} and \ref{fig:jets:nClus}\subref{fig:jets:nClus:3:ov}. 
The peak at $\fhardestmop \approx \unit{60}{\%}$ in this distribution is consistent with jets with $\Nclusjet = 1$, and the leftmost shoulder indicates contributions  from jets with $\Nclusjet = 2$, with the region in between populated by jets with $\Nclusjet > 2$.  
All distributions of \fhardest{} are modelled well in the \MC{} simulations with fully simulated \pu.

\subsubsection{\PU{} dependence of leading \topo{} signal features} \label{\thislabel:hardest:pu}

\begin{figure}[t!] \centering
	\subfloat[]{\includegraphics[width=\fighalfwidth]{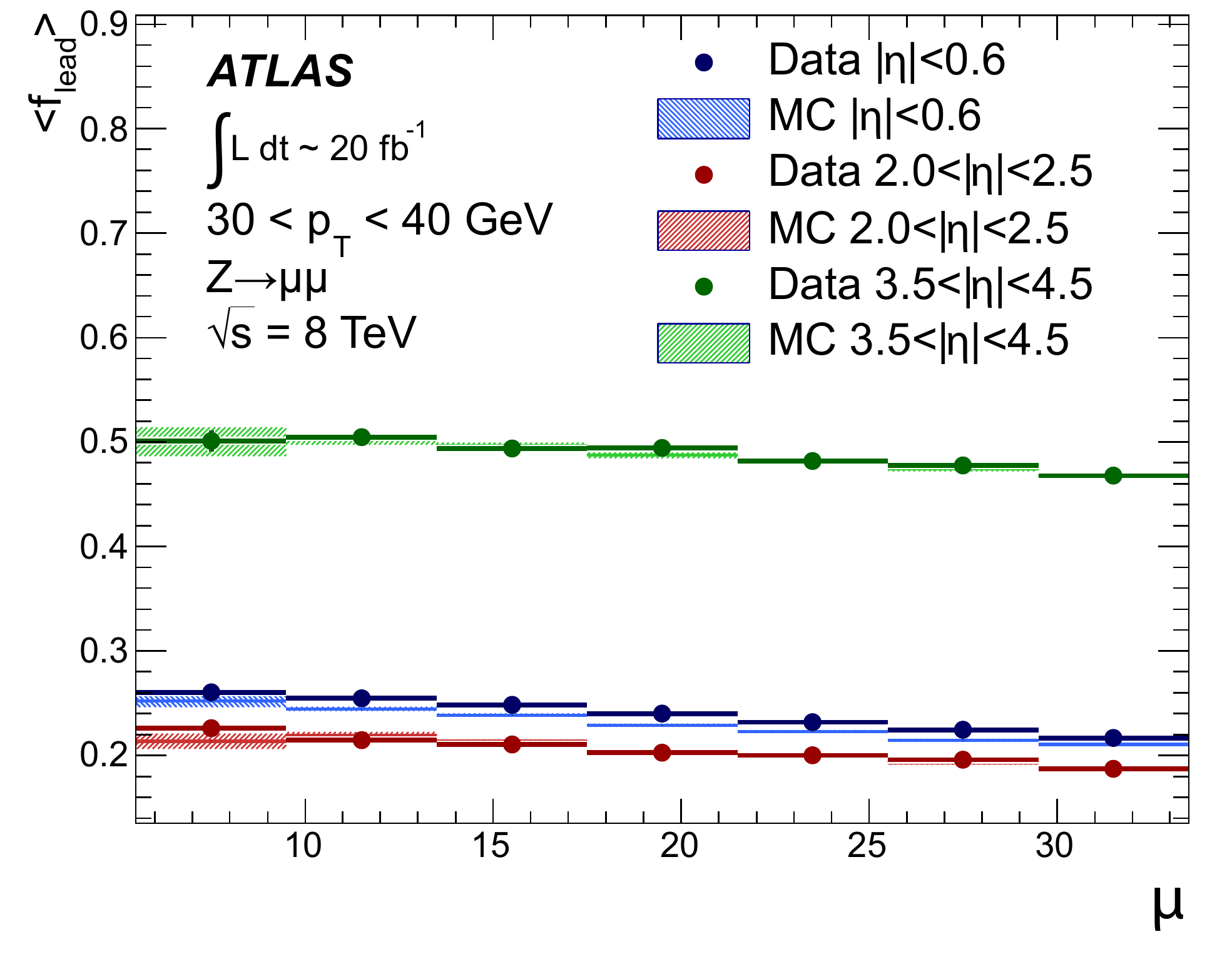}\label{fig:jet:lead:pu:fHard}}
	\subfloat[]{\includegraphics[width=\fighalfwidth]{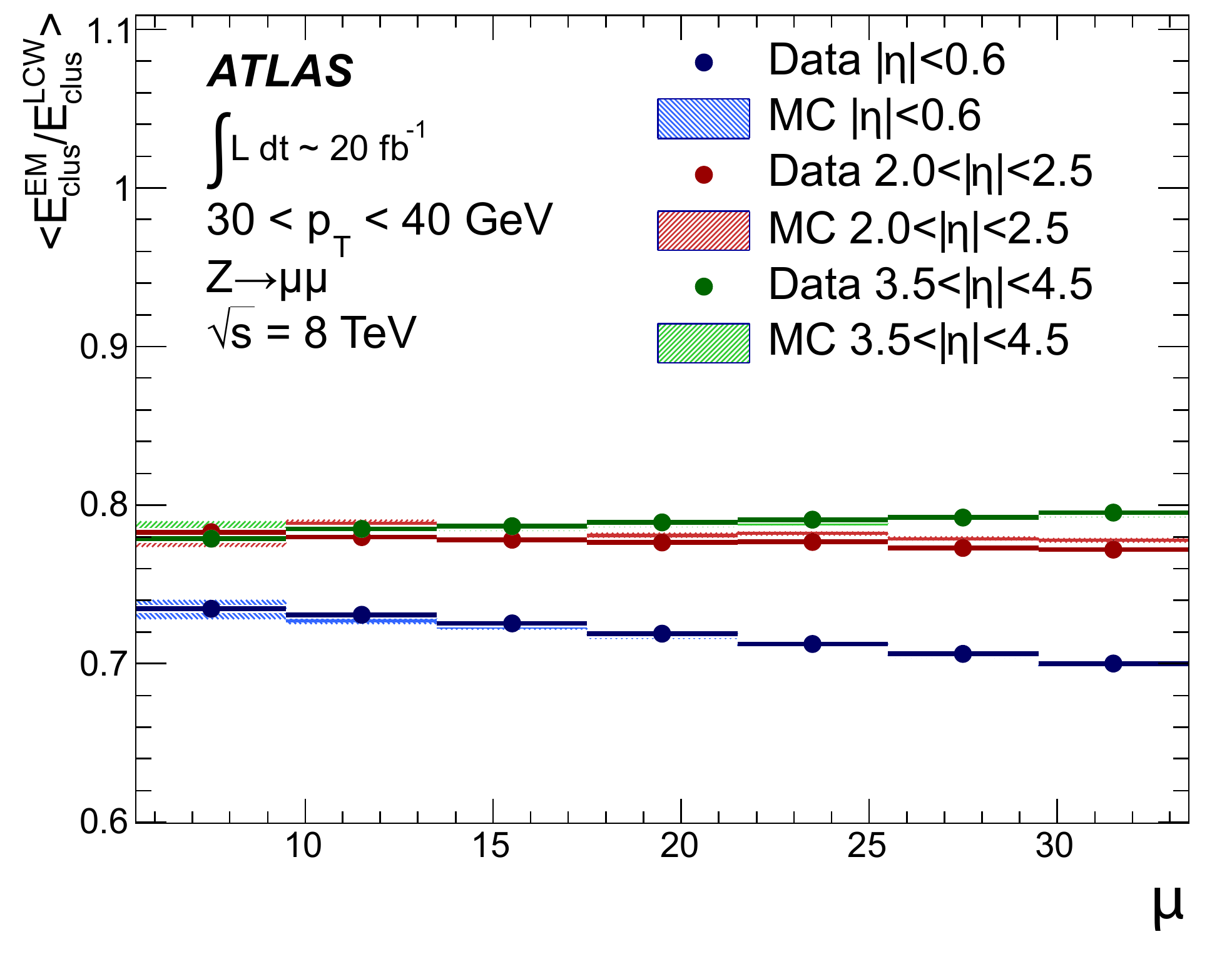}\label{fig:jet:lead:pu:LC}}
        \\
	\subfloat[]{\includegraphics[width=\fighalfwidth]{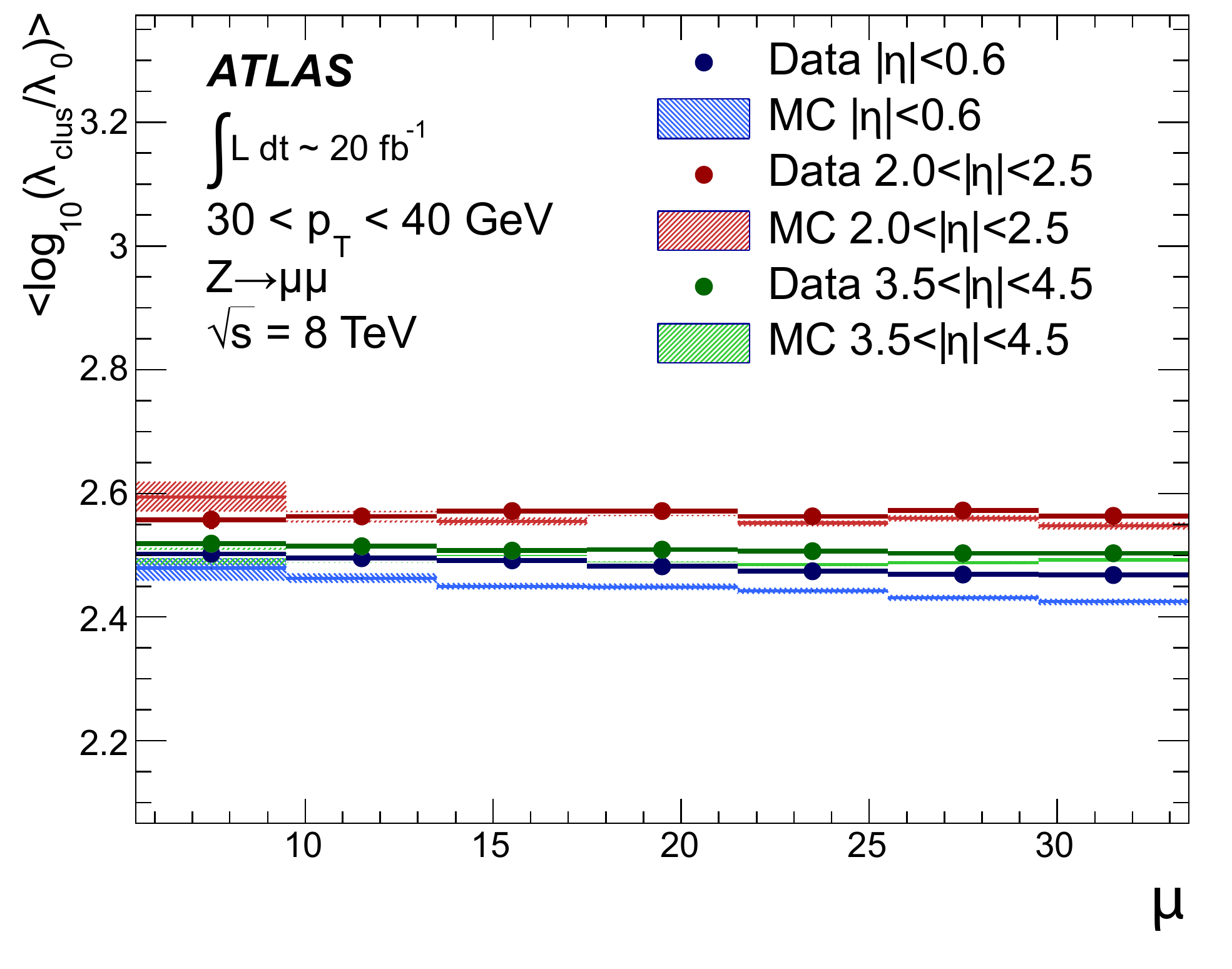}\label{fig:jet:lead:pu:lam}}
	\caption[]{The \pu{} dependence of \subref{fig:jet:lead:pu:fHard} \fhardest{} defined in \eqRef{eq:fhardest}, \subref{fig:jet:lead:pu:LC} $\eclusem/\ecluslcw$, and \subref{fig:jet:lead:pu:lam} the depth location \lamctr{} of the leading \topo{} in fully calibrated \antikt{} jets reconstructed with $R = 0.4$ and with $\unit{30}{\GeV} < \ptjetlcwjes < \unit{40}{\GeV}$ in \Zmumu{} events in 2012 data and \MC{} simulations with fully simulated \pu. The reference scale for \lamctr{} is $\lambda_{0} = \unit{1}{\mm}$. The \pu{} activity is measured in terms of the number of \pu{} interactions $\mu$.
The shaded bands shown for the results obtained from \MC{} simulations indicate statistical uncertainties.} 
	\label{fig:jet:lead:pu}
\end{figure}

The \pu{} dependence of the average leading cluster signal fraction \AVE{\fhardest}, the average \AVE{\eclusem/\ecluslcw} ratio, and the average depth location of the leading \topo{} are displayed in 
\figRef{fig:jet:lead:pu}. The \pu{} activity is measured in terms of $\mu$ for this evaluation. A small linear drop of $\AVE{\fhardest}(\mu)$ is observed for increasing $\mu$ in all three detector regions in \figRef{fig:jet:lead:pu}\subref{fig:jet:lead:pu:fHard}. 
This signal loss of the leading \topo{} can arise from two effects. First, the increase of the \opu{} contributions due to the rising $\mu$ reduces 
the signal due to the bipolar signal shaping function employed in the \ATLAS{} \LAr{} calorimeters (see discussion in \secRef{sec:atlas:data:pu}). 
Second, the increasing \ipu{} contributions at higher $\mu$ and the increased noise introduced by more \opu{} leads to additional splitting in the \topo{} formation, which can take signal away from the leading cluster in the jets. 

\FigRef{fig:jet:lead:pu}\subref{fig:jet:lead:pu:LC} shows that the overall \LCW{} calibration applied to the leading \topo, measured by the average ratio \AVE{\eclusem/\ecluslcw}, in the \EndCap{} and forward detector regions is stable against increasing \pu{} activity. 
A slight drop can be observed with increasing $\mu$ in the central detector region, which indicates changes in the \topo{} properties relevant to the \LCW{} calibration introduced by increasing \pu. 
One possible reason for that may be effects on the \topo{} splitting in this region, as \pu{} can induce spatial energy distributions leading to modifications
in the splitting even for hard signal clusters.\footnote{In particular, \pu{} can introduce an additional signal maximum at the boundary of a relatively dense leading \topo, which can have a significant effect 
on e.g. \rhoclus{} and other cluster properties pertinent to the
\LCW{} calibration.} The depth location \lamctr, which enters the
\LCW{} calibration in the classification step discussed in
\secRef{sec:lcw:classification}, is found to be rather stable against
\pu, as shown in \figRef{fig:jet:lead:pu}\subref{fig:jet:lead:pu:lam}. The pile-up
dependence of the leading \topo{} features discussed here are found to
be well modelled in \MC{} simulations with fully simulated \pu.

\subsubsection{Leading \topo{} geometry and shapes}\label{\thislabel:hardest:geometry}

\begin{figure}[tp!] \centering
        \sfcompress
	\subfloat[]{\includegraphics[width=\figsixpanelwidth]{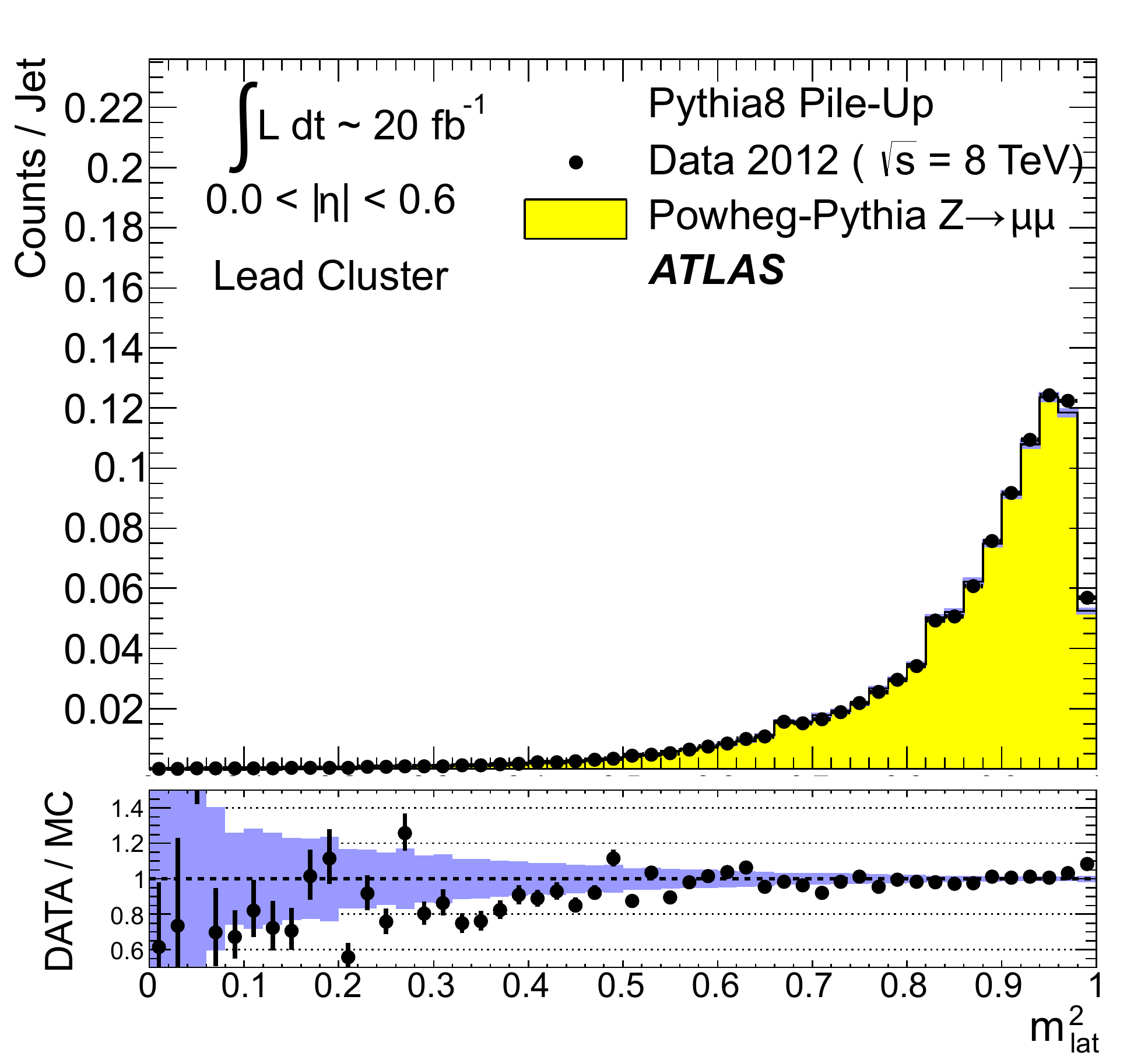}\label{fig:jet:lead:lat:1:mc}}  \qquad 
	\subfloat[]{\includegraphics[width=\figsixpanelwidth]{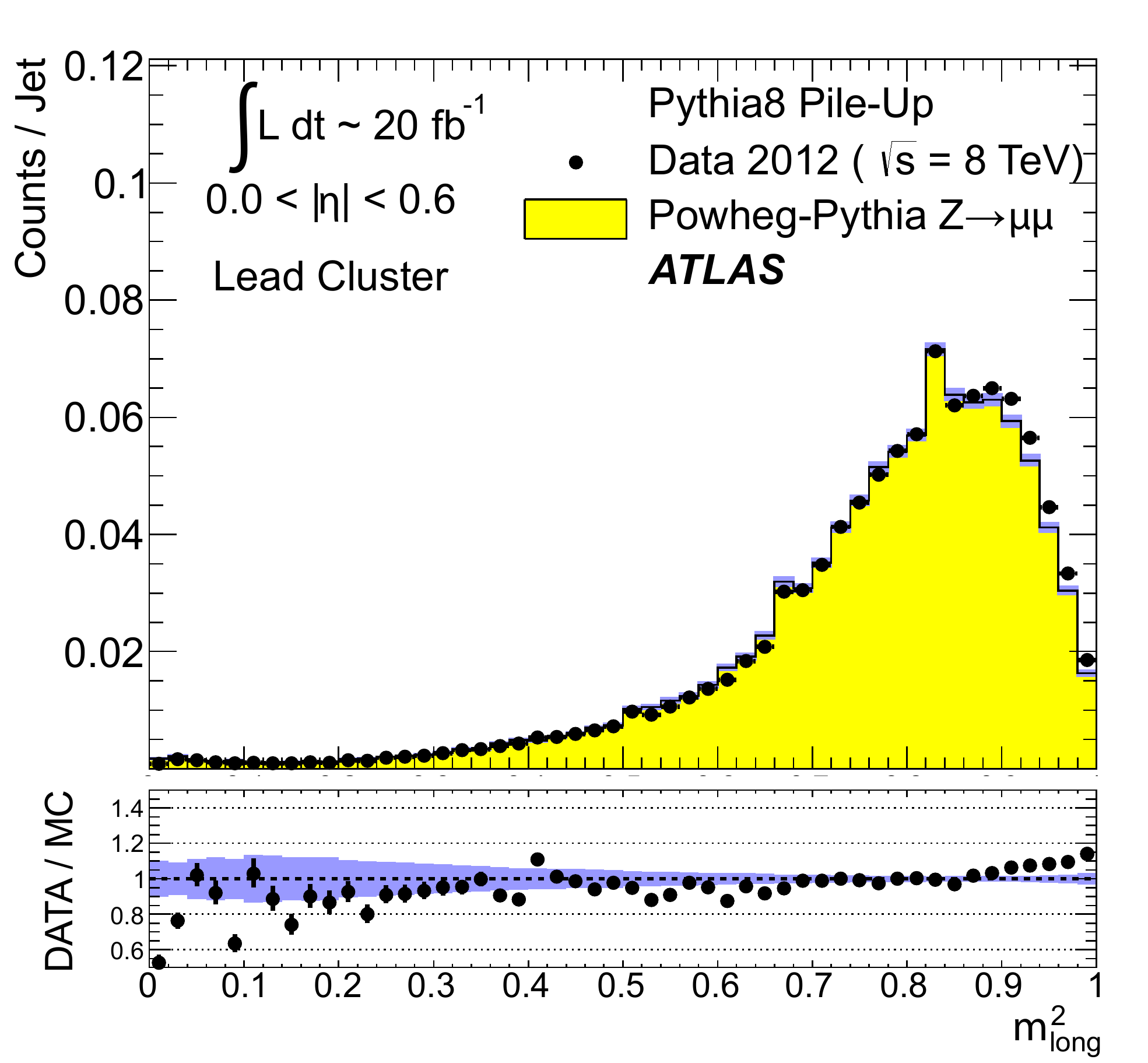}\label{fig:jet:lead:long:1:mc}}
        \\      
	\subfloat[]{\includegraphics[width=\figsixpanelwidth]{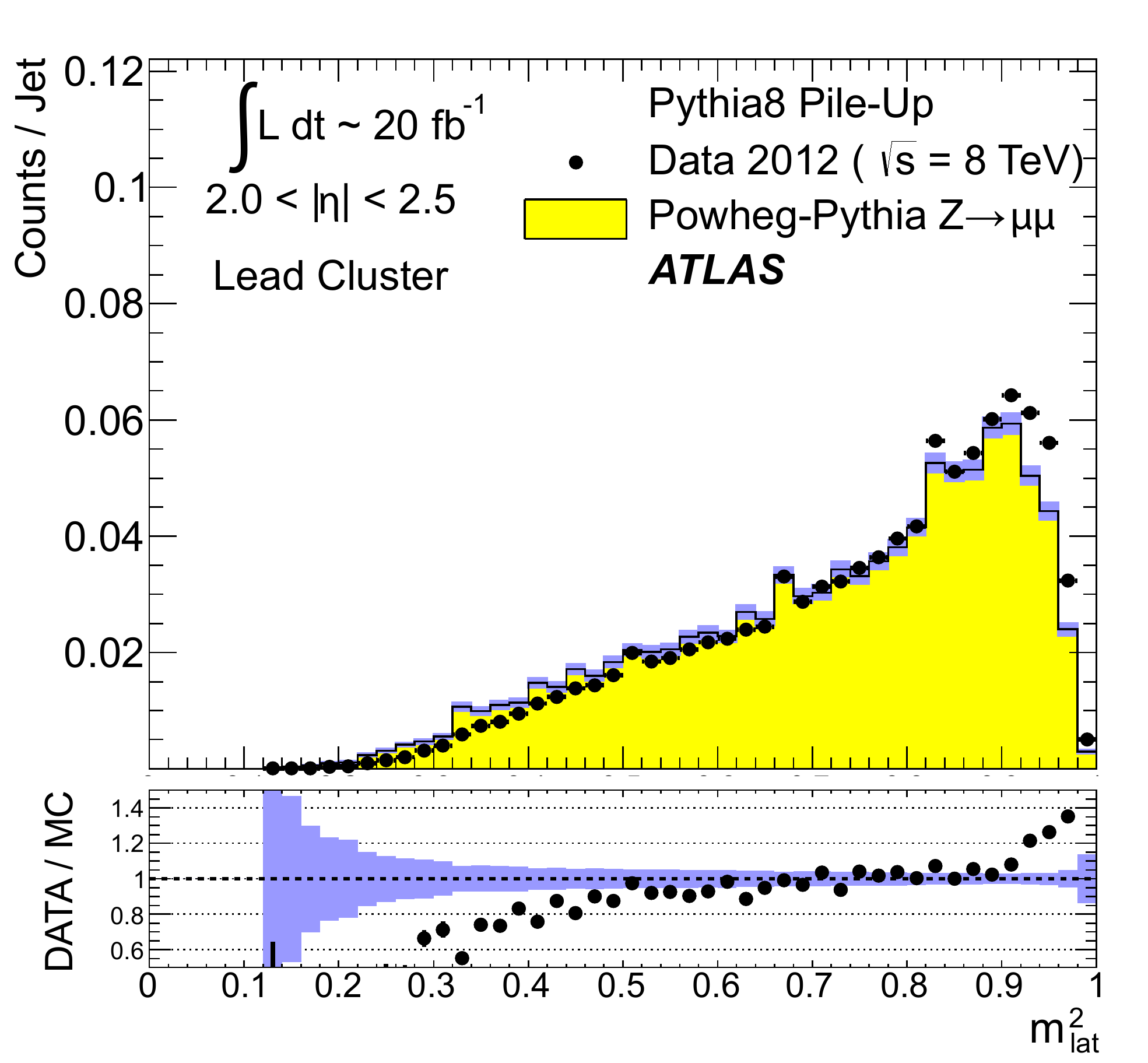}\label{fig:jet:lead:lat:2:mc}}  \qquad
	\subfloat[]{\includegraphics[width=\figsixpanelwidth]{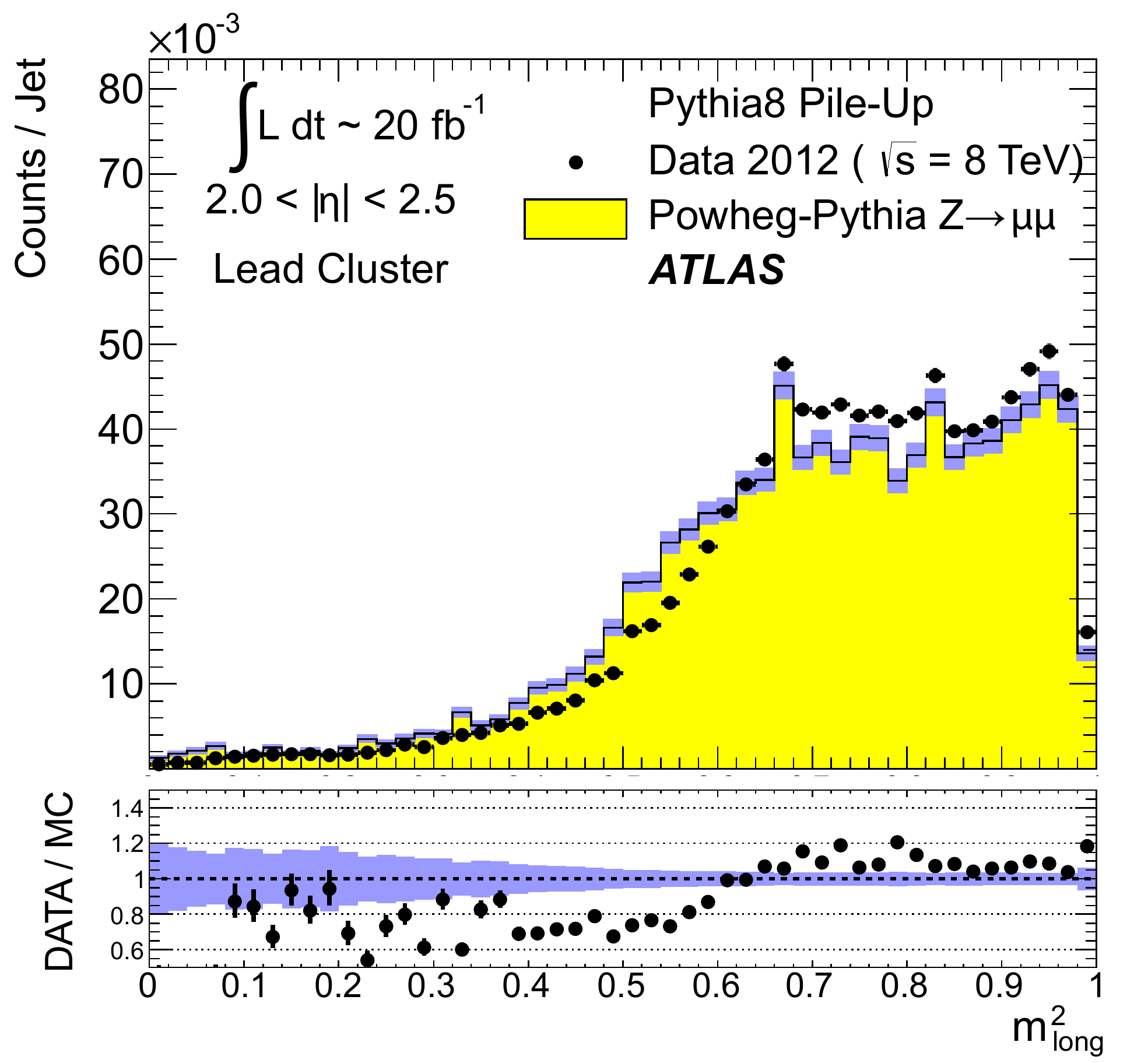}\label{fig:jet:lead:long:2:mc}}
        \\
	\subfloat[]{\includegraphics[width=\figsixpanelwidth]{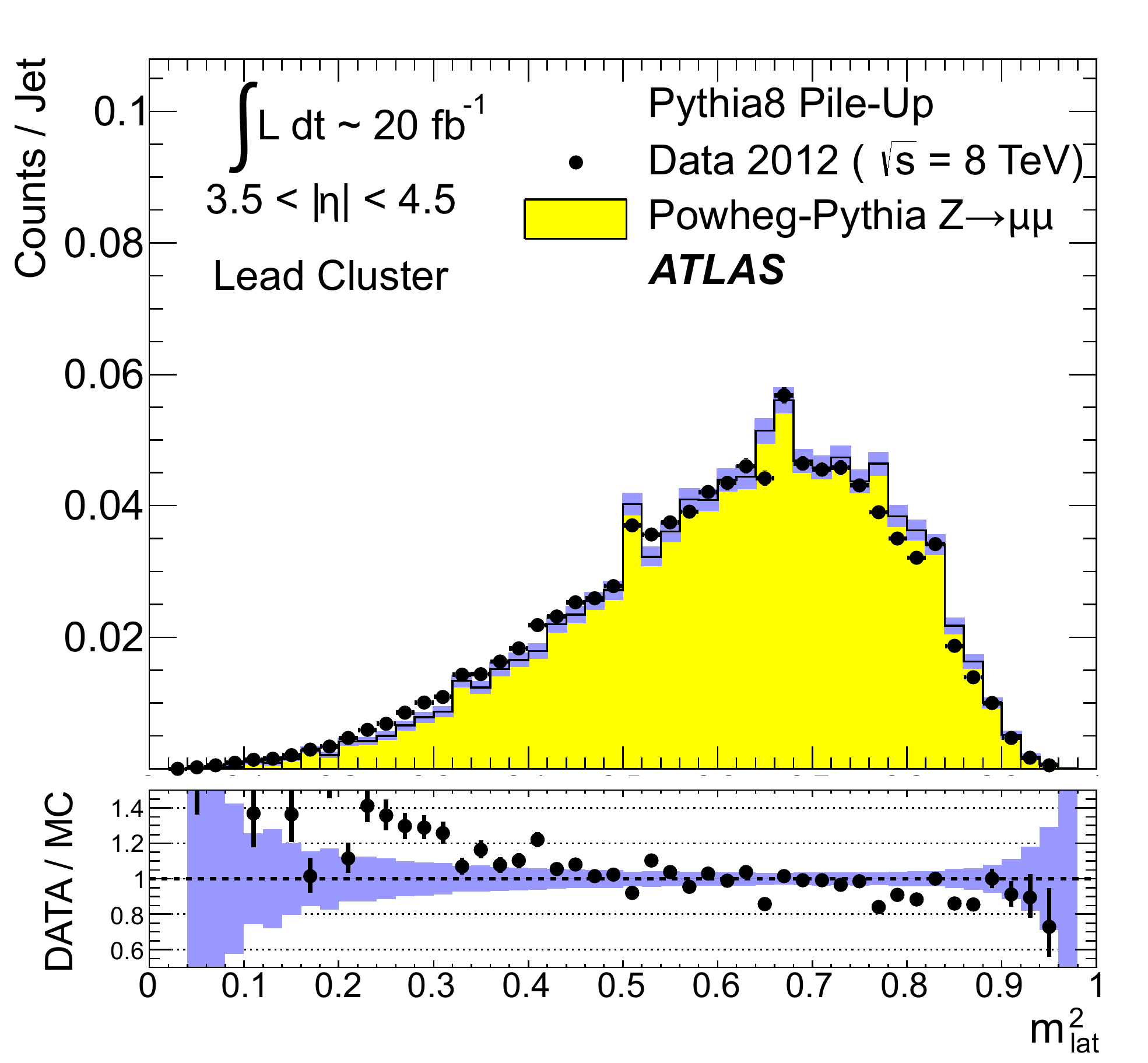}\label{fig:jet:lead:lat:3:mc}}  \qquad
	\subfloat[]{\includegraphics[width=\figsixpanelwidth]{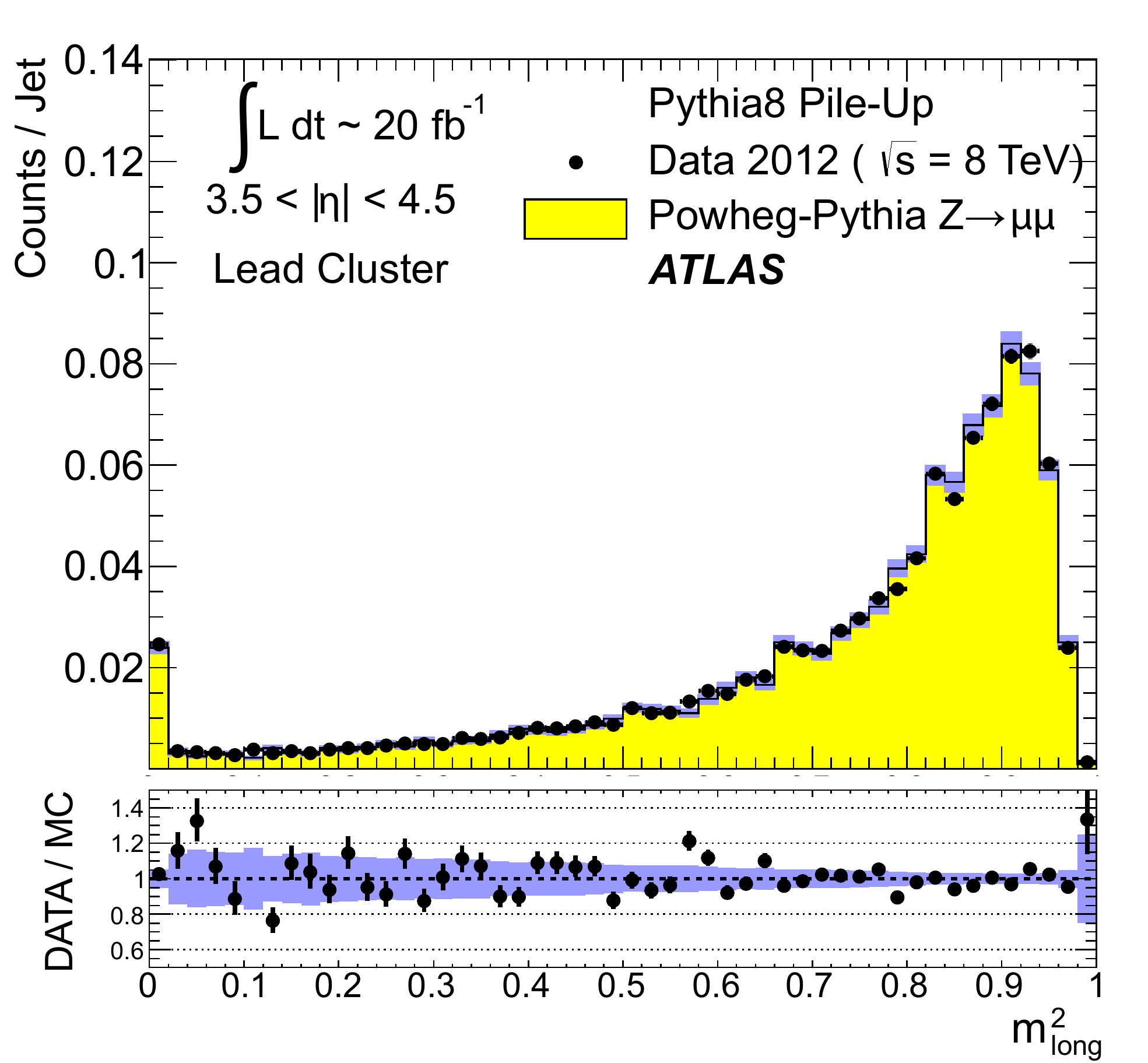}\label{fig:jet:lead:long:3:mc}}
	\caption[]{The distribution of the normalised (\subref{fig:jet:lead:lat:1:mc}, \subref{fig:jet:lead:lat:2:mc}, and \subref{fig:jet:lead:lat:3:mc}) lateral (\LAT) and (\subref{fig:jet:lead:long:1:mc}, \subref{fig:jet:lead:long:2:mc}, and \subref{fig:jet:lead:long:3:mc}) longitudinal (\LONG) extension measures of the leading \topo{} in fully calibrated \antikt{} jets with $R = 0.4$ and $\unit{30}{\GeV} < \ptjetlcwjes < \unit{40}{\GeV}$ in \Zmumu{} events in 2012 data and \MC{} simulations with fully simulated \pu, for jets in the (\subref{fig:jet:lead:lat:1:mc}, \subref{fig:jet:lead:long:1:mc}) central ($|\eta|<0.6$), the (\subref{fig:jet:lead:lat:2:mc}, \subref{fig:jet:lead:long:2:mc}) \EndCap{} ($2.0<|\eta|<2.5$), and the (\subref{fig:jet:lead:lat:3:mc}, \subref{fig:jet:lead:long:3:mc}) forward detector region ($3.5<|\eta|<4.5$) of \ATLAS. The ratios of data and \MC{} simulation distributions are shown below the plots. The shaded bands shown for the distributions obtained from \MC{} simulations indicate statistical uncertainties and the corresponding uncertainty bands in the ratio plots.}
	\label{fig:jet:lead:shape:norm}
\end{figure}

The spatial extensions of the leading \topo{} in a jet are calculated as described in \secRef{sec:moments:geometry}. 
The distributions of the normalised lateral energy dispersion \LAT{} given in \eqRef{eq:radial2} and the normalised longitudinal energy dispersion \LONG{} given in \eqRef{eq:long2}
are shown in \figRef{fig:jet:lead:shape:norm} for the leading \topo{} in jets reconstructed with the \antikt{} algorithm with  $R = 0.4$ and $\unit{30}{\GeV} < \ptjetlcwjes < \unit{40}{\GeV}$, in \Zmumu{} events in 2012 data and \MC{} simulations with fully simulated \pu.
The lateral extensions represented by \LAT{} are reasonably well modelled in all three detector regions, with some residual discrepancies in particular in the low-value tails and upper edges of the spectra in the \EndCap{} and forward regions. 
The longitudinal extensions measured by \LONG{} are modelled well in the central and forward detector regions, but their modelling shows some deficiencies in the \EndCap{} region. 

\begin{figure}[t!]\centering	
	\subfloat[]{\includegraphics[width=\fighalfwidth]{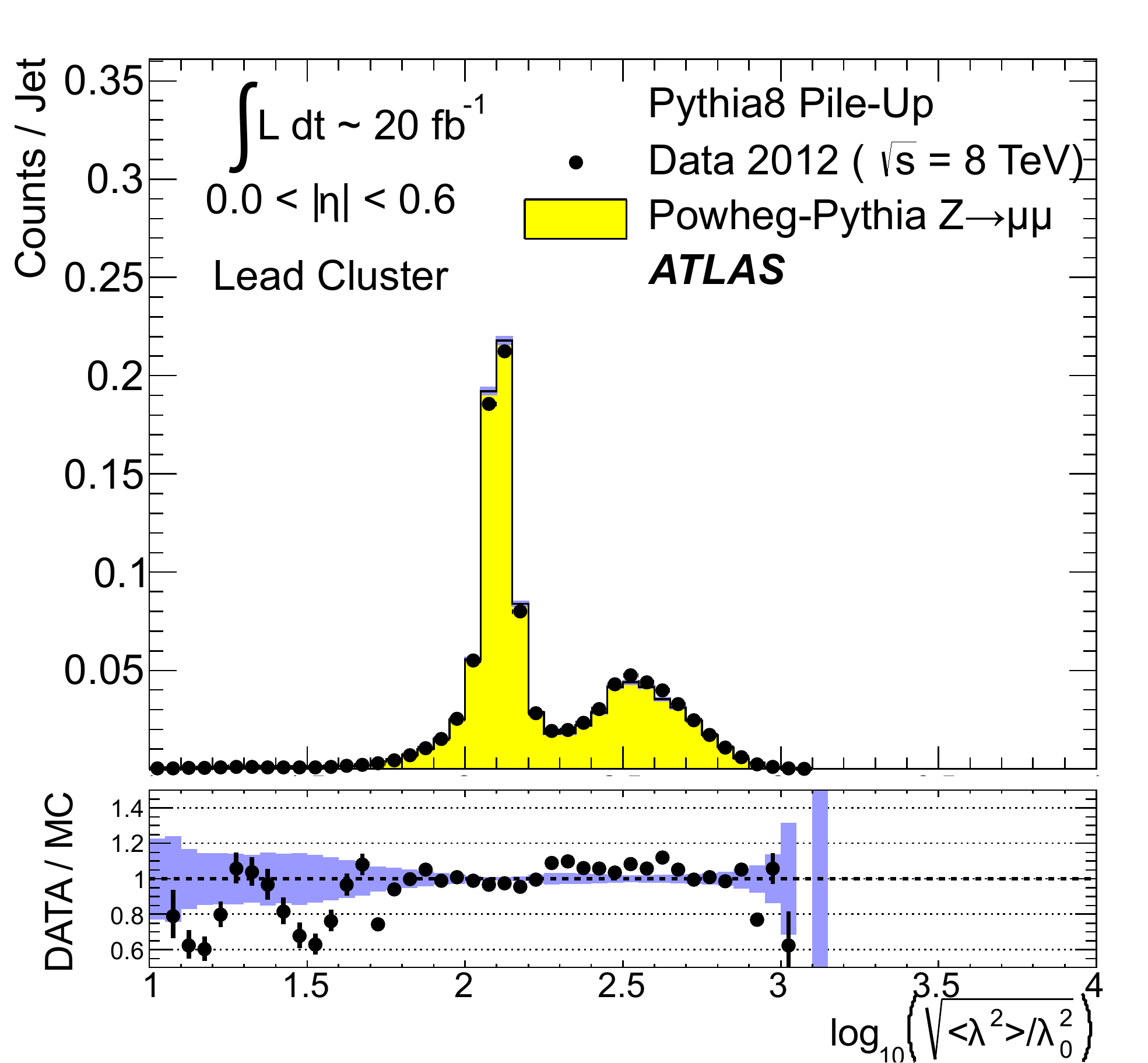}\label{fig:jet:lead:geo:long:1:mc}} 
	\subfloat[]{\includegraphics[width=\fighalfwidth]{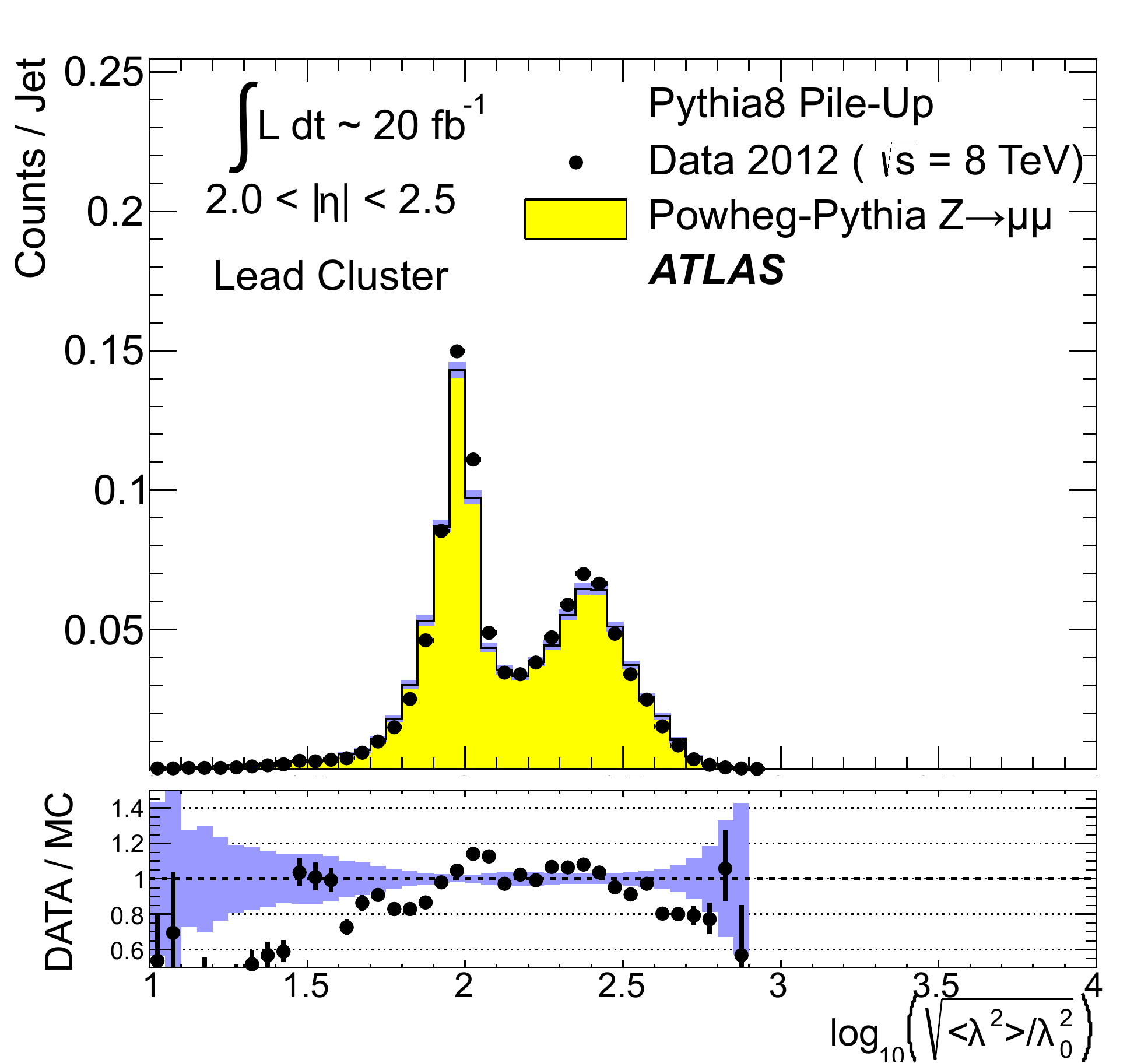}\label{fig:jet:lead:geo:long:2:mc}}
	\\
	\subfloat[]{\includegraphics[width=\fighalfwidth]{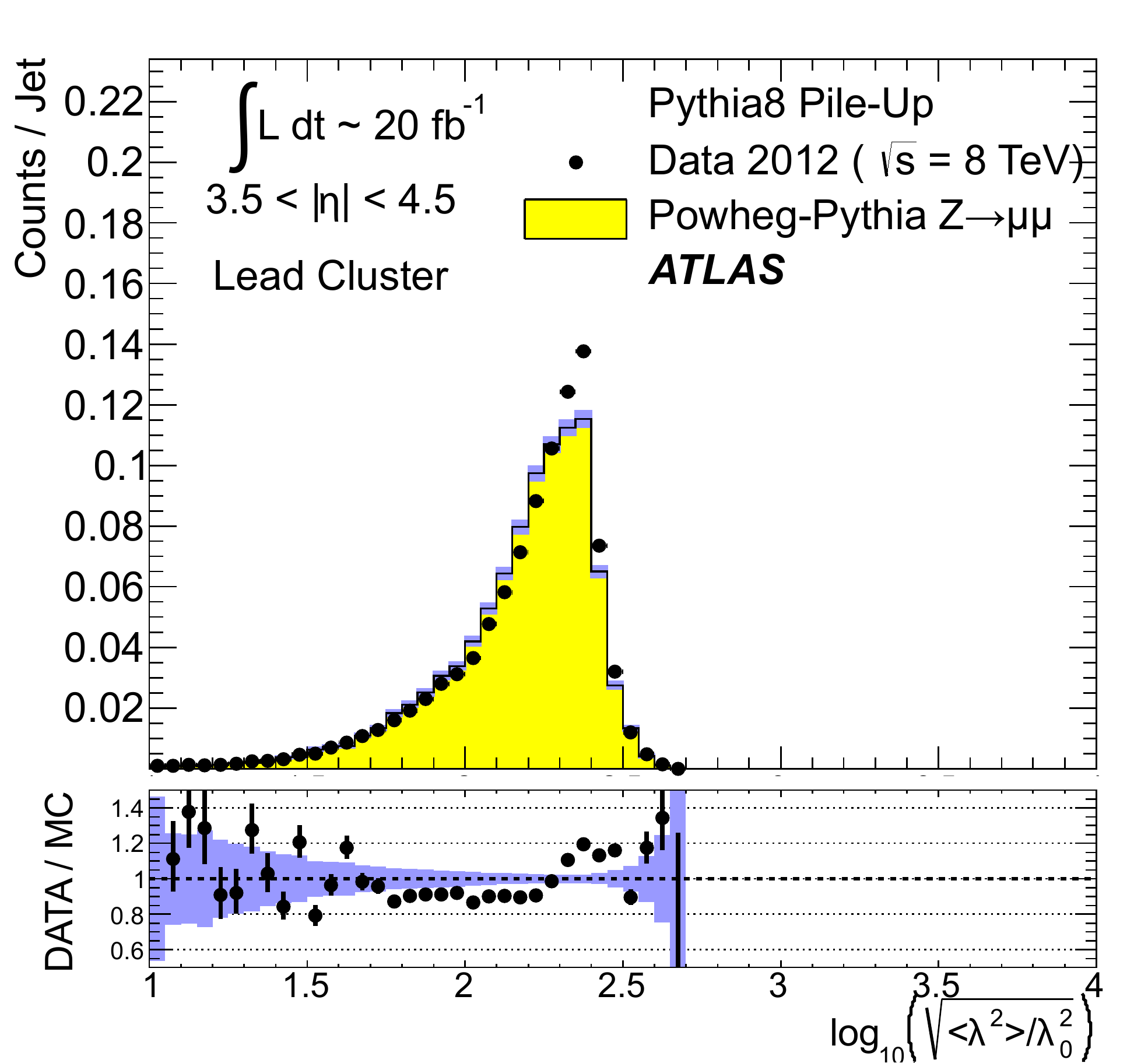}\label{fig:jet:lead:geo:long:3:mc}} 
	\caption[]{The length of the leading \topo, measured in terms of the longitudinal spread (second moment) \AVE{\lambda^{2}}{} of the cell coordinates along the principal cluster axis by $\sqrt{\AVE{\lambda^{2}}/\lambda_{0}^{2}}$, in \antikt{} jets reconstructed with $R = 0.4$ and $\unit{30}{\GeV} < \ptjetlcwjes < \unit{40}{\GeV}$ in \Zmumu{} events in 2012 data and \MC{} simulations with fully simulated \pu. Distributions are shown for jets in the \subref{fig:jet:lead:geo:long:1:mc} central ($|\eta|<0.6$), the \subref{fig:jet:lead:geo:long:2:mc} \EndCap{} ($2.0<|\eta|<2.5$), and the \subref{fig:jet:lead:geo:long:3:mc} forward detector region ($3.5<|\eta|<4.5$). The normalisation of the longitudinal spread is given by $\lambda_{0} = \unit{1}{\mm}$.
	The ratios of data-to-\MC{} simulations are shown below the distributions.
The shaded bands indicate statistical uncertainties of the distributions from \MC{} simulations and the resulting uncertainty bands in the ratio plots.}  
\label{fig:jet:lead:geo:long}		
\end{figure}

The distribution of the leading \topo{} length measure $\sqrt{\AVE{\lambda^{2}}}$ defined in \secRef{sec:moments:geometry:size} in the three detector regions is shown in \figMultiRefLabel~\ref{fig:jet:lead:geo:long}\subref{fig:jet:lead:geo:long:1:mc} to \ref{fig:jet:lead:geo:long}\subref{fig:jet:lead:geo:long:3:mc}. 
The \MC{} simulations reproduce the shape of the $\sqrt{\AVE{\lambda^{2}}}$ distributions from data well in the central and forward regions, with some deficiencies observed in the \EndCap{} region. 
The shapes in the central and \EndCap{} region are due to leading \topos{} contained in the electromagnetic calorimeters populating the left peak of the distribution (short clusters) and leading \topos{} in the hadronic calorimeters populating the right peak with longer clusters. 
The shape of the length distribution in the forward region shown in \figRef{fig:jet:lead:geo:long}\subref{fig:jet:lead:geo:long:3:mc} is characterised by a sharp drop on the right of the spectrum, which corresponds to the half-depth of cells (\unit{225}{\mm}) in the \LArFCAL{} modules. 
This shows that in this detector region the leading \topo{} rarely extends into all three \LArFCAL{} modules, as indicated by only few \topos{} with $\sqrt{\AVE{\lambda^{2}}} > \unit{225}{\mm}$. 
The leading cluster is more likely to share its energy between the first two modules \LArFCALN{0}{} and \LArFCAL{1}, with $\sqrt{\AVE{\lambda^{2}}} \approx \unit{225}{\mm}$ indicating a near equal share and $\sqrt{\AVE{\lambda^{2}}} < \unit{225}{\mm}$ indicating that most of the cluster energy is in \LArFCALN{0}.   

\begin{figure}[t!]\centering	
	\subfloat[]{\includegraphics[width=\fighalfwidth]{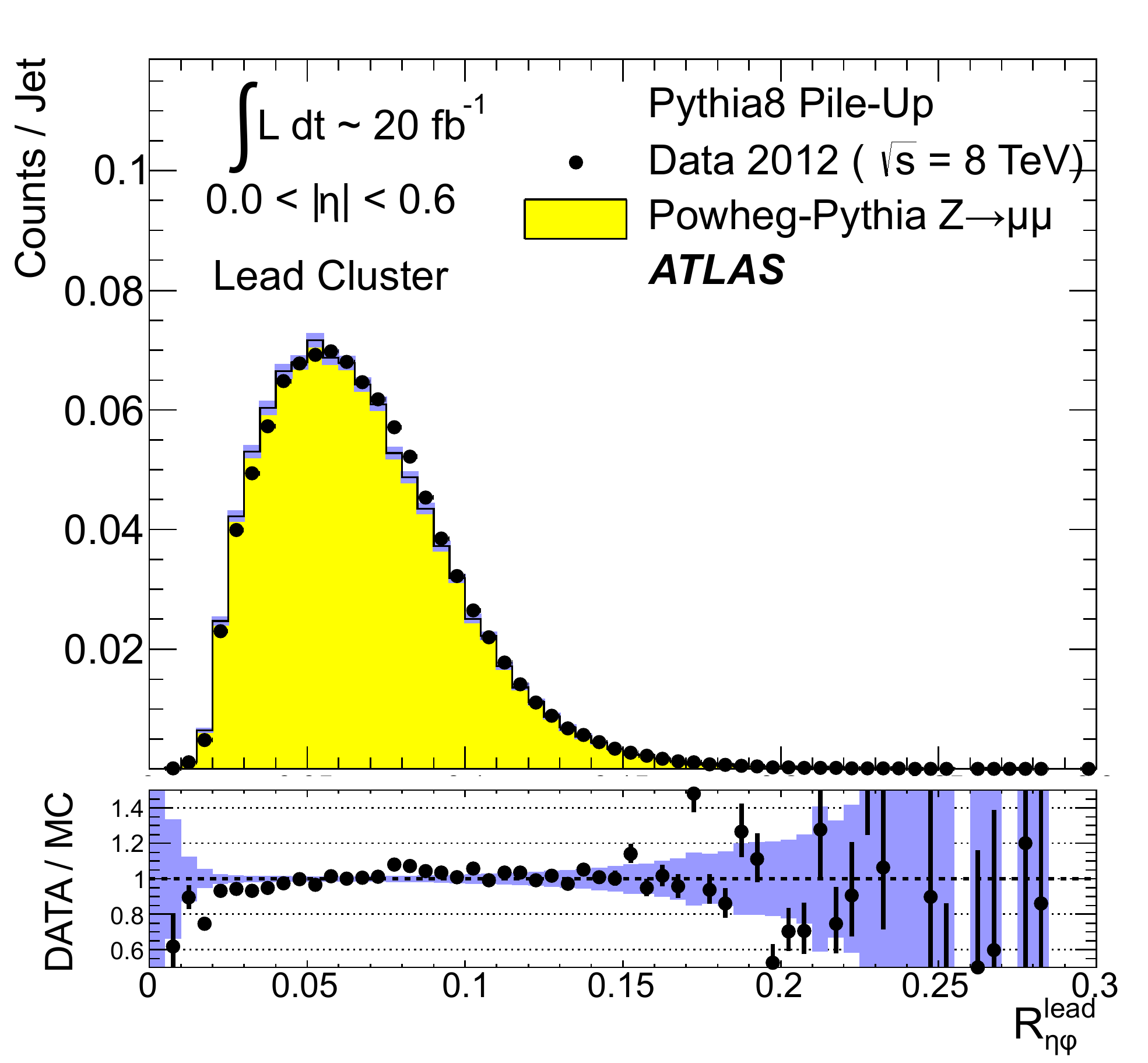}\label{fig:jet:lead:geo:dR:1:mc}} 
	\subfloat[]{\includegraphics[width=\fighalfwidth]{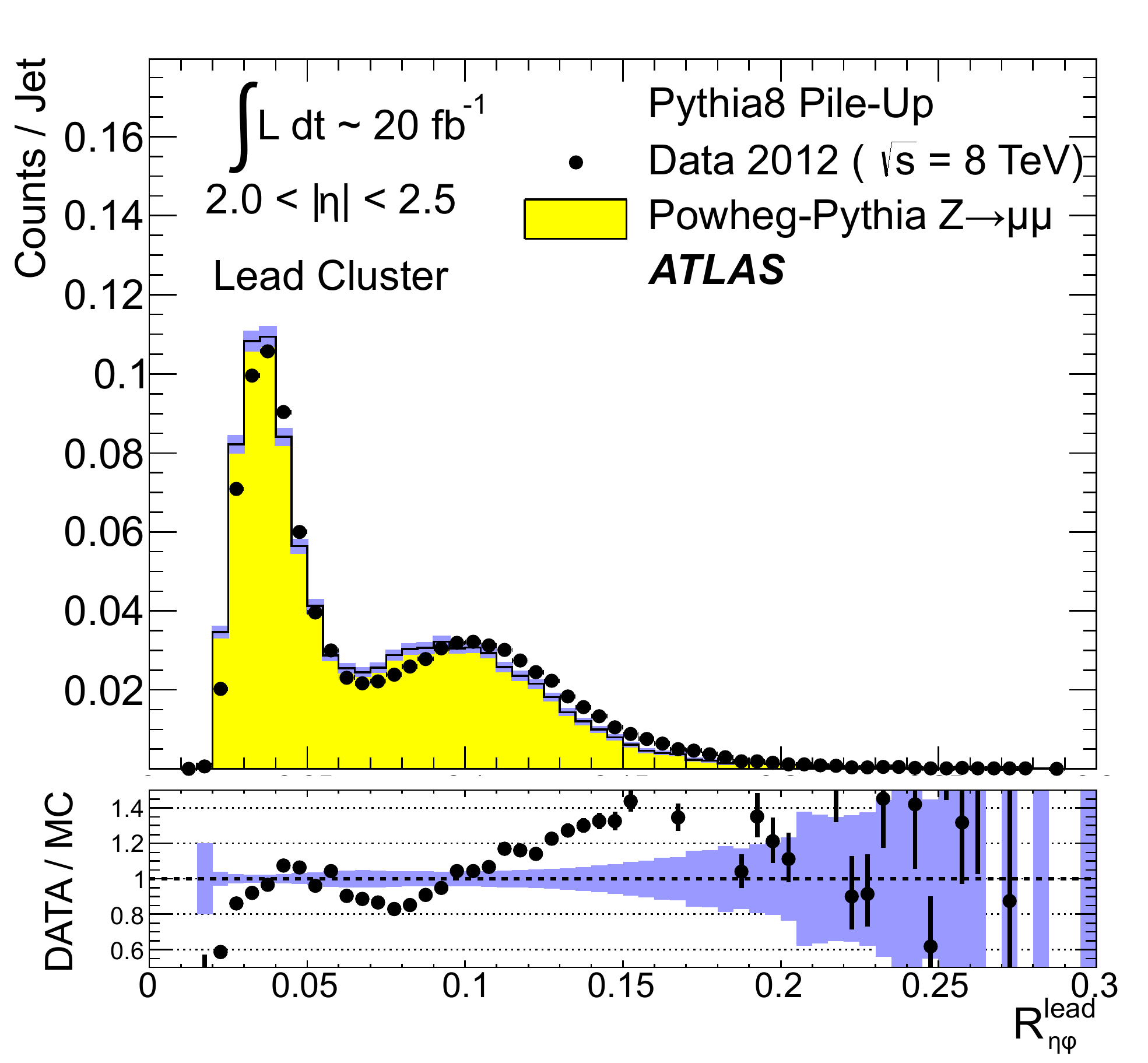}\label{fig:jet:lead:geo:dR:2:mc}}
	\\
	\subfloat[]{\includegraphics[width=\fighalfwidth]{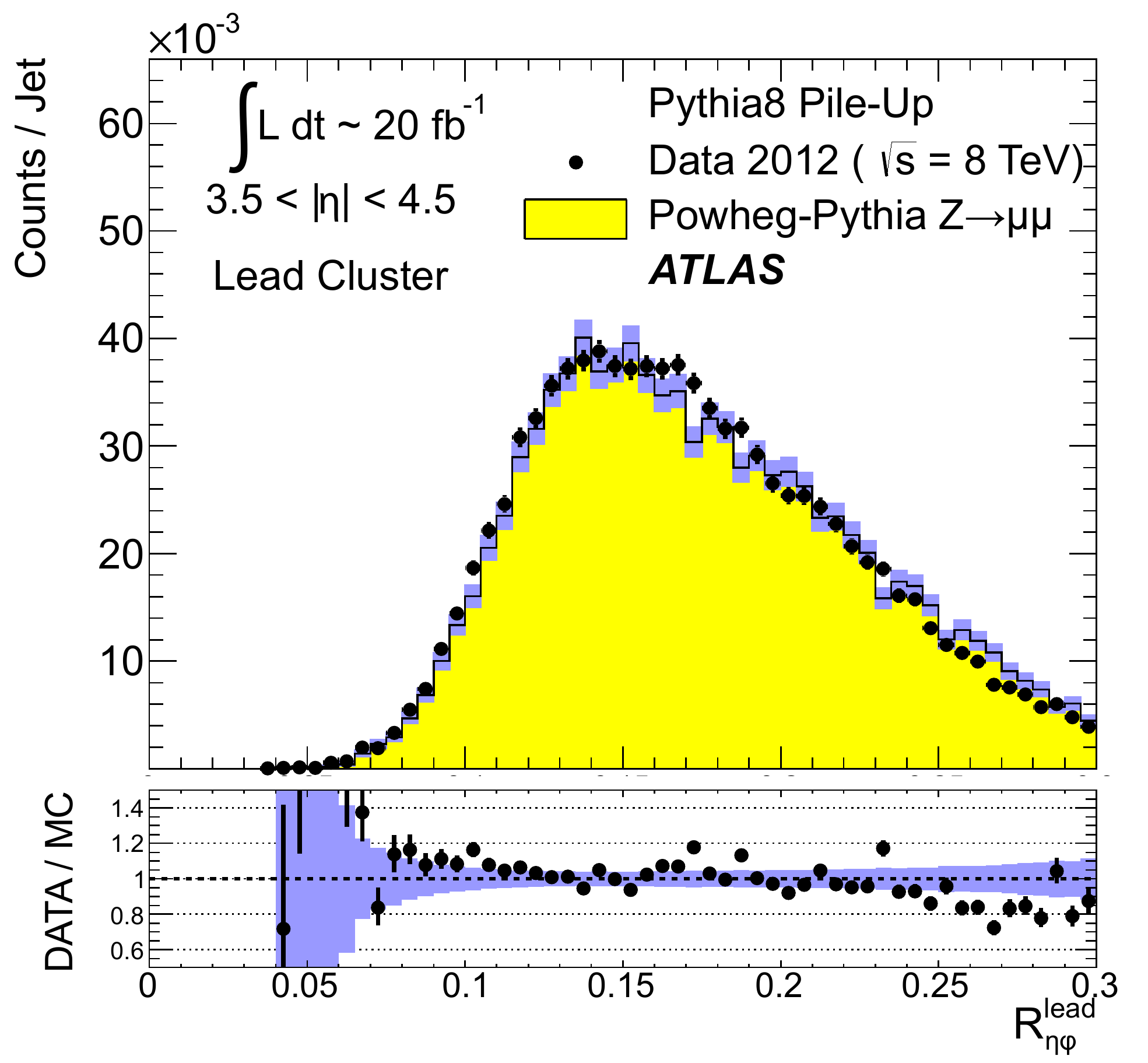}\label{fig:jet:lead:geo:dR:3:mc}} 
	\caption[]{The size \clussize{}
	of the leading \topo{} in $(\eta,\phi)$ space, measured using \eqRef{eq:cluswidth}, 
	in \antikt{} jets reconstructed with $R = 0.4$ and with $\unit{30}{\GeV} < \ptjetlcwjes < \unit{40}{\GeV}$ in \Zmumu{} events in 2012 data and \MC{} simulations with fully simulated \pu. Distributions are shown for jets in the \subref{fig:jet:lead:geo:dR:1:mc} central ($|\eta|<0.6$), the \subref{fig:jet:lead:geo:dR:2:mc} \EndCap{} ($2.0<|\eta|<2.5$), and the \subref{fig:jet:lead:geo:dR:3:mc} forward detector region ($3.5<|\eta|<4.5$) in \ATLAS.
	The ratios of data to \MC{} simulations are shown below the distributions.
The shaded bands shown for the distributions obtained from \MC{} simulations indicate statistical uncertainties and the corresponding uncertainty bands in the ratio plots.} 
\label{fig:jet:lead:geo:dR}		
\end{figure}

The size \clussize{} of the  leading \topo{} in $(\eta,\phi)$ space is calculated from the respective cluster width estimates $\sigma_{\eta(\phi)}$ given in \eqRef{eq:cluswidth}.
Its distributions in various calorimeter regions are shown in \figRef{fig:jet:lead:geo:dR}. 
The \clussize{} distribution in the central region in \figRefLabel~\ref{fig:jet:lead:geo:dR}\subref{fig:jet:lead:geo:dR:1:mc} is consistent with \topos{} in a calorimeter with a fine and regular \readout{} granularity. 
The double-peak structure in the \EndCap{} region in \figRefLabel~\ref{fig:jet:lead:geo:dR}\subref{fig:jet:lead:geo:dR:2:mc} shows contributions from leading \topos{} extending beyond $|\eta| = |\etajet| = 2.5$, where the cell granularity drops sharply by about a factor of four.  
This generates the right peak in the distribution.\footnote{The location of this peak is consistent with the change of the cell size  
in sampling layers \LArEMEN{1} and \LArEMEN{2} of the electromagnetic \EndCap{} calorimeter at $|\eta| = 2.5$, see \tabRef{tab:readout}.} 
The \clussize{} distribution in the forward detector region displayed in \figRefLabel~\ref{fig:jet:lead:geo:dR}\subref{fig:jet:lead:geo:dR:3:mc} is consistent with a non-pointing calorimeter \readout{} segmentation with smooth transitions in the granularity from about $\Delta\eta\times\Delta\phi \approx 0.15\times 0.15$ at $|\eta| = 3.5$ to $\Delta\eta\times\Delta\phi \approx 0.3\times 0.3$ for $|\eta| = 4.5$. 

\begin{figure}[t!] \centering
	\subfloat[]{\includegraphics[width=\fighalfwidth]{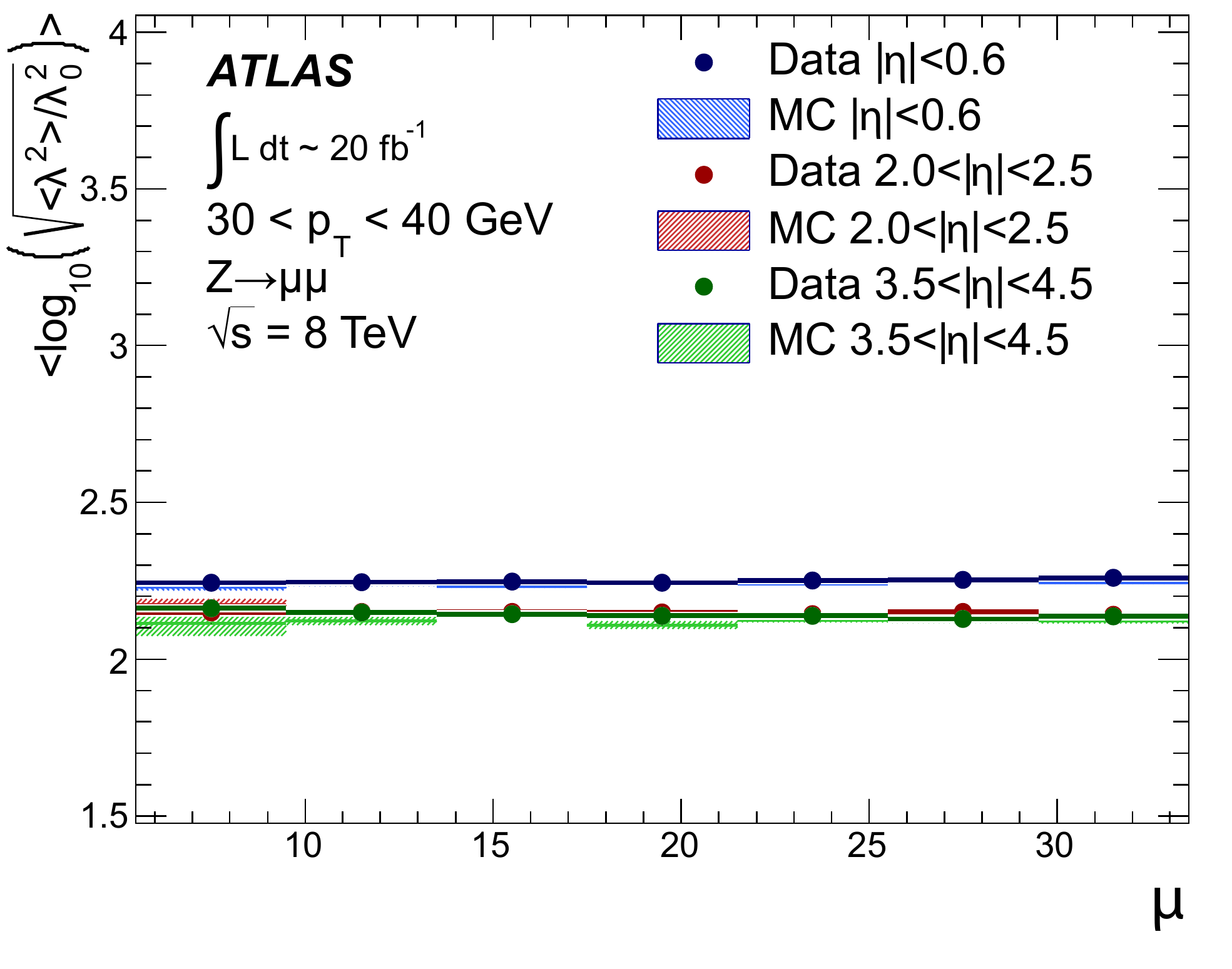}\label{fig:jet:lead:geo:pu:long2}} 
	\subfloat[]{\includegraphics[width=\fighalfwidth]{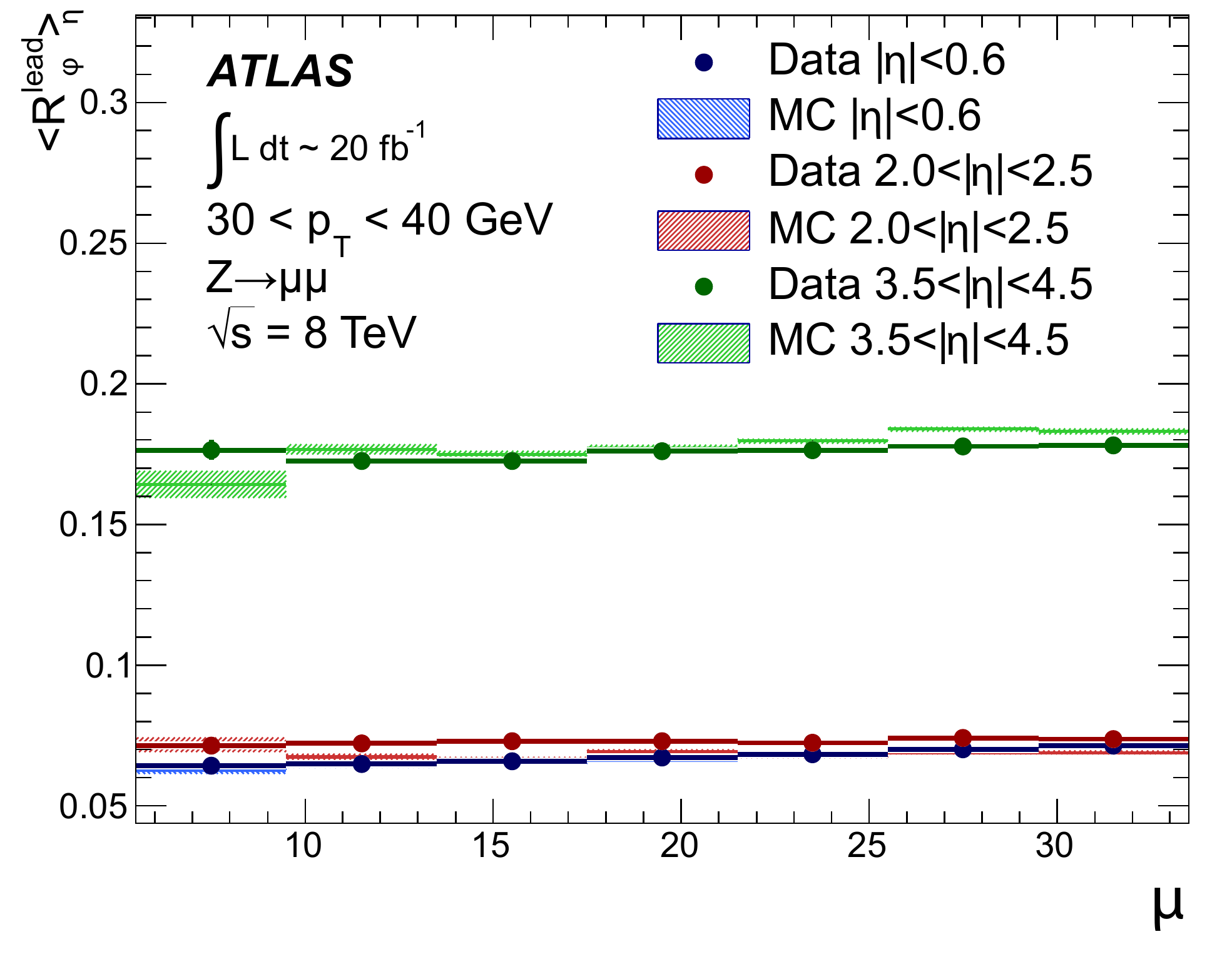}\label{fig:jet:lead:geo:pu:dR}}
        \\      
	\subfloat[]{\includegraphics[width=\fighalfwidth]{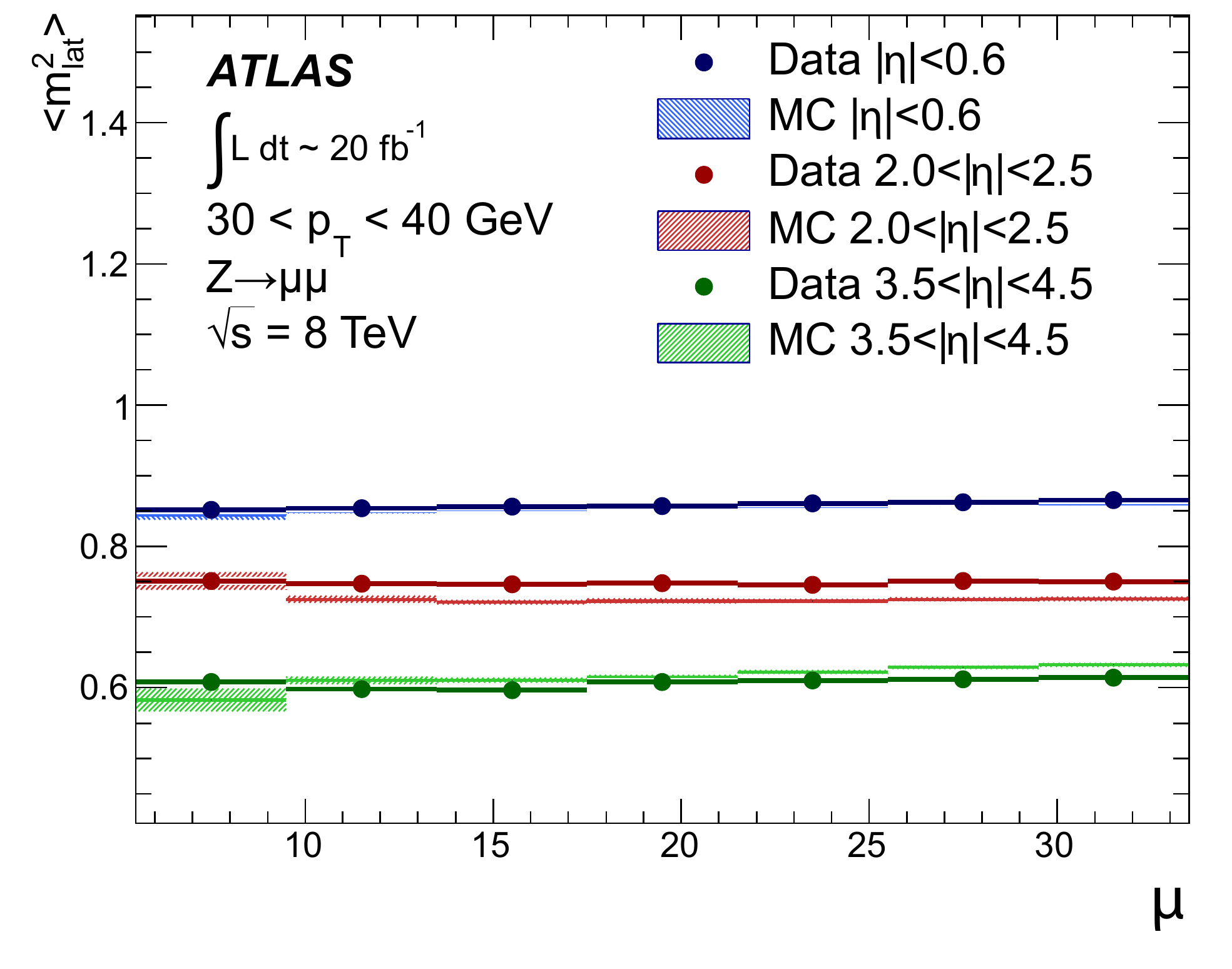}\label{fig:jet:lead:geo:pu:rad}} 
	\caption[]{The average \pu{} dependence of various geometric observables reconstructed from the leading \topo{} in \antikt{} jets reconstructed with $R = 0.4$ and $\unit{30}{\GeV} < \ptjetlcwjes < \unit{40}{\GeV}$ in \Zmumu{} events in 2012 data and \MC{} simulations with fully simulated \pu. 
	The average cluster length, represented by $\log_{10}(\AVE{\lambda^{2}/\lambda_{0}^{2}}^{1/2})$ with the reference scale $\lambda_{0} = \unit{1}{\mm}$, is shown as a function of $\mu$ in \subref{fig:jet:lead:geo:pu:long2}, for three detector regions.
	The average size $\langle\clussize\rangle$ of the leading \topo{} in $(\eta,\phi)$ space is displayed for the same detector regions and as a function of $\mu$ in \subref{fig:jet:lead:geo:pu:dR}. 
	 The average normalised lateral energy dispersion \AVE{\LAT}{} of the cluster, as a function of $\mu$ for the three detector regions, is shown in \subref{fig:jet:lead:geo:pu:rad}.
The shaded bands shown around the results obtained from \MC{} simulations indicate statistical uncertainties.}  
	\label{fig:jet:lead:geo:pu}
\end{figure}

\subsubsection{\PU{} dependence of leading \topo{} geometry and shapes}\label{\thislabel:hardest:geometry:pu}

The dependence of the geometry and shape of the leading \topo{} in a jet on the \pu{} activity measured by $\mu$ is shown in \figRef{fig:jet:lead:geo:pu}. No significant dependence is observed for the average longitudinal extension of this cluster shown in \figRef{fig:jet:lead:geo:pu}\subref{fig:jet:lead:geo:pu:long2}, the average size of this cluster in $(\eta,\phi)$ space 
in \figRef{fig:jet:lead:geo:pu}\subref{fig:jet:lead:geo:pu:dR}, and its average lateral energy dispersion, defined in \eqRef{eq:radial2} 
and displayed in \figRefLabel~\ref{fig:jet:lead:geo:pu}\subref{fig:jet:lead:geo:pu:rad}.    

The \datatomc{} comparison of the average \pu{} dependences shows generally acceptable agreement, but also suggests some residual deficiencies likely related to the simulation of the longitudinal and lateral (hadronic) shower shapes. 
Corresponding observations are reported in \citMultiRef{Adragna:2010zz,Abat:2010zz,Abat:2010zza,Abat:1900zz} in the context of detailed comparisons of \ATLAS{} test-beam data with simulations.

\FloatBarrier

\section{Conclusion} \label{sec:concl}
Topological cell signal clusters (\topos) provide a well-understood and calibrated signal definition for hadronic final-state reconstruction in the \ATLAS{} calorimeters. 
The principal algorithm generating these \topos{} includes a noise-suppression scheme based on signal-significance patterns which is similar to applications in previous experiments.
The innovative approach developed for the \ATLAS{} calorimeters not only employs a highly refined implementation of this algorithm in a high-energy, high-luminosity hadron collider environment characterised by significant collision backgrounds introduced by \pu, but also uses the \topos{} as a signal base for a local hadronic calibration (\LCW) in a non-compensating calorimeter. 

Both the \topo{} formation and the \LCW{} calibration have been validated in collisions without \pu{} recorded in 2010, and in the more active \pu{} environments observed in 2011 and 2012 operations. 
The residual effects of \pu{} on cluster kinematics and observables in data are well controlled in that they can be reproduced with sufficient precision in \MC{} simulations for \topos{} either inside or outside jets. 
The largest observed \datamc{} differences mainly arise from imperfect modelling of the soft collision physics affecting \pu. 
Overlaying \pu{} from data on generated hard-scatter interactions in \MC{} simulations yields significantly better agreement for most kinematic variables and \topo{} moments.

From the \LHC{} \runone{} experience, \topos{} are now established as a well-performing signal base for jet and transverse missing momentum (\met) reconstruction in \ATLAS. 
They provide noise suppression important for a high-quality calorimeter signal, and in this reduce the amount of data needed to represent the final state in the detector.   
Their spatial resolution allows not only detailed analysis of the energy flow in the \pp{} collision events as needed for \met{} reconstruction but also analysis of more localised energy-flow structures inside jets. 
This is done routinely in boosted-object reconstruction techniques applied in jet substructure analysis, with recent examples from \ATLAS{} discussed in \citMultiRefLabel{} \cite{Aad:2013gja,Aad:2012meb,Aad:2012dpa,Aad:2015owa}.

\section*{Acknowledgements}
\addcontentsline{toc}{section}{Acknowledgements}

We thank CERN for the very successful operation of the LHC, as well as the
support staff from our institutions without whom ATLAS could not be
operated efficiently.

We acknowledge the support of ANPCyT, Argentina; YerPhI, Armenia; ARC, Australia; BMWFW and FWF, Austria; ANAS, Azerbaijan; SSTC, Belarus; CNPq and FAPESP, Brazil; NSERC, NRC and CFI, Canada; CERN; CONICYT, Chile; CAS, MOST and NSFC, China; COLCIENCIAS, Colombia; MSMT CR, MPO CR and VSC CR, Czech Republic; DNRF and DNSRC, Denmark; IN2P3-CNRS, CEA-DSM/IRFU, France; SRNSF, Georgia; BMBF, HGF, and MPG, Germany; GSRT, Greece; RGC, Hong Kong SAR, China; ISF, I-CORE and Benoziyo Center, Israel; INFN, Italy; MEXT and JSPS, Japan; CNRST, Morocco; NWO, Netherlands; RCN, Norway; MNiSW and NCN, Poland; FCT, Portugal; MNE/IFA, Romania; MES of Russia and NRC KI, Russian Federation; JINR; MESTD, Serbia; MSSR, Slovakia; ARRS and MIZ\v{S}, Slovenia; DST/NRF, South Africa; MINECO, Spain; SRC and Wallenberg Foundation, Sweden; SERI, SNSF and Cantons of Bern and Geneva, Switzerland; MOST, Taiwan; TAEK, Turkey; STFC, United Kingdom; DOE and NSF, United States of America. In addition, individual groups and members have received support from BCKDF, the Canada Council, CANARIE, CRC, Compute Canada, FQRNT, and the Ontario Innovation Trust, Canada; EPLANET, ERC, ERDF, FP7, Horizon 2020 and Marie Sk{\l}odowska-Curie Actions, European Union; Investissements d'Avenir Labex and Idex, ANR, R{\'e}gion Auvergne and Fondation Partager le Savoir, France; DFG and AvH Foundation, Germany; Herakleitos, Thales and Aristeia programmes co-financed by EU-ESF and the Greek NSRF; BSF, GIF and Minerva, Israel; BRF, Norway; CERCA Programme Generalitat de Catalunya, Generalitat Valenciana, Spain; the Royal Society and Leverhulme Trust, United Kingdom.

The crucial computing support from all WLCG partners is acknowledged gratefully, in particular from CERN, the ATLAS Tier-1 facilities at TRIUMF (Canada), NDGF (Denmark, Norway, Sweden), CC-IN2P3 (France), KIT/GridKA (Germany), INFN-CNAF (Italy), NL-T1 (Netherlands), PIC (Spain), ASGC (Taiwan), RAL (UK) and BNL (USA), the Tier-2 facilities worldwide and large non-WLCG resource providers. Major contributors of computing resources are listed in Ref.~\cite{ATL-GEN-PUB-2016-002}.

\bibliographystyle{bibtex/bst/atlasBibStyleWithTitle}
\bibliography{clusternote,Acknowledgements}

\newpage
\begin{flushleft}
{\Large The ATLAS Collaboration}

\bigskip

G.~Aad$^\textrm{\scriptsize 86}$,
B.~Abbott$^\textrm{\scriptsize 114}$,
J.~Abdallah$^\textrm{\scriptsize 152}$,
O.~Abdinov$^\textrm{\scriptsize 11}$,
R.~Aben$^\textrm{\scriptsize 108}$,
M.~Abolins$^\textrm{\scriptsize 91}$,
O.S.~AbouZeid$^\textrm{\scriptsize 159}$,
H.~Abramowicz$^\textrm{\scriptsize 154}$,
H.~Abreu$^\textrm{\scriptsize 153}$,
R.~Abreu$^\textrm{\scriptsize 117}$
% ...
\bigskip
\\
% ...
$^{11}$ Institute of Physics, Azerbaijan Academy of Sciences, Baku, Azerbaijan\\
% ...
$^{86}$ CPPM, Aix-Marseille Universit{\'e} and CNRS/IN2P3, Marseille, France\\
% ...
$^{91}$ Department of Physics and Astronomy, Michigan State University, East Lansing MI, United States of America\\
% ...
$^{108}$ Nikhef National Institute for Subatomic Physics and University of Amsterdam, Amsterdam, Netherlands\\
% ...
$^{114}$ Homer L. Dodge Department of Physics and Astronomy, University of Oklahoma, Norman OK, United States of America\\
% ...
$^{117}$ Center for High Energy Physics, University of Oregon, Eugene OR, United States of America\\
% ...
$^{152}$ Institute of Physics, Academia Sinica, Taipei, Taiwan\\
$^{153}$ Department of Physics, Technion: Israel Institute of Technology, Haifa, Israel\\
$^{154}$ Raymond and Beverly Sackler School of Physics and Astronomy, Tel Aviv University, Tel Aviv, Israel\\
% ...
$^{159}$ Department of Physics, University of Toronto, Toronto ON, Canada\\
% ...
\end{flushleft}


\end{document}